\PassOptionsToPackage{svgnames}{xcolor}
\documentclass[twocolumn]{aastex62}

\usepackage{fixfoot}
\usepackage[clockwise, figuresright]{rotating}
\usepackage{amsmath}
\usepackage{color,soul}
\usepackage{multirow}
\usepackage{IEEEtrantools}
\usepackage{bm}
\usepackage{textcomp}

\received{2018 May 18}
\revised{2018 October 19}
\accepted{2018 November 13}
\published{2019 March 6, \href{https://iopscience.iop.org/article/10.3847/1538-4357/aaf3b8/meta}{\textcolor{magenta}{\apj}}, \href{http://adsabs.harvard.edu/cgi-bin/bib_query?arXiv:1810.04887}{873, 85}}

\shorttitle{The $M_{\rm BH}$--$M_{\rm *,sph}$ Relation for Spiral Galaxies}
\shortauthors{Davis, Graham, and Cameron}

\begin{document}

\title{Black Hole Mass Scaling Relations for Spiral Galaxies. I. $M_{\rm BH}$--$M_{\rm *,sph}$}

\correspondingauthor{Benjamin L. Davis}
\email{benjamindavis@swin.edu.au}

\author[0000-0002-4306-5950]{Benjamin L. Davis}
\affil{Centre for Astrophysics and Supercomputing, Swinburne University of Technology, Hawthorn, Victoria 3122, Australia}

\author[0000-0002-6496-9414]{Alister W. Graham}
\affil{Centre for Astrophysics and Supercomputing, Swinburne University of Technology, Hawthorn, Victoria 3122, Australia}

\author[0000-0001-8311-1491]{Ewan Cameron}
\affil{Oxford Big Data Institute, University of Oxford, Oxford OX3 7LF, United Kingdom}


\begin{abstract}
The (supermassive black hole mass, $M_\text{BH}$)--(bulge stellar mass, $M_{\rm*,sph}$) relation is, obviously, derived using two quantities. We endeavor to provide accurate values for the latter via detailed multicomponent galaxy decompositions for the current full sample of 43 spiral galaxies having directly measured $M_\text{BH}$ values; 35 of these galaxies have been alleged to contain pseudobulges, 21 have water maser measurements, and three appear bulgeless. This more than doubles the previous sample size of spiral galaxies with a finessed image analysis. We have analyzed near-infrared images, accounting for not only the bulge, disk (exponential, truncated, or inclined), and bar but also for spiral arms and rings and additional central components (active galactic nuclei (AGNs), etc.). A symmetric Bayesian analysis finds $\log\left(M_\text{BH}/M_{\sun}\right)=\left(2.44_{-0.31}^{+0.35}\right)\log\left\{M_{\rm*,sph}/[\upsilon(1.15\times10^{10}\,M_{\sun})]\right\}+(7.24\pm0.12)$, with $\upsilon$ a stellar mass-to-light ratio term. The level of scatter equals that about the $M_{\rm BH}$--$\sigma_*$ relation. The nonlinear slope rules out the idea that many mergers, coupled with the central limit theorem, produced this scaling relation, and it corroborates previous observational studies and simulations, which have reported a near-quadratic slope at the low-mass end of the $M_\text{BH}$--$M_{\rm*,sph}$ diagram. Furthermore, bulges with AGNs follow this relation; they are not offset by an order of magnitude, and models that have invoked AGN feedback to establish a linear $M_{\rm BH}$--$M_{\rm*,sph}$ relation need revisiting. We additionally present an updated $M_\text{BH}$--(S{\'e}rsic index, $n_\text{sph}$) relation for spiral galaxy bulges with a comparable level of scatter and a new $M_{\rm*,sph}$--(spiral-arm pitch angle, $\phi$) relation.

%\textcolor{red}{This Abstract has 245 words out of a maximum allowed 250 words.}
\end{abstract}

\section{Introduction}

The connection between supermassive black hole (SMBH) mass, $M_{\rm BH}$, and host bulge/spheroid stellar mass, $M_{\rm *,sph}$, has been a topic of discussion and publication for nearly three decades, since \citet{Dressler:1989}, see also \citet{Yee:1992}, suggested that a linear relation exists. Most of the first generation of observational studies of the $\log{M_{\rm BH}}$--$\log{M_{\rm *,sph}}$ relationship \citep{Kormendy:Richstone:1995,Franceschini:1998,Magorrian:1998,McLure:Dunlop:2002,Marconi:Hunt:2003,Haring:Rix:2004} reported quasi-linear (i.e., approximately linear) $\log$(black hole mass)--$\log$(spheroid mass) scaling relations based on luminosities and masses from predominantly high-mass, early-type galaxies, supporting the notion of \citet{Dressler:1989}. This period also introduced the correlation between black hole mass and the spheroid stellar velocity dispersion, $\sigma_{*}$ \citep{Ferrarese:2000,Gebhardt:2000,Merritt:2000}.

The second generation of studies \citep[e.g.,][]{Ferrarese:Ford:2005,Graham:2007,Lauer:2007,Gultekin:2009,Sani:2011,Beifiori:2012,Erwin:Gadotti:2012,Bosch:2012,Vika:2012,McConnell:Ma:2013} continued to recover a near-linear $M_{\rm BH}$--$M_{\rm *,sph}$ relation. However, the inclusion of more low-mass spheroids revealed departures and produced a notably steeper distribution than was seen at the high-mass end. \citet{Laor:1998,Laor:2001}, \citet{Wandel:1999}, and \citet{Ryan:2007} were the first to realize this, and their work suggested a steeper (single) power law, with a slope of $1.53\pm0.14$, presented in \citep{Laor:2001}. \citet{Salucci:2000} reported that the $M_{\rm BH}$--$M_{\rm *,sph}$ relation might be significantly steeper for spiral galaxies than for (massive) elliptical galaxies based on their study of black holes with upper-limit estimates for their masses.

\citet{Graham:2012} highlighted that the (stellar luminosity, $L$)--(stellar velocity dispersion, $\sigma_{*}$) relation for low-luminosity early-type galaxies was inconsistent with the $M_{\rm BH}$--$L$ and $M_{\rm BH}$--$\sigma_{*}$ relations of the day.  With $L\propto\sigma_{*}^2$ at $B$-band absolute magnitudes $\mathfrak{M}_B \gtrsim -20.5$ (Vega) mag \citep[][see also \citealt{Graham:2016b} and references therein]{Davies:1983}, and the literature reporting $M_{\rm BH}\propto\sigma_{*}^4$--$\sigma_{*}^6$, one must have that $M_{\rm BH}\propto L^2$--$L^3$, where $\mathfrak{M}_B \gtrsim -20.5\,\text{mag}$.  This is much steeper than the $M_{\rm BH}\propto L^1$ relation that had typically been reported in the literature, and it explained the apparent departures at low magnitudes in many past $M_{\rm BH}$--$L$ diagrams \citep[e.g.,][]{Sani:2011,Bosch:2012,McConnell:Ma:2013}. \citet{Graham:2012} therefore advocated a broken $M_{\rm BH}$--$M_{\rm *,sph}$ power-law relation with a near-linear slope at the high-mass end and a near-quadratic slope (or slope of $\approx2.5$) at the low-mass end.

As discussed in a recent review article by \citet{Graham:2016b} regarding the various bulge--(black hole) scaling relations, the consequences of this steeper relation are numerous and far-reaching. It has implications ranging from theories of galaxy--(black hole) coevolution \citep{Graham:Scott:2013} to the design of space-based gravitational-wave detectors \citep{Mapelli:2012}. \citet{Graham:2012}, \citet{Scott:2013}, and \citet{Graham:Scott:2013,Graham:Scott:2015} offered an interpretation for the bent $M_{\rm BH}$--$M_{\rm *,sph}$ relation involving core-S{\'e}rsic \citep{Graham:2003} and S{\'e}rsic spheroids at the high- and low-mass ends of the diagram, respectively.  An alternative proposal was offered in \citet{Savorgnan:2016:II} in terms of a red (early-type galaxy) and blue (late-type galaxy) sequence.

While most studies of the $M_{\rm BH}$--$M_{\rm *,sph}$ scaling relation have been dominated by early-type galaxies with high masses ($M_{\rm BH}\gtrsim10^7\,M_{\sun}$), here we will focus on spiral galaxies. We endeavor to provide further insight into the true slope of the low-mass end of the $M_{\rm BH}$--$M_{\rm *,sph}$ relation. We do this, in part, by performing a detailed but justifiable decomposition of the galaxy light into its constituent structural components. Our work builds upon the study of \citet{Savorgnan:2016}, which included 18 spiral galaxies\footnote{\citet{Savorgnan:2016} counted only 17 spiral galaxies because they did not include NGC~4594 in their subsample of late-type galaxies. We do consider it to be a spiral galaxy \citep[see][]{Gadotti:2012}.} with eight galaxies possessing SMBHs less massive than $\approx10^7\,M_{\sun}$. Using an array of different regression methods, their data did not enable them to discriminate between slopes of 2 or 3 for the spiral galaxy $M_{\rm BH}$--$M_{\rm *,sph}$ relation. We have more than doubled the sample size, increasing it to 40 spiral galaxies with bulges, including 17 galaxies having $M_{\rm BH}<10^7\,M_{\sun}$. This will enable us to better constrain the $M_{\rm BH}$--$M_{\rm *,sph}$ relation for spiral galaxies, thereby contributing to the discussion as to whether early- and late-type galaxies follow a different relation, possibly indicative of a different formation channel.

In addition to the care that has gone into acquiring the bulge masses that are tabulated here, we have applied a sophisticated Bayesian linear regression to obtain the optimal $M_{\rm BH}$--$M_{\rm *,sph}$ relation. We have also included the results when using the more familiar \textsc{bces} regression from \citet{BCES} and the modified \textsc{fitexy} routine \citep{Press:1992,Tremaine:2002}. We also briefly present the $M_{\rm BH}$--$n_{\rm maj}$ relation from a sample of only spiral galaxy bulges. Furthermore, given the strong relation between the black hole mass and the disk's spiral-arm pitch angle, $\phi$ \citep{Seigar:2008,Berrier:2013,Davis:2017}, we have investigated the relation between pitch angle and bulge stellar mass.

In what follows, we first introduce our sample of spiral galaxies and their image sources. We then discuss our surface brightness profile decomposition technique and the conversion of luminosity into stellar mass (see Section~\ref{DM}). In Section~\ref{sec:compare}, we compare our bulge measurements with those of similar studies in the literature. Next, we provide our primary analysis of the $M_{\rm BH}$--$M_{\rm *,sph}$ scaling relation (see Section~\ref{AR}), plus relations with the S{\'e}rsic index and the spiral-arm pitch angle. We discuss our results and provide a comparison with scaling relations from massive early-type galaxies with core-S{\'e}rsic profiles in Section~\ref{DI}.  Finally, we summarize the overall outcome of this paper in Section~\ref{END}. In the appendices, we detail the statistical modeling framework of our Bayesian linear regression analysis, and we provide photometric decompositions for all of the galaxies, along with commentary (see Appendices \ref{App1}, \ref{App2}, and \ref{App3}).

We adopt a spatially flat $\Lambda$CDM cosmology with the best-fit Planck TT $+$ lowP $+$ lensing cosmographic parameters estimated by the Planck mission \citep{Planck:2015}: $\Omega_{\rm M} = 0.308$, $\Omega_\Lambda = 0.692$, and $h_{67.81} = h/0.6781 = H_0/(67.81\,\text{km\,s}^{-1}\,\text{Mpc}^{-1}) \equiv 1$. Throughout this paper (unless otherwise stated), all printed errors and plotted error bars represent $1\,\sigma$ ($\approx 68.3\%$) confidence levels, and all magnitudes are expressed in the absolute (AB) system \citep{Oke:1974}. We use the terms ``spheroid'' and ``bulge'' interchangeably.

\section{Data and Methodology}\label{DM}

\subsection{Sample with Directly Measured Black Hole Masses}

Our sample consists of what we believe is the current complete sample of spiral galaxies with directly measured SMBH masses. We do not include black hole masses estimated from reverberation mapping or other secondary estimators calibrated to the $M_{\rm BH}$--$\sigma_{*}$ or $M_{\rm BH}$--$M_{\rm *,sph}$ relations. Literature searches on these criteria yield a total of 44 galaxies that have been classified as spiral types (including one ambiguously classified spiral: Cygnus~A). Our sample of 44 is ultimately culled down by removing four galaxies: Cygnus A (probably an early-type galaxy) and three bulgeless galaxies. This yields our primary sample of 40 spiral galaxies with bulges. All black hole masses have been adjusted from their originally published masses to conform with our fiducial distances. Correspondingly, the error bars on the SMBH masses have been adjusted considering the uncertainty in the distance to their host galaxies. All references for our adopted distances and black hole masses are compiled in \citet{Davis:2017}.

\subsection{Imaging Data}

Following \citet{Sani:2011} and \citet{Savorgnan:2016}, we have used $3.6\,\micron$ imaging from the \textit{Spitzer Space Telescope} for our bulge/disk/etc.\ light profile decompositions due to the lack of significant dust extinction at this wavelength.\footnote{Optical images, especially blue images, can be biased by a small percentage (of the stellar mass) of young hot stars because of the way their near-blackbody spectral energy distributions (SEDs) peak at these wavelengths, while the SEDs of older, cooler stars peak at redder wavelengths \citep[e.g.,][]{MacArthur:2009}.} The $3.6\,\micron$ imaging data are obtained primarily from the \textit{Spitzer} Survey of Stellar Structure in Galaxies \citep[$\rm{S^4G}$;][]{Sheth:2010},\footnote{\url{http://irsa.ipac.caltech.edu/data/SPITZER/S4G/}} which provides large mosaicked, processed images and masks for all of their galaxies. For galaxies not part of the $\rm{S^4G}$, we obtained processed images from the \textit{Spitzer} Heritage Archive (SHA).\footnote{\url{http://sha.ipac.caltech.edu}} When both of these sources failed to include imaging of a target galaxy, or the resolution was not adequate to quantify small bulges, alternative processed images were collected from the \textit{Hubble Space Telescope} (\textit{HST})\footnote{\url{https://mast.stsci.edu/}} or the Two Micron All Sky Survey (2MASS) Large Galaxy Atlas (LGA)\footnote{\citet{Jarrett:2003}} when the field of view (FoV) was not sufficient to accurately measure the sky background. Overriding this ranking, F814W \textit{HST} imaging was preferred (due to the FoV, image availability, and our desire to minimize the number of bands requiring a stellar mass-to-light ratio) for all of the smaller bulges with effective radii approaching the full width at half maximum (FWHM) of the point spread function (PSF) of the \textit{Spitzer Space Telescope}. Given that the S$^4$G only observed galaxies with recessional velocities less than $3000\,\text{km\,s}^{-1}$, and bulges naturally appear smaller in more distant galaxies, this resulted in us using \textit{HST} images for 11 of the 13 galaxies more distant than $40\,\text{Mpc}$. From the reduced sample of 40, 26 have \textit{Spitzer} data, 11 have \textit{\textit{HST}} data, two have 2MASS data, and the Milky Way's stellar bulge mass was obtained from \citet{Licquia:Newman:2015}.\footnote{\citet{Licquia:Newman:2015} determined the stellar mass of the Galactic bulge from a meta-analysis of literature measurements using Bayesian hierarchical modeling.} Results for the additional four galaxies are also provided. Photometric parameters for all imaging are provided in Table~\ref{table:photo}.

\begin{deluxetable*}{llclll}
\tablecolumns{6}
\tabletypesize{\normalsize}
\tablecaption{Filters and Photometric Calibrations\label{table:photo}}
\tablehead{
\colhead{Source} & \colhead{$\lambda$} & \colhead{ps} & \colhead{zp} & \colhead{$\mathfrak{M}_\sun$} & \colhead{$\Upsilon_*$} \\
\colhead{} & \colhead{($\micron$)} & \colhead{(arcsec)} & \colhead{} & \colhead{(mag)} & \colhead{($M_{\sun}/L_{\sun}$)} \\
\colhead{(1)} & \colhead{(2)} & \colhead{(3)} & \colhead{(4)} & \colhead{(5)} & \colhead{(6)}
}
\startdata
\textit{Spitzer} IRAC1: $\rm{S^4G}$ & 3.550 & 0.75 & 21.097\tablenotemark{a} & 6.02\tablenotemark{b} & $0.60\pm0.09$\tablenotemark{c} \\
\textit{Spitzer} IRAC1: SHA & 3.550 & 0.60 & 21.581 & 6.02\tablenotemark{b} & $0.60\pm0.09$\tablenotemark{c} \\
\textit{HST} WFC3 UVIS2 F814W & 0.8024 & 0.04 & 25.110\tablenotemark{d} & 4.52 & $1.88\pm0.40$ \\
\textit{HST} WFPC2/WFC F814W & 0.8012 & 0.10 & 24.211\tablenotemark{e} & 4.52 & $1.88\pm0.40$ \\
2MASS LGA $K_s$ & 2.159 & 1.00 & Image-specific\tablenotemark{f} & 5.08 & $0.62\pm0.08$ \\
\enddata
\tablecomments{Columns:
(1) Image source.
(2) Effective wavelength midpoint.
(3) Pixel size.
(4) AB magnitude photometric zero-point.
(5) Solar absolute AB magnitude (from \url{http://mips.as.arizona.edu/~cnaw/sun.html}).
(6) (Stellar mass)-to-(stellar light) ratio.
}
\tablenotetext{a}{From \citet[][Equation~(13)]{Salo:2015}. \citet{zp} provided a zero-point of $20.472$ for the $\rm{S^4G}$ surface brightness maps with the same $0\farcs75$ pixel size but have combined the zero-point and pixel size into one constant such that $20.472=21.097+5\log({0\farcs75})$.}
\tablenotetext{b}{From \citet{Oh:2008}, after applying a $3.6\,\micron$ Vega-to-AB magnitude conversion: $\mathfrak{m}_{\rm AB}=\mathfrak{m}_{\rm Vega}+2.78\,\text{mag}$.}
\tablenotetext{c}{From \citet{Meidt:2014}, assuming a \citet{Chabrier:2003} IMF and a \citet{Bruzual:Charlot:2003} SSP with exponentially declining SFHs for a range of metallicities.}
\tablenotetext{d}{\url{http://www.stsci.edu/hst/wfc3/documents/ISRs/WFC3-2017-14.jpg}}
\tablenotetext{e}{\url{http://www.stsci.edu/hst/acs/documents/handbooks/currentDHB/acs_Ch52.html\#94716}}
\tablenotetext{f}{Photometric zero-points have been converted to the AB system via $\mathfrak{m}_{K_{s},{\rm AB}} = \mathfrak{m}_{K_{s},{\rm Vega}} + 1.85\,\text{mag}$ \citep{Blanton:2005}.}
\end{deluxetable*}

%For a spectrograph with some fixed limiting spatial resolution, and a black hole of arbitrary constant mass, the ability to spatially resolve the black hole's gravitational sphere-of-influence --- given by $r_{\rm infl} = G M_{\rm BH} / \sigma_*^2$ \citep[e.g.,][]{Peebles:1972} --- in the bulges of spiral galaxies will decrease linearly with the distance to, and the luminosity of, the host galaxy because $L\propto \sigma_*^2$ in spiral galaxies \citep[see][]{Merritt:Ferrarese:2001}. 

\subsection{Masking and Sky Subtraction}

It is important to isolate the light of the target galaxy by masking foreground stars and background galaxies and subtracting the sky background, which is particularly important for our decomposition method (see Section~\ref{Sec_SBF}). We used the \textsc{iraf} routines \textsc{objmasks} and \textsc{mskregions} to identify and mask contaminating sources, respectively. We also took care to manually identify contaminating sources coincident with the target galaxy. For $\rm{S^4G}$ images, we began with their provided mask and then manually masked finer sources of contamination. We determined the sky background and its associated uncertainty from the image's histogram of pixel intensities \citep[e.g.,][]{Almoznino:1993}. After subtracting the median sky value, we added and subtracted the uncertainty in the sky background to determine the radial extent of the surface brightness profile that is largely unaffected by the uncertainty in this sky background. The public \textit{Spitzer} and 2MASS images were already mosaicked to provide a sufficient FoV for one to do this, while the \textit{HST} images have a $162\arcsec\times162\arcsec$ and $150\arcsec\times150\arcsec$ FoV for the WFC3\footnote{Here we are referring to the UVIS2 channel; we do not use the IR instrument due to its smaller FoV ($136\arcsec\times123\arcsec$).} and WFPC2\footnote{The actual collecting area of the WFPC2 CCD is less due to one diminished quadrant.} cameras, respectively.  None of our galaxies imaged by \textit{HST} has a semiminor axis (radius)---as defined by the isophote where their $B$-band surface brightness equals $25\,\text{mag\,arcsec}^{-2}$---greater than 75$\arcsec$; their semiminor axes are smaller than 42$\arcsec$ for all but two galaxies (NGC~2273 and NGC~3393) and smaller than 34$\arcsec$ for all but three galaxies.

\subsection{Isophote Fitting}

We then used the software packages \textsc{isofit} and \textsc{cmodel} \citep{Ciambur:2015}, which respectively fit and model the isophotal structure of galaxies. These routines are improvements upon the standard \textsc{iraf} packages \textsc{ellipse} and \textsc{bmodel}, respectively. Importantly, as discussed by \citet{Ciambur:2015}, these improvements include using eccentric anomalies ($\psi$) instead of azimuthal/plane-polar angles ($\theta$) for the angular metric of elliptical isophotes. The two quantities are related via
\begin{equation}
\psi = -\arctan\left ( \frac{\tan\theta}{1-\epsilon} \right ),
\label{anomaly}
\end{equation}
with the ellipticity $\epsilon=1-(b/a)$, where $a$ is the major axis length and $b$ is the minor-axis length of an isophote. Use of eccentric anomalies enables a more accurate representation of the light when analyzing perturbations to quasi-elliptical isophotes as a function of the angle $\psi$, via a Fourier series decomposition,
\begin{equation}
I(\psi)=I_\text{ell}+\sum_{m}\left [ A_m\sin(m\psi)+B_m\cos(m\psi) \right ],
\label{harmonics}
\end{equation}
where $I(\psi)$ is the intensity profile along the isophote, expressed as a function of the eccentric anomaly; $I_\text{ell}$ is the median intensity of the purely elliptical path; and the summation represents Fourier harmonic perturbations to $I_\text{ell}$, with $m$ being the harmonic (integer) order.\footnote{For additional information, see \citet{Ciambur:2015}.}

As noted by \citet{Lasker:2014a}, triaxial spheroids can have observed projections (on the plane of the sky) that display ellipticity and positional angle twists that are not captured by 2D studies that use a S{\'e}rsic bulge model with fixed ellipticity and positional angle, whereas the series of 1D profiles within \textsc{isofit} captures these radial changes and others, such as $B_4$ and $B_6$. In this work, we extracted a set of isophotal profiles that included the radial gradient of the Fourier harmonic orders $m$ = 2, 3, 4, 6, 8, and 10.  Although 2D modeling can use (a radially constant) set of Fourier harmonic terms to describe perturbations to otherwise elliptical isophotes, \citet{Savorgnan:2016} demonstrated greater success with 1D modeling techniques than with  2D modeling of their galaxy sample.

\subsection{Galaxy Components}

In this subsection, we describe all of the various components that we considered and fit, as needed, to the galaxies in our sample. Here we list their names, as well as the defining parameters for each functional form that were solved for in our decompositions. For a thorough listing of their mathematical forms, plots, and descriptions, we refer readers to \citet{Ciambur:2016}.

We have modeled bulges using the \citet{Sersic:1963} function (see \citealt{Caon:1993} and \citealt{Graham:Driver:2005} for additional equations and a discussion in English). These are parameterized by three quantities: the effective ``half-light'' radius ($R_e$), the effective surface brightness at $R_e$ ($\mu_e$), and the S{\'e}rsic index ($n$), which describes the radial concentration of the function. Alternatively, bulges that display a central deficit\footnote{The PSF-convolved profiles of galaxies with cores that are smaller than the PSF still display a central deficit.} are modeled using the core-S{\'e}rsic function \citep{Graham:2003}. These are parameterized by six quantities: the break (transition) radius ($R_b$), the half-light radius ($R_e$), the inner profile slope ($\gamma$), the smoothness of the transition ($\alpha$),\footnote{In order to reduce computational time and degeneracy, we set $\alpha\equiv\infty$ to define a sharp transition between the inner power-law and outer-S{\'e}rsic regimes.} the S{\'e}rsic index ($n$), and the surface brightness term ($\mu'$). We tabulate all of the S{\'e}rsic profile parameters for each galaxy and present them in Table~\ref{table:Sersic}. 

\begin{longrotatetable}
\begin{deluxetable*}{llrrrrrrrrrr}
\tabletypesize{\scriptsize}
\tablecolumns{12}
\tablecaption{S{\'e}rsic Bulge Model Profile Parameters\label{table:Sersic}}
\tablehead{
\colhead{Galaxy Name} & \colhead{Type} & \colhead{$d_L$} & \colhead{$i$} & \colhead{$R_{e\rm, maj}$} & \colhead{$R_{e\rm, maj}$} & \colhead{$\mu_{e\rm, maj}$} & \colhead{$n_{\rm maj}$} & \colhead{$R_{e\rm, eq}$} & \colhead{$R_{e\rm, eq}$} & \colhead{$\mu_{e\rm, eq}$} & \colhead{$n_{\rm eq}$} \\
\colhead{} & \colhead{} & \colhead{(Mpc)} & \colhead{(deg)} & \colhead{(arcsec)} & \colhead{(kpc)} & \colhead{($\text{mag\,arcsec}^{-2}$)} & \colhead{} & \colhead{(arcsec)} & \colhead{(kpc)} & \colhead{($\text{mag\,arcsec}^{-2}$)} & \colhead{} \\
\colhead{(1)} & \colhead{(2)} & \colhead{(3)} & \colhead{(4)} & \colhead{(5)} & \colhead{(6)} & \colhead{(7)} & \colhead{(8)} & \colhead{(9)} & \colhead{(10)} & \colhead{(11)} & \colhead{(12)}
}
\startdata
\object{Circinus} & SABb & $4.21\pm0.76$	& $66.9\pm0.9$ &	 $33.26\pm1.99$ 	&	 $0.68\pm0.04$ 	&	 $18.29\pm0.12$ 	&	 $2.21\pm0.56$ 	&	 $23.13\pm1.22$ 	&	 $0.47\pm0.03$ 	&	 $18.06\pm0.11$ 	&	 $1.80\pm0.60$ 	\\
\object{Cygnus~A} & S	& $258.4\pm3.9$\tablenotemark{a} & $33.9\pm3.8$ &	 $19.56\pm0.80$ 	&	 $21.97\pm0.90$ 	&	 $22.22\pm0.07$ 	&	 $1.45\pm0.10$ 	&	 $46.48\pm12.1$ 	&	 $52.21\pm13.94$ 	&	 $23.74\pm0.41$ 	&	 $2.44\pm0.21$ 	\\
\object{ESO~558-G009} & Sbc	& $115.4\pm1.7$\tablenotemark{a} & $73.4\pm1.6$ &	 $0.62\pm0.01$ 	&	 $0.33\pm0.01$ 	&	 $18.17\pm0.04$ 	&	 $1.28\pm0.03$ 	&	 $0.68\pm0.03$ 	&	 $0.36\pm0.01$ 	&	 $18.62\pm0.06$ 	&	 $1.63\pm0.05$ 	\\
\object{IC~2560} & SBb	& $31\pm13$ & $54.9\pm1.8$ &	 $7.15\pm1.19$ 	&	 $4.21\pm0.70$ 	&	 $19.64\pm0.31$ 	&	 $2.27\pm0.84$ 	&	 $3.92\pm0.26$ 	&	 $0.59\pm0.04$ 	&	 $19.07\pm0.15$ 	&	 $0.68\pm0.20$ 	\\
\object[SDSS J043703.67+245606.8]{J0437+2456}\tablenotemark{b} & SB 	& $72.8\pm1.1\tablenotemark{a}$ & $54.0\pm0.6$ &	 $1.22\pm0.13$ 	&	 $0.42\pm0.04$ 	&	 $19.42\pm0.15$ 	&	 $1.73\pm0.12$ 	&	 $0.87\pm0.15$ 	&	 $0.30\pm0.05$ 	&	 $19.40\pm0.25$ 	&	 $1.97\pm0.23$ 	\\
\object{Milky~Way}\tablenotemark{c} & SBbc 	& $7.86\pm0.15\,\text{kpc}$ & \nodata &	 \nodata 	&	 $1.04\pm0.06$ 	&	 \nodata 	&	 $1.30\pm0.10$ 	&	 \nodata 	&	 $1.04\pm0.06$ 	&	 \nodata 	&	 $1.3\pm0.1$ 	\\
\object{Mrk~1029} & S	& $136.9\pm2.1$\tablenotemark{a} & $45.9\pm5.8$ &	 $0.47\pm0.00$ 	&	 $0.30\pm0.00$ 	&	 $16.53\pm0.02$ 	&	 $1.15\pm0.02$ 	&	 $0.28\pm0.00$ 	&	 $0.17\pm0.00$ 	&	 $16.29\pm0.02$ 	&	 $1.07\pm0.02$ 	\\
\object{NGC~0224}\tablenotemark{d} & SBb	& $0.75\pm0.02$ & $64.3\pm3.5$ &	 $418.6$ 	&	 $1.51$ 	&	 $19.58$	&	 $2.2\pm0.3$	&	 $173.6$	&	 $0.62$ 	&	 $18.41$ 	&	 $1.3\pm0.2$ 	\\
\object{NGC~0253} & SABc 	& $3.47\pm0.11$ & $75.3\pm2.0$ &	 $55.55\pm1.82$ 	&	 $0.93\pm0.03$ 	&	 $19.22\pm0.06$ 	&	 $2.53\pm0.08$ 	&	 $27.89\pm0.71$ 	&	 $0.47\pm0.01$ 	&	 $18.82\pm0.05$ 	&	 $2.33\pm0.08$ 	\\
\object{NGC~1068} & SBb 	& $10.1\pm1.8$ & $37.2\pm2.6$ &	 $10.52\pm0.71$ 	&	 $0.51\pm0.03$ 	&	 $16.17\pm0.31$ 	&	 $0.71\pm0.14$ 	&	 $8.29\pm0.79$ 	&	 $0.41\pm0.04$ 	&	 $16.14\pm0.35$ 	&	 $0.87\pm0.23$ 	\\
\object{NGC~1097} & SBb 	& $24.9\pm1.0$ & $48.4\pm9.2$ &	 $15.72\pm1.93$ 	&	 $1.90\pm0.23$ 	&	 $18.71\pm0.24$ 	&	 $1.95\pm0.26$ 	&	 $11.39\pm1.88$ 	&	 $1.37\pm0.23$ 	&	 $18.27\pm0.33$ 	&	 $1.52\pm0.33$ 	\\
\object{NGC~1300} & SBbc 	& $14.5\pm2.5$ & $49.6\pm10.1$ &	 $24.37\pm13.93$ 	&	 $1.72\pm0.98$ 	&	 $21.97\pm0.94$ 	&	 $4.20\pm0.48$ 	&	 $7.39\pm2.36$ 	&	 $0.52\pm0.17$ 	&	 $19.99\pm0.65$ 	&	 $2.83\pm0.38$ 	\\
\object{NGC~1320} & Sa	& $37.7\pm16.8$ & $65.8\pm1.6$ &	 $3.35\pm0.32$ 	&	 $0.61\pm0.06$ 	&	 $17.93\pm0.19$ 	&	 $3.08\pm0.12$ 	&	 $2.23\pm0.14$ 	&	 $0.41\pm0.03$ 	&	 $17.40\pm0.13$ 	&	 $2.87\pm0.10$ 	\\
\object{NGC~1398} & SBab	& $24.8\pm4.5$ & $43.3\pm2.7$ &	 $17.53\pm2.03$ 	&	 $2.11\pm0.24$ 	&	 $19.75\pm0.26$ 	&	 $3.44\pm0.23$ 	&	 $10.38\pm0.84$ 	&	 $1.25\pm0.10$ 	&	 $19.04\pm0.17$ 	&	 $3.00\pm0.17$ 	\\
\object{NGC~2273} & SBa	& $31.6\pm6.2$ & $50.1\pm3.3$ &	 $2.99\pm0.09$ 	&	 $0.46\pm0.01$ 	&	 $18.13\pm0.05$ 	&	 $2.24\pm0.04$ 	&	 $3.15\pm0.11$ 	&	 $0.48\pm0.02$ 	&	 $18.52\pm0.06$ 	&	 $2.49\pm0.05$ 	\\
\object{NGC~2748} & Sbc 	& $18.2\pm4.2$ & $62.4\pm10.7$ &	 $5.74\pm0.79$ 	&	 $0.51\pm0.07$ 	&	 $19.91\pm0.17$ 	&	 $1.59\pm0.11$ 	&	 $8.29\pm0.38$ 	&	 $0.73\pm0.03$ 	&	 $20.15\pm0.08$ 	&	 $1.71\pm0.06$ 	\\
\object{NGC~2960} & Sa	& $71.1\pm26.8$ & $51.5\pm9.2$ &	 $2.35\pm0.27$ 	&	 $0.81\pm0.09$ 	&	 $18.04\pm0.20$ 	&	 $2.59\pm0.12$ 	&	 $2.19\pm0.24$ 	&	 $0.76\pm0.08$ 	&	 $18.30\pm0.19$ 	&	 $2.86\pm0.11$ 	\\
\object{NGC~2974} & SB	& $21.5\pm2.5$ & $48.1\pm2.6$ &	 $9.21\pm0.45$ 	&	 $0.96\pm0.05$ 	&	 $18.49\pm0.08$ 	&	 $1.56\pm0.09$ 	&	 $6.53\pm0.14$ 	&	 $0.68\pm0.01$ 	&	 $18.12\pm0.04$ 	&	 $1.17\pm0.06$ 	\\
\object{NGC~3031}\tablenotemark{e} & SABab	& $3.48\pm0.13$ & $54.4\pm2.3$ &	 $36.19\pm1.43$ 	&	 $0.61\pm0.02$ 	&	 \nodata 	&	 $2.81\pm0.11$ 	&	 $42.98\pm0.74$ 	&	 $0.73\pm0.01$ 	&	 \nodata 	&	 $3.46\pm0.06$ 	\\
\object{NGC~3079} & SBcd	& $16.5\pm2.9$ & $75.0\pm3.9$ &	 $5.91\pm0.57$ 	&	 $0.47\pm0.05$ 	&	 $16.79\pm0.25$ 	&	 $0.52\pm0.21$ 	&	 $4.35\pm0.54$ 	&	 $0.35\pm0.04$ 	&	 $17.13\pm0.35$ 	&	 $0.58\pm0.47$ 	\\
\object{NGC~3227} & SABa	& $21.1\pm3.0$ & $59.3\pm3.9$ &	 $17.91\pm3.31$ 	&	 $1.83\pm0.34$ 	&	 $20.26\pm0.32$ 	&	 $2.60\pm0.44$ 	&	 $8.34\pm0.60$ 	&	 $0.85\pm0.06$ 	&	 $19.32\pm0.13$ 	&	 $1.90\pm0.28$ 	\\
\object{NGC~3368} & SABa	& $10.7\pm0.6$ & $46.2\pm3.8$ &	 $5.98\pm0.31$ 	&	 $0.31\pm0.02$ 	&	 $17.07\pm0.08$ 	&	 $1.19\pm0.09$ 	&	 $4.83\pm0.14$ 	&	 $0.25\pm0.01$ 	&	 $16.92\pm0.04$ 	&	 $1.00\pm0.05$ 	\\
\object{NGC~3393} & SBa	& $55.8\pm0.8$\tablenotemark{a} & $31.8\pm8.5$ &	 $1.64\pm0.02$ 	&	 $0.43\pm0.01$ 	&	 $17.27\pm0.05$ 	&	 $1.14\pm0.07$ 	&	 $1.77\pm0.09$ 	&	 $0.47\pm0.02$ 	&	 $17.63\pm0.15$ 	&	 $1.36\pm0.13$ 	\\
\object{NGC~3627} & SBb	& $10.6\pm0.6$ & $59.2\pm3.0$ &	 $11.07\pm1.54$ 	&	 $0.57\pm0.08$ 	&	 $18.44\pm0.24$ 	&	 $3.17\pm0.19$ 	&	 $3.92\pm0.57$ 	&	 $0.20\pm0.03$ 	&	 $16.98\pm0.29$ 	&	 $2.10\pm0.31$ 	\\
\object{NGC~4151} & SABa	& $19.0\pm2.5$ & $46.7\pm3.6$ &	 $6.23\pm0.35$ 	&	 $0.57\pm0.03$ 	&	 $17.75\pm0.12$ 	&	 $2.24\pm0.33$ 	&	 $6.00\pm0.34$ 	&	 $0.55\pm0.03$ 	&	 $17.77\pm0.05$ 	&	 $1.85\pm0.27$ 	\\
\object{NGC~4258} & SABb	& $7.60\pm0.17$ & $63.3\pm2.9$ &	 $41.80\pm6.71$ 	&	 $1.54\pm0.25$ 	&	 $20.14\pm0.27$ 	&	 $3.21\pm0.31$ 	&	 $26.40\pm3.90$ 	&	 $0.97\pm0.14$ 	&	 $19.73\pm0.25$ 	&	 $2.60\pm0.28$ 	\\
\object{NGC~4303} & SBbc	& $12.3\pm0.6$ & $32.3\pm4.5$ &	 $2.28\pm0.09$ 	&	 $0.14\pm0.01$ 	&	 $16.51\pm0.10$ 	&	 $1.02\pm0.13$ 	&	 $2.16\pm0.09$ 	&	 $0.13\pm0.01$ 	&	 $15.78\pm0.11$ 	&	 $0.90\pm0.13$ 	\\
\object{NGC~4388} & SBcd	& $17.8\pm4.1$ & $71.6\pm1.9$ &	 $21.68\pm0.54$ 	&	 $1.87\pm0.05$ 	&	 $19.83\pm0.08$ 	&	 $0.89\pm0.13$ 	&	 $14.30\pm0.55$ 	&	 $1.23\pm0.05$ 	&	 $19.82\pm0.10$ 	&	 $1.15\pm0.09$ 	\\
\object{NGC~4395} & SBm	& $4.76\pm0.02$ & $47.7\pm4.6$ &	 \nodata 	&	 \nodata 	&	 \nodata 	&	 \nodata 	&	 \nodata 	&	 \nodata 	&	 \nodata 	&	 \nodata 	\\
\object{NGC~4501} & Sb	& $11.2\pm0.1$ & $58.7\pm2.9$ &	 $21.22\pm1.22$ 	&	 $1.15\pm0.21$ 	&	 $19.53\pm0.26$ 	&	 $2.33\pm0.23$ 	&	 $20.35\pm2.79$ 	&	 $1.10\pm0.15$ 	&	 $19.91\pm0.21$ 	&	 $2.83\pm0.20$ 	\\
\object{NGC~4594}\tablenotemark{e} & Sa	& $9.55\pm0.44$ & $47.9\pm4.9$ &	 $44.94\pm2.88$ 	&	 $2.08\pm0.13$ 	&	 \nodata 	&	 $6.14\pm0.54$ 	&	 $41.36\pm1.94$ 	&	 $1.92\pm0.09$ 	&	 \nodata 	&	 $4.24\pm0.20$ 	\\
\object{NGC~4699}\tablenotemark{e} & SABb	& $23.7\pm4.8$ & $32.1\pm4.9$ &	 $24.44\pm0.46$ 	&	 $2.80\pm0.05$ 	&	 \nodata 	&	 $5.35\pm0.28$ 	&	 $29.75\pm0.22$ 	&	 $3.41\pm0.03$ 	&	 \nodata 	&	 $6.77\pm0.08$ 	\\
\object{NGC~4736} & SABab 	& $4.41\pm0.08$ & $41.4\pm8.0$ &	 $9.79\pm0.10$ 	&	 $0.21\pm0.00$ 	&	 $16.17\pm0.03$ 	&	 $0.93\pm0.02$ 	&	 $9.65\pm0.13$ 	&	 $0.21\pm0.00$ 	&	 $16.31\pm0.03$ 	&	 $1.03\pm0.02$ 	\\
\object{NGC~4826} & Sab 	& $5.55\pm1.28$ & $55.2\pm4.1$ &	 $13.89\pm0.19$ 	&	 $0.37\pm0.01$ 	&	 $17.86\pm0.03$ 	&	 $0.73\pm0.07$ 	&	 $11.93\pm0.29$ 	&	 $0.32\pm0.01$ 	&	 $17.98\pm0.05$ 	&	 $0.76\pm0.05$ 	\\
\object{NGC~4945} & SABc	& $3.72\pm0.19$ & $77.0\pm1.7$ &	 $26.33\pm5.17$ 	&	 $0.47\pm0.09$ 	&	 $18.48\pm0.34$ 	&	 $3.40\pm0.28$ 	&	 $13.93\pm2.81$ 	&	 $0.25\pm0.05$ 	&	 $17.99\pm0.35$ 	&	 $3.19\pm0.29$ 	\\
\object{NGC~5055} & Sbc 	& $8.87\pm0.39$ & $52.5\pm1.3$ &	 $55.12\pm4.56$ 	&	 $2.37\pm0.20$ 	&	 $20.09\pm0.12$ 	&	 $2.02\pm0.13$ 	&	 $43.52\pm3.15$ 	&	 $1.87\pm0.14$ 	&	 $19.85\pm0.10$ 	&	 $1.76\pm0.11$ 	\\
\object{NGC~5495} & SBc	& $101.1\pm1.5$\tablenotemark{a} & $32.8\pm7.5$ &	 $3.75\pm0.34$ 	&	 $1.76\pm0.16$ 	&	 $20.21\pm0.14$ 	&	 $2.60\pm0.12$ 	&	 $3.99\pm0.27$ 	&	 $1.87\pm0.13$ 	&	 $20.23\pm0.11$ 	&	 $2.46\pm0.12$ 	\\
\object{NGC~5765b} & SABb	& $133.9\pm11.6$ & $42.4\pm1.8$ &	 $1.11\pm0.05$ 	&	 $0.72\pm0.03$ 	&	 $18.72\pm0.06$ 	&	 $1.46\pm0.04$ 	&	 $1.00\pm0.05$ 	&	 $0.65\pm0.03$ 	&	 $18.83\pm0.07$ 	&	 $1.51\pm0.05$ 	\\
\object{NGC~6264} & SBb	& $153.9\pm19.0$ & $59.6\pm7.0$ &	 $1.13\pm0.03$ 	&	 $0.84\pm0.02$ 	&	 $19.23\pm0.04$ 	&	 $1.04\pm0.05$ 	&	 $1.05\pm0.06$ 	&	 $0.78\pm0.05$ 	&	 $19.37\pm0.08$ 	&	 $1.35\pm0.09$ 	\\
\object{NGC~6323} & SBab	& $116.9\pm36.0$\tablenotemark{a} & $69.0\pm1.2$ &	 $1.53\pm0.10$ 	&	 $0.83\pm0.05$ 	&	 $20.39\pm0.17$ 	&	 $2.09\pm0.20$ 	&	 $1.71\pm0.24$ 	&	 $0.92\pm0.13$ 	&	 $19.98\pm0.12$ 	&	 $1.15\pm0.13$ 	\\
\object{NGC~6926} & SBc	& $87.6\pm3.0$ & $58.0\pm7.8$ &	 $0.57\pm0.17$ 	&	 $0.24\pm0.07$ 	&	 $17.52\pm1.17$ 	&	 $1.60\pm1.01$ 	&	 $0.86\pm0.24$ 	&	 $0.36\pm0.10$ 	&	 $18.69\pm0.78$ 	&	 $2.33\pm0.53$ 	\\
\object{NGC~7582} & SBab	& $19.9\pm0.9$ & $64.3\pm5.2$ &	 $5.33\pm1.26$ 	&	 $0.51\pm0.12$ 	&	 $17.04\pm0.52$ 	&	 $2.20\pm0.54$ 	&	 $4.55\pm1.26$ 	&	 $0.44\pm0.12$ 	&	 $17.66\pm0.54$ 	&	 $2.21\pm0.56$ 	\\
\object{UGC~3789} & SABa	& $49.6\pm5.1$ & $41.4\pm17.7$ &	 $1.60\pm0.04$ 	&	 $0.38\pm0.01$ 	&	 $18.38\pm0.05$ 	&	 $2.37\pm0.05$ 	&	 $3.11\pm0.10$ 	&	 $0.75\pm0.02$ 	&	 $19.03\pm0.06$ 	&	 $2.67\pm0.05$ 	\\
\object{UGC~6093} & SBbc	& $152.8\pm10.8$\tablenotemark{a} & $37.7\pm6.5$ &	 $1.84\pm1.17$ 	&	 $1.36\pm0.13$ 	&	 $19.27\pm0.14$ 	&	 $1.55\pm0.20$ 	&	 $1.27\pm0.09$ 	&	 $0.94\pm0.07$ 	&	 $18.87\pm0.10$ 	&	 $1.41\pm0.16$ 	\\
\enddata
\tablecomments{Columns:
(1) Galaxy name.
(2) Morphological type from HyperLeda or NED. The presence of a bar has been updated according to findings in our surface brightness profile decompositions.
(3) Luminosity distance from \citet{Davis:2017} and references therein.
(4) Disk inclination angle (in degrees), where $i=\arccos(1-\epsilon_{\rm disk})$ and $\epsilon_{\rm disk}$ is the average ellipticity for the outer disk of the galaxy.
(5) and (6) Bulge major axis effective half-light radius (in arcsec and kpc, respectively).
(7) Bulge effective surface brightness (in AB $\text{mag\,arcsec}^{-2}$) at the radius listed in columns 5 and 6 (and at the same corresponding wavelengths noted in Table~\ref{table:Sample}).
(8) Bulge major-axis S{\'e}rsic index.
(9)--(12) Similar to columns 5--8, but from an independent decomposition of the equivalent axis surface brightness profile.
\emph{Note:} Uncertainties on parameters in columns 5--12 are formal estimates from \textsc{profiler}; they should be considered minimum estimates.
}
\tablenotetext{a}{Hubble flow distance.}
\tablenotetext{b}{SDSS J043703.67+245606.8}
\tablenotetext{c}{S{\'e}rsic profile parameters were obtained from \citet{Okamoto:2013}.}
\tablenotetext{d}{S{\'e}rsic profile parameters were obtained from \citet{Savorgnan:2016}.}
\tablenotetext{e}{The core-S{\'e}rsic model was applied for this galaxy.}
\end{deluxetable*}
\end{longrotatetable}

For three galaxies from our sample (NGC 3031, NGC 4594, and NGC 4699; see Figures \ref{NGC3031_plot}, \ref{NGC4594_plot}, and \ref{NGC4699_plot}, respectively), we used the core-S{\'e}rsic model to describe their spheroids. In each of these galaxies, the \textit{Spitzer} image displays a small deficit of light in their cores. This is an odd occurrence, given that legitimate partially depleted cores have (to date) been observed exclusively in massive early-type galaxies. Therefore, for these three galaxies, we consulted \textit{HST} visible-light images with higher spatial resolution. We do not detect evidence of partially depleted cores and do not classify these galaxies as canonical core-S{\'e}rsic spheroids.\footnote{Therefore, the core-S{\'e}rsic parameters $R_b$, $\gamma$, and $\alpha$ are not physical quantities for these three galaxies.} This situation does not, however, produce a problem. We could have, for instance, excluded the questionable inner deficit from these galaxies' \textit{Spitzer} light profiles (and fit a S{\'e}rsic function rather than a core-S{\'e}rsic function), effectively obtaining the same decomposition. The central flux deficit is but $\approx1\%$ of the (S{\'e}rsic) bulge light.

Nuclear point sources were modeled by us using the image's PSF and are characterized by one parameter, the central surface brightness of the PSF ($\mu_0$).

While bars can be modeled using a low-$n$ S{\'e}rsic profile, we used a Ferrers function \citep{Ferrers:1877,Sellwood:Wilkinson:1993} with a similarly flat core. These are parameterized by four quantities: the central brightness ($\mu_0$), the cutoff radius ($R_{\rm end}$), and two parameters that control the inner slope ($\alpha$) and break sharpness ($\beta$). However, since $\beta>0$ causes a cusp in the central parts of the bar profile, we permanently set $\beta\equiv0$ in all of our models.\footnote{Examining potentially legitimate upturns in the inner light profile of bars is beyond the scope of the present paper and left for future work.}

Most disks were described by us using an exponential model with two parameters: the central surface brightness ($\mu_0$) and exponential scale length ($h$). Truncated disks were modeled using a broken exponential model with four parameters: the central surface brightness ($\mu_0$), the break radius ($R_b$), and the inner and outer scale lengths, $h_1$ and $h_2$, respectively. Inclined disks with close to edge-on orientations were fit with an edge-on disk model \citep{Kruit:1981}. Additional features that cause ``bumps'' in the light profiles (e.g., rings, spiral arms, ansae, etc.) could usually be modeled using a Gaussian profile centered at the radius of the ``bump.'' These Gaussian components were parameterized by three quantities: the radius of the bump ($R_{\rm r}$), its peak surface brightness ($\mu_0$), and its FWHM.

\subsection{Surface Brightness Profiles}\label{Sec_SBF}

The study of bulge masses  requires a decomposition of a galaxy's total light into its separate components. The generation of bulge mass estimates in haste by automated pipelines (i.e., without human guidance) can at times be misleading; there are substantial uncertainties in automated decompositions \citep[see, e.g.,][]{Benson:2007,Tasca:2011}. This is especially true when using a predetermined number of S{\'e}rsic components \citep[as noted by][see also \citealt{Huang:2013}]{Peng:2010}, or when a single exponential disk is used to model every disk, or when an intermediate-scale disk is treated as if it were a large-scale disk \citep{Liller:1966,Savorgnan:Graham:2016}. Simply adding S{\'e}rsic components without recourse to the physical components in a galaxy can also lead one astray as to the mass of the spheroid \citep[e.g.,][in the case of NGC~1277]{Graham:2016c}.

Spiral galaxies frequently contain multiple components such as bars, spiral arms, rings, etc., which, if not accounted for in the modeling process, can bias the estimate of the bulge magnitude, and often considerably so (as noted by \citealt{Lasker:2014a} and \citealt{Savorgnan:2016}). In addition, galaxy-centric weighting schemes \citep[e.g.,][used a Poisson error weighting: see their Equation~(1)]{Peng:2002,Peng:2010}, used by minimization routines to fit a model to the  data, can be undermined by the presence of an active galactic nucleus (AGN) or a nuclear disk that was not accounted for in the model, or because of central dust, or by a poorly represented PSF. It is also important to use an adequate radial extent of the surface brightness profile in order to diagnose the contribution from the disk and thereby determine what model should be used (e.g., single, truncated, or inclined exponential) and thus how it extrapolates into the inner regions of the galaxy. All of this can hinder the analysis of not only of individual galaxies but also of pipeline surveys that subsequently apply the black hole mass scaling relations to their bulge luminosity function \citep[e.g.,][]{Marconi:2004} in order to construct the black hole mass function (BHMF).

\citet{Savorgnan:2016} investigated the published decompositions of 18 spiral galaxies and identified where improvements could be made. Their decomposition figures and captions, along with comparisons to the literature, provide valuable insight for those attempting this type of work. Here we continue the methodology employed by \citet{Savorgnan:2016}.

Although we model the 1D light profile, this is not simply obtained from a 1D cut through a galaxy.  Rather, the information from a set of 1D profiles---including the ellipticity profile, the position angle (PA) profile, and the radially changing Fourier harmonic terms describing isophotal deviations from pure ellipses---are all effectively folded into the final 1D surface brightness profile \citep[see the Appendix of][for more details]{Ciambur:2015}, and we refer to these profiles for our user-guided decompositions.\footnote{It is difficult to provide a formula that automated pipelines could follow; the methodology is complex and requires human intervention in consultation with all of the available information, such as ellipticity profiles, PA profiles, Fourier harmonic profiles, the surface brightness profiles, care with the sky background, kinematical data (if available), viewing the image at a range of contrasts, and knowledge of second-order components like ansae, rings, spirals, etc.} We model the surface brightness profiles using the decomposition routine \textsc{profiler} \citep{Ciambur:2016}, which convolves the fitted galaxy model with the PSF but intentionally\footnote{Signal-to-noise weighted fitting schemes are not always ideal in practice. They can be significantly hampered by central galaxy components and features, such as dust and AGNs, which are present in the data but not present in one's fitted model.} does not use a Poisson or galaxy-centric weighting scheme in its minimization routine, and therefore making it important to carefully measure the sky background. We determine the PSF by sampling (with the \textsc{iraf} task \textsc{imexamine}) numerous bright foreground stars in the image\footnote{\citet{Savorgnan:2016} demonstrated that it is necessary to measure the PSF from ``real'' stars in the image, because the \textit{Spitzer} instrument point-response function is systematically smaller than the PSF of ``real'' stars.} and represent it with a \citep{Moffat:1969} profile via its equivalent FWHM and $\beta$-value.\footnote{\textsc{profiler} can be run using the exact PSF rather than a Moffat function. For bulges that are considerably larger than the PSF, the difference in PSF model has a negligible effect on the analysis of bulges. Here $R_e/(\text{PSF FWHM})>3$ for 75\% of our sample. Tests between exact and mean PSFs for the galaxies in our sample with the smallest $R_e/(\text{PSF FWHM})$ ratios yield discrepancies $\lesssim0.02\,\text{dex}$ ($\lesssim5\%$) in stellar spheroid masses. The median PSF FWHMs for our images are $2\farcs00$, $1\farcs90$, $0\farcs09$, $0\farcs19$, and $3\farcs04$ for our $\rm{S^4G}$, SHA, WFC3, WFPC2, and 2MASS images, respectively.}

We examine the surface brightness profile along both the major and ``equivalent'' (geometric mean of the major and minor) axes. Given an ellipse with semimajor and semiminor axis lengths ``$a$'' and ``$b$,'' the area of the ellipse is equivalent to that of a circle with a geometric mean radius equal to $\sqrt{ab}$. Moreover, the same is true when using the semimajor and semiminor axis lengths of quasi-elliptical isophotes from \textsc{isofit}, which include the Fourier harmonic terms (for a derivation, see the Appendix of \citealt{Ciambur:2015}). Thus, by mapping the semimajor-axis radius ($R_{\rm maj}$) to the equivalent axis radius ($R_{\rm eq}$),\footnote{$R_{\rm eq} = R_{\rm maj}\sqrt{1-\epsilon(R_{\rm maj})}$} we convert an isophote into the equivalent circle that conserves the original surface area of the isophote. This allows for simple, analytical computation of component luminosity from what can be thought of as a circularly symmetric surface brightness profile (see Section~\ref{AbsMags}).

\begin{sidewaysfigure*}
\includegraphics[clip=true,trim= 11mm 1mm 3mm -15cm,width=0.25\textwidth]{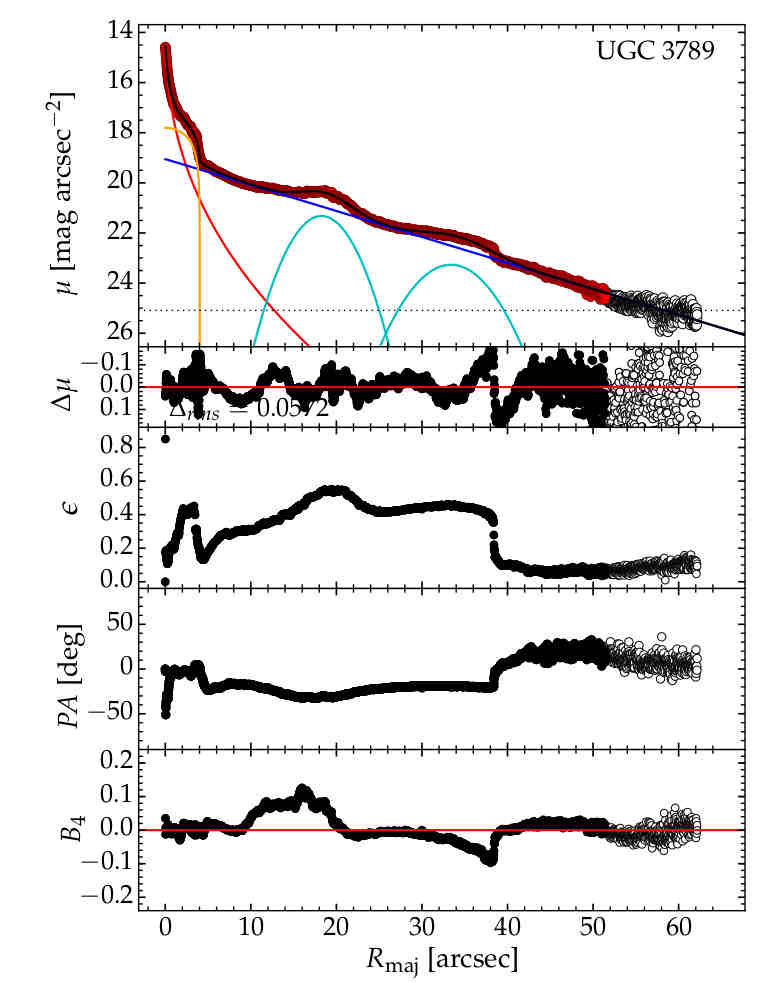}
\includegraphics[clip=true,trim= 11mm 1mm 3mm -15cm,width=0.25\textwidth]{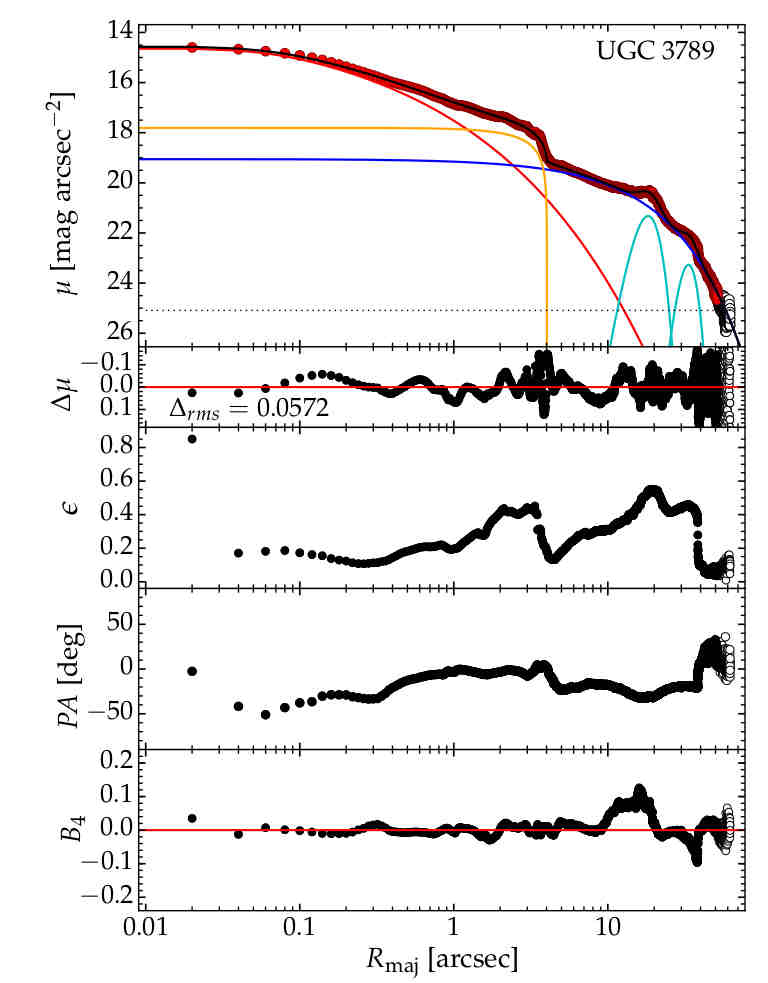}
\includegraphics[clip=true,trim= 11mm 1mm 3mm -15cm,width=0.25\textwidth]{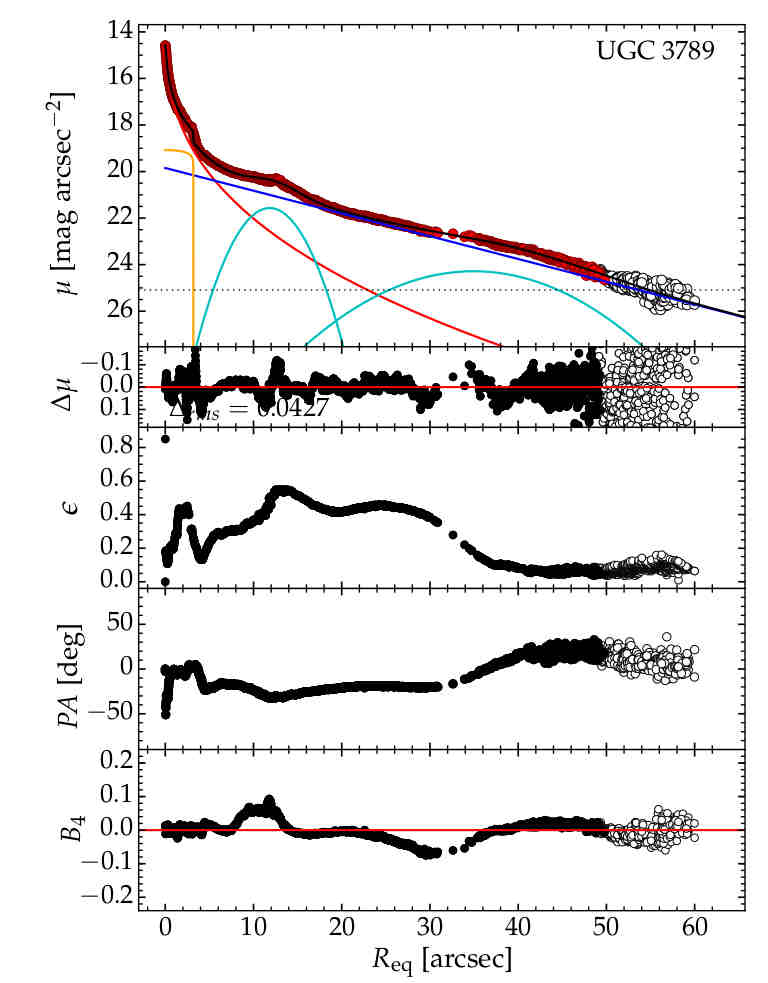}
\includegraphics[clip=true,trim= 11mm 1mm 3mm -15cm,width=0.25\textwidth]{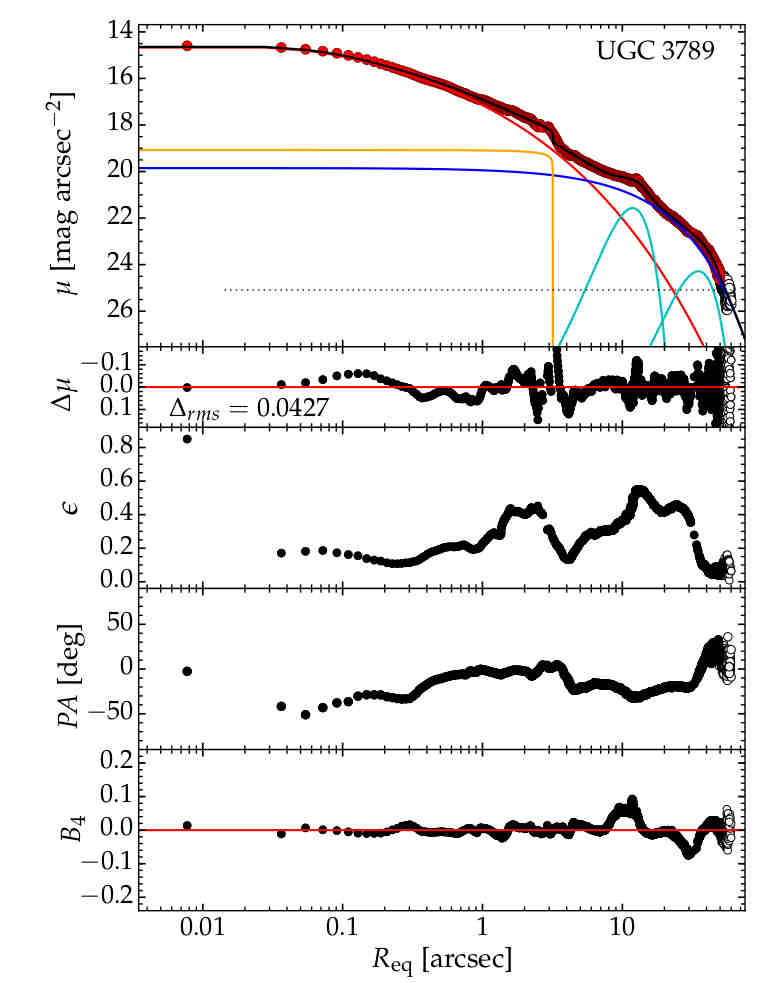}
\caption{\textit{HST} WFC3 UVIS2 F814W surface brightness profile for the major axis (left two panels) and the equivalent axis (right two panels) of UGC~3789 after masking, sky subtraction, and PSF convolution with the model. At the estimated distance of UGC~3789, the physical sizes can be converted via 0.2405$\,\text{kpc}\,\text{arcsec}^{-1}$. \textbf{Left two panels}---The model represents $0\arcsec \leq R_{\rm maj} \leq 51\farcs45$ with $\Delta_{\rm rms}=0.0572\,\text{mag\,arcsec}^{-2}$, and additional unfit data are plotted for $51\farcs45 < R_{\rm maj} \leq 62\farcs18$. \underline{S{\'e}rsic Profile Parameters:} \textcolor{red}{$R_e=1\farcs60\pm0\farcs04$, $\mu_e=18.38\pm0.05\,\text{mag\,arcsec}^{-2}$, and $n=2.37\pm0.05$.} \underline{Ferrers Profile Parameters:} \textcolor{Orange}{$\mu_0 = 17.80\pm0.02\,\text{mag\,arcsec}^{-2}$, $R_{\rm end} = 4\farcs02\pm0\farcs01$, and $\alpha = 1.67\pm0.05$.} \underline{Exponential Profile Parameters:} \textcolor{blue}{$\mu_0 = 19.05\pm0.01\,\text{mag\,arcsec}^{-2}$ and $h = 10\farcs47\pm0\farcs02$.} \underline{Additional Parameters:} two Gaussian components added at: \textcolor{cyan}{$R_{\rm r}=18\farcs26\pm0\farcs03$ \& $33\farcs39\pm0\farcs05$; with $\mu_0 = 21.32\pm0.01$ \& $23.27\pm0.01\,\text{mag\,arcsec}^{-2}$; and FWHM = $6\farcs03\pm0\farcs07$ \& $7\farcs97\pm0\farcs09$, respectively.} \textbf{Right two panels}---The model represents $0\arcsec \leq R_{\rm eq} \leq 49\farcs83$ with $\Delta_{\rm rms}=0.0427\,\text{mag\,arcsec}^{-2}$, and additional unfit data are plotted for $49\farcs83 < R_{\rm maj} \leq 60\farcs32$. \underline{S{\'e}rsic Profile Parameters:} \textcolor{red}{$R_e=3\farcs11\pm0\farcs10$, $\mu_e=19.03\pm0.06\,\text{mag\,arcsec}^{-2}$, and $n=2.67\pm0.05$.} \underline{Ferrers Profile Parameters:} \textcolor{Orange}{$\mu_0 = 19.08\pm0.06\,\text{mag\,arcsec}^{-2}$, $R_{\rm end} = 3\farcs18\pm0\farcs05$, and $\alpha = 0.31\pm0.10$.} \underline{Exponential Profile Parameters:} \textcolor{blue}{$\mu_0 = 19.85\pm0.04\,\text{mag\,arcsec}^{-2}$ and $h = 11\farcs10\pm0\farcs21$.} \underline{Additional Parameters:} two Gaussian components added at: \textcolor{cyan}{$R_{\rm r}=11\farcs84\pm0\farcs05$ \& $34\farcs94\pm0\farcs26$; with $\mu_0 = 21.57\pm0.02$ \& $24.28\pm0.08\,\text{mag\,arcsec}^{-2}$; and FWHM = $5\farcs93\pm0\farcs10$ \& $18\farcs52\pm0\farcs88$, respectively.}}
\label{UGC3789_plot}
\end{sidewaysfigure*}

We present four light profiles for each of the spiral galaxies in our sample. An example is given in Figure~\ref{UGC3789_plot}. Each of the four light profiles is presented in five-paneled plots, each with a common horizontal axis (abscissa) representing either the major axis or the geometric mean axis, equivalent to a circularized axis and referred to as the equivalent axis, in units of arcseconds. Moving from left to right between the four five-panel plots, we present surface brightness profiles for the major axis on a linear scale, the major axis on a logarithmic scale (required to see the features at smaller radii), the equivalent axis on a linear scale, and the equivalent axis on a logarithmic scale. For all panels, open \textbf{black} circles represent datapoints that were omitted from the fitting process. We include the major axis light profile because this is what is usually shown in the literature and will enable reader comparison with past work. We include the circularized profile because this is what we used to derive the spheroid stellar flux and mass.

The individual isophotal surface brightness datapoints, in $\text{mag\,arcsec}^{-2}$, are depicted by filled \textcolor{red}{red} circles. Model components are depicted by unique colors: \textcolor{LimeGreen}{lime green} for point sources, \textcolor{blue}{blue} lines for exponential (as well as broken exponential and near edge-on) disks, \textcolor{red}{red} curves for S{\'e}rsic (and core-S{\'e}rsic) spheroids, \textcolor{Orange}{orange} ``shelf'' shapes for Ferrers bars, and \textcolor{cyan}{cyan} lines for Gaussians, which perform well at capturing excess flux due to features like spiral arms and rings. The integrated surface brightness profile model (summation of all the individual components), after convolution with the PSF, is depicted as a solid \textbf{black} line that contours (as best as possible) to the pattern of \textcolor{red}{red} circles. The horizontal \textbf{black} dotted line (if shown) represents the ``threshold surface brightness,'' which depicts our level of uncertainty on the sky background.

The second row of panels depicts the residual surface brightness profile: $\Delta\mu(R) = \mu_{\rm data}(R) - \mu_{\rm model}(R)$ in $\text{mag\,arcsec}^{-2}$. Additionally, a \textcolor{red}{red} line at $\Delta\mu = 0$ is printed as a reference, and the root mean square (rms) scatter ($\Delta_{\rm rms}$) about the line is provided. The middle (third) row of panels depicts the ellipticity ($\epsilon$) profile of the target galaxy. The fourth row of panels depicts the PA profile of the target galaxy, with PA in degrees (east of north). The bottom (fifth) row of panels depicts the isophotal ``boxyness/diskyness'' profile, as quantified by the Fourier harmonic coefficient, $B_4$ (fourth cosine harmonic amplitude from Equation~(\ref{harmonics})).\footnote{The $B_4$ coefficient is the relative difference in radius along the major axis between that of a perfect ellipse and the modified isophote. ``Disky'' profiles have positive values, and ``boxy'' profiles have negative values \citep[for additional information, see][]{Ciambur:2015}.} These latter three panels can be helpful in identifying substructure in the galaxy, especially in combination with visual inspection of the image.

Figure~\ref{UGC3789_plot} presents the surface brightness profile plots for UGC~3789. The modeled surface brightness profiles of the other galaxies appear at the end of the paper in Appendix~\ref{App3}, preceded by textual descriptions of the galaxies' components in Appendix~\ref{App2}. In Figure~\ref{fig:annotate}, we provide a visual aid for our component fitting for UGC~3789.

\begin{figure}
\includegraphics[clip=true,trim= 1mm 1mm 1mm 1mm,width=\columnwidth]{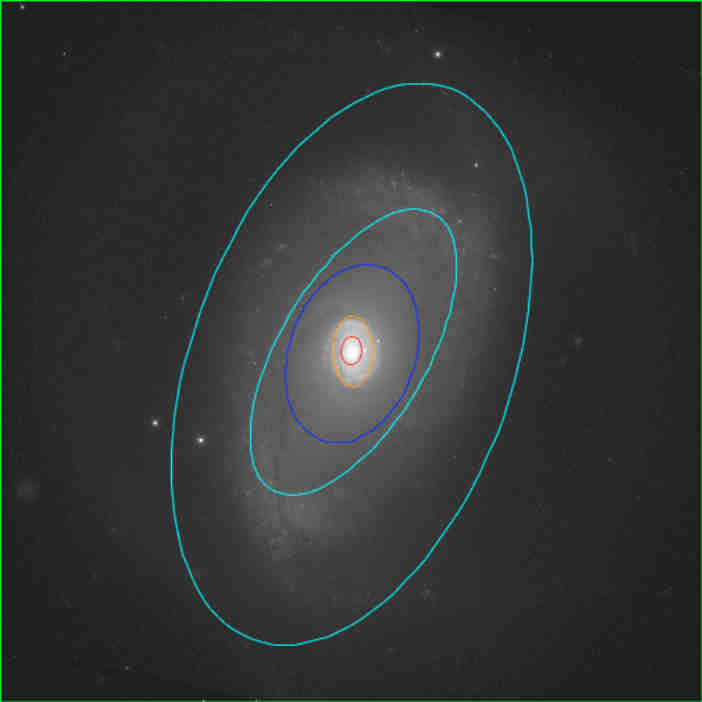}
\caption{The \textit{HST} WFC3 UVIS2 F814W image of UGC~3789 with overlaid components, as detailed in Figure~\ref{UGC3789_plot}, with a matching color scheme. Each component is represented by an elliptical isophote with a matching PA, ellipticity, and semimajor axis as the relative location of each component in the major axis surface brightness profile. The S{\'e}rsic component isophote (\textcolor{red}{red}) is positioned at $R_{\rm maj}\equiv R_e$. The Ferrers component isophote (\textcolor{orange}{orange}) is positioned at $R_{\rm maj}\equiv R_{\rm end}$, with an ellipticity equal to the maximum bar ellipticity. The exponential profile isophote (\textcolor{blue}{blue}) is positioned at $R_{\rm maj}\equiv h$. The Gaussian component isophotes (\textcolor{cyan}{cyan}) are positioned at their peak locations with $R_{\rm maj}\equiv R_{\rm r}$. Note that the image is a square with $80\arcsec$ ($19.24\,\text{kpc}$) on a side; north is up and east is left. The black pixel values are set to $\mu=26\,\text{mag\,arcsec}^{-2}$, and the white pixel values are set to $\mu=14\,\text{mag\,arcsec}^{-2}$ with a logarithmic contrast stretch.}
\label{fig:annotate}
\end{figure}

\subsection{Magnitudes}\label{AbsMags}

Using the parameters from the \citet{Sersic:1963} bulge model fit to the equivalent axis, we are able to analytically determine the apparent magnitude, $\mathfrak{m}$, of the bulge\footnote{Core-S{\'e}rsic bulge apparent magnitudes are determined from equations presented in Appendix~A of \citet{Trujillo:2004}. Specifically, their Equations (A20) and (A6) are used instead of our Equations (\ref{apparent_mag}) and (\ref{b_n}), respectively.} using the relation
\begin{equation}
\mathfrak{m} = \mu_e - 5\log{R_e} - 2.5\log \left[2\pi n\frac{e^{b_n}}{(b_n)^{2n}}\Gamma(2n)\right],
\label{apparent_mag}
\end{equation}
where $b_n$ is a constant such that
\begin{equation}
\Gamma(2n) = 2\gamma(2n,\, b_n).
\label{b_n}
\end{equation} 
A derivation of this, and related equations pertaining to the \citet{Sersic:1963} model, can be found in \citet{Graham:Driver:2005}.
Equation~(\ref{apparent_mag}) does not need to include any ellipticity terms because the ellipticity profile is already accounted for within the equivalent axis light profile, as are the departures from a pure ellipse as described by the Fourier harmonic terms.

We then calculate the corrected bulge absolute magnitude via
\begin{equation}
\mathfrak{M} = \mathfrak{m} - 5\log{d_L} -25 - A_{\lambda} - K_{\lambda}(z) -10\log(1+z),
\label{absolute_mag}
\end{equation}
where $d_L$ is the luminosity distance (in Mpc), $A_{\lambda}$ is the Galactic extinction (in mag) at the observed wavelength ($\lambda$) due to dust attenuation in the Milky Way, $K_\lambda(z)$ is the rest-frame $K$-correction\footnote{See \citet{Hogg:2002} for a pedagogical discussion about $K$-corrections. The $3.6\,\micron$ $K$-corrections are estimated via the template of \citet{Glikman:2006}, $K$-corrections for the other imaging sources are obtained from the NASA-\textit{Sloan} Atlas (\url{http://nsatlas.org}). This research made use of the ``$K$-corrections calculator'' service available at \url{http://kcor.sai.msu.ru/} \citep{Chilingarian:2010,Chilingarian:2012}. The median $|K_{\lambda}(z)|=0.003\,\text{mag}$ for our sample.} (in mag) at the observed wavelength and redshift ($z$), and the final term corrects for the $(1+z)^4$ cosmological surface brightness dimming \citep{Tolman:1930,Tolman:1934}.\footnote{Our median cosmological surface brightness dimming correction is $0.02\,\text{mag}$. However, for our eight most distant galaxies, the corrections are all $\geq0.10\,\text{mag}$.}

Uncertainties on bulge magnitudes are determined from $10^4$~Monte~Carlo samplings per galaxy of Equations (\ref{apparent_mag}) and (\ref{b_n}) by allowing $R_e$, $\mu_e$, and $n$ to vary based on their joint distribution from the decomposition analysis. For each sample, the above variables are randomly selected from their joint normal distribution, which comes from the \textsc{profiler} software and is possibly an underestimate due to potentially misdiagnosed galaxy substructure. After $10^4$ samples, we compute the rms error for $\mathfrak{m_{\rm sph}}$. The numerically determined uncertainties on $\mathfrak{m_{\rm sph}}$\footnote{As explored by \citet{Savorgnan:2016}, precise errors on the bulge apparent magnitude are difficult to ascertain due to unknown levels of degeneracy amongst the different components (i.e., adding or subtracting a component can significantly affect the bulge apparent magnitude). From the result of many tests of the degeneracy for our sample of galaxies, we decided to qualitatively restrict our uncertainties on $\mathfrak{m_{\rm sph}}$ to the range $0.13\,\text{mag} \leq \delta\mathfrak{m_{\rm sph}} \leq 0.45\,\text{mag}$.} are then propagated, along with uncertainties with the distance, to calculate uncertainties on $\mathfrak{M_{\rm sph}}$, such that
\begin{equation}
\delta\mathfrak{M_{\rm sph}}=\sqrt{\delta\mathfrak{m_{\rm sph}}^2+\left[\frac{5(\delta d_L)}{d_L\ln(10)}\right]^2}.
\label{spheroid_abs_mag_error}
\end{equation}

\subsection{Mass-to-light Ratios}

At $3.6\,\micron$, the thermal glow of dust in spiral galaxies contributes, on
average, one-third as much flux as the stars.  That is, $\approx25\%$ of the observed $3.6,\micron$ flux comes from the dust \citep[][their Figures 8 and 9]{Querejeta:2015}.\footnote{In galaxies with rather high specific star formation rates (SSFR), with
$\log({\rm SSFR/yr}) > -9.75$, roughly one-third of the observed flux comes from the
glow of dust.} Figure~10 in \citet{Querejeta:2015} presents a slight trend
such that the (stellar mass)-to-(observed $3.6\,\micron$ luminosity) decreases as the
observed, i.e., the dust-affected $[3.6\,\micron]-[4.5\,\micron]$ color becomes
redder.  The explanation is that more dust will result in a higher nonstellar
(i.e., dust) luminosity, and the dust glows more brightly at $4.5\,\micron$ than at
$3.6\,\micron$.  However, complicating matters is that the (stellar
mass)-to-(stellar $3.6\,\micron$ luminosity) ratio ($\Upsilon_{\rm *,IRAC1}$) also varies with
the stellar population, as traced by the dust-free $[3.6\,\micron]-[4.5\,\micron]$ color
\citep[][their Figure~4]{Meidt:2014}. This relation has the opposite sense,
such that $\Upsilon_{\rm *,IRAC1}$ increases as the stellar population becomes redder.

Given this complication and given that we do not have the \emph{stellar} $[3.6\,\micron]-[4.5\,\micron]$ color for most of our galaxies, we proceed by adopting a median $\log(M_*/L_{\rm obs,IRAC1})$ ratio equal to $-0.35$ from \citet[][their
Figure~10]{Querejeta:2015}. This median value arises from the use of a color-independent $\Upsilon_{\rm *,IRAC1}$ ratio of
$0.60\pm0.09$ from \citet{Meidt:2014}, which is based on a \citet{Chabrier:2003} initial mass function (IMF)\footnote{\citet{Dokkum:2017} asserted that the stellar IMF in the centers of massive early-type galaxies is bottom-heavy. Thus, these massive galaxies are better represented by a \citet{Salpeter:1955} IMF than a \citet{Kroupa:2001} or \citet{Chabrier:2003} IMF. Specifically, a \citet{Salpeter:1955} IMF implies stellar masses that are higher by a factor of 1.6.} and a \citet{Bruzual:Charlot:2003} synthesized stellar population (SSP) with exponentially declining star formation histories (SFHs) for a range of metallicities, coupled with a median $(L_*/L_{\rm obs})_{\rm IRAC1}$ ratio of $0.755\pm0.042$ \citep[][derived from the data in their
Figures 8 and 9]{Querejeta:2015}, giving us a median ratio $M_*/L_{\rm obs,IRAC1}=0.453\pm0.072$ that we apply
to the observed $3.6\,\micron$ luminosities to obtain the stellar masses of
the galaxies observed with the \textit{Spitzer Space Telescope}.

According to \citet{Meidt:2014}, $\Upsilon_{\rm *,IRAC1}=0.60\pm0.09$, is fairly constant\footnote{\citet{Savorgnan:2016:II} had reported that use of this constant $\Upsilon_{\rm *,IRAC1}$ yielded consistent results for the $M_{\rm BH}$--$M_{\rm *,sph}$ relation as obtained when using the $[3.6\,\micron]-[4.5\,\micron]$ color-dependent $\Upsilon_{\rm *,IRAC1}$ \citep[][their
Figure~4 and Equation~(4)]{Meidt:2014}.} for stellar populations with ages $\lesssim10$ Gyr. To be consistent, we select corresponding $\Upsilon_*$ values for all other filters that yield equivalent stellar masses as predicted by our \textit{Spitzer} imaging. Specifically, we define the other stellar mass-to-light ratios such that
\begin{equation}
\frac{M_*}{L_{\rm F814W}}=\left(\frac{L_{\rm IRAC1}}{L_{\rm F814W}}\right)\left(\frac{M_*}{L_{\rm IRAC1}}\right)=1.88\pm0.40
\label{M/L_i}
\end{equation}
and
\begin{equation}
\frac{M_*}{L_{K_s}}=\left(\frac{L_{\rm IRAC1}}{L_{K_s}}\right)\left(\frac{M_*}{L_{\rm IRAC1}}\right)=0.62\pm0.08,
\label{M/L_i}
\end{equation}
where $M_*/L_{\rm F814W}$ is the calibrated stellar mass-to-light ratio for our F814W imaging, $M_*/L_{K_s}$ is the calibrated stellar mass-to-light ratio for our $K_s$ imaging, $L_{\rm IRAC1}$ is the observed luminosity in the IRAC1 filter, $L_{\rm F814W}$ is the observed luminosity in the F814W filter, $L_{K_s}$ is the observed luminosity in the $K_s$ filter, and $M_*/L_{\rm IRAC1}=0.453$. 
All $\Upsilon_*$ values, as well as solar absolute magnitude values, are listed in Table~\ref{table:photo}. Spheroid absolute magnitudes and stellar masses are provided in Table~\ref{table:Sample}. Uncertainties on stellar mass are defined as
\begin{IEEEeqnarray}{rCl}
&&\delta \log{M_{*,\rm sph}}=\nonumber\\
&&\sqrt{\left(\frac{\delta\mathfrak{m_{\rm sph}}}{2.5}\right)^2+\left[\frac{2(\delta d_L)}{d_L\ln(10)}\right]^2+\left[\frac{\delta\Upsilon_*}{\Upsilon_*\ln(10)}\right]^2}.
\label{spheroid_mass_error}
\end{IEEEeqnarray}

\startlongtable
\begin{deluxetable*}{lllrlrrrr}
\tabletypesize{\small}
\tablecolumns{9}
\tablecaption{Galaxy Sample and Masses\label{table:Sample}}
\tablehead{
\colhead{Galaxy Name} & \colhead{$\lambda$} & \colhead{$A_{\lambda}$} & \colhead{$K_{\lambda}(z)$} & \colhead{$\log(M_{\rm BH}/M_{\sun})$} & \colhead{$|\phi|$} & \colhead{$\mathfrak{m}_{\lambda,{\rm sph}}$} & \colhead{$\mathfrak{M}_{\lambda,{\rm sph}}$} & \colhead{$\log(M_{\rm *,sph}/M_{\sun})$} \\
\colhead{} & \colhead{($\micron$)} & \colhead{(mag)} & \colhead{(mag)} & \colhead{} & \colhead{(deg)} & \colhead{(mag)} & \colhead{(mag)} & \colhead{} \\
\colhead{(1)} & \colhead{(2)} & \colhead{(3)} & \colhead{(4)} & \colhead{(5)} & \colhead{(6)} & \colhead{(7)} & \colhead{(8)} & \colhead{(9)}
}
\startdata
\object{Circinus} 		&	3.550	&	0.265  &	 $-0.001$ 	&	 $6.25^{+0.10}_{-0.12}$ 	&	 $17.0\pm3.9$ 	& $8.26\pm0.27$ &	 $-19.83\pm0.48$ 	&	 $10.12\pm0.20$	\\
\object{Cygnus~A} 		&	0.8012	&	0.067 &	 $-0.017$ 	&	 $9.44^{+0.11}_{-0.14}$ 	&	 $2.7\pm0.2$ 	& $12.28\pm0.45$ &	 $-25.69\pm0.45$ 	&	 $12.36\pm0.20$	\\
\object{ESO~558-G009} 		&	0.8024	&	 $0.610$  &	 $0.033$ 	&	 $7.26^{+0.03}_{-0.04}$ 	&	 $16.5\pm1.3$ 	& $16.54\pm0.13$ &	 $-19.52\pm0.13$ 	&	 $9.89\pm0.11$	\\
\object{IC~2560} 		&	3.550	&	 $0.017$  &	 $-0.006$ 	&	 $6.49^{+0.19}_{-0.21}$ 	&	 $22.4\pm1.7$ 	& $13.59\pm0.28$ &	 $-18.62\pm0.96$ 	&	 $9.63\pm0.39$		\\
\object[SDSS J043703.67+245606.8]{J0437+2456}\tablenotemark{a} 		&	0.8024	&	1.821  &	 $-0.012$ 	&	 $6.51^{+0.04}_{-0.05}$ 	&	 $16.9\pm4.1$ 	& $16.67\pm0.45$ &	 $-19.54\pm0.45$ 	&	 $9.90\pm0.20$	\\
\object{Milky~Way} 	 	&	0.7625	&	 \nodata  &	 \nodata 	&	 $6.60\pm0.02$ 	&	 $13.1\pm0.6$ 	& \nodata &	 $-19.9\pm0.3$\tablenotemark{b} 	&	 $9.96\pm0.05$\tablenotemark{c}\\
\object{Mrk~1029} 		&	0.8024	&	0.064  &	0.033	&	 $6.33^{+0.10}_{-0.13}$ 	&	 $17.9\pm2.1$ 	& $16.36\pm0.13$ &	 $-19.55\pm0.13$ 	&	 $9.90\pm0.11$	\\
\object{NGC~0224} 		&	3.550	&	0.124  &	0.001	&	 $8.15^{+0.22}_{-0.11}$ 	&	 $8.5\pm1.3$ 	& $4.39\pm0.15$\tablenotemark{d} &	 $-19.80\pm0.17$\tablenotemark{e} 	&	 $10.11\pm0.09$\tablenotemark{e}	\\
\object{NGC~0253} 		&	3.550	&	0.003  &	 $-0.001$ 	&	 $7.00\pm0.30$ 	&	 $13.8\pm2.3$ 	& $8.47\pm0.13$ &	 $-18.93\pm0.16$ 	&	 $9.76\pm0.09$	\\
\object{NGC~1068} 		&	2.159	&	0.010  &	 $-0.019$ 	&	 $6.75\pm0.08$ 	&	 $17.3\pm1.9$ 	& $8.92\pm0.44$ &	 $-21.11\pm0.59$ 	&	 $10.27\pm0.24$	\\
\object{NGC~1097} 		&	3.550	&	0.005 &	 $-0.003$ 	&	 $8.38^{+0.03}_{-0.04}$ 	&	 $9.5\pm1.3$ 	& $10.09\pm0.45$ &	 $-21.61\pm0.46$ 	&	 $10.83\pm0.20$	\\
\object{NGC~1300} 		&	3.550	&	0.005 &	 $-0.003$ 	&	 $7.71^{+0.19}_{-0.14}$ 	&	 $12.7\pm2.0$ 	& $12.43\pm0.45$ &	 $-18.10\pm0.59$ 	&	 $9.42\pm0.25$	\\
\object{NGC~1320} 		&	3.550	&	0.008  &	 $-0.006$ 	&	 $6.78^{+0.24}_{-0.34}$ 	&	 $19.3\pm2.0$ 	& $12.44\pm0.19$ &	 $-20.17\pm0.99$ 	&	 $10.25\pm0.40$	\\
\object{NGC~1398} 		&	3.550	&	0.002  &	 $0.023$ 	&	 $8.03\pm0.11$ 	&	 $9.7\pm0.7$ 	& $10.72\pm0.25$ &	 $-20.96\pm0.47$ 	&	 $10.57\pm0.20$	\\
\object{NGC~2273} 		&	0.8024	&	0.107  &	 $0.004$ 	&	 $6.97\pm0.09$ 	&	 $15.2\pm3.9$ 	& $12.89\pm0.13$ &	 $-19.75\pm0.45$ 	&	 $9.98\pm0.20$	\\
\object{NGC~2748} 		&	0.8012	&	0.041 &	 $0.001$ 	&	 $7.54^{+0.17}_{-0.25}$ 	&	 $6.8\pm2.2$ 	& \nodata &	 \nodata 	&	 \nodata\\
\object{NGC~2960} 		&	3.550	&	0.008  &	 $-0.009$ 	&	 $7.06^{+0.16}_{-0.17}$ 	&	 $14.9\pm1.9$ 	& $13.39\pm0.30$ &	 $-20.64\pm0.87$ 	&	 $10.44\pm0.36$	\\
\object{NGC~2974} 		&	3.550	&	0.010  &	 $-0.004$ 	&	 $8.23^{+0.07}_{-0.08}$ 	&	 $10.5\pm2.9$ 	& $11.27\pm0.13$ &	 $-20.11\pm0.29$ 	&	 $10.23\pm0.13$	\\
\object{NGC~3031} 	&	3.550	&	0.014  &	0.000	&	 $7.83^{+0.11}_{-0.07}$ 	&	 $13.4\pm2.3$ 	& $7.48\pm0.19$\tablenotemark{f} &	 $-19.94\pm0.22$\tablenotemark{f} 	&	 $10.16\pm0.11$	\\
\object{NGC~3079} 		&	3.550	&	0.002  &	 $-0.003$ 	&	 $6.38^{+0.11}_{-0.13}$ 	&	 $20.6\pm3.8$ 	& $11.48\pm0.45$ &	 $-19.32\pm0.59$ 	&	 $9.92\pm0.25$	\\
\object{NGC~3227} 		&	3.550	&	0.004  &	 $-0.003$ 	&	 $7.88^{+0.13}_{-0.14}$ 	&	 $7.7\pm1.4$ 	& $11.70\pm0.22$ &	 $-19.63\pm0.38$ 	&	 $10.04\pm0.17$	\\
\object{NGC~3368} 		&	3.550	&	0.004  &	 $-0.002$ 	&	 $6.89^{+0.08}_{-0.10}$ 	&	 $14.0\pm1.4$ 	& $10.80\pm0.13$ &	 $-19.06\pm0.19$ 	&	 $9.81\pm0.10$	\\
\object{NGC~3393} 		&	0.8024	&	0.116  &	 $0.009$ 	&	 $7.49^{+0.05}_{-0.16}$ 	&	 $13.1\pm2.5$ 	& $13.54\pm0.19$ &	 $-20.37\pm0.19$ 	&	 $10.23\pm0.12$	\\
\object{NGC~3627} 		&	3.550	&	0.006  &	 $-0.002$ 	&	 $6.95\pm0.05$ 	&	 $18.6\pm2.9$ 	& $10.95\pm0.44$ &	 $-18.88\pm0.46$ 	&	 $9.74\pm0.20$		\\
\object{NGC~4151} 		&	3.550	&	0.005  &	 $-0.002$ 	&	 $7.68^{+0.15}_{-0.58}$ 	&	 $11.8\pm1.8$ 	& $10.88\pm0.15$ &	 $-20.22\pm0.33$ 	&	 $10.27\pm0.15$	\\
\object{NGC~4258} 		&	3.550	&	0.003  &	 $-0.001$ 	&	 $7.60\pm0.01$ 	&	 $13.2\pm2.5$ 	& $9.45\pm0.42$ &	 $-19.65\pm0.42$ 	&	 $10.05\pm0.18$	\\
\object{NGC~4303} 		&	3.550	&	0.004  &	 $-0.003$ 	&	 $6.58^{+0.07}_{-0.26}$ 	&	 $14.7\pm0.9$ 	& $12.09\pm0.16$ &	 $-18.08\pm0.20$ 	&	 $9.42\pm0.10$\\
\object{NGC~4388} 		&	3.550	&	0.006  &	 $-0.005$ 	&	 $6.90\pm0.11$ 	&	 $18.6\pm2.6$ 	& $11.28\pm0.13$ &	 $-19.70\pm0.52$ 	&	 $10.07\pm0.22$	\\
\object{NGC~4395} 		&	3.550	&	0.003 &	 $-0.001$ 	&	 $5.64^{+0.22}_{-0.12}$ 	&	 $22.7\pm3.6$ 	& \nodata &	 \nodata 	&	 \nodata \\
\object{NGC~4501} 		&	3.550	&	0.007  &	 $-0.005$ 	&	 $7.13\pm0.08$ 	&	 $12.2\pm3.4$ 	& $10.15\pm0.37$ &	 $-19.81\pm0.38$ 	&	 $10.11\pm0.16$	\\
\object{NGC~4594} 		&	3.550	&	0.009 &	 $-0.002$ 	&	 $8.81\pm0.03$ 	&	 $5.2\pm0.4$ 	& $8.06\pm0.45$\tablenotemark{f} &	 $-21.56\pm0.47$\tablenotemark{f} 	&	 $10.81\pm0.20$	\\
\object{NGC~4699} 		&	3.550	&	0.006 &	 $-0.003$ 	&	 $8.34\pm0.10$ 	&	 $5.1\pm0.4$ 	& $9.26\pm0.45$\tablenotemark{f} &	 $-22.33\pm0.63$\tablenotemark{f} 	&	 $11.12\pm0.26$	\\
\object{NGC~4736} 		&	3.550	&	0.003  &	 $-0.001$ 	&	 $6.78^{+0.09}_{-0.11}$ 	&	 $15.0\pm2.3$ 	& $8.67\pm0.13$ &	 $-19.25\pm0.15$ 	&	 $9.89\pm0.09$	\\
\object{NGC~4826} 		&	3.550	&	0.007  &	 $-0.001$ 	&	 $6.07^{+0.14}_{-0.16}$ 	&	 $24.3\pm1.5$ 	& $10.02\pm0.13$ &	 $-18.40\pm0.52$ 	&	 $9.55\pm0.22$	\\
\object{NGC~4945} 		&	2.159	&	0.055  &	 $-0.009$ 	&	 $6.15\pm0.30$ 	&	 $22.2\pm3.0$ 	& $9.00\pm0.45$ &	 $-18.91\pm0.47$ 	&	 $9.39\pm0.19$		\\
\object{NGC~5055} 		&	3.550	&	0.003 &	 $-0.001$ 	&	 $8.94^{+0.09}_{-0.11}$ 	&	 $4.1\pm0.4$ 	& $8.69\pm0.19$ &	 $-20.76\pm0.22$ 	&	 $10.49\pm0.11$	\\
\object{NGC~5495} 		&	0.8024	&	0.089  &	 $0.023$ 	&	 $7.04^{+0.08}_{-0.09}$ 	&	 $13.3\pm1.4$ 	& $14.09\pm0.18$ &	 $-21.15\pm0.19$ 	&	 $10.54\pm0.12$	\\
\object{NGC~5765b} 		&	0.8024	&	0.057  &	 $0.025$ 	&	 $7.72\pm0.05$ 	&	 $13.5\pm3.9$ 	& $15.95\pm0.13$ &	 $-19.89\pm0.23$ 	&	 $10.04\pm0.13$	\\
\object{NGC~6264} 		&	0.8024	&	0.100  &	 $0.055$ 	&	 $7.51\pm0.06$ 	&	 $7.5\pm2.7$ 	& $16.43\pm0.16$ &	 $-19.81\pm0.31$ 	&	 $10.01\pm0.15$\\
\object{NGC~6323} 		&	0.8024	&	0.026  &	 $0.036$ 	&	 $7.02^{+0.13}_{-0.14}$ 	&	 $11.2\pm1.3$ 	& $16.06\pm0.34$ &	 $-19.46\pm0.75$ 	&	 $9.86\pm0.31$	\\
\object{NGC~6926} 		&	3.550	&	0.029  &	 $-0.011$ 	&	 $7.74^{+0.26}_{-0.74}$ 	&	 $9.1\pm0.7$ 	& \nodata &	 \nodata 	&	 \nodata\\
\object{NGC~7582} 		&	3.550	&	0.002  &	 $-0.003$ 	&	 $7.67^{+0.09}_{-0.08}$ 	&	 $10.9\pm1.6$ 	& $11.29\pm0.45$ &	 $-19.92\pm0.47$ 	&	 $10.15\pm0.20$	\\
\object{UGC~3789} 		&	0.8024	&	0.100  &	 $0.008$ 	&	 $7.06\pm0.05$ 	&	 $10.4\pm1.9$ 	& $13.39\pm0.13$ &	 $-20.24\pm0.26$ 	&	 $10.18\pm0.14$	\\
\object{UGC~6093} 		&	0.8024	&	0.041 &	 $0.051$ 	&	 $7.41^{+0.04}_{-0.03}$\tablenotemark{g} 	&	 $10.2\pm0.9$ 	& $15.50\pm0.19$ &	 $-20.67\pm0.25$ 	&	 $10.35\pm0.14$	\\
\enddata
\tablecomments{Columns:
(1) Galaxy name.
(2) Filter wavelength (see Table~\ref{table:photo} for the source of the image).
(3) Galactic extinction (in mag) due to dust attenuation in the Milky Way at the reference wavelength listed in column~2, from \citet{Schlafly:Finkbeiner:2011}.
(4) Rest-frame $K$-correction for the wavelength listed in column~2.
(5) Black hole mass listed in \citet{Davis:2017}, compiled from references therein.
(6) Logarithmic spiral-arm pitch angle (absolute value in degrees) from \citet{Davis:2017}.
(7) Bulge apparent magnitude (in AB mag) for the wavelength listed in column~2 (calculated via Equations (\ref{apparent_mag})--(\ref{b_n})).
(8) Fully corrected bulge absolute magnitude (in AB mag) for the wavelength listed in column~2 (calculated via Equation~(\ref{absolute_mag})); \textit{Spitzer} images are additionally corrected for dust emission.
(9) Bulge stellar mass (from the bulge absolute magnitude in column~8, converted to a mass via the appropriate solar absolute magnitude and stellar mass-to-light ratio from Table~\ref{table:photo}).
}
\tablenotetext{a}{SDSS J043703.67+245606.8}
\tablenotetext{b}{From \citet{Okamoto:2013}.} 
\tablenotetext{c}{From \citet{Licquia:Newman:2015}.}
\tablenotetext{d}{From \citet{Savorgnan:2016}.}
\tablenotetext{e}{From \citet{Savorgnan:2016:II}.}
\tablenotetext{f}{Determined from the core-S{\'e}rsic model.}
\tablenotetext{g}{From \citet{Zhao:2018}.}
\end{deluxetable*}

\subsubsection{Conversion to Alternate Mass-to-light Ratios}

In cosmology, ``$h$ is a dimensionless number parameterizing our ignorance'' \citep{Hogg:1999} of the true value of $H_0$. The use of $h$, e.g., $h_{70}=h/0.70=H_0/(70\,{\rm km\,s^{-1}\,Mpc^{-1}})\equiv1$, in published cosmological studies allows readers to easily convert to their own preferred cosmographic parameter. Just as redshift-dependent distances are determined by the assumed value of $H_0$, our values of $M_{\rm *,sph}$ are dependent on our assumption of stellar mass-to-light ratios, $\Upsilon_*$. Whereas modern cosmology has defined the value of $H_0$ to within $\approx 2\%$ and thus rendered $h$ a trifle constant \citep{Croton:2013}, the value of $\Upsilon_*$ is far less certain. Given that individual choices of IMFs, metallicities, ages, and SFHs can alter $\Upsilon_*$ by a factor of two at $3.6\,\micron$, our ignorance of $\Upsilon_*$ hinders the measurement of absolute extragalactic stellar masses and requires calibration to the same system when comparing masses derived from different studies.  Therefore, when using the equations presented here for the prediction of black hole masses, one needs to ensure that one's galaxies' stellar masses are consistent with the $\Upsilon_*$ ratios used here.

To help facilitate this, and with inspiration from $h$, we define a new (passband-dependent) parameter, denoted by the lowercase upsilon $\upsilon$, such that
\begin{equation}
\upsilon_{\rm *,IRAC1,0.453} = \frac{\Upsilon_{\rm *,IRAC1}}{0.453},
\label{upsilon}
\end{equation}
where 0.453 is our adopted stellar mass-to-light ratio\footnote{This value of 0.453 is derived from a \citet{Chabrier:2003} IMF and a \citet{Bruzual:Charlot:2003} SSP having a range of metallicities and exponentially declining SFHs that are, by and large, compatible with $\Upsilon_{\rm *,IRAC1}=0.60\pm0.09$ for a dust-free stellar population (and $\Upsilon_{\rm *,IRAC1}=0.453\pm0.072$ for the total observed light from dusty late-type galaxies) taken from \citet[][see their Figure~1]{Meidt:2014} and \citet{Querejeta:2015}.} and $\Upsilon_{\rm *,IRAC1}$ is an alternative stellar mass-to-light ratio that someone may use to derive their stellar masses. Hereafter, for simplicity, we shall drop the subscripts from $\upsilon$, and we include it as a coefficient in our scaling relations to remind users of this necessary conversion.  This enables one to more readily compare with other $\Upsilon_*$-dependent black hole mass scaling relations and, more generally, to work with studies that have adopted a different $\Upsilon_*$ to define their galaxies' stellar masses.  Throughout this paper, $\upsilon\equiv1$.

\section{Comparison of $3.6\,\micron$ Magnitudes}\label{sec:compare}

\subsection{Savorgnan \& Graham (2016a)}\label{Sav_compare}

\begin{figure}
\includegraphics[clip=true,trim= 0mm 0mm 0mm 0mm,width=\columnwidth]{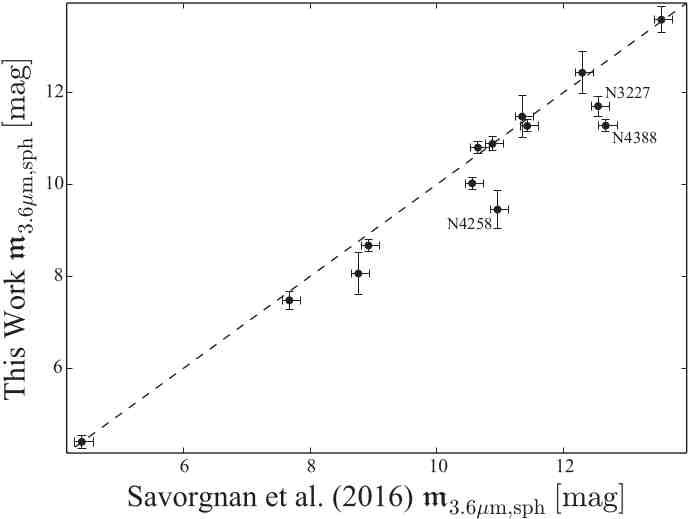}
\caption{Comparison of the spheroid $3.6\,\micron$ apparent magnitudes (with a 1:1 dashed line) for data from 14 spiral galaxies that are in common with \citet{Savorgnan:2016}, yields $\Delta_{\rm rms,\perp}=0.27\,\text{mag}$. Note that the magnitudes from \citet{Savorgnan:2016} have been converted to the AB magnitude system.}
\label{SG16_compare_plot}
\end{figure}

Our study is an expansion of that of \citet{Savorgnan:2016}, who analyzed a sample of 66 galaxies with directly measured SMBH masses (including 18 spiral galaxies when we additionally count NGC~4594). They performed careful photometric decompositions of $3.6\,\micron$ \textit{Spitzer} images. They additionally provided a detailed comparison of their decompositions with those of many earlier studies, pointing out where things had changed. Since our methodology is similar to that of \citet{Savorgnan:2016}, we expect our results to resemble one another for the 18 spiral galaxies in common (see Figure~\ref{SG16_compare_plot}). However, it should be noted that we have not used the \textit{Spitzer} images that were reduced, mosaicked, and masked by \citet{Savorgnan:2016}; thus, our images from the online archives (see Table~\ref{table:photo}) will be slightly different.

Furthermore, \citet{Savorgnan:2016} relied on the \textsc{iraf} task \textsc{ellipse} \citep{ellipse} to fit quasi-elliptical isophotes to their galaxy images. However, \citet{Ciambur:2015} pointed out several inaccuracies with this popular algorithm and created updated software (\textsc{isofit} and \textsc{cmodel}) that we used for our work. In addition, we use the new surface brightness profile decomposition code \textsc{profiler} \citep{Ciambur:2016}, rather than \textsc{profiterol} \citep{Savorgnan:2016}, which better deals with the PSF convolution. Additionally, \citet{Savorgnan:2016} did not model broken exponential profiles. Instead, they elected to truncate the outer surface brightness profile prior to the breaks. They also did not use the inclined disk model that we used but instead used $n<1$ S{\'e}rsic functions to model inclined disks. While \citet{Savorgnan:2016} did, therefore, allow for the influence of nonexponential disks, we have treated these features differently.

Figure~\ref{SG16_compare_plot} shows that our bulge apparent magnitudes match well with \citet{Savorgnan:2016}, with $\Delta_{\rm rms,\perp}=0.27\,\text{mag}$\footnote{In this and subsequent 1:1 comparisons in this work, we analyze the agreement by calculating the raw orthogonal rms scatter ($\Delta_{\rm rms,\perp}$) about the 1:1 line, with $\Delta_{\rm rms,\perp} = \Delta_{\rm rms}/\sqrt{2}.$} for the comparison of spheroid absolute magnitudes. We found that the offset of points from the 1:1 line can generally be explained because our surface brightness profiles extended to greater radii, allowing for a more complete analysis of the disk light, which can influence the determination of the bulge luminosities.

Among the outliers, we find NGC 3227 to have a higher luminosity because we were able to fit the bar in the equivalent axis profile (see Appendix~\ref{NGC3227} and Figure~\ref{NGC3227_plot}). This had the effect of lowering the disk central surface brightness and increasing the disk scale length, which ultimately increased the spheroid luminosity. Again, in NGC 4258, we fit a Ferrers function for the bar (see Appendix~\ref{NGC4258} and Figure~\ref{NGC4258_plot}), whereas \citet{Savorgnan:2016} ignored it (along with the inner $6\farcs1$ of the profile) and fit the profile out to only $R_{\rm maj}\approx 130\arcsec$; we fit the profile out to $R_{\rm maj}=431\arcsec$. For NGC 4388, \citet{Savorgnan:2016}'s exclusion of data in the range $35\arcsec\lesssim R_{\rm maj}\lesssim65\arcsec$ led them to overestimate the contribution of the bar and underestimate the contribution of the bulge (see Appendix~\ref{NGC4388} and Figure~\ref{NGC4388_plot}).

\subsection{S$^4$G}

\begin{figure}
\includegraphics[clip=true,trim= 0mm 0mm 0mm 0mm,width=\columnwidth]{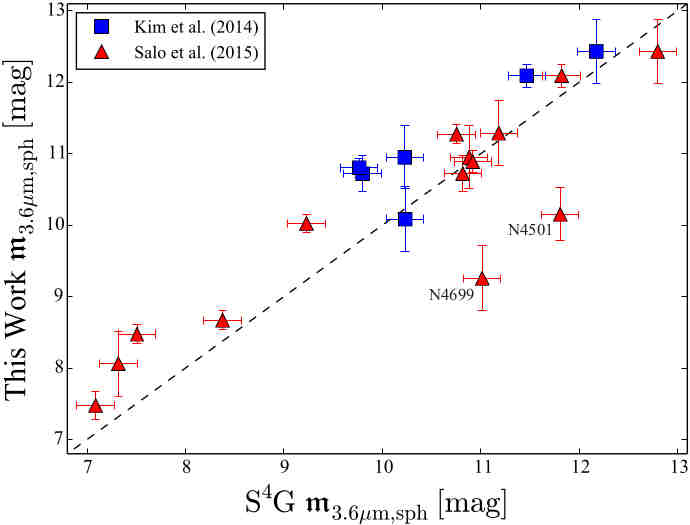}
\caption{Comparison of the spheroid $3.6\,\micron$ apparent magnitudes (with 1:1 dashed line) for data from 14 spiral galaxies that are in common with the S$^4$G sample from \citet{Salo:2015} plus six from \citet{Kim:2014}. The agreement is such that $\Delta_{\rm rms,\perp}=0.20\,\text{mag}$ \citep{Kim:2014} and $\Delta_{\rm rms,\perp}=0.34\,\text{mag}$ \citep{Salo:2015}. Note that the S$^4$G does not provide error estimates, so we have added error bars equivalent to our median error.}
\label{S4G_compare}
\end{figure}

We have also compared our $3.6\,\micron$ bulge apparent magnitudes to those from the S$^4$G \citep{Salo:2015}, which also provided multicomponent photometric decompositions that yield spheroid components for 14 galaxies in common with our sample (see Figure~\ref{S4G_compare}). It is beneficial to compare these galaxies because, in both our work and theirs, identical $3.6\,\micron$ images were analyzed. As part of the S$^4$G pipeline, \citet{Salo:2015} used \textsc{galfit} \citep{Peng:2002,Peng:2010} to perform automated 2D surface brightness decompositions for 2352 galaxies. They provided one-component S{\'e}rsic fits and two-component S{\'e}rsic bulge + exponential disk fits. When determined to be necessary, they initiated human-supervised multicomponent decompositions with additional components, such as a central point source and bar.

We also compare our work to six common galaxies from \citet{Kim:2014}, who similarly modeled S$^4$G data, except that they used an alternate decomposition software \textsc{budda} \citep{BUDDA:2004,Gadotti:2008,Gadotti:2009}. Their decompositions differ slightly from those of \citet{Salo:2015} because they take into account disk breaks using a broken exponential model. This additional consideration brings their spheroid analyses into closer agreement with ours than the analyses of \citet{Salo:2015} because treating all disks as purely exponential will introduce errors that ultimately affect the determination of the spheroid luminosities.

For the most part, we find a good agreement with the decompositions from the S$^4$G analyses. Notable outliers in Figure~\ref{S4G_compare} include NGC~4501 and NGC~4699. In the case of NGC~4501, because the S$^4$G analysis did not model the prominent spiral arms, which increases the brightness of the model (exponential) disk (see Appendix~\ref{NGC4501} and Figure~\ref{NGC4501_plot}), the bulge mass was underestimated by S$^4$G. The S$^4$G decomposition by \citet{Salo:2015} for NGC~4699 included an additional inner exponential disk component (for which we do not see evidence; see Appendix~\ref{NGC4699} and Figure~\ref{NGC4699_plot}), which came at the expense of the bulge light.

\subsection{Other}

Several other studies have presented multicomponent decompositions for some of our galaxies' surface brightness profiles.  \citet{Savorgnan:2016} have closely scrutinized many of these, providing qualitative and quantitative comparisons (see their Figure~11) with the decompositions from \citet{Graham:Driver:2007}, \citet{Laurikainen:2010}, \citet{Sani:2011}, \citet{Beifiori:2012}, \citet{Vika:2012}, \citet{Rusli:2013}, and \citet{Lasker:2014a}. We do not repeat these detailed comparisons here; instead, we have transitively compared to these works through our comparison with \citet{Savorgnan:2016}.

\citet{Lasker:2016} appeared after \citet{Savorgnan:2016} and provided detailed, multicomponent  decompositions for eight spiral galaxies\footnote{Plus one lenticular galaxy, NGC~1194.} that are in our sample.  Their bulge magnitudes are derived from higher spatial resolution but dust-affected $H$-band images. For all but three of these eight galaxies, the spheroid masses reported by \citet{Lasker:2016} are contained within the $1\,\sigma$ uncertainty assigned to our spheroid masses, and for the remaining three galaxies, the difference only exceeds our $1\,\sigma$ uncertainty by $0.06$--$0.08\,{\rm dex}$. Moreover, the two mass estimates for all eight bulges have overlapping error bars. We also find a good overall agreement when comparing our  decompositions with theirs, although some disagreement arises in our use of broken or inclined exponential models for the large-scale disk, whereas \citet{Lasker:2016} only used an exponential model. This resulted in them sometimes using an ``envelope'' component to capture what we consider to be the outer disk.
  
%  For example, adding an "envelope" to the Sb galaxy NGC~4388, \citet{Lasker:2016} report a bulge-to-disk size ratio, $R_{\rm e}/h$, equal to 0.7, notably higher than the expected value of $0.21^{+0.15}_{-.07}$ \citep{Graham:Worley:2008}. 

\section{Regression Analyses}\label{AR}

The focus of this paper is to determine the slope of the $M_{\rm BH}$--$M_{\rm *,sph}$ relation for spiral galaxies, enabling greater clarity at low masses in the $M_{\rm BH}$--$M_{\rm *,sph}$ diagram. For this task, we have employed a robust (i.e., stable against outlying data) Bayesian analysis \citep[e.g.,][or see a review by \citealt{Andreon:2013}]{Barnes:2003,Wyithe:2006,Kelly:2007,Andreon:2010,Shetty:2013,Souza:2015,HyperFit,Sereno:2016,Pihajoki:2017} with a \emph{symmetric} treatment of the data and allowing for errors in both coordinates, as well as a \emph{conditional} analysis that optimizes the prediction of black hole mass given an input stellar bulge mass (see Appendix~\ref{App1} for additional details). As pointed out in \citet{Novak:2006}, since there is no natural division into ``dependent'' and ``independent'' variables in the black hole mass scaling relations when constructed for comparison with theory, we represent our scaling relations with a symmetric treatment of the variables.\footnote{The ordinary least-squares linear regression of $X$ on $Y$, denoted here by ($X|Y$), minimizes the residual offset, in the direction parallel to the $X$-coordinate axis of the data, about the fitted line (typically resulting in a steeper slope). The ordinary least-squares linear regression of $Y$ on $X$, denoted here by ($Y|X$), minimizes the residual offset in the direction parallel to the $Y$-coordinate axis (typically resulting in a shallower slope). The \textit{bisector} linear regression bisects the angle between the ordinary least-squares ($X|Y$) and the ordinary least squares ($Y|X$) fits.}

The Bayesian analysis was used to check on our primary scaling relation, the $M_{\rm BH}$--$M_{\rm *,sph}$ relation, which was also derived here using two linear regressions that are more commonly employed for black hole scaling relations. First, we used the \textit{bisector} line from the \textsc{bces} (Bivariate Correlated Errors and intrinsic Scatter)\footnote{The \textsc{bces} routine \citep{BCES} was run via the \textsc{python} module written by Rodrigo Nemmen \citep{Nemmen:2012}, which is available at \url{https://github.com/rsnemmen/BCES}.} regression method \citep{BCES}, which takes into account measurement error in both coordinates and allows for intrinsic scatter in the data. Second, we used the \textsc{mpfitexy} routine \citep{Press:1992,Tremaine:2002,Bedregal:2006,Novak:2006,Markwardt:2009,MPFIT,Williams:2010}, which also takes into account measurement error in both coordinates and intrinsic scatter. The \textit{bisector} line from the \textsc{mpfitexy} routine is obtained, for example, by using ${\rm X} = \log{M_{\rm *,sph}}$ and ${\rm Y} = \log{M_{\rm BH}}$, and then repeating the regression using ${\rm X} = \log{M_{\rm BH}}$ and ${\rm Y} = \log{M_{\rm *,sph}}$, and finding the line that bisects these two lines.

We derive the spiral galaxy $M_{\rm BH}$--$M_{\rm *,sph}$ relation, excluding the ambiguously classified galaxy Cygnus~A and the three (potentially) bulgeless galaxies NGC~4395 \citep{Sandage:1981,Brok:2015}, NGC~2748 \citep{Salo:2015},\footnote{We note that, like us, \citet{Lasker:2014a,Lasker:2014} fit a S{\'e}rsic component to NGC 2478. Both their and our S{\'e}rsic components account for approximately one-third of the total galaxy luminosity. However, they considered it to be a \textit{bona fide} bulge, whereas \citet{Salo:2015} did not.} and NGC~6926 (see Appendix~\ref{NGC6926} and Figure~\ref{NGC6926_plot}). Moreover, NGC~2748 has a discontinuous light profile (see Figure~\ref{NGC2748_plot}). Its pitch-angle measurement is also questionable due to its inclination and irregular spiral shapes. As noted in Section~\ref{DM}, this reduced our sample from 44 to 40 galaxies. For consistency between scaling relations, we analyze the same subsample of 40 galaxies for additional relations presented in this work.

\vspace{3mm}

\subsection{Relations with Black Hole Mass ($M_{\rm BH}$)}

\subsubsection{The $M_{\rm BH}$--$M_{\rm *,sph}$ Relation}\label{sec:M_BH-M_sph}

\begin{figure}
\includegraphics[clip=true,trim= 1mm 1mm 0mm 0mm,width=\columnwidth]{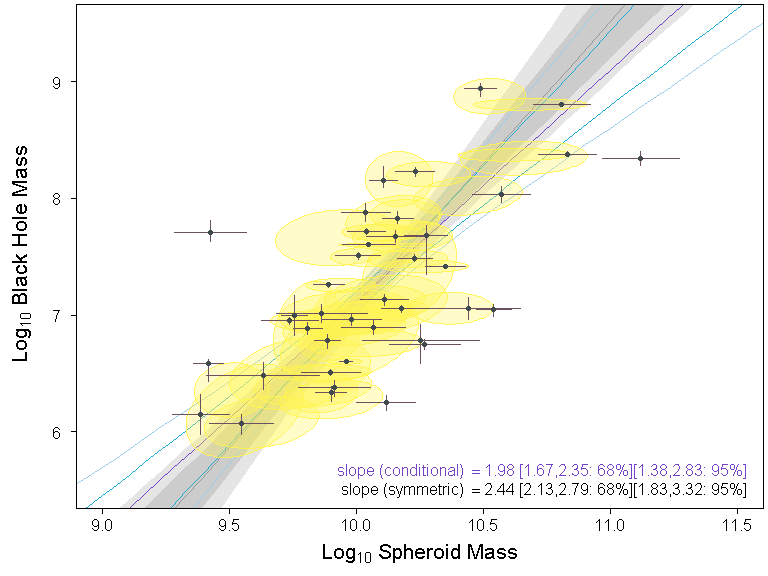}
\caption{The \textit{symmetric} line of best fit (see Equation~(\ref{Ewan_M_BH-M_sph_eqn})) is presented as its pointwise median with $\pm68\%$ and $\pm95\%$ (shaded) intervals, while the $\pm68\%$ posterior estimates of the true stellar spheroid and black hole mass of each galaxy are highlighted in yellow. The \textit{conditional} (purple) line of best fit is additionally supplied with similar (cyan) error intervals. Masses are in units of solar masses.}
\label{Ewan_M_BH-M_sph_plot}
\end{figure}

Our $\left(\log {M_{\rm *,sph}},\,\log{M_{\rm BH}}\right)$ data set has a Pearson correlation coefficient $r = 0.66$, with a $p$-value probability of $4.49\times10^{-6}$ that the null hypothesis is true. The Spearman rank-order correlation coefficient $r_s=0.62$, with a $p_s$-value probability of $2.38\times10^{-5}$ that the null hypothesis is true. Our \textit{symmetric} Bayesian analysis, detailed in Appendix \ref{App1}, yields the following equation:
\begin{IEEEeqnarray}{rCl}
\log\left( \frac{M_{\rm BH}}{M_{\sun}}\right ) & = & \left(2.44_{-0.31}^{+0.35}\right)\log\left[ \frac{M_{\rm *,sph}}{\upsilon(1.15\times10^{10}\,M_{\sun})}\right] \nonumber \\ 
&& +\> (7.24\pm0.12).
\label{Ewan_M_BH-M_sph_eqn}
\end{IEEEeqnarray}
The data have an rms scatter about this line in the $\log{M_{\rm BH}}$ direction of $\Delta_{\rm rms} = 0.70\,\text{dex}$, with an associated intrinsic scatter of $\epsilon=0.51\,\text{dex}$.\footnote{Intrinsic scatter naturally increases with increasing slope; this complicates the simple comparison of intrinsic scatter across scaling relations with different slopes. In contrast, correlation coefficients (i.e., Pearson and Spearman) do not account for measurement error. However, they are independent of slope, unlike intrinsic scatter, and are arguably a more informative quantity.} This relationship is illustrated in Figure~\ref{Ewan_M_BH-M_sph_plot}. Equation~(\ref{Ewan_M_BH-M_sph_eqn}) and its correlation parameters are listed at the top of Table~\ref{table:bent}, along with all subsequent linear regressions of interest, i.e., \textsc{bces} and \textsc{mpfitexy}. We note that this seemingly high level of scatter for the \textit{symmetric} linear regression is significantly diminished when the \textit{conditional} linear regression is employed, resulting in $\Delta_{\rm rms} = 0.60\,\text{dex}$ and $\epsilon=0.47\,\text{dex}$.

\begin{deluxetable*}{lcllccccrrrr}
\tabletypesize{\normalsize}
\tablecolumns{11}
\tablecaption{Linear Regressions\label{table:bent}}
\tablehead{
\colhead{Regression} & \colhead{Minimization} & \colhead{$\alpha$} & \colhead{$\beta$} & \colhead{$\epsilon$} & \colhead{$\Delta_{\rm rms}$} & \colhead{} & \colhead{$r$} & \colhead{$\log{p}$} & \colhead{$r_s$} & \colhead{$\log{p_s}$} \\
\colhead{} & \colhead{} & \colhead{} & \colhead{(dex)} & \colhead{(dex)} & \colhead{(dex)} & \colhead{} & \colhead{} & \colhead{(dex)} & \colhead{} & \colhead{(dex)} \\
\colhead{(1)} & \colhead{(2)} & \colhead{(3)} & \colhead{(4)} & \colhead{(5)} & \colhead{(6)} & \colhead{} & \colhead{(7)} & \colhead{(8)} & \colhead{(9)} & \colhead{(10)}
}
\startdata
\multicolumn{11}{c}{\textbf{40 Late-Type Galaxies with S{\'e}rsic Bulges}} \\
\hline
\multicolumn{11}{c}{\textbf{The} $\bm{M_{\rm BH}}$\textbf{--}$\bm{M_{\rm *,sph}}$ \textbf{Relation:} $\bm{\log(M_{\rm BH}/M_{\sun})=\alpha\log(M_{\rm *,sph}/[\upsilon(1.15\times10^{10}\,M_{\sun})])+\beta$}} \\
\hline
Bayesian & \textit{Symmetric} & $2.44_{-0.31}^{+0.35}$ & $7.24\pm0.12$ & 0.51 & 0.70 & \multirow{8}{*}{$\left\}\begin{tabular}{@{}l@{}} \\ \\ \\ \\ \\ \\ \\ \\ \end{tabular}\right.$} & \multirow{8}{*}{$0.66$} & \multirow{8}{*}{$-5.35$} & \multirow{8}{*}{$0.62$} & \multirow{8}{*}{$-4.62$} \\
Bayesian & $M_{\rm BH}$ & $1.98_{-0.31}^{+0.37}$ & $7.24\pm0.12$ & 0.47 & 0.60 & & & & \\
\textsc{bces} & \textit{Symmetric} & $2.17\pm0.32$ & $7.21\pm0.10$ & $0.48$ & $0.64$ & & & & \\
\textsc{bces} & $M_{\rm BH}$ & $1.69\pm0.35$ & $7.22\pm0.09$ & $0.47$ & $0.56$ & & & & \\
\textsc{bces} & $M_{\rm *,sph}$ & $2.90\pm0.55$ & $7.19\pm0.13$ & $0.59$ & $0.82$ & & & & \\
\textsc{mpfitexy} & \textit{Symmetric} & $2.23\pm0.36$ & $7.24\pm0.10$ & $0.49$ & $0.65$ & & & & \\
\textsc{mpfitexy} & $M_{\rm BH}$ & $1.74\pm0.27$ & $7.25\pm0.09$ & $0.46$ & $0.57$ & & & & \\
\textsc{mpfitexy} & $M_{\rm *,sph}$ & $3.00\pm0.53$ & $7.24\pm0.13$ & $0.61$ & $0.85$ & & & & \\
\hline
\multicolumn{11}{c}{\textbf{The} $\bm{M_{\rm BH}}$\textbf{--}$\bm{n_{\rm sph, maj}}$ \textbf{Relation:} $\bm{\log(M_{\rm BH}/M_{\sun})=\alpha\log(n_{\rm sph, maj}/2.20)+\beta}$} \\
\hline
\textsc{bces} & \textit{Symmetric} & $2.69\pm0.33$ & $7.43\pm0.12$ & $0.66$ & $0.70$ & \multirow{6}{*}{$\left\}\begin{tabular}{@{}l@{}} \\ \\ \\ \\ \\ \\ \end{tabular}\right.$} & \multirow{6}{*}{$0.46$} & \multirow{6}{*}{$-2.58$} & \multirow{6}{*}{$0.39$} & \multirow{6}{*}{$-1.88$} \\
\textsc{bces} & $M_{\rm BH}$ & $1.60\pm0.38$ & $7.36\pm0.11$ & $0.62$ & $0.64$ & & & & \\
\textsc{bces} & $n_{\rm sph, maj}$ & $6.44\pm2.24$ & $7.67\pm0.24$ & $1.23$ & $1.35$ & & & & \\
\textsc{mpfitexy} & \textit{Symmetric} & $2.76\pm0.70$ & $7.45\pm0.13$ & $0.66$ & $0.71$ & & & \\
\textsc{mpfitexy} & $M_{\rm BH}$ & $1.67\pm0.43$ & $7.38\pm0.11$ & $0.62$ & $0.64$ & & & & \\
\textsc{mpfitexy} & $n_{\rm sph, maj}$ & $6.33\pm1.91$ & $7.66\pm0.24$ & $1.21$ & $1.33$ & & & & \\
\hline
\multicolumn{11}{c}{\textbf{The} $\bm{M_{\rm *,sph}}$\textbf{--}$\bm{\phi}$ \textbf{Relation:} $\bm{\log(M_{\rm *,sph}/M_{\sun})=\alpha\left[|\phi|-13\fdg4 \right ]{\rm deg}^{-1}+\beta+\log(\upsilon)}$} \\
\hline
\textsc{bces} & \textit{Symmetric} & $-0.078\pm0.013$ & $10.11\pm0.05$ & $0.22$ & $0.32$ & \multirow{6}{*}{$\left\}\begin{tabular}{@{}l@{}} \\ \\ \\ \\ \\ \\ \end{tabular}\right.$} & \multirow{6}{*}{$-0.63$} & \multirow{6}{*}{$-4.89$} & \multirow{6}{*}{$-0.56$} & \multirow{6}{*}{$-3.78$} \\
\textsc{bces} & $M_{\rm *,sph}$ & $-0.063\pm0.012$ & $10.11\pm0.05$ & $0.20$ & $0.30$ & & & & \\
\textsc{bces} & $|\phi|$ & $-0.093\pm0.020$ & $10.11\pm0.05$ & $0.25$ & $0.35$ & & & & \\
\textsc{mpfitexy} & \textit{Symmetric} & $-0.079\pm0.013$ & $10.06\pm0.05$ & $0.21$ & $0.32$ & & & \\
\textsc{mpfitexy} & $M_{\rm *,sph}$ & $-0.060\pm0.011$ & $10.07\pm0.05$ & $0.20$ & $0.30$ & & & & \\
\textsc{mpfitexy} & $|\phi|$ & $-0.097\pm0.015$ & $10.05\pm0.06$ & $0.25$ & $0.37$ & & & & \\
\hline
\multicolumn{11}{c}{\textbf{21\tablenotemark{a} Early-Type Galaxies with Core-S{\'e}rsic Bulges}} \\
\hline
\multicolumn{11}{c}{\textbf{The} $\bm{M_{\rm BH}}$\textbf{--}$\bm{M_{\rm *,sph}}$ \textbf{Relation:} $\bm{\log(M_{\rm BH}/M_{\sun})=\alpha\log(M_{\rm *,sph}/[\upsilon(2.10\times10^{11}\,M_{\sun})])+\beta}$} \\
\hline
\textsc{bces} & \textit{Symmetric} & $1.28\pm0.26$ & $9.23\pm0.10$ & $0.43$ & $0.46$ & \multirow{6}{*}{$\left\}\begin{tabular}{@{}l@{}} \\ \\ \\ \\ \\ \\ \end{tabular}\right.$} & \multirow{6}{*}{$0.61$} & \multirow{6}{*}{$-2.45$} & \multirow{6}{*}{$0.56$} & \multirow{6}{*}{$-2.10$} \\
\textsc{bces} & $M_{\rm BH}$ & $0.88\pm0.32$ & $9.18\pm0.11$ & $0.38$ & $0.41$ & & & & \\
\textsc{bces} & $M_{\rm *,sph}$ & $1.94\pm0.41$ & $9.31\pm0.11$ & $0.59$ & $0.63$ & & & & \\
\textsc{mpfitexy} & \textit{Symmetric} & $1.20\pm0.25$ & $9.21\pm0.09$ & $0.41$ & $0.44$ & & & \\
\textsc{mpfitexy} & $M_{\rm BH}$ & $0.74\pm0.22$ & $9.16\pm0.09$ & $0.37$ & $0.41$ & & & & \\
\textsc{mpfitexy} & $M_{\rm *,sph}$ & $2.04\pm0.64$ & $9.31\pm0.17$ & $0.62$ & $0.66$ & & & & \\
\enddata
\tablecomments{Late-type galaxies are from this work, and early-type galaxies are from \citet{Savorgnan:2016:II}. The calculation of the total rms scatter ($\Delta_{\rm rms}$), the correlation coefficients ($r$ and $r_s$), and their associated probabilities do not take into account the uncertainties on the datapoints.
Columns:
(1) Regression software used.
(2) Variable that had its offsets from the regression line minimized.
(3) Slope.
(4) Intercept.
(5) Intrinsic scatter in the $Y$-coordinate direction \citep[][their Equation~(1)]{Graham:Driver:2007}.
(6) The rms scatter in the $Y$-coordinate direction.
(7) Pearson correlation coefficient.
(8) Logarithm of the Pearson correlation probability value.
(9) Spearman rank-order correlation coefficient.
(10) Logarithm of the Spearman rank-order correlation probability value.
}
\tablenotetext{a}{This number was 22 in \citet{Savorgnan:2016:II} because they considered NGC 4594 to have a core-S{\'e}rsic bulge (and not to be a spiral galaxy).}
\end{deluxetable*}

\begin{figure}
\includegraphics[clip=true,trim= 0mm 0mm 0mm 0mm,width=\columnwidth]{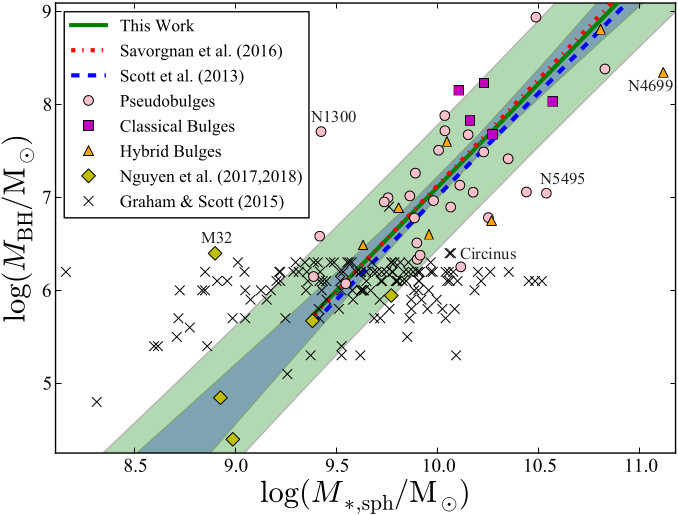}
\caption{Similar to Figure~\ref{Ewan_M_BH-M_sph_plot}, except here we present our black hole vs.\ spheroid stellar mass data combined with the data for an additional five low-mass, early-type galaxies \citep{Nguyen:2017a,Nguyen:2017} and 139 low-mass AGNs \citep{Jiang:2011,Graham:Scott:2015}. Additionally, our data have been labeled as possessing a pseudobulge, classical bulge, or both (classifications from \citealt{Davis:2017}, and references therein). Instead of the Bayesian regression shown in Figure~\ref{Ewan_M_BH-M_sph_plot}, here we plot our \textsc{mpfitexy} \textit{bisector} regression (not fitting the additional data), represented by the solid green line. The dark green band shows the $\pm1$\,$\sigma$ uncertainty on the slope and the intercept from the regression, while the light green band delineates the $\pm1$\,$\sigma$ scatter of the data about the regression line. We also plot the comparable best-fit linear regressions from \citet{Scott:2013} and \citet{Savorgnan:2016:II}. Note that all data and regressions from other works have been adjusted to conform with the stellar mass-to-light ratios used in this work. The 139 AGN black hole masses $\lesssim10^6\,M_{\sun}$ were derived from reverberation-mapping techniques with a greater level of uncertainty than directly measured masses. Error bars have been omitted for clarity.}
\label{spheroid_plot2}
\end{figure}

The Bayesian analyses agree with the respective \textsc{bces} and \textsc{mpfitexy} analyses, with slopes that are slightly steeper than the latter regressions. Figure~\ref{spheroid_plot2} shows the \textsc{mpfitexy} \textit{bisector} regression and provides additional data (not plotted in Figure~\ref{Ewan_M_BH-M_sph_plot}) that are plotted here but not included in the regression analysis. These additional data include five low-mass early-type galaxies from \citet{Nguyen:2017a,Nguyen:2017} and 139 low-mass AGNs \citep{Jiang:2011,Graham:Scott:2015}. Our best-fit trend line cuts through the cloud of points from the low-mass AGN sample, and its extrapolation coincides with the least massive black hole (in NGC 205) from \citet{Nguyen:2017}. Here our galaxies are additionally labeled as possessing either a pseudobulge, classical bulge, or a hybrid mix of both \citep[see][and references therein]{Davis:2017}; no obvious trends appear amongst these subsamples. Also in Figure~\ref{spheroid_plot2}, we compare our \textsc{mpfitexy} \textit{bisector} regression with the comparable linear regression from the spiral galaxy sample in \citet{Savorgnan:2016:II} and the S{\'e}rsic galaxy sample in \citet{Scott:2013}. Their slopes both match well with ours and are consistent within our $\pm\Delta_{\rm rms}$ scatter.

We compare our $M_{\rm BH}$--$M_{\rm *,sph}$ relations with \citet{Savorgnan:2016:II}, beginning with the \textsc{bces} \textit{bisector} results. We find a slope of $2.17\pm0.32$ compared with their slope of $3.00\pm1.30$. This is consistent at the level of $0.51\,\sigma$.\footnote{Here $n\,\sigma=|\mu_1-\mu_2|/(\sigma_1+\sigma_2)$, where observation 1 has a mean ($\mu_1$) with standard deviation ($\sigma_1$), observation 2 has a mean ($\mu_2$) with standard deviation ($\sigma_2$), and the left-hand side of the equation ($n\,\sigma$) represents the joint number of standard deviations at which both observations agree.} Next, we compare our \textsc{mpfitexy} \textit{bisector} slope ($2.23\pm0.36$) to theirs ($2.28^{+1.67}_{-1.01}$), which agrees at the level of $0.04\,\sigma$. Finally, we compare our \textit{symmetric} Bayesian slope ($2.44_{-0.31}^{+0.35}$) with their Bayesian \textsc{linmix\_err} \citep{Kelly:2007} \textit{bisector} slope ($1.94\pm1.24$), which agrees with ours at the level of $0.32\,\sigma$. Given the previous wide range of slopes and considerable uncertainty on those slopes, which were derived using a sample of only 17 spiral galaxies, our sample of 40 spiral galaxies has allowed us to finally narrow down the slope to a more precise level.

\subsubsection{The $M_{\rm BH}$--$n_{\rm sph, maj}$ Relation}\label{sec:bh-sersic}

\begin{figure}
\includegraphics[clip=true,trim= 0mm 0mm 0mm 0mm,width=\columnwidth]{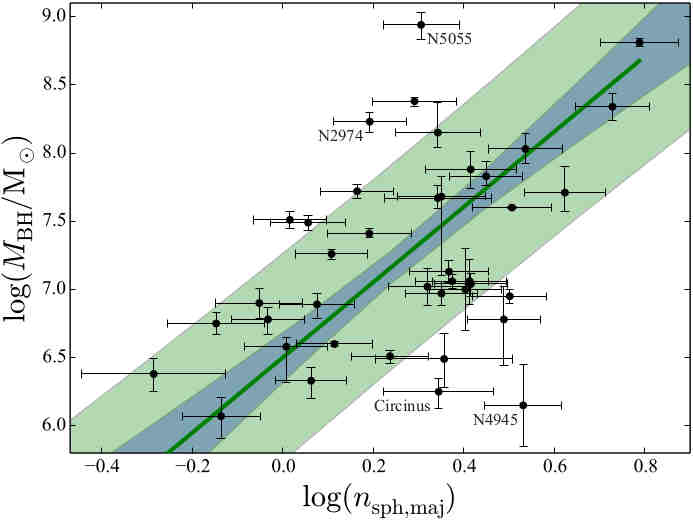}
\caption{Black hole mass versus the spheroid major axis S{\'e}rsic index for our sample of 40 spiral galaxies. The \textsc{mpfitexy} \textit{bisector} linear regression (Equation~(\ref{bh-sersic_eqn})) is presented.}
\label{sersic_plot}
\end{figure}

The $M_{\rm BH}$--$n_{\rm sph, maj}$ relation \citep{Graham:2001,Graham:2003a,Graham:Driver:2007}, where $n_{\rm sph, maj}$ is the major axis S{\'e}rsic index of the spheroidal component, has been shown to be a reliable predictor of SMBH mass, with a level of scatter similar to that of the $M_{\rm BH}$--$\sigma_{*}$ relation when using a sample dominated by massive spheroids. Here we explore, from a sample of only spiral galaxies, how well this relation holds up at lower masses. The $\left(\log  n_{\rm sph, maj},\, \log M_{\rm BH}\right)$ data set is defined by $r = 0.46$, $p=2.61\times10^{-3}$, $r_s=0.39$, and $p_s=1.32\times10^{-2}$. We find from the \textsc{mpfitexy} \textit{bisector} linear regression that
\begin{IEEEeqnarray}{rCl}
\log\left( \frac{M_{\rm BH}}{M_{\sun}}\right ) & = & \left(2.76\pm0.70\right)\log\left(\frac{n_{\rm sph, maj}}{2.20}\right) \nonumber \\
&& +\> (7.45\pm0.13),
\label{bh-sersic_eqn}
\end{IEEEeqnarray}
with $\Delta_{\rm rms} = 0.71\,\text{dex}$ and $\epsilon=0.66\,\text{dex}$ (see Figure~\ref{sersic_plot}).\footnote{The formal errors from column~8 of Table~\ref{table:Sersic} are increased by adding 20\% in quadrature in an attempt to better represent the unknown influence of component degeneracy. Alternatively, if we do not add 20\% in quadrature, our \textit{symmetric} \textsc{bces} slope is $2.51\pm0.31$. Instead, if we add a rather large 40\% in quadrature to the formal \textsc{profiler} errors, our \textit{symmetric} \textsc{bces} slope becomes $3.44\pm0.50$.} Amongst the outliers, we find that NGC 5055 continues to be deviant, as it was in the $M_{\rm BH}$--$\sigma_*$ relation from \citet{Davis:2017}, with a possible overmassive black hole. Performing a regression that minimizes the scatter in the $\log M_{\rm BH}$ direction, we find $\Delta_{\rm rms} = 0.64$ (see Table~\ref{table:bent}).

\subsection{The $M_{\rm *,sph}$--$\phi$ Relation}

\begin{figure}
\centering
\includegraphics[clip=true,trim= 0mm 0mm 0mm 0mm,width=\columnwidth]{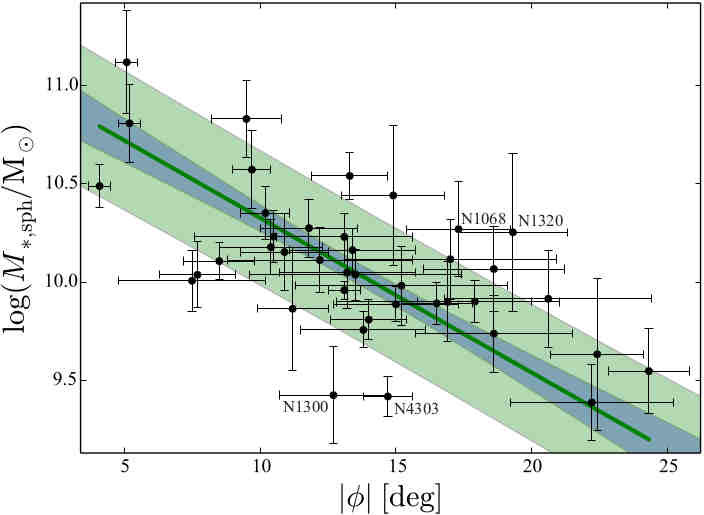}
\caption{Logarithmic spiral-arm pitch angle absolute value versus the spheroid stellar mass; Equation~(\ref{phi-M_sph_eqn}) is presented.}
\label{phi-M_sph_plot}
\end{figure}

Since logarithmic spiral-arm pitch angle ($\phi$) has been shown to correlate well with black hole mass \citep{Seigar:2008,Berrier:2013,Davis:2017}, and we are showing in this work that black hole mass correlates with the spheroid stellar mass, it is prudent to check on the $M_{\rm *,sph}$--$\phi$ relation. We find from the \textsc{mpfitexy} \textit{bisector} linear regression
\begin{IEEEeqnarray}{rCl}
\log\left( \frac{M_{\rm *,sph}}{M_{\sun}}\right ) & = & -\left(0.079\pm0.013\right)\left[|\phi|-13\fdg4\right]{\rm deg}^{-1} \nonumber \\
&& +\> \left(10.06\pm0.05\right)+\log(\upsilon),
\label{phi-M_sph_eqn}
\end{IEEEeqnarray}
with $\Delta_{\rm rms} = 0.32\,\text{dex}$ and $\epsilon=0.21\,\text{dex}$ in the $\log{M_{\rm *,sph}}$ direction; the data set is described by $r = -0.63$, $p=1.28\times10^{-5}$, $r_s=-0.56$, and $p_s=1.66\times10^{-4}$ (see Figure~\ref{phi-M_sph_plot}). We note that NGC~1300 has been an outlier ($\geq2$ $\Delta_{\rm rms}$) in most of the relations explored thus far, and so it may be worth revisiting its black hole mass, although this is beyond the scope of this study. However, NGC~1300 does stand out as having perhaps the strongest, most well-defined bar in our sample.

\section{Discussion}\label{DI}

\subsection{The $M_{\rm BH}$--$M_{\rm *,sph}$ Relation}

With three different linear regression analysis techniques, we find consistent results for the $M_{\rm BH}$--$M_{\rm *,sph}$ relation, indicating that $M_{\rm BH}\propto M_{\rm *,sph}^{\alpha}$ with $\alpha>2$. The agreement in slope between the \textit{symmetric} Bayesian analysis ($2.44_{-0.31}^{+0.35}$) and the \textsc{bces} \textit{bisector} regression ($2.17\pm0.32$) is at the level of $0.43\,\sigma$, and at the level of $0.32\,\sigma$ when compared with the \textsc{mpfitexy} \textit{bisector} slope $(2.23\pm0.36)$. Additionally, we find that the agreement in slope between the \textit{conditional} Bayesian analysis ($1.98_{-0.31}^{+0.37}$) and the \textsc{bces} ($Y|X$) regression ($1.69\pm0.35$) is at the level of $0.43\,\sigma$, and at the level of $0.42\,\sigma$ when compared with the \textsc{mpfitexy} ($Y|X$) slope $(1.74\pm0.27)$. These high levels of agreement across three independent regression analyses instill confidence that the best fit for the \textit{symmetric} slope of the $M_{\rm BH}$--$M_{\rm *,sph}$ relation lies in the range of $\approx2.3\pm0.2$. We note that \citet{Savorgnan:2016:II} had also measured a similar slope, but with over 50\% uncertainty.

We compare our result with our past results that have attempted to quantify the slope of the $M_{\rm BH}$--$M_{\rm *,sph}$ relation at low masses, by which we mean the departure from a near-linear relation in $\log$--$\log$ space. We find good agreement between our slope for spiral galaxies (which have S{\'e}rsic bulges) and slopes in the literature for any type of galaxy with a S{\'e}rsic bulge.  In particular, \citet{Scott:2013} revealed an apparent dichotomy in the $M_{\rm BH}$--$M_{\rm *,sph}$ diagram between S{\'e}rsic and core-S{\'e}rsic galaxies, with the former having a significantly steeper slope (of $2.22\pm0.58$) obtained using 23 spiral plus 26 early-type galaxies. \citet{Savorgnan:2016:II} had also reported a  similarly steep but notably less certain slope of 2--3 using the (S{\'e}rsic) bulges of 17 spiral galaxies.

\citet{Graham:Scott:2015} showed that the inclusion of galaxies with AGNs, having black hole masses between $2\times10^5\,M_{\odot}$ and $2\times10^6\,M_{\odot}$, followed the near-quadratic $M_{\rm BH}$--$M_{\rm *,sph}$ relation down to $M_{\rm BH}\approx10^5\,M_\sun$, the lower limit of the black hole masses derived from reverberation mapping. In Figure~\ref{spheroid_plot2}, we repeat \citet{Graham:Scott:2015}'s comparison to 139 AGNs from \citet{Jiang:2011}. We also find that an extrapolation of our $M_{\rm BH}$--$M_{\rm *,sph}$ relation coincides nicely with the low-mass data down to $M_{\rm BH}\approx10^5\,M_\sun$.\footnote{Many works have observed this cloud or ``plume'' of data at the low-mass end of the $M_{\rm BH}$--$\sigma_*$ relation \citep{Greene:2006,Jiang:2011,Mezcua:2017,Navarro:2018} and attribute it to an asymptotic flattening of the $M_{\rm BH}$--$\sigma_*$ relation at low masses due to a direct collapse-formation scenario of $\sim10^5\,M_\sun$ black hole seeds \citep{Volonteri:2009,Volonteri:2010,Wassenhove:2010}.} While it may appear that any relation would intersect the cloud of data pertaining to the AGN sample, three points should probably be borne in mind: (i) larger measurement errors on the spheroid masses of the AGN sample will broaden the distribution at a given black hole mass but not alter the median of the distribution, (ii) our relation passes through the center of the cloud, and (iii) the larger AGN sample (139 vs.\ 40) will naturally populate the $2\,\sigma$ and $3\,\sigma$ wings of the distribution more fully.

Recently, \citet{Nguyen:2017a,Nguyen:2017} studied five nearby low-mass early-type galaxies (M32, NGC 205, NGC 404, NGC 5102, and NGC 5206). They also found that these galaxies (except for M32) match (see their Figure~10, left panel) the low-mass end of the $M_{\rm BH}$--$M_{\rm *,sph}$ relation of \citet{Scott:2013}. We plot these five galaxies in Figure~\ref{spheroid_plot2}, finding similar agreement down to the mass $M_{\rm BH}\approx2.5\times10^4\,M_\sun$ in NGC 205. The rare ``compact elliptical'' galaxy M32 appears as a notable outlier \citep[see][and references therein]{Graham:Spitler:2009}, perhaps due to stripping of its outer stellar layers by M31.

The implications of this near-quadratic scaling relation for the bulges of spiral galaxies are deep and wide-ranging. Indeed, two decades of research, simulations, theory, and publications have mistakenly thought that the black hole mass is linearly proportional to the host spheroid mass. However, this linear coupling is only the case in massive early-type galaxies that had previously dominated our data sets.

Here we simply list some of the areas of astronomical research that are affected. These include (i) black hole mass predictions, (ii) estimates of the local BHMF and mass density based on local spheroid luminosity functions, and (iii) evolutionary studies of the $M_{\rm BH}/M_{\rm *,sph}$ mass ratio over different cosmic epochs. From our \textit{conditional} Bayesian $M_{\rm BH}$--$M_{\rm *,sph}$ relation, the ratio $M_{\rm BH}/(\upsilon M_{\rm *,sph})$ equals $0.043\%_{-0.023\%}^{+0.049\%}$ at $\log{M_{\rm BH}}=9.5$ and $0.71\%_{-0.45\%}^{+1.25\%}$ at $\log{M_{\rm BH}}=10.75$, near the low- and high-mass ends of our relation, respectively. For comparison, \citet{Savorgnan:2016:II} reported a ratio of $(0.68\%\pm0.04\%)$ from their 45 early-type galaxies.

Additional areas affected by a modification to the $M_{\rm BH}$--$M_{\rm *,sph}$ relation include (i) galaxy/black hole formation theories, which extend to (ii) AGN feedback models, (iii) predictions for space-based gravitational-wave detections, (iv) connections with nuclear star cluster scaling relations, (v) derivations of past quasar accretion efficiency as a function of mass, (vi) searches for fundamental, rather than secondary, black hole mass scaling relations, and (vii) calibrations matching (predominantly) inactive galaxy samples with low-mass AGN data to determine the optimal virial factor, $f$, for measuring black hole masses in AGNs. \citet{Graham:Scott:2013} and \citet{Graham:2016b} have already discussed these many implications of the steeper $M_{\rm BH}$--$M_{\rm *,sph}$ relation, and we refer readers to those works (and references therein) if they would like further details.

Finally, we remind readers that the intercept of the $M_{\rm BH}$--$M_{\rm *,sph}$ relation and thus the above $M_{\rm BH}/M_{\rm *,sph}$ ratios depend on the adopted IMF and SFH that one uses to derive their stellar masses. Obviously, one should not derive the stellar mass of the bulge of a galaxy using a stellar mass-to-light ratio based on a different IMF and SFH than that used here and then expect to be able to predict the central black hole mass without adjusting either the $M_{\rm BH}$--$M_{\rm *,sph}$ relation presented here or adjusting the stellar mass of one's bulge. Perhaps surprisingly, this important point is often overlooked, which has led us to include the $\upsilon$ term in Equation~(\ref{upsilon}).

\subsubsection{Pseudobulges}

Following \citet{Graham:2008} and \citet{Hu:2008}, pseudobulges have often been accused of not following black hole mass scaling relations. For example, \citet{Sani:2011} claimed that pseudobulges with small black holes are significantly displaced from (black hole)--bulge scaling relations. What \citet{Graham:2007,Graham:2008} and \citet{Hu:2008} found is that barred galaxies, possibly containing pseudobulges, are offset to higher spheroid stellar velocity dispersions and/or lower black hole masses than nonbarred galaxies in the $M_{\rm BH}$--$\sigma_{*}$ diagram. \citet[][see also \citealt{Debattista:2013} and \citealt{Monari:2014}]{Hartmann:2014} demonstrated via simulations that the evolution of bars results in elevated spheroid stellar velocity dispersions to a degree that fully explained the observations. That is, the observed offset is not thought to be because their black hole masses might be low but rather because of their modified dynamics.

The existence of many galaxies hosting both a classical bulge and a pseudobulge \citep{Ganda:2006,Peletier:2007} could undermine the exercise of trying to bin galaxies into either the classical or pseudobulge category. This classification dilemma is somewhat bypassed by observing in Figure~\ref{spheroid_plot2} that all of the bulges appear to co-define the same $M_{\rm BH}$--$M_{\rm *,sph}$ relation for spiral galaxies. As such, pseudobulges are not discrepant outliers with low black hole masses, supporting the view that the offset seen in the $M_{\rm BH}$--$\sigma_{*}$ diagram was due to the increased velocity dispersions of the barred galaxies alleged to have pseudobulges.

Identifying ``pseudobulges'' is an imprecise science due to a host of ambiguous demarcations between pseudo- and classical bulges \citep{Graham:2015a,Fisher:Drory:2016}. Recently, \citet{Costantin:2018} went as far as to say that most kinematic or photometric properties of bulges, particularly S{\'e}rsic index, are poor indicators of whether a bulge is pseudo or classical. They advocate that the intrinsic shape of bulges is the best single characteristic that correlates with bulge type, and we leave that exercise for others. Nonetheless, our decompositions identify a single bulge for each galaxy plus, in some instances, an inner disk or bar that some may consider to be the pseudobulge. The idea that the bulges identified here require further subdivision into a classical bulge and a pseudobulge, such that the classical bulge would follow the near-linear $M_{\rm BH}$--$M_{\rm *,sph}$ scaling relation defined by the spheroids in (massive) early-type galaxies \citep{Nowak:2010}, seems fanciful.

\subsubsection{Potential Overmassive Black Holes}

Figure~\ref{spheroid_plot2} reveals that NGC 1300 is $>2.5\,\Delta_{\rm rms}$ above the $M_{\rm BH}$--$M_{\rm *,sph}$ line. Perhaps its spheroid mass is lower than expected, or its black hole mass is higher than expected. We have been able to check on the spheroid mass by converting the absolute spheroid magnitude from \citet{Lasker:2014a} into a stellar mass. After adjusting their adopted distance to match our distance, we find their stellar mass is $0.18\,\text{dex}$ smaller\footnote{\citet{Lasker:2014a} used a single exponential disk model, whereas we used a broken exponential disk model as required for galaxies with bar/peanut-shell-shaped structures \citep{Saha:2018}. Consequently, this resulted in \citet{Lasker:2014a} underestimating the luminosity of the bulge due to an overluminous disk central surface brightness.} than our value, which had a $1\,\sigma$ uncertainty of $0.25\,\text{dex}$. \citet{Savorgnan:Graham:2016} demonstrated that early-type galaxies that possess intermediate-scale disks can sometimes \textit{appear} to have overmassive black holes---a term used to describe systems with unusually high $M_{\rm BH}/M_{\rm *,sph}$ mass ratios---in the $M_{\rm BH}$--$M_{\rm *,sph}$ diagram if a large-scale disk is erroneously fit to the galaxy, resulting in an underestimation of the bulge mass. However, the correct determination and accurate modeling of their intermediate-scale disks can rectify the situation.

It is possible that the bright nuclear star cluster in NGC~1300 \citep{Lasker:2014a} could be contaminating the central gravitational potential, confounding the SMBH's actual dynamical influence. With a very strong bar and nuclear spiral arms, NGC~1300 is additionally unique.\footnote{Notably, \citet{Lasker:2014a} describe NGC 1300 as being the ``most complex'' galaxy in their sample.} However, the uncertainty on the stellar mass of the bulge brings NGC~1300 to within $2\,\Delta_{\rm rms}$ of the best-fit line, and therefore all may be fine. There is thus little evidence for overmassive black holes in spiral galaxies.

\subsection{The $M_{\rm BH}$--$n_{\rm sph, maj}$ Relation}

In Section~\ref{sec:bh-sersic}, we presented the interesting result of a correlation (albeit a weak one) between $M_{\rm BH}$ and $n_{\rm sph, maj}$ for spiral galaxies. Previous studies have derived the $M_{\rm BH}$--$n_{\rm sph, maj}$ relation from samples of predominantly early-type galaxies. Here we first compare our results to those of \citet{Graham:Driver:2007}, who studied a sample of 27 galaxies that included only three spiral galaxies (the Milky Way, NGC 3031, and NGC 4258). They found a strong correlation with a Pearson correlation coefficient of $r=0.88$ and a Spearman rank-order correlation coefficient of $r_s=0.95$. From our sample of 40 spiral galaxies, we find a much weaker correlation with $r=0.46$ and $r_s=0.39$. We note that traditionally, $n_{\rm sph, maj}$, rather than $n_{\rm sph, eq}$ (spheroid equivalent axis S{\'e}rsic index), has been used as a predictor of black hole mass. We find that if instead $n_{\rm sph, eq}$ is used instead, it yields a consistent relation with $r=0.43$ and $r_s=0.42$. Despite our weaker correlation, our slope is consistent with the \textit{linear} $M_{\rm BH}$--$n_{\rm sph, maj}$ relation in \citet{Graham:Driver:2007};\footnote{\citet{Graham:Driver:2007} advocated for a non-(log-linear) relation between $\log{M_{\rm BH}}$ and $\log{n_{\rm sph, maj}}$, such that it is steeper at lower black hole masses.} our \textsc{bces} \textit{bisector} slope ($2.69\pm0.33$) is consistent at $0.22\,\sigma$ with their \textsc{bces} \textit{bisector} slope ($2.85\pm0.40$).

\citet{Savorgnan:2013} increased the sample size to 48 galaxies, which included a subsample of 21 core-S\'ersic galaxies and 27 galaxies with S\'ersic bulges, 15 of which were spiral galaxies. They reported a steeper \textit{symmetric} \textsc{bces} slope of $4.11\pm0.72$ for the 27 S\'ersic galaxies, with $r_s=0.60$. We also compare our result with the $M_{\rm BH}$--$n_{\rm sph, maj}$ relation, derived from a sample of 17 spiral galaxies by \citet{Savorgnan:2016:III}, who reported a \textsc{bces} \textit{bisector} slope of $6.06\pm3.66$, which is steeper but consistent at the level of $0.84\,\sigma$ with our slope due to the large uncertainties on their measurement. As was the case with the $M_{\rm BH}$--$M_{\rm *,sph}$ relation, our larger sample of spiral galaxies has allowed us to more precisely quantify the slope of the $M_{\rm BH}$--$n_{\rm sph, maj}$ relation. With an rms scatter of $0.70$--$0.71\,\text{dex}$ in the $\log{M_{\rm BH}}$ direction about the $M_{\rm BH}$--$n_{\rm sph, maj}$ relation (derived using the \textit{symmetric} \textsc{bces} and \textsc{mpfitexy} regressions), it remains competitive with the $M_{\rm BH}$--$\sigma_*$ relation \citep[see][]{Davis:2017} and the $M_{\rm BH}$--$M_{\rm *,sph}$ relation.

We note that our tests indicate that the slope of the $M_{\rm BH}$--$n_{\rm sph, maj}$ relation is highly sensitive to the assumed uncertainties on the $n_{\rm sph, maj}$ measurements; the steepness of the slope increases when increasing the errors on $n_{\rm sph, maj}$. \citet{Savorgnan:2016:III} assumed a much larger (39\%) median relative error than we do (21\%).\footnote{Our assumed uncertainties on the $n_{\rm sph, maj}$ measurements are the formal \textsc{profiler} errors plus adding 20\% in quadrature.} \citet{Savorgnan:2016:III} derived the uncertainties based on comparison with past measurements in the literature. If we instead adjust the median relative error of \citet{Savorgnan:2016:III}'s data to match our lower value, the \textit{symmetric} \textsc{bces} regression gives a slope of $3.03\pm0.46$, which is now consistent at the level of $0.43\,\sigma$ with our slope.

\subsection{The $M_{\rm *,sph}$--$\phi$ Relation}

The Hubble-Jeans sequence of galaxies \citep{Jeans:1919,Jeans:1928,Hubble:1926,Hubble:1936} famously established a qualitative connection between the apparent prominence of bulges and the tightness of winding for their spiral arms. Thus, our $M_{\rm *,sph}$--$\phi$ relation (Figure~\ref{phi-M_sph_plot}) is roughly a quantitative representation of the Hubble--Jeans spiral galaxy sequence.\footnote{Bulge-to-total flux ratio versus $|\phi|$ would be a more accurate representation of the spiral sequence, although it has been shown that the bulge-to-total flux ratio versus morphological-type correlation is primarily driven by the bulge flux \citep[][and references therein]{Graham:Worley:2008}. Moreover, it should be remembered that the nature of the spiral arms is the primary criterion in establishing the morphological type, with the prominence of the bulge a secondary criterion \citep{Sandage:1961}.} We find that the pitch angle is indeed a good predictor of bulge mass with relatively low scatter.

\citet{Davis:2015} demonstrated that the addition of a third parameter (disk density) will significantly tighten the $M_{\rm *,sph}$--$\phi$ relation. Specifically, galaxies with large bulges may have loosely wound spiral arms if they possess dense disks, and galaxies with small bulges may have tightly wound spiral arms if they possess rarified disks. Exploring this is, however, beyond the desired scope of this current paper.

\vspace{6mm}

\subsection{Morphology-dependent $M_{\rm BH}$--$M_{\rm *,sph}$ Relations}\label{sec:bent}

\begin{figure}
\includegraphics[clip=true,trim= 10mm 6mm 20mm 15mm,width=\columnwidth]{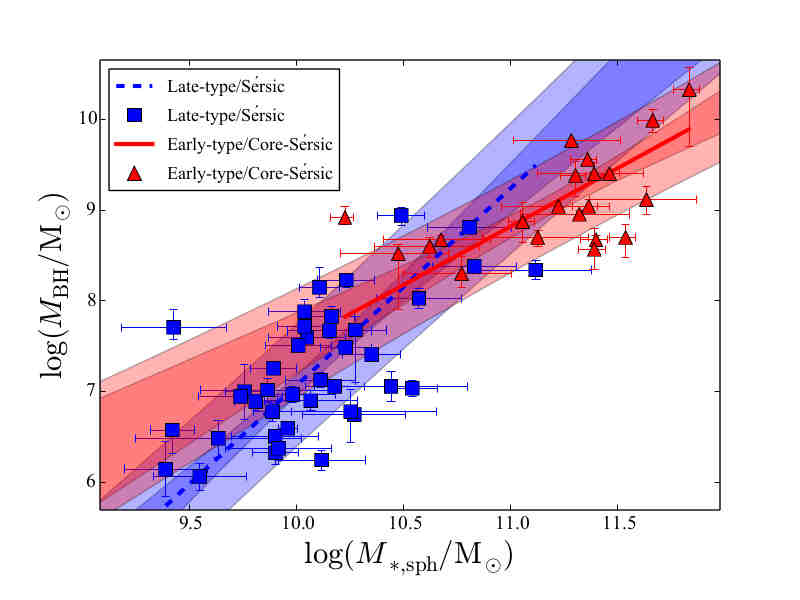}
\caption{Comparison plot of $M_{\rm BH}$ vs.\ $M_{\rm *,sph}$ for our 40 late-type/S{\'e}rsic and 21 early-type/core-S{\'e}rsic galaxies from \citet{Savorgnan:2016:II}. Note that all trend lines are from the \textsc{bces} \textit{bisector} routine.}
\label{bent_plots}
\end{figure}

For a more complete look at the $M_{\rm BH}$--$M_{\rm *,sph}$ diagram, we contrast the distribution of our 40 late-type galaxies (all having S{\'e}rsic spheroids) with 21 early-type galaxies having core-S{\'e}rsic spheroids \citep{Savorgnan:2016:II}.\footnote{We reduced the \citet{Savorgnan:2016:II} stellar spheroid masses by $24.5\%$ according to the dust emission estimation of \citet{Querejeta:2015}. Additionally, \citet{Savorgnan:2016:II} counted 22 early-type galaxies with core-S{\'e}rsic spheroids because they considered NGC 4594 to have a core-S{\'e}rsic bulge (and not to be a spiral galaxy).} The results of the regression involving the early-type galaxies are given in Table~\ref{table:bent} and can be seen in Figure~\ref{bent_plots}. The late-type sample defines an $M_{\rm BH}$--$M_{\rm *,sph}$ relation with a slope that is approximately double the slope of the early-type sample, indicating the existence of a red (early-type) and blue (late-type) sequence, which in this instance also reflects a core-S{\'e}rsic and a S{\'e}rsic sequence.

Figure~\ref{bent_plots} shows a dichotomy between the slopes of the early- and late-type galaxy samples. If we compare the \textsc{bces} \textit{bisector} slopes of the $M_{\rm BH}$--$M_{\rm *,sph}$ relation for the early-type ($1.28\pm0.26$) and late-type ($2.17\pm0.32$) galaxies, we find that they are statistically different, agreeing only at the level of $1.53\,\sigma$. This illustrates that the two samples are fundamentally different. Fitting a single power law to the combined sample of 61 galaxies yields a slope of $1.71\pm0.10$ (according to the \textsc{bces} \textit{bisector} routine), which almost exactly bisects the early- and late-type sample slopes; it is unclear whether this has any physical meaning.

Moreover, the ratio of core-S{\'e}rsic to S{\'e}rsic early-type galaxies in one's sample, or the ratio of ``slow'' and ``fast'' rotating early-type galaxies, or the numbers of high- and low-mass early-type galaxies will dictate the slope of the red sequence if fit with a single power law. For that reason, we elect to only compare with the core-S{\'e}rsic spheroids, excluding (for now) early-type galaxies with S{\'e}rsic spheroids. This population will be quantified in N.\ Sahu et al.\ (2019, submitted), who are performing a photometric analysis of $\approx80$ early-type galaxies with directly measured black hole masses. We note that \citet[][see their Figure~5]{Savorgnan:2016:II} have already revealed that early-type galaxies follow a different relation than late-type galaxies at the low-mass end of their diagram ($3\times10^9 \lesssim M_{\rm *,sph}/M_{\odot} \lesssim 3\times10^{10}$), ruling out the notion of either a single curved relation for all galaxy types or a division in this diagram based on spheroid stellar mass.

To date, the $M_{\rm BH}$--$M_{\rm *,sph}$ relation has been studied using samples of predominantly massive early-type galaxies, yielding linear relations. Two decades ago, \citet[][see also \citealt{Franceschini:1998}]{Magorrian:1998} presented a linear $M_{\rm BH}$--$M_{\rm *,sph}$ relationship. Almost immediately (within 4 months), \citet{Laor:1998} found that $M_{\rm BH}\propto M_{\rm *,sph}^{1.5-1.8}$. \citet{Laor:1998} remarked that a better agreement was found, consistent with a steeper-than-linear slope, when using the low-mass inactive galaxies from \citet{Magorrian:1998}. \citet{Wandel:1999} also reported a steeper-than-linear slope from a sample of Seyfert galaxies with $M_{\rm BH}$ predominantly less than $10^8\,M_{\sun}$, a population not well sampled from other studies of the period. \citet{Salucci:2000} was the first to specifically suggest that the $M_{\rm BH}$--$M_{\rm *,sph}$ relation is significantly steeper for spiral galaxies than for (massive) elliptical galaxies, based on hints from a study of black holes with upper limits for their masses. These studies during the last couple of years of the previous millennium were largely ignored; a score of years later, studies still adhere to the belief of a linear relation.

\subsection{Fundamental Planes with Black Hole Mass}

\citet{Bosch:2016} has claimed that black hole masses correlate strongly with the stellar velocity dispersion, $\sigma_*$, but only weakly with bulge mass.  However, we have found that upon increasing the accuracy of the bulge masses and performing a regression that minimizes the scatter in the $\log M_{\rm BH}$ direction, the level of scatter for spiral galaxies is $0.56$--$0.60\,{\rm dex}$ for both of these relations \citep[see][for the $M_{\rm BH}$--$\sigma_*$ diagram]{Davis:2017}. Similar conclusions regarding the equality of scatter or lower levels of scatter about the $M_{\rm BH}$--$M_{\rm *,sph}$ relation have repeatedly been uncovered in the past when the quality of the galaxy decompositions has been improved \citep[e.g.,][]{McLure:Dunlop:2002, Graham:2007}. Considering $\sigma_*$ to be a galaxy parameter and abandoning bulges, \citet{Bosch:2016} went on to advocate for a plane involving the galaxy stellar mass, the galaxy half-light radius, and the black hole mass, for which he reported a level of scatter comparable to that about the $M_{\rm BH}$--$\sigma_*$ relation. However, the total rms scatter of $0.41$--$0.46\,{\rm dex}$ about the $M_{\rm BH}$--$\phi$ relation for spiral galaxies is notably smaller \citep{Davis:2017}, as is the scatter of $0.41$--$0.48\,{\rm dex}$ about the $M_{\rm BH}$--$M_{\rm *,sph}$ relation for early-type galaxies (e.g., \citealt{Lasker:2014,Savorgnan:2016:II}; Table \ref{table:bent}; N.\ Sahu et al.\ 2019, submitted), challenging the notion that either the proposed plane or the velocity dispersion define the fundamental relation.  Moreover, the focus by \citet{Bosch:2016} on the scatter in just the $\log M_{\rm BH}$ direction may be misleading, and we suggest that one should consider taking into account the slopes of the various relations and planes and using something more akin to the orthogonal scatter.

One needs to be careful if using both early-type and spiral galaxies together, because their different distributions in the $M_{\rm BH}$--$M_{\rm *,sph}$ diagram may lead one to construct an artificial ``fundamental plane'' that only serves as a correction for the mixing of morphological types. For our spiral galaxy sample, we do not find the need for an $M_{\rm BH}$--$M_{\rm *,sph}$--$R_{e\rm, maj}$ relation; i.e., there is no correlation between the $M_{\rm BH}$ residuals from our $M_{\rm BH}$--$M_{\rm *,sph}$ relation and their associated bulge half-light radii. For now, we postpone a discussion of and search for a ``fundamental plane'' involving three parameters. Such a plane was first explored by \citet{Marconi:Hunt:2003}, and attempts to this day have continued \citep[see][for a review of developments and concerns since 2003]{Graham:2016b}.

\subsection{Coevolution of Galaxies and Black Holes}

As \citet{Graham:Scott:2013} noted, the steep nonlinear slope implies that the low-mass (mostly spiral) galaxies, which grow via accretion of gas and/or ``wet'' mergers, tend to grow at a slower fractional rate than their central black hole, assuming growth \emph{along} the redshift $z\approx0$ relations (which need not be the case). In contrast, high-mass SMBHs, which reside in older, more massive, gas-poor early-type galaxies, will nowadays grow primarily via major ``dry'' merging events, causing the black holes to grow at the same relative rate as their host spheroid, preserving their $M_{\rm BH}/M_{\rm *,sph}$ ratio, and producing a near-linear relation. \citet{Savorgnan:2016:II} modified this picture, with data suggesting that the early-type galaxies follow a near-linear relation and the spiral galaxies follow a steeper relation with a log-linear slope between two and three.

Maintaining the steeper-than-linear mass scaling relation between
a galaxy's stellar mass and its central black hole reveals that as gas becomes
available, the fractional mass gain of the black hole must grow dramatically compared
to the fractional mass gain of the galaxy's stellar mass.
This has also been borne out in observations of black
hole accretion rate versus star formation rate \citep[e.g.,][]{Seymour:2012,LaMassa:2013,Yang:2018}. The $M_{\rm BH}/M_{\rm *,sph}$
ratio increases from the observed values of $10^{-3}$ at low masses to
several times $10^{-2}$ at high masses (see Figure~\ref{bent_plots} and \citealt{Graham:Scott:2015}). Obviously,
switching the AGN on/off, according to the AGN duty cycle, will not
instantly result in this mass ratio changing by an order of magnitude; rather, over many
cycles, the ratio will steadily increase as the black hole and galaxy coevolve.
Uncertainties do, however, remain. The near-quadratic, 
$z=0$, $M_{\rm BH}$--$M_{\rm *,sph}$ relation represents the state of affairs to which
the universe has evolved to today. However, galaxies may not have always evolved \emph{along}
this relation to higher masses. That is, the relation itself may also have
evolved with time.

The coevolution of galaxies and their central black holes has often been
presented as a ``chicken-and-egg'' problem. Which came first, the black hole or
the galaxy? In high-mass systems, perhaps a (seed) black hole formed before the galaxy, while
in lower-mass systems, perhaps the galaxy formed before the black hole. With
the myriad of potential pathways to create intermediate-mass black holes \citep[IMBHs; e.g.,][]{Hirano:2017,Regan:2017}, searches for a single answer to the chicken-and-egg
problem may therefore prove fruitless, as there may be a mass-dependent
answer or multiple possibilities valid at the same mass.

\subsection{Simulations}

For over a decade now, some simulations have illustrated a characteristic bend, such that the $M_{\rm BH}$--$M_{\rm *,sph}$ relation is steeper than linear at lower masses \citep{Cirasuolo:2005,Fontanot:2006,Dubois:2012,Khandai:2012,Bonoli:2014,Neistein:2014}. Recently, \citet{FIRE:2017} presented simulations showing that the $M_{\rm BH}$--$M_{\rm *,sph}$ diagram (see their Figure~2) displays a steep slope for stellar bulge masses between $\approx10^{10}\,M_{\sun}$ and $\approx10^{11}\,M_{\sun}$ that gradually becomes shallower at higher masses. This trend roughly follows the path connecting our (low-mass) late-type galaxies with the (high-mass) core-S\'ersic early-type galaxies in our Figure~\ref{bent_plots}.

Nonetheless, most well-cited simulations and BHMFs have tied themselves to a linear scaling between black hole mass and spheroid mass. \citet{Marconi:2004} generated a BHMF via the linear black hole mass scaling relations presented in \citet{Marconi:Hunt:2003}. The Millennium Simulation \citep{Springel:2005} normalized the feedback process of black holes to reproduce the scaling relation of \citet{Magorrian:1998}. \citet{Croton:2006} implemented semi-analytic models on the output of the Millennium Simulation, incorporating radio-mode feedback from AGNs, in order to simulate the growth of SMBHs and their host galaxies; however, they tuned their models to reproduce a linear $M_{\rm BH}$--$M_{\rm *,sph}$ scaling relation \citep{Magorrian:1998,Marconi:Hunt:2003,Haring:Rix:2004}. \citet{Hopkins:2006} derived an analytical model by assuming that black hole mass growth is proportional to the inflowing gas mass in its host galaxy core, which reproduces the \citet{Magorrian:1998} linear $M_{\rm BH}$--$M_{\rm *,sph}$ relation. \citet{Booth:2009} placed an observational constraint of $M_{\rm BH}\approx0.006M_{\rm *,sph}$ \citep{Magorrian:1998} on the growth of their black holes. More recently, the EAGLE project \citep{Schaye:2015}, which is another suite of simulations designed to track the formation of galaxies and SMBHs, closely followed the method for AGN feedback described in \citet{Booth:2009}. Given that their SMBH accretion was adjusted to reproduce the local galaxy stellar mass function, and the $\mathfrak{M}_B$--$R_{\rm e}$ relation, then the efficiency of the AGN feedback will be in error if the assigned SMBH masses were not correct. Thus, with so many influential studies anchored by the simplistic assumption of a universal linear rate of growth between the bulges and their black holes, our work echoes the voices of change that have existed since \citet{Salucci:2000}, that a steeper-than-linear relation should be implemented in studies that are derivative of the $M_{\rm BH}$--$M_{\rm *,sph}$ relation.

\subsection{Predicting Black Hole Masses}

Here we expound the synergy that exists between the use of bulge mass or pitch angle for SMBH mass prediction. These two quantities ($M_{\rm *,sph}$ and $\phi$) are naturally complementary to each other. The $M_{\rm BH}$--$M_{\rm *,sph}$ relation is not applicable for bulgeless galaxies, and the $M_{\rm BH}$--$\phi$ relation is not applicable for galaxies that lack visible spiral structure (e.g., elliptical, lenticular, ellicular, or edge-on spiral galaxies). Together, these two general characteristics of galaxies ($M_{\rm *,sph}$ and $\phi$) can estimate central SMBH masses in most observable galaxies. For the intersection of galaxies with both measurable spiral structure and a bulge, the relations provide two independent predictions and thus, serve as an important double-check for SMBH mass.

Finally, our newly defined relations allow us to estimate which spheroids/galaxies might potentially harbor IMBHs ($10^2\leq M_{\rm BH}/M_{\sun}\leq10^5$). The \textit{symmetric} Bayesian analysis between $M_{\rm BH}$ and $M_{\rm *,sph}$ predicts that spheroids with $M_{\rm *,sph}\leq\upsilon(1.39\times10^9)\,M_{\sun}$ should possess IMBHs. It is our hope that this target mass, along with corresponding pitch angles ($|\phi|\geq26\fdg7$) and spheroid stellar velocity dispersions \citep[$\sigma_*\leq57.5\,\text{km\,s}^{-1}$, from][]{Davis:2017}, can help future studies identify galaxies that may host IMBHs.

For the above extrapolations to low black hole masses, we advocate the \emph{symmetric} regressions, not the \emph{conditional} regressions. While our \emph{conditional} regression of $M_{\rm BH}$ on some galaxy property will result in the lowest level of scatter in the $\log M_{\rm BH}$ direction over the interpolated data range, this is not ideal when extrapolating to lower masses. Due to the diminished slope of a \textit{conditional}, ($Y|X$), or ordinary least-squares regression, this will lead to an overprediction of black hole masses when extrapolated below the range of values used to define the relation.

%\vspace{6mm}

\section{Conclusions}\label{END}

We have extracted and modeled the surface brightness profiles for the current complete sample of 43 spiral galaxies (plus Cygnus A) with directly measured SMBH masses, more than doubling the sample size of recent work that provided similarly accurate decompositions. Our multicomponent decomposition of galaxy light was based upon evidence for real substructure (e.g., bars, rings, spiral arms, etc.) in galaxy images and ellipticity, light, PA, and $B_4$ profiles. The results of this work are presented in Appendix~\ref{App3}, with an example in Figures \ref{UGC3789_plot} and \ref{fig:annotate}. We provide improved spheroid stellar magnitudes and masses in Table~\ref{table:Sample}. We find the following key results
\begin{enumerate}
\item For spiral galaxies derived from a \textit{symmetric} Bayesian analysis, $\log{M_{\rm BH}}\propto\left(2.44_{-0.31}^{+0.35}\right )\log{M_{\rm *,sph}}$, which is approximately double the slope for early-type galaxies with core-S{\'e}rsic spheroids.
\item Fitting a single power law to varying ratios of early- and late-type galaxies in the $M_{\rm BH}$--$M_{\rm *,sph}$ diagram (Figure~\ref{bent_plots}) will result in varying slopes that more reflect one's sample selection more than anything physically meaningful. We are in the process of adding $\approx60$ early-type galaxies to the 21 shown in Figure~\ref{bent_plots}, and will report on how the ``red sequence'' of early-type galaxies bends as a function of core-S{\'e}rsic versus S{\'e}rsic galaxies, as well as a function of other properties, such as the presence of a disk (i.e., fast rotator versus slow rotator).
\item The $M_{\rm BH}$--$n_{\rm sph, maj}$ relation, when derived from a sample of only spiral galaxies, remains consistent with previous evaluations of the relation. Furthermore, its scatter remains competitive with other black hole mass scaling relations.
\item In Figure~\ref{phi-M_sph_plot}, we provide the relation between the spiral-arm pitch angle $\phi$ and the stellar mass of the bulge. Given the strong correlation between $\phi$ and $M_{\rm BH}$ \citep[e.g.,][]{Davis:2017}, this correlation draws strong parallels with the $M_{\rm BH}$--$M_{\rm *,sph}$ relation.
\end{enumerate}

We bring further clarity to the results of previous studies that have suggested the existence of a bend in the slope of the $\log M_{\rm BH}$ vs.\ $\log M_{\rm *,sph}$ diagram between the populations of late- and early-type galaxies. We were able to greatly narrow down the uncertainty on the slope of the $M_{\rm BH}$--$M_{\rm *,sph}$ relation for spiral galaxies. We promote the use of spheroid stellar mass along with logarithmic spiral-arm pitch angle as important properties of galaxies that can be used to produce accurate central black hole mass estimates in spiral galaxies. 

\acknowledgments

We thank Nandini Sahu for her helpful comments and insights, which helped improve this paper. A.W.G.\ was supported under the Australian Research Council's funding scheme DP17012923. Parts of this research were conducted by the Australian Research Council Centre of Excellence for Gravitational Wave Discovery (OzGrav) through project number CE170100004. This research has made use of NASA's Astrophysics Data System. This research has made use of the NASA/IPAC Infrared Science Archive. We acknowledge the usage of the HyperLeda database \citep{HyperLeda}, \url{http://leda.univ-lyon1.fr}. This research has made use of the NASA/IPAC Extragalactic Database (NED). Some of the data presented in this paper were obtained from the Mikulski Archive for Space Telescopes (MAST). This publication makes use of data products from the Two Micron All Sky Survey, which is a joint project of the University of Massachusetts and the Infrared Processing and Analysis Center/California Institute of Technology. Error propagation calculations were performed via the \textsc{python} package \textsc{uncertainties} (\url{http://pythonhosted.org/uncertainties/}).

\bibliography{bibliography}

\appendix

\section{Statistical Modeling Framework}\label{App1}

Two aspects of the linear regression analysis have shaped our statistical modeling approach: specifically, that (i) there
are substantial uncertainties associated with the measurement of \textit{both} the spheroid log-masses
and black hole log-masses, and (ii) we wish for a symmetric treatment of the
relationship between these two variables. For these reasons, we do not pursue an ordinary
regression approach, in which the variables are described as one ``dependent'' and the other
``independent'' and the model structured toward estimation of the mean of the former
conditional on the latter. Instead, it was decided to treat the statistical challenge as one of
joint density estimation, in which a bivariate normal density is used to represent the
distribution of latent (``true'') spheroid and black hole log-masses that might occur in our
sample. This is conceptually equivalent to the generative framework sketched by \citet{Hogg:2010}, in which the observed datapoints are imagined to be
drawn from a distribution centered around the ``line of best fit,'' except that here we allow
Bayesian ``shrinkage'' by estimating the underlying distribution along the line rather than
keeping this as an improper uniform prior. For a bivariate normal with marginal standard
deviations, $\sigma_{\rm{*,sph}}$ and $\sigma_{\text{BH}}$, and correlation coefficient, $\rho$, the corresponding symmetric and conditional slopes are simply 
\begin{equation}
\beta_{\rm symmetric}=\frac{\sigma_{\text{BH}}}{\sigma_{\rm{*,sph}}}\;\text{and}\;\beta_{\rm conditional}=\rho\frac{\sigma_{\text{BH}}}{\sigma_{\rm{*,sph}}}.
\end{equation}

One technical point is that we are fitting our statistical model separately to the step at which observational data are compared against the physical models from which our ``observed'' spheroid and black hole log-masses are derived.  This means that instead of using an ordinary sampling distribution to form our likelihood function, we must invert the uncertainty distributions compiled in our data table as summaries of the likelihood for each datapoint, imagining improper uniform priors to have effectively been adopted during the original physical model comparison.  In the case of the spheroid log-masses, we have ordinary normal distributions of known mean and standard deviation.  In this case, the distinction between the above and the ordinary likelihood function is purely theoretical due to the symmetry of the normal distribution.  However, uncertainties in the black hole log-masses are fundamentally asymmetric and come as upper and lower ``standard deviations,'' which we represent via skew normal distributions matching the suggested quantiles as close as possible. Our reason for choosing this particular representation rather than, e.g., joining two normal densities at zero, is that it is everywhere differentiable and hence amenable to sampling via hybrid Monte Carlo \citep[here implemented with the \textsc{stan} package in \textsc{r};][]{RStan:2016}.

In the notation of hierarchical Bayesian statistics, our model may be written as below, complete with our chosen priors.  Note that the ``$\sim$'' symbol means ``is distributed as'' and the index $i$ runs from 1 to 40, referencing each object in our sample:

\begin{equation}
L(\text{data}|M_{\rm{*,sph},i}^{\text{true}})\propto \text{Normal}(M_{\rm{*,sph},i}^{\text{true}}|M_{\rm{*,sph},i}^{\text{obs.}},[\sigma_{\rm{*,sph},i}^{\text{obs.}}]^2)
\end{equation}

\begin{equation}
L(\text{data}|M_{\text{BH},i}^{\text{true}})\propto \text{SkewNormal}(M_{\text{BH},i}^{\text{true}}|M_{\text{BH},i}^{\text{obs.}},[\sigma_{\text{BH},i}^{\text{obs.}}]^2,\alpha_{\text{BH},i}^{\text{obs.}})
\end{equation}

\begin{equation}
\{M_{\rm{*,sph},i}^{\text{true}},M_{\text{BH},i}^{\text{true}}\}\sim \text{BivariateNormal}(\mu,\Sigma)
\end{equation}

\begin{equation}
\mu=\{\mu_{\rm{*,sph}},\mu_{\text{BH}}\},\ \Sigma=\begin{Bmatrix}
\sigma_{\text{*,sph dist.}}^2 & \rho\sigma_{\rm{*,sph}}\sigma_{\text{BH}} \\ 
\rho\sigma_{\rm{*,sph}}\sigma_{\text{BH}} & \sigma_{\text{BH dist.}}^2
\end{Bmatrix}
\end{equation}

\begin{equation}
\mu_{\rm{*,sph}}\sim \text{Normal}(10.5,2),\ \mu_{\text{BH}}\sim \text{Normal}(7,2), \sigma_{\rm{*,sph}}\sim \text{Gamma}(1,1),\ \sigma_{\text{BH}}\sim \text{Gamma}(1,1),\ \rho\sim\beta(10,1)
\end{equation}

We have summarized the results of fitting this model against the observational data set in Table~\ref{table:priors}. In particular, we report the estimated quantiles at 2.5\%, 16\%, 50\%, 84\%, and 97.5\% for each parameter (or composition of parameters, in the case of the slope); from these can be read the median, 68\% (``$\pm1$ $\sigma$''), and 95\% (``$\pm2$ $\sigma$'') credible intervals.  An illustration of our fit is also presented in Figure~\ref{Ewan_M_BH-M_sph_plot}. From inspection of Table~\ref{table:priors}, it is evident that our priors are strongly updated by the data; that is, our solution is not hamstrung by the choice of priors.

\begin{deluxetable}{lrrrrr|rrrrr}
\tabletypesize{\normalsize}
\tablecolumns{11}
\tablecaption{Fitting Results of Our Model against the Observational Data Set $\left(\log{M_{\rm *,sph}},\, \log{M_{\rm BH}}\right)$\label{table:priors}}
\tablehead{
\colhead{} & \multicolumn{5}{c}{Prior} & \multicolumn{5}{c}{Posterior}
}
\startdata
Quantile & 2.5\% & 16\% & 50\% & 84\% & 97.5\% & 2.5\% & 16\% & 50\% & 84\% & 97.5\% \\
\hline
\textit{Symmetric} slope & 0.03 & 0.19 & 1.00 & 5.21 & 38.89 & 1.83 & 2.13 & 2.44 & 2.79 & 3.32 \\
\textit{Conditional} ($Y|X$) slope & 0.02 & 0.17 & 0.90 & 4.69 & 34.23 & 1.38 & 1.67 & 1.98 & 2.35 & 2.83 \\
\textit{Symmetric} $M_{\rm BH}$ scatter (dex)  & 0.01 & 0.05 & 0.22 & 0.66 & 1.56 & 0.61 & 0.69 & 0.77 & 0.87 & 0.99 \\
\textit{Conditional} $M_{\rm BH}$ scatter (dex) & 0.01 & 0.05 & 0.22 & 0.69 & 1.57 & 0.30 & 0.37 & 0.43 & 0.51 & 0.58 \\
Normalized $X$-intercept, $X_0$ & 6.58 & 8.51 & 10.50 & 12.49 & 14.42 & 9.96 & 10.01 & 10.06 & 10.12 & 10.18 \\
Normalized $Y$-intercept, $Y_0$ & 3.08 & 5.01 & 7.00 & 8.99 & 10.92 & 6.99 & 7.12 & 7.24 & 7.36 & 7.47 \\
\enddata
\end{deluxetable}

\section{Notes on Individual Galaxies}\label{App2}

Here we provide a detailed accounting of the components we identified and implemented in the decompositions of the galaxies in our sample. In Table~\ref{table:comps}, we provide a tabular list of the components fit to each galaxy.

\begin{sidewaysfigure}
\includegraphics[clip=true,trim= 11mm 1mm 1mm 5mm,width=0.249\textwidth]{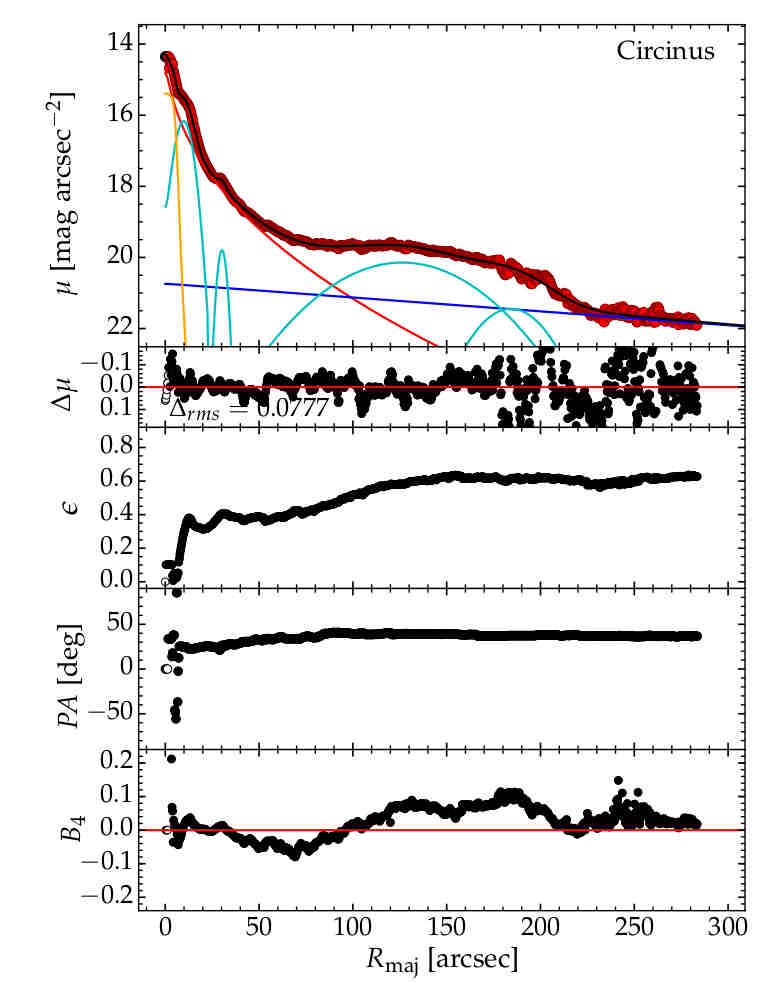}
\includegraphics[clip=true,trim= 11mm 1mm 1mm 5mm,width=0.249\textwidth]{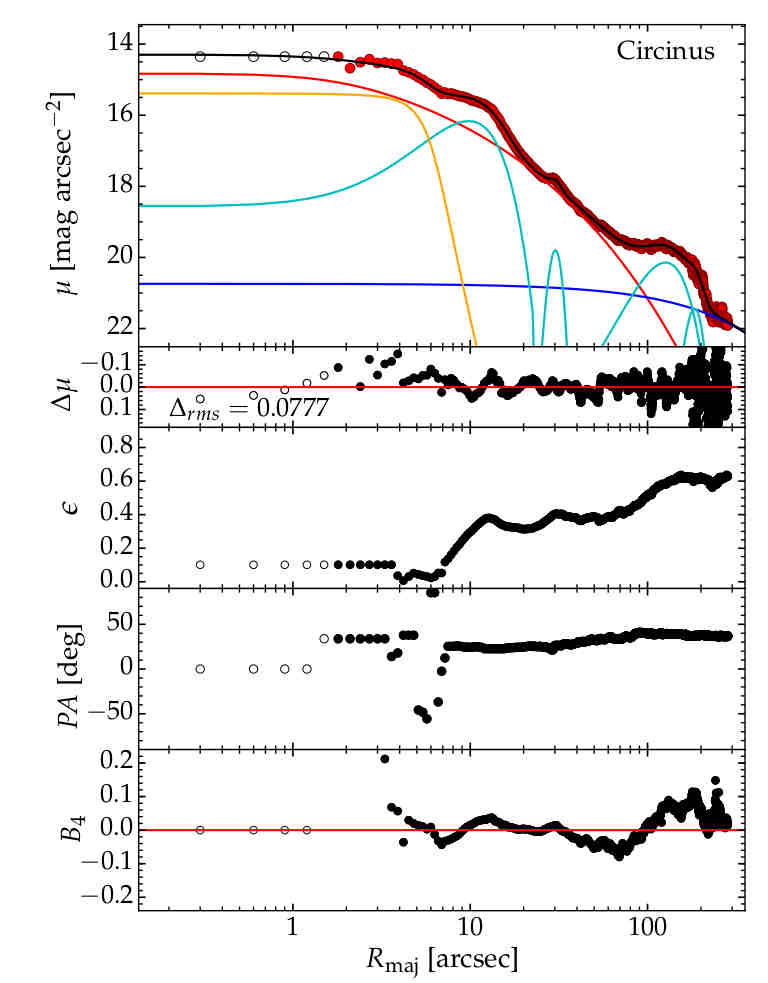}
\includegraphics[clip=true,trim= 11mm 1mm 1mm 5mm,width=0.249\textwidth]{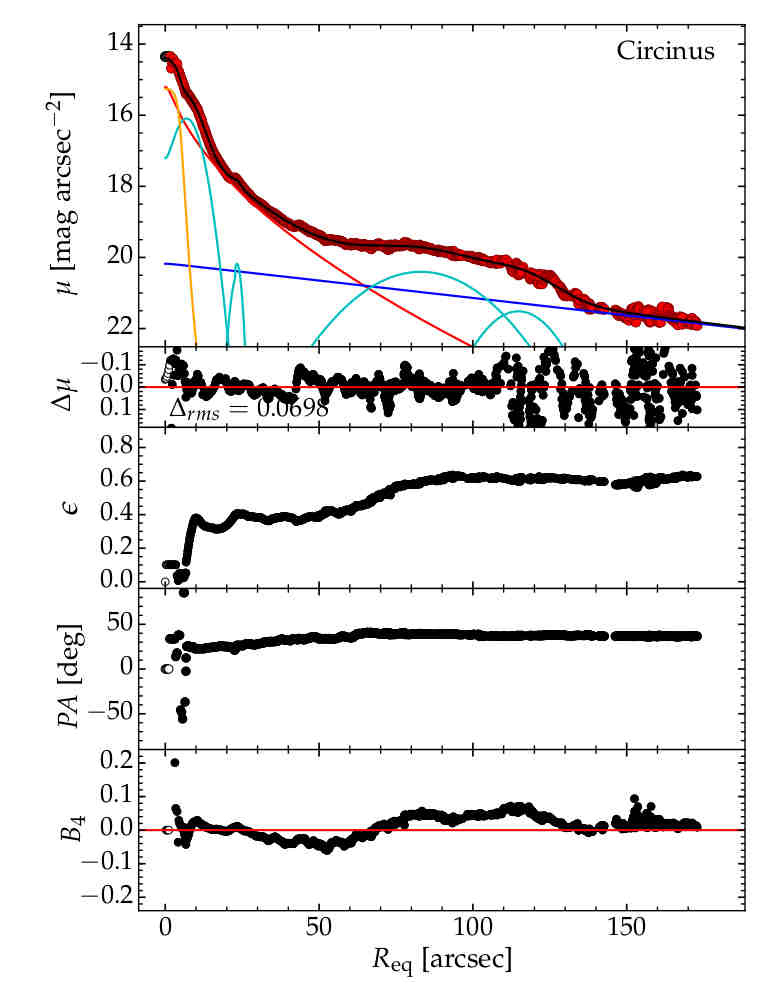}
\includegraphics[clip=true,trim= 11mm 1mm 1mm 5mm,width=0.249\textwidth]{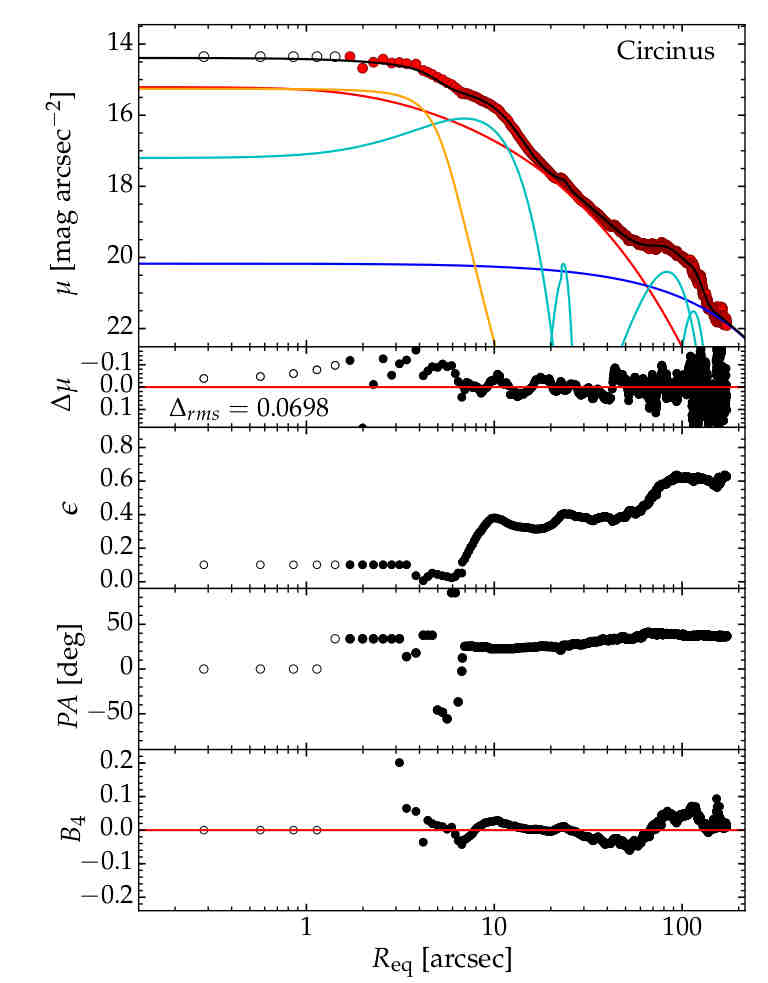}
\caption{\textit{Spitzer} $3.6\,\micron$ surface brightness profile for the Circinus galaxy, with a physical scale of 0.0204$\,\text{kpc}\,\text{arcsec}^{-1}$. \textbf{Left two panels}---The model represents $1\farcs9 \leq R_{\rm maj} \leq 284\arcsec$ with $\Delta_{\rm rms}=0.0777\,\text{mag\,arcsec}^{-2}$. \underline{S{\'e}rsic Profile Parameters:} \textcolor{red}{$R_e=33\farcs26\pm1\farcs99$, $\mu_e=18.29\pm0.12\,\text{mag\,arcsec}^{-2}$, and $n=2.21\pm0.56$.} \underline{Ferrers Profile Parameters:} \textcolor{Orange}{$\mu_0 = 15.36\pm0.64\,\text{mag\,arcsec}^{-2}$, $R_{\rm end} = 5\farcs40\pm0\farcs00$, and $\alpha = 0.23\pm1.41$.} \underline{Exponential Profile Parameters:} \textcolor{blue}{$\mu_0 = 20.74\pm0.18\,\text{mag\,arcsec}^{-2}$ and $h = 278\farcs28\pm39\farcs19$.} \underline{Additional Parameters:} four Gaussian components added at: \textcolor{cyan}{$R_{\rm r}=9\farcs82\pm0\farcs49$, $30\farcs23\pm0\farcs38$, $126\farcs27\pm1\farcs07$, \& $184\farcs21\pm1\farcs27$; with $\mu_0 = 16.10\pm0.10$, $19.80\pm0.17$, $20.14\pm0.04$, \& $21.44\pm0.16\,\text{mag\,arcsec}^{-2}$; and FWHM = $8\farcs84\pm0\farcs74$, $5\farcs12\pm0\farcs95$, $80\farcs09\pm5\farcs57$, \& $40\farcs64\pm2\farcs93$, respectively.} \textbf{Right two panels}---The model represents $1\farcs9 \leq R_{\rm eq} \leq 173\arcsec$ with $\Delta_{\rm rms}=0.0698\,\text{mag\,arcsec}^{-2}$. \underline{S{\'e}rsic Profile Parameters:} \textcolor{red}{$R_e=23\farcs13\pm1\farcs22$, $\mu_e=18.06\pm0.11\,\text{mag\,arcsec}^{-2}$, and $n=1.80\pm0.60$.} \underline{Ferrers Profile Parameters:} \textcolor{Orange}{$\mu_0 = 15.21\pm0.56\,\text{mag\,arcsec}^{-2}$, $R_{\rm end} = 4\farcs67\pm0\farcs00$, and $\alpha = 0.26\pm1.32$.} \underline{Exponential Profile Parameters:} \textcolor{blue}{$\mu_0 = 20.16\pm0.27\,\text{mag\,arcsec}^{-2}$ and $h = 110\farcs93\pm13\farcs98$.} \underline{Additional Parameters:} four Gaussian components added at: \textcolor{cyan}{$R_{\rm r}=7\farcs00\pm0\farcs75$, $23\farcs30\pm0\farcs31$, $83\farcs01\pm0\farcs69$, \& $114\farcs77\pm1\farcs00$; with $\mu_0 = 16.03\pm0.19$, $20.18\pm0.24$, $20.40\pm0.05$, \& $21.51\pm0.19\,\text{mag\,arcsec}^{-2}$; and FWHM = $8\farcs63\pm0\farcs96$, $3\farcs00\pm0\farcs83$, $42\farcs51\pm3\farcs61$, \& $25\farcs22\pm1\farcs82$, respectively.} Given our focus on isolating the bulge light, we have allowed degeneracy among the components which dominate at large radii (whose parameters are therefore neither stable nor reliable) when this appears to not compromise the bulge.}
\label{Circinus_plot}
\end{sidewaysfigure}

\subsection{Circinus (Figure~\ref{Circinus_plot})}

The Circinus galaxy possesses a Type 2 Seyfert AGN \citep{Maiolino:1998}. Due to corrupt pixels in the nucleus of the \textit{Spitzer} image of the Circinus galaxy, we have excluded modeling of the surface brightness profile for $R_{\rm maj} = R_{\rm eq} < 1\farcs9$. We have added a nuclear Ferrers component to account for residual AGN light. We have also added four Gaussian components at $R_{\rm maj} \approx 10\arcsec$ ($R_{\rm eq}\approx 7\arcsec$), $R_{\rm maj} \approx 30\arcsec$ ($R_{\rm eq}\approx 23\arcsec$), $R_{\rm maj} \approx 126\arcsec$ ($R_{\rm eq}\approx 83\arcsec$), and $R_{\rm maj} \approx 184\arcsec$ ($R_{\rm eq}\approx115\arcsec$) to account for a nuclear ring and three spiral-arm contributions in the surface brightness profile, respectively.

\begin{sidewaysfigure}
\includegraphics[clip=true,trim= 11mm 1mm 1mm 5mm,width=0.249\textwidth]{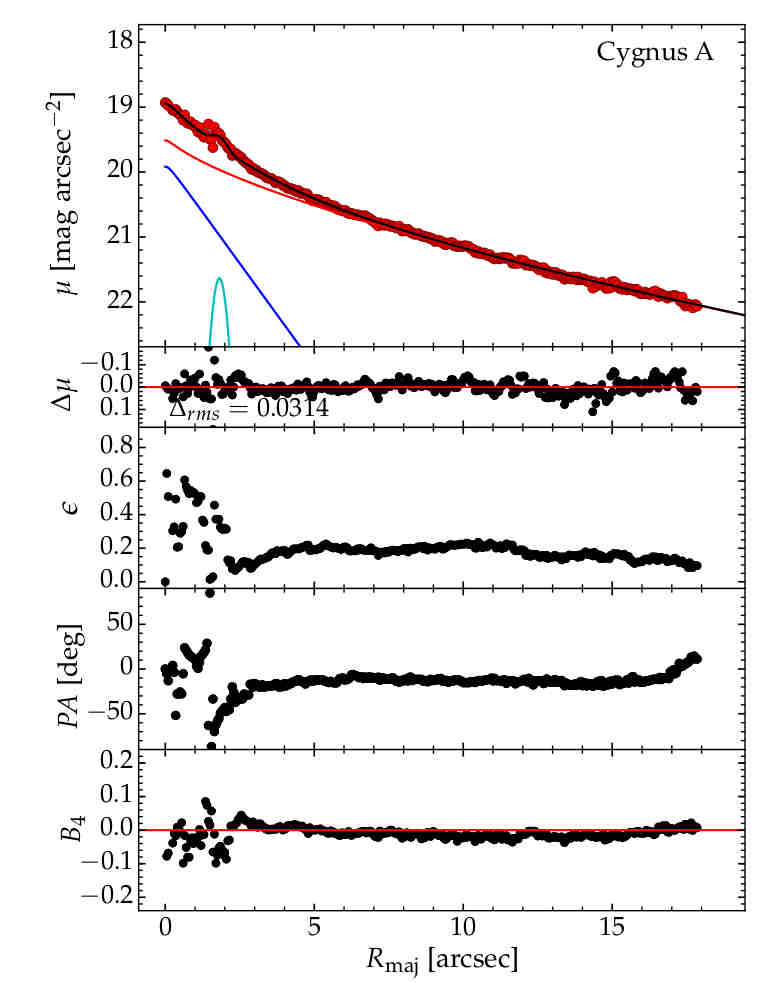}
\includegraphics[clip=true,trim= 11mm 1mm 1mm 5mm,width=0.249\textwidth]{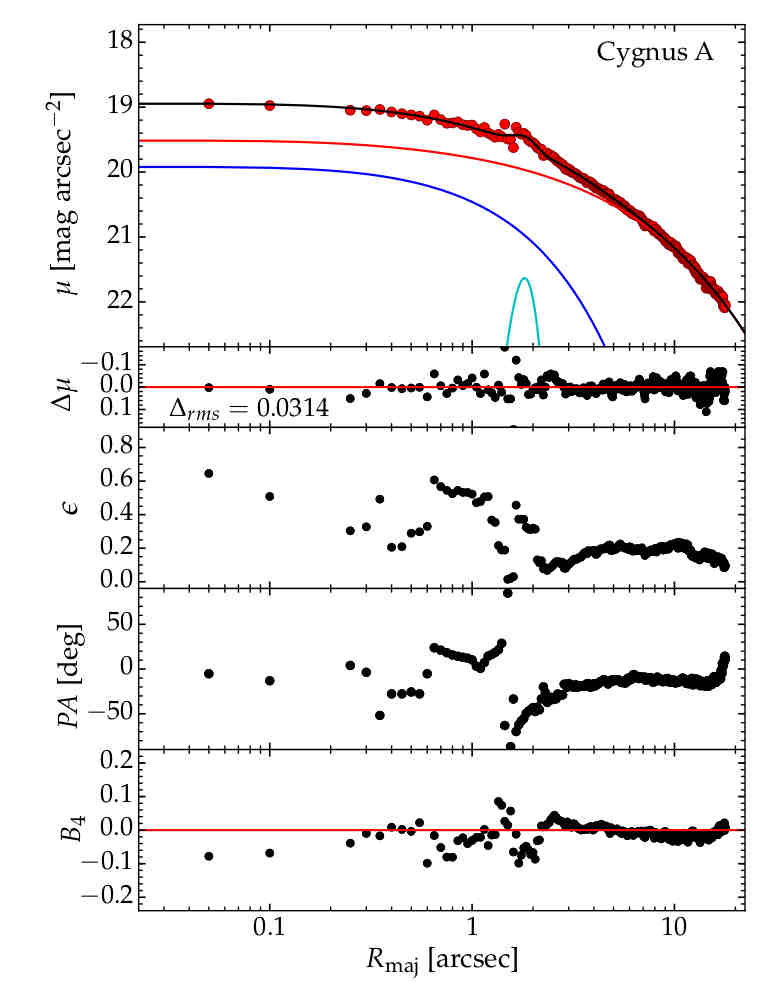}
\includegraphics[clip=true,trim= 11mm 1mm 1mm 5mm,width=0.249\textwidth]{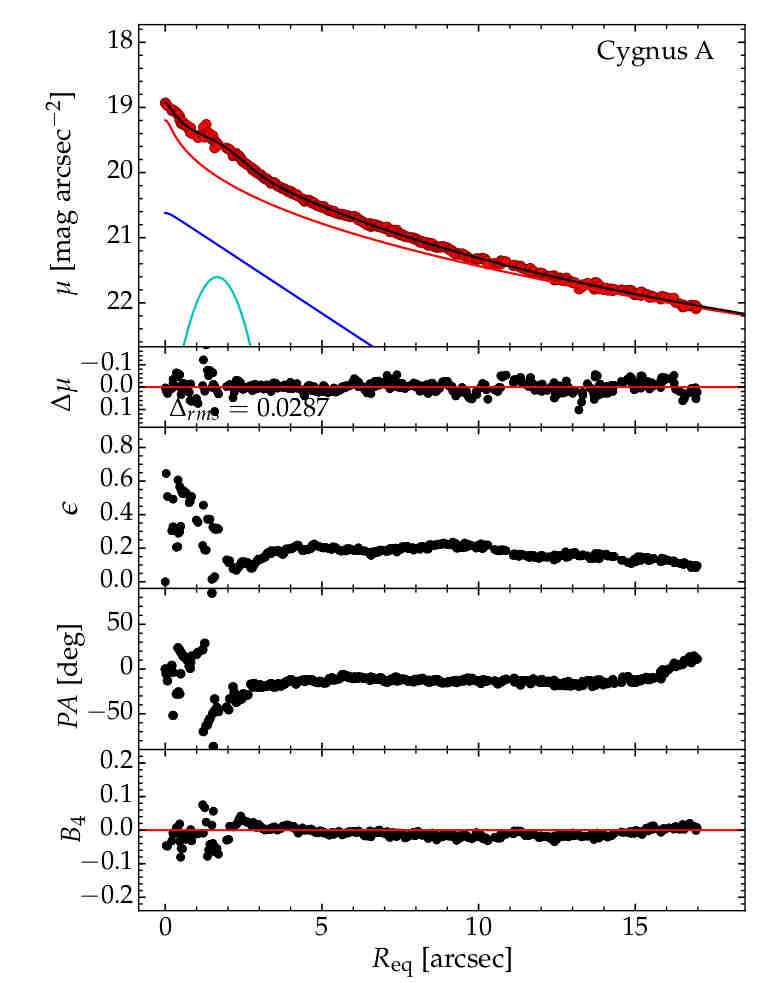}
\includegraphics[clip=true,trim= 11mm 1mm 1mm 5mm,width=0.249\textwidth]{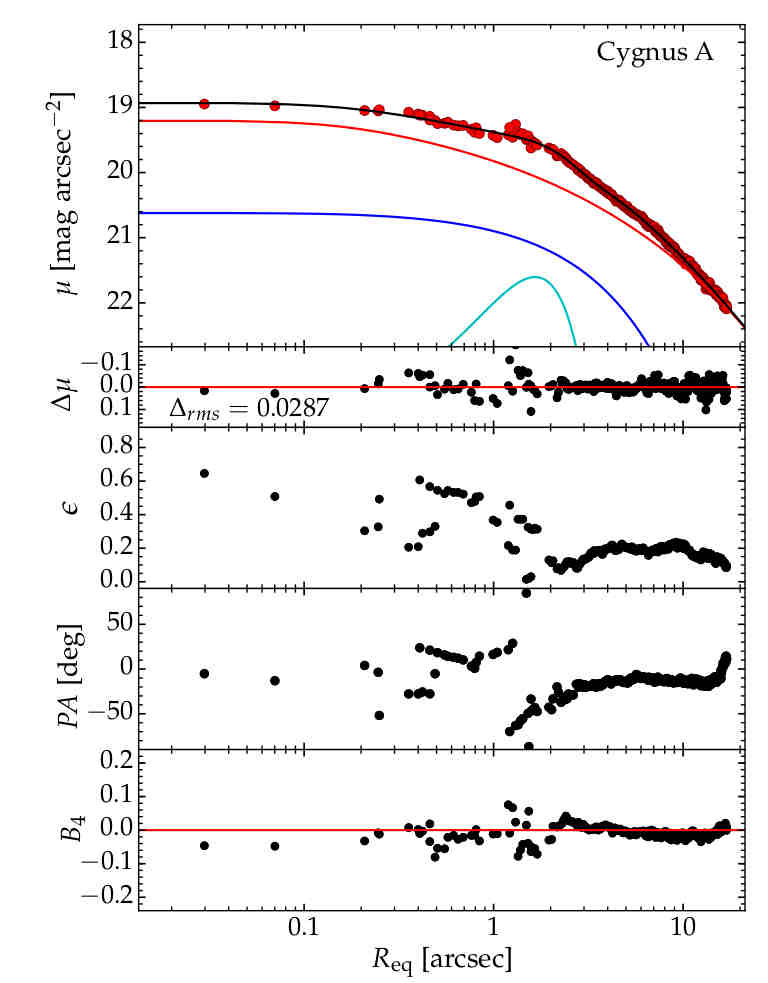}
\caption{\textit{HST} WFPC2/WFC F814W surface brightness profile for Cygnus~A, with a physical scale of 1.1232$\,\text{kpc}\,\text{arcsec}^{-1}$. \textbf{Left two panels}---The model represents $0\arcsec \leq R_{\rm maj} \leq 17\farcs9$ with $\Delta_{\rm rms}=0.0314\,\text{mag\,arcsec}^{-2}$. \underline{S{\'e}rsic Profile Parameters:} \textcolor{red}{$R_e=19\farcs56\pm0\farcs80$, $\mu_e=22.22\pm0.07\,\text{mag\,arcsec}^{-2}$, and $n=1.45\pm0.10$}. \underline{Exponential Profile Parameters:} \textcolor{blue}{$\mu_0 = 19.83\pm0.14\,\text{mag\,arcsec}^{-2}$ and $h = 1\farcs72\pm0\farcs05$.} \underline{Additional Parameters:} one Gaussian component added at: \textcolor{cyan}{$R_{\rm r}=1\farcs82\pm0\farcs02$ with $\mu_0 = 21.51\pm0.11\,\text{mag\,arcsec}^{-2}$ and FWHM = $0\farcs50\pm0\farcs06$.} \textbf{Right two panels}---The model represents $0\arcsec \leq R_{\rm eq} \leq 17\farcs0$ with $\Delta_{\rm rms}=0.0287\,\text{mag\,arcsec}^{-2}$. \underline{S{\'e}rsic Profile Parameters:} \textcolor{red}{$R_e=46\farcs48\pm12\farcs41$, $\mu_e=23.74\pm0.41\,\text{mag\,arcsec}^{-2}$, and $n=2.44\pm0.21$}. \underline{Exponential Profile Parameters:} \textcolor{blue}{$\mu_0 = 20.58\pm0.12\,\text{mag\,arcsec}^{-2}$ and $h = 3\farcs44\pm0\farcs37$.} \underline{Additional Parameters:} one Gaussian component added at: \textcolor{cyan}{$R_{\rm r}=1\farcs66\pm0\farcs07$ with $\mu_0 = 21.59\pm0.10\,\text{mag\,arcsec}^{-2}$ and FWHM = $1\farcs78\pm0\farcs16$.}}
\label{Cygnus_A_plot}
\end{sidewaysfigure}

\subsection{Cygnus~A (Figure~\ref{Cygnus_A_plot})}

It may be that Cygnus~A is an early-type galaxy with an intermediate-scale disk hosting a spiral; see CG 611 \citep{Graham:2017}. We have added a Gaussian component to model spiral structure at $R_{\rm maj} = R_{\rm eq} \approx2\arcsec$.

\begin{sidewaysfigure}
\includegraphics[clip=true,trim= 11mm 1mm 3mm 5mm,width=0.249\textwidth]{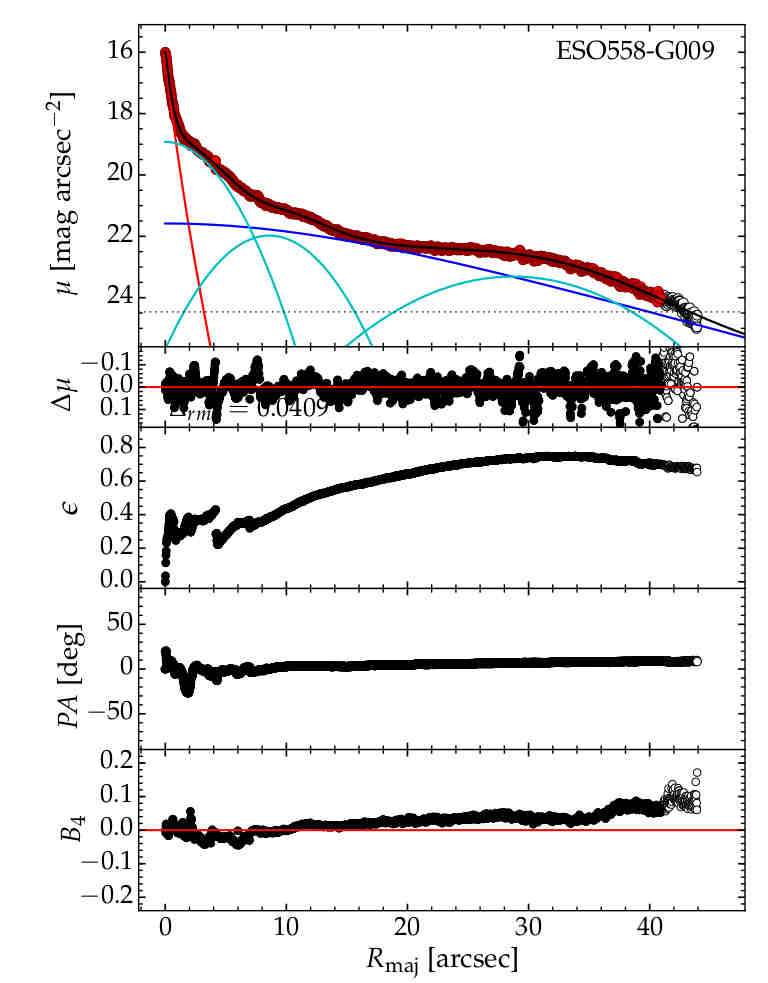}
\includegraphics[clip=true,trim= 11mm 1mm 3mm 5mm,width=0.249\textwidth]{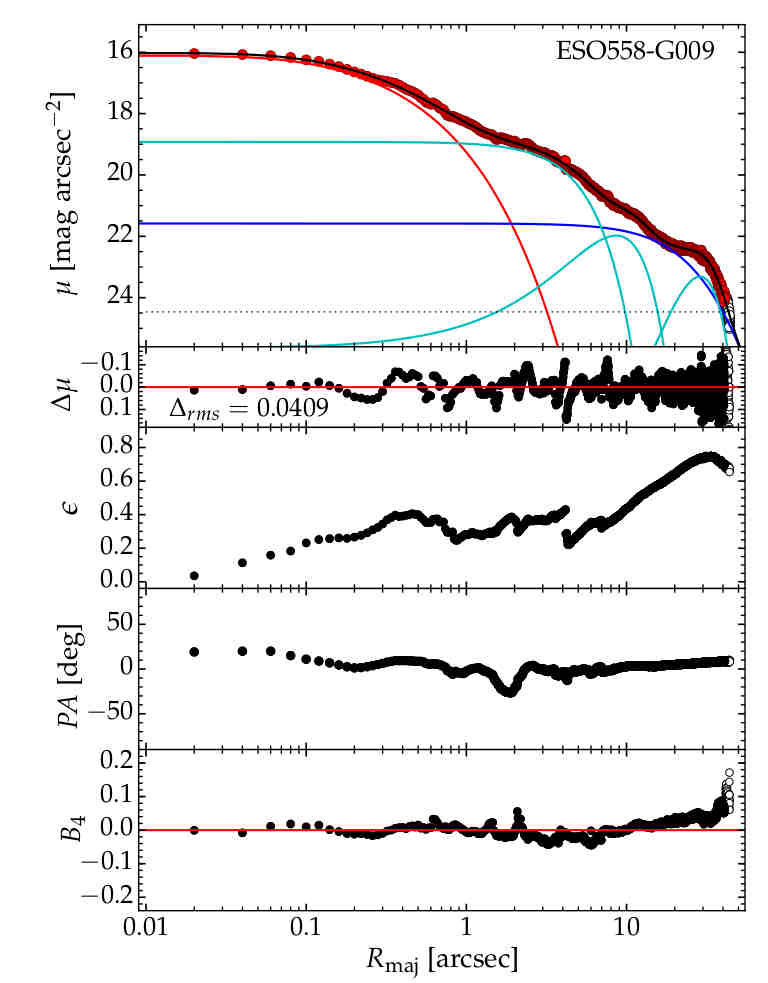}
\includegraphics[clip=true,trim= 11mm 1mm 3mm 5mm,width=0.249\textwidth]{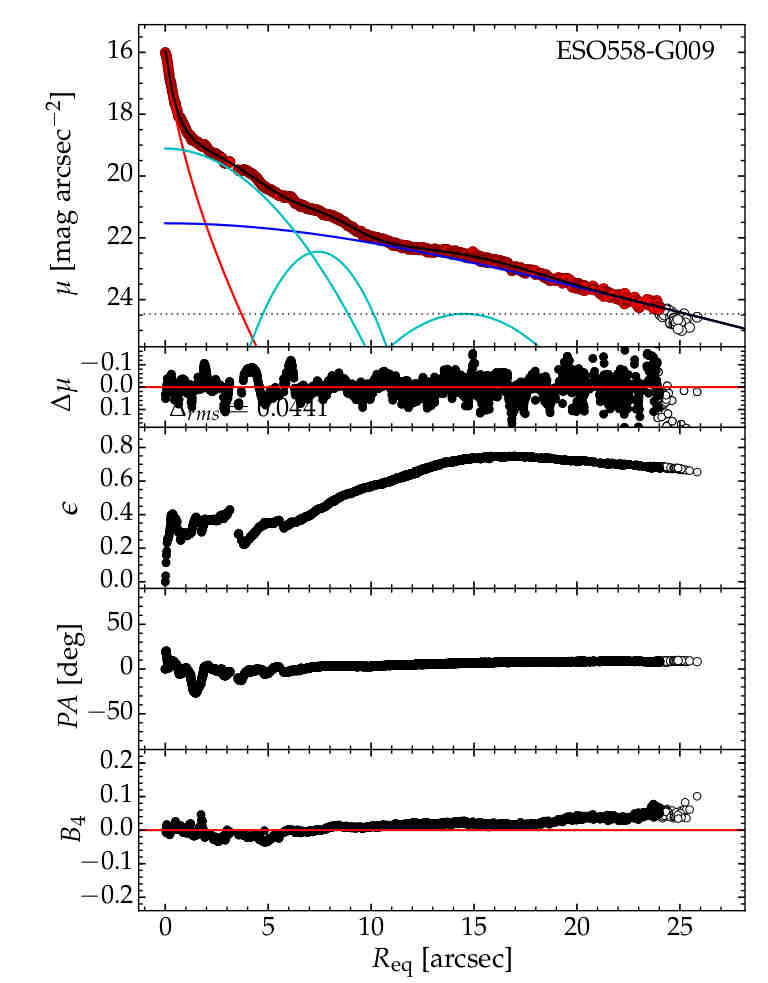}
\includegraphics[clip=true,trim= 11mm 1mm 3mm 5mm,width=0.249\textwidth]{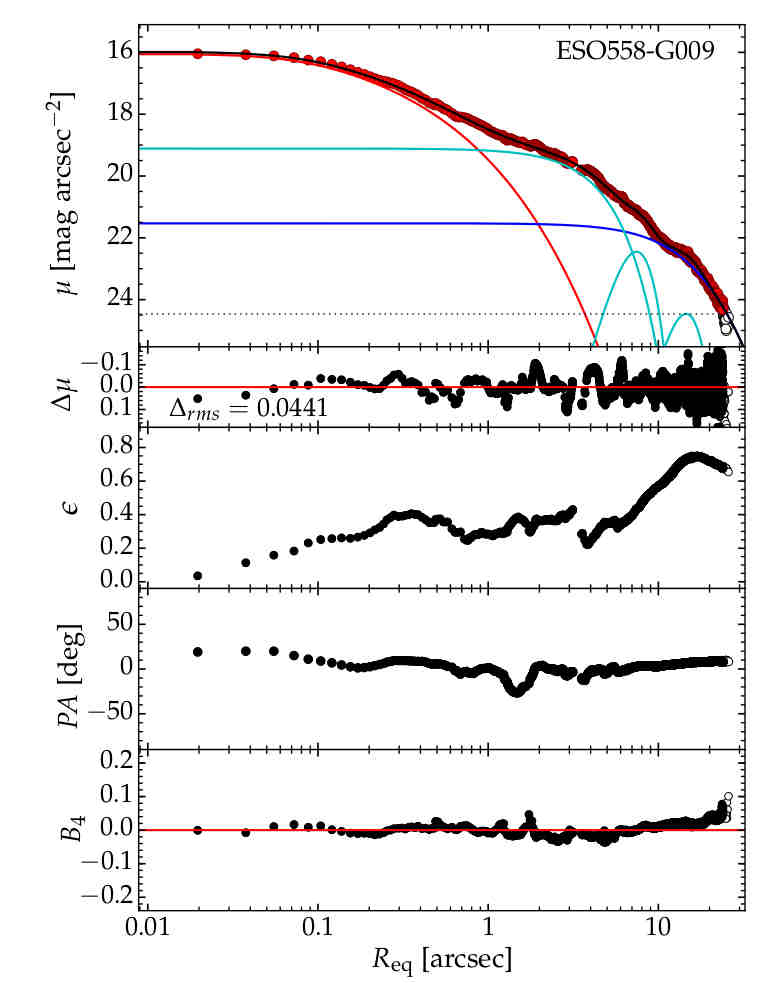}
\caption{\textit{HST} WFC3 UVIS2 F814W surface brightness profile for ESO558-G009, with a physical scale of 0.5318$\,\text{kpc}\,\text{arcsec}^{-1}$. \textbf{Left two panels}---The model represents $0\arcsec \leq R_{\rm maj} \leq 40\farcs90$ with $\Delta_{\rm rms}=0.0409\,\text{mag\,arcsec}^{-2}$ and additional data from $40\farcs90 < R_{\rm maj} \leq 43\farcs92$ is plotted, but not modeled. \underline{S{\'e}rsic Profile Parameters:} \textcolor{red}{$R_e=0\farcs62\pm0\farcs01$, $\mu_e=18.17\pm0.04\,\text{mag\,arcsec}^{-2}$, and $n=1.28\pm0.03$}. \underline{Edge-on Disk Model Parameters:} \textcolor{blue}{$\mu_0 = 21.58\pm0.02\,\text{mag\,arcsec}^{-2}$ and $h_z = 19\farcs95\pm0\farcs41$.} \underline{Additional Parameters:} three Gaussian components added at: \textcolor{cyan}{$R_{\rm r}=0\arcsec$, $8\farcs60\pm0\farcs18$, \& $28\farcs72\pm0\farcs12$; with $\mu_0 = 18.92\pm0.01$, $21.98\pm0.04$, \& $23.31\pm0.04\,\text{mag\,arcsec}^{-2}$; and FWHM = $7\farcs22\pm0\farcs07$, $7\farcs75\pm0\farcs25$, \& $15\farcs87\pm0\farcs43$, respectively.} \textbf{Right two panels}---The model represents $0\arcsec \leq R_{\rm eq} \leq 24\farcs38$ with $\Delta_{\rm rms}=0.0441\,\text{mag\,arcsec}^{-2}$ and additional data from $24\farcs38 < R_{\rm maj} \leq 25\farcs84$ is plotted, but not modeled. \underline{S{\'e}rsic Profile Parameters:} \textcolor{red}{$R_e=0\farcs68\pm0\farcs03$, $\mu_e=18.62\pm0.06\,\text{mag\,arcsec}^{-2}$, and $n=1.63\pm0.05$}. \underline{Edge-on Disk Model Parameters:} \textcolor{blue}{$\mu_0 = 21.53\pm0.05\,\text{mag\,arcsec}^{-2}$ and $h_z = 12\farcs50\pm0\farcs17$.} \underline{Additional Parameters:} three Gaussian components added at: \textcolor{cyan}{$R_{\rm r}=0\arcsec$, $7\farcs43\pm0\farcs06$, \& $14\farcs53\pm0\farcs28$; with $\mu_0 = 19.11\pm0.02$, $22.45\pm0.07$, \& $24.46\pm0.13\,\text{mag\,arcsec}^{-2}$; and FWHM = $6\farcs66\pm0\farcs06$, $3\farcs29\pm0\farcs15$, \& $6\farcs05\pm0\farcs47$, respectively.}}
\label{ESO558-G009_plot}
\end{sidewaysfigure}

\subsection{ESO~558-G009 (Figure~\ref{ESO558-G009_plot})}

A near edge-on galaxy, ESO~558-G009 is inclined by a maximum of $73\fdg4\pm1\fdg6$ with respect to the plane of the sky. Because of this, we use the edge-on disk model. A central Gaussian component has been added to account for the influence from a potential nuclear disk. Two additional Gaussian components have been added at $R_{\rm maj} \approx9\arcsec$ ($R_{\rm eq}\approx7\arcsec$) and $R_{\rm maj} \approx29\arcsec$ ($R_{\rm eq}\approx15\arcsec$) to capture the very broad and elongated influence of the spiral arms. We note that the surface brightness profile for this galaxy, as well as for the other $\rm H_2O$ megamaser host galaxies in our sample, can be compared with those presented in the parent sample of \citet{Greene:2013} and \citet{Pjanka:2017}.\footnote{However, the scale of the semi-major axis length appears to be 50\% too small throughout \citet{Pjanka:2017}.}

\begin{sidewaysfigure}
\includegraphics[clip=true,trim= 11mm 1mm 3mm 5mm,width=0.249\textwidth]{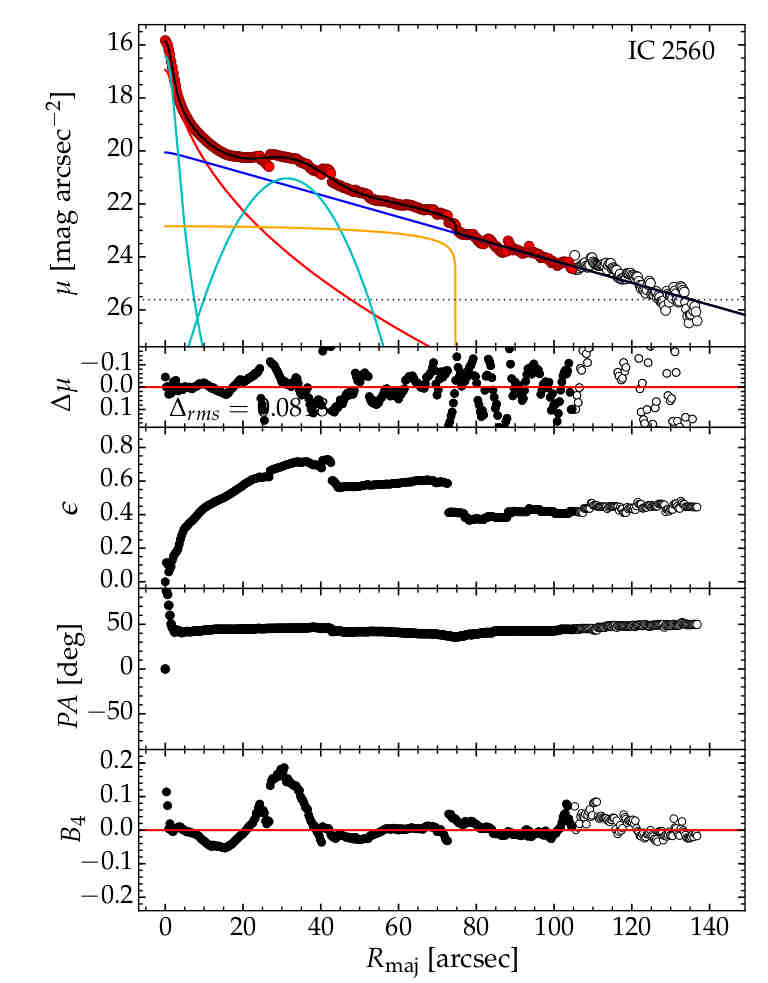}
\includegraphics[clip=true,trim= 11mm 1mm 3mm 5mm,width=0.249\textwidth]{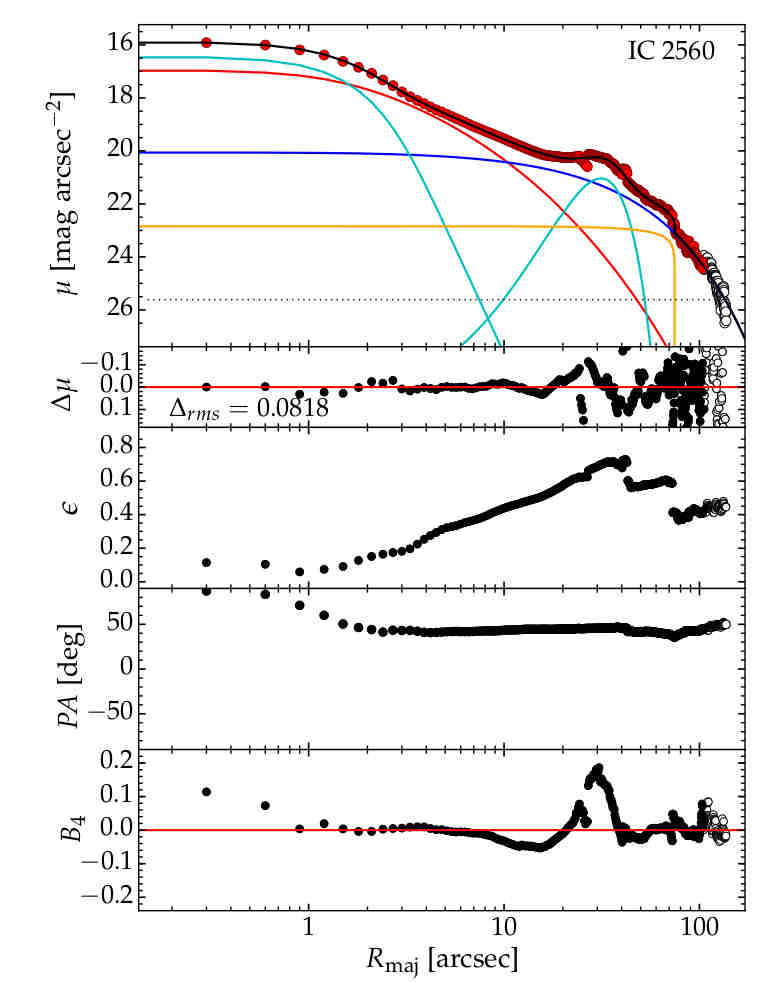}
\includegraphics[clip=true,trim= 11mm 1mm 3mm 5mm,width=0.249\textwidth]{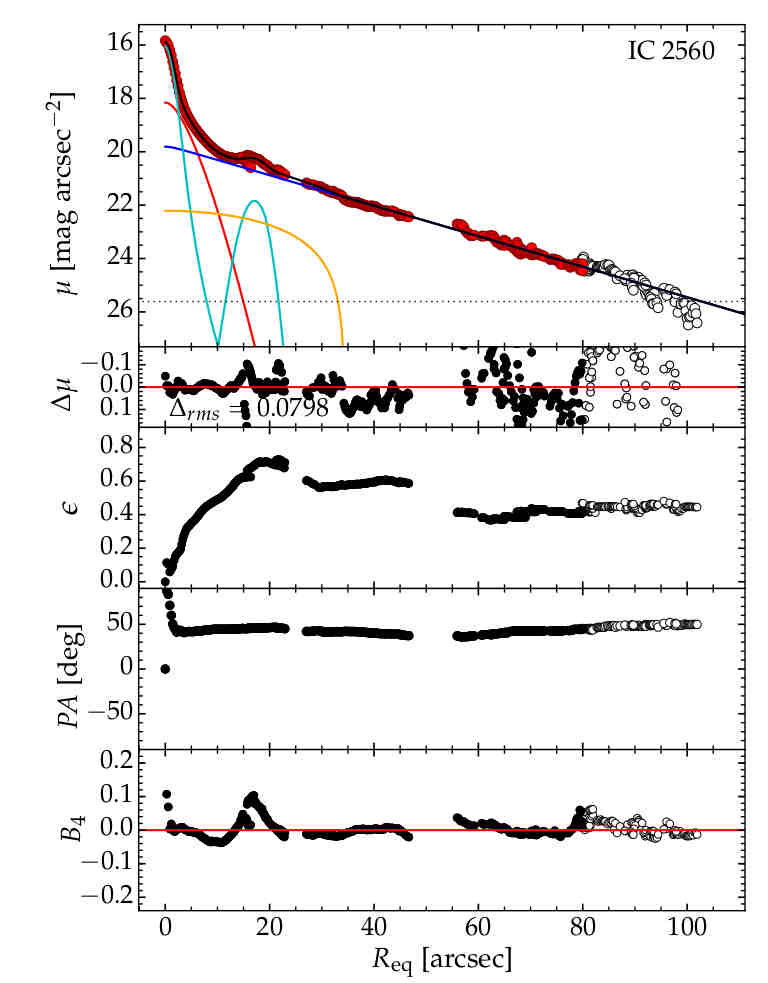}
\includegraphics[clip=true,trim= 11mm 1mm 3mm 5mm,width=0.249\textwidth]{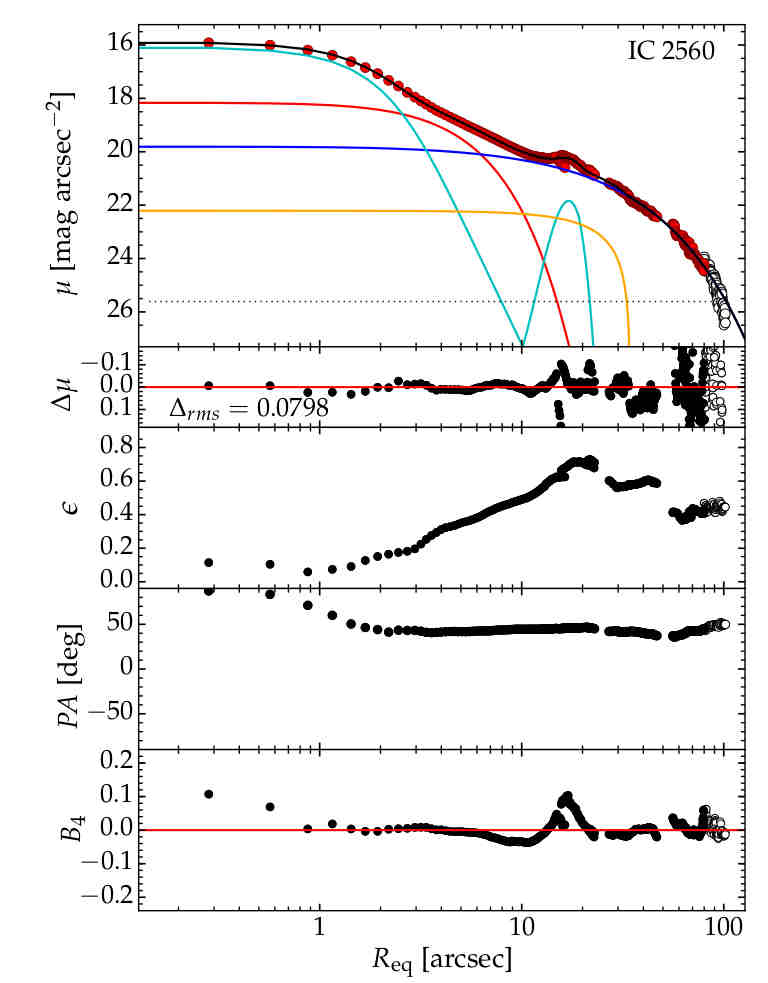}
\caption{\textit{Spitzer} $3.6\,\micron$ surface brightness profile for IC~2560, with a physical scale of 0.1503$\,\text{kpc}\,\text{arcsec}^{-1}$. \textbf{Left two panels}---The model represents $0\arcsec \leq R_{\rm maj} \leq 105\arcsec$ with $\Delta_{\rm rms}=0.0818\,\text{mag\,arcsec}^{-2}$ and additional data from $105\arcsec < R_{\rm maj} \leq 137\arcsec$ is plotted, but not modeled. \underline{S{\'e}rsic Profile Parameters:} \textcolor{red}{$R_e=7\farcs15\pm1\farcs19$, $\mu_e=19.64\pm0.31\,\text{mag\,arcsec}^{-2}$, and $n=2.27\pm0.84$}. \underline{Ferrers Profile Parameters:} \textcolor{Orange}{$\mu_0 = 22.84\pm0.09\,\text{mag\,arcsec}^{-2}$, $R_{\rm end} = 74\farcs75\pm0\farcs10$, and $\alpha = 0.63\pm0.09$}. \underline{Exponential Profile Parameters:} \textcolor{blue}{$\mu_0 = 20.00\pm0.11\,\text{mag\,arcsec}^{-2}$ and $h = 26\farcs15\pm0\farcs68$.} \underline{Additional Parameters:} two Gaussian components added at: \textcolor{cyan}{$R_{\rm r}=0\arcsec$ \& $31\farcs32\pm0\farcs31$; with $\mu_0 = 15.27\pm0.30$ \& $21.04\pm0.04\,\text{mag\,arcsec}^{-2}$; and FWHM = $1\farcs71\pm0\farcs18$ \& $17\farcs08\pm0\farcs64$, respectively.} \textbf{Right two panels}---The model represents $0\arcsec \leq R_{\rm eq} \leq 80\arcsec$ with $\Delta_{\rm rms}=0.0520\,\text{mag\,arcsec}^{-2}$ and additional data from $80\arcsec < R_{\rm maj} \leq 102\arcsec$ is plotted, but not modeled. \underline{S{\'e}rsic Profile Parameters:} \textcolor{red}{$R_e=3\farcs92\pm0\farcs26$, $\mu_e=19.07\pm0.15\,\text{mag\,arcsec}^{-2}$, and $n=0.68\pm0.20$}. \underline{Ferrers Profile Parameters:} \textcolor{Orange}{$\mu_0 = 22.21\pm0.40\,\text{mag\,arcsec}^{-2}$, $R_{\rm end} = 34\farcs38\pm2\farcs31$, and $\alpha = 3.00\pm1.56$}. \underline{Exponential Profile Parameters:} \textcolor{blue}{$\mu_0 = 19.74\pm0.03\,\text{mag\,arcsec}^{-2}$ and $h = 19\farcs0\pm0\farcs15$.} \underline{Additional Parameters:} two Gaussian components added at: \textcolor{cyan}{$R_{\rm r}=0\arcsec$ \& $17\farcs07\pm0\farcs25$; with $\mu_0 = 15.00\pm0.12$ \& $21.66\pm0.14\,\text{mag\,arcsec}^{-2}$; and FWHM = $1\farcs63\pm0\farcs16$ \& $4\farcs06\pm0\farcs74$, respectively.}  We note that a plausible alternative for the decomposition can be seen in Figure~8 of \citet{Saha:2018}, but the bulge magnitude would be similar in either scenario.}
\label{IC2560_plot}
\end{sidewaysfigure}

\subsection{IC~2560 (Figure~\ref{IC2560_plot})}

The galaxy IC~2560 possesses a Type 2 Seyfert AGN \citep{Veron-Cetty:2006}, which we have modeled with a central Gaussian. We have added an additional Gaussian at $R_{\rm maj} \approx31\arcsec$ ($R_{\rm eq}\approx17\arcsec$) to account for a thickened bar or ``peanut.''

\begin{sidewaysfigure}
\includegraphics[clip=true,trim= 11mm 1mm 3mm 5mm,width=0.249\textwidth]{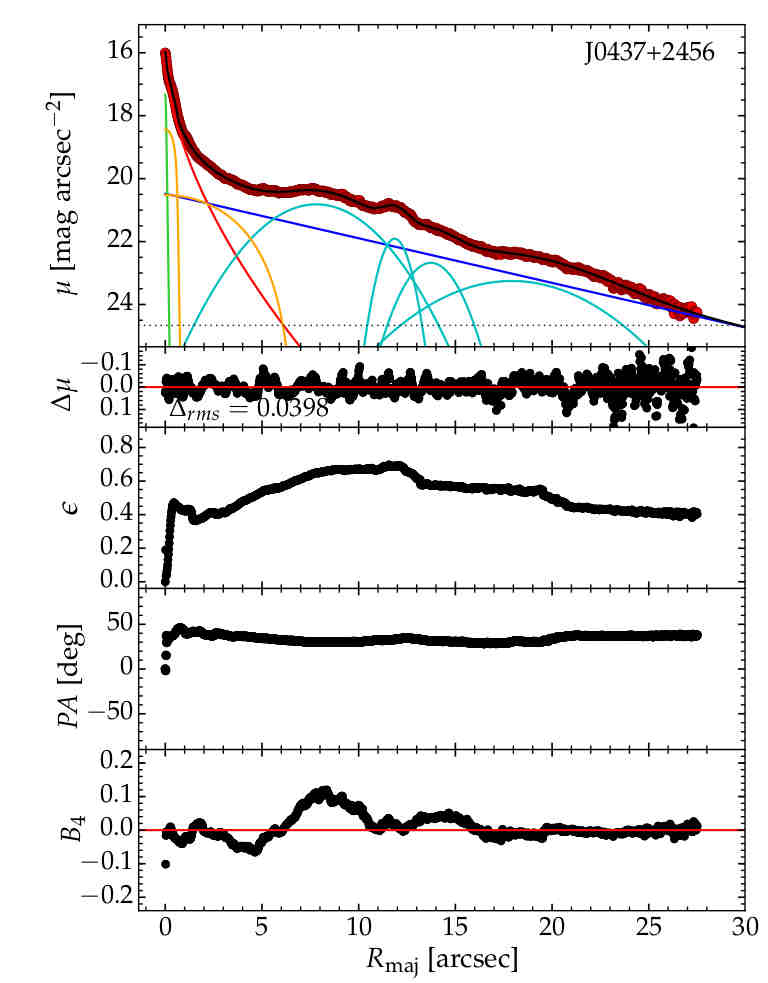}
\includegraphics[clip=true,trim= 11mm 1mm 3mm 5mm,width=0.249\textwidth]{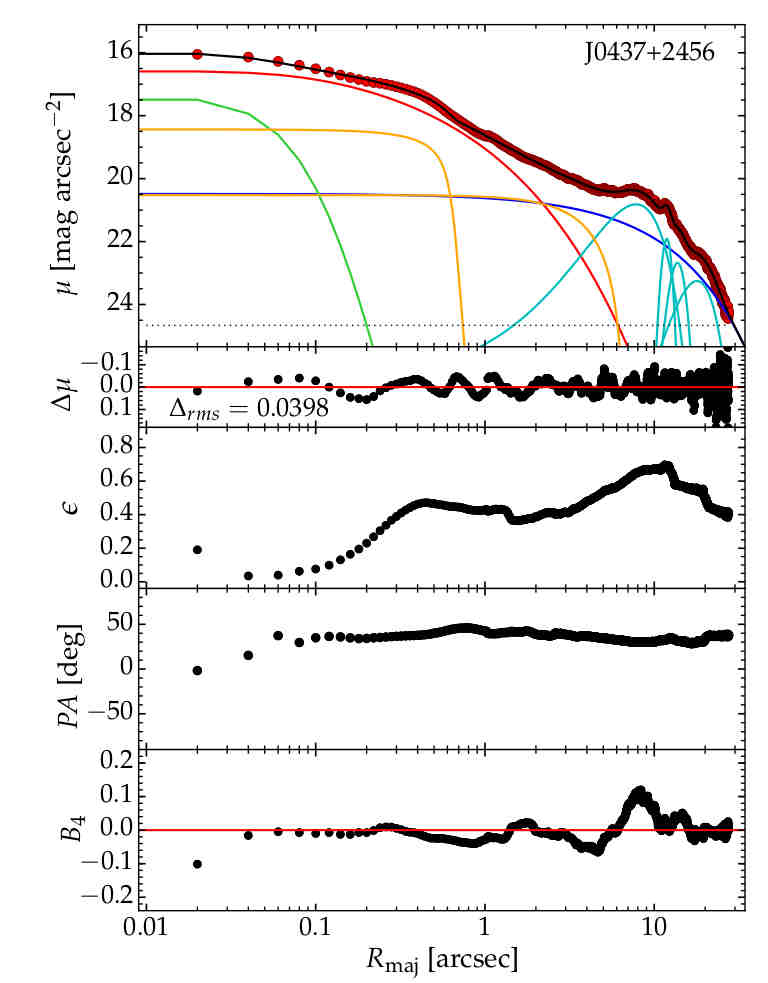}
\includegraphics[clip=true,trim= 11mm 1mm 3mm 5mm,width=0.249\textwidth]{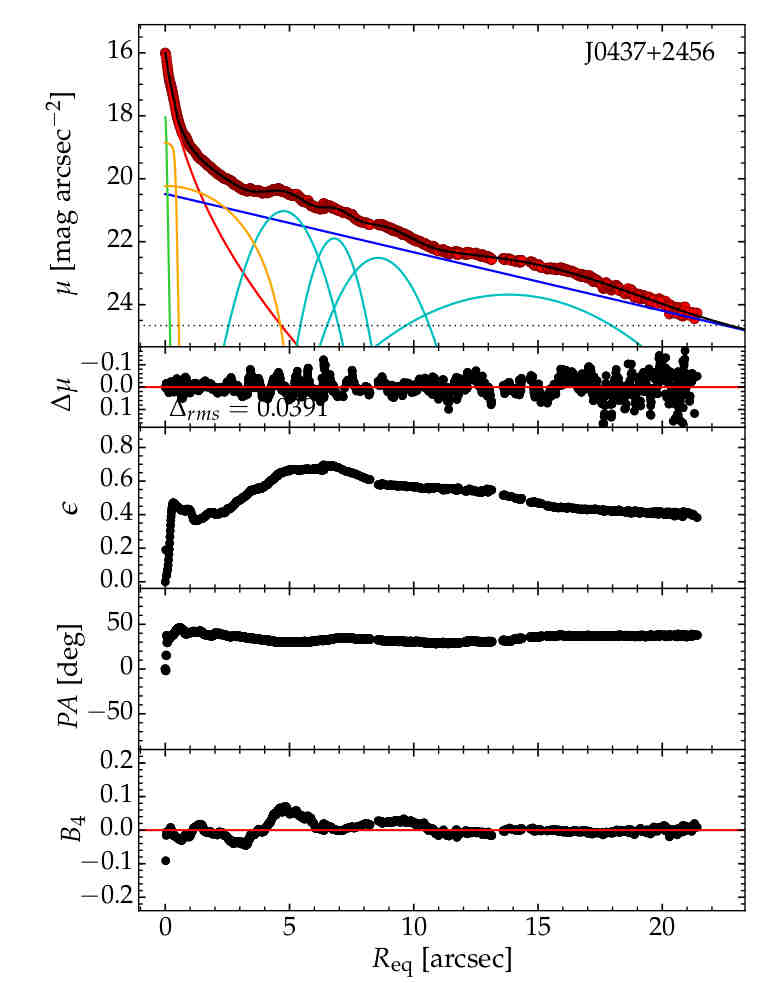}
\includegraphics[clip=true,trim= 11mm 1mm 3mm 5mm,width=0.249\textwidth]{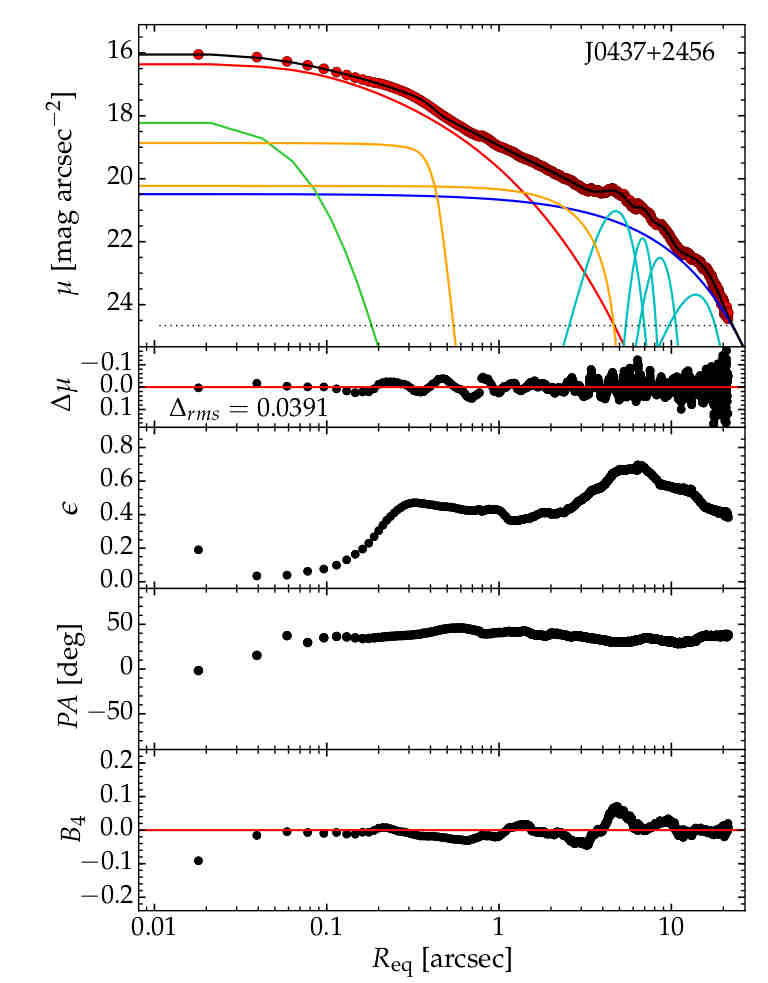}
\caption{\textit{HST} WFC3 UVIS2 F814W surface brightness profile for J0437+2456, with a physical scale of 0.3417$\,\text{kpc}\,\text{arcsec}^{-1}$. \textbf{Left two panels}---The model represents $0\arcsec \leq R_{\rm maj} \leq 27\farcs5$ with $\Delta_{\rm rms}=0.0398\,\text{mag\,arcsec}^{-2}$. \underline{Point Source:} \textcolor{LimeGreen}{$\mu_0 = 17.33\pm0.17\,\text{mag\,arcsec}^{-2}$.} \underline{S{\'e}rsic Profile Parameters:} \textcolor{red}{$R_e=1\farcs22\pm0\farcs13$, $\mu_e=19.42\pm0.15\,\text{mag\,arcsec}^{-2}$, and $n=1.73\pm0.12$}. \underline{Ferrers Profile Parameters:} \textcolor{Orange}{$\mu_0 = 18.43\pm0.18$ \& $20.52\pm0.16\,\text{mag\,arcsec}^{-2}$; $R_{\rm end} = 0\farcs63\pm0\farcs04$ \& $6\farcs76\pm0\farcs44$; and $\alpha = 1.63\pm0.80$ \& $5.73\pm1.18$, respectively.} \underline{Exponential Profile Parameters:} \textcolor{blue}{$\mu_0 = 20.48\pm0.14\,\text{mag\,arcsec}^{-2}$ and $h = 7\farcs66\pm0\farcs03$.} \underline{Additional Parameters:} four Gaussian components added at: \textcolor{cyan}{$R_{\rm r}=7\farcs80\pm0\farcs04$, $11\farcs86\pm0\farcs02$, $13\farcs73\pm0\farcs07$, \& $17\farcs91\pm0\farcs09$; with $\mu_0 = 20.82\pm0.01$, $21.91\pm0.04$, $22.68\pm0.04$, \& $23.25\pm0.02\,\text{mag\,arcsec}^{-2}$; and FWHM = $5\farcs60\pm0\farcs13$, $1\farcs48\pm0\farcs05$, $2\farcs77\pm0\farcs12$, \& $8\farcs40\pm0\farcs17$, respectively.} \textbf{Right two panels}---The model represents $0\arcsec \leq R_{\rm maj} \leq 21\farcs5$ with $\Delta_{\rm rms}=0.0391\,\text{mag\,arcsec}^{-2}$. \underline{Point Source:} \textcolor{LimeGreen}{$\mu_0 = 18.05\pm0.51\,\text{mag\,arcsec}^{-2}$.} \underline{S{\'e}rsic Profile Parameters:} \textcolor{red}{$R_e=0\farcs87\pm0\farcs15$, $\mu_e=19.40\pm0.25\,\text{mag\,arcsec}^{-2}$, and $n=1.97\pm0.23$}. \underline{Ferrers Profile Parameters:} \textcolor{Orange}{$\mu_0 = 18.86\pm0.47$ \& $20.23\pm0.22\,\text{mag\,arcsec}^{-2}$; $R_{\rm end} = 0\farcs41\pm0\farcs37$ \& $5\farcs25\pm1\farcs15$; and $\alpha = 0.67\pm2.42$ \& $6.76\pm3.55$, respectively.} \underline{Exponential Profile Parameters:} \textcolor{blue}{$\mu_0 = 20.48\pm0.14\,\text{mag\,arcsec}^{-2}$ and $h = 5\farcs84\pm0\farcs17$.} \underline{Additional Parameters:} four Gaussian components added at: \textcolor{cyan}{$R_{\rm r}=4\farcs76\pm0\farcs08$, $6\farcs79\pm0\farcs03$, $8\farcs57\pm0\farcs08$, \& $13\farcs82\pm0\farcs30$; with $\mu_0 = 21.03\pm0.09$, $21.90\pm0.06$, $22.51\pm0.11$, \& $23.68\pm0.13\,\text{mag\,arcsec}^{-2}$; and FWHM = $2\farcs01\pm0\farcs10$, $1\farcs41\pm0\farcs07$, $2\farcs44\pm0\farcs16$, \& $7\farcs25\pm0\farcs43$, respectively.}}
\label{J0437+2456_plot}
\end{sidewaysfigure}

\subsection{J0437+2456 (Figure~\ref{J0437+2456_plot})}

We have added four Gaussian components at $R_{\rm maj} \approx 8\arcsec$ ($R_{\rm eq}\approx 5\arcsec$), $R_{\rm maj} \approx12\arcsec$ ($R_{\rm eq}\approx 7\arcsec$), $R_{\rm maj} \approx 14\arcsec$ ($R_{\rm eq}\approx 9\arcsec$), and $R_{\rm maj} \approx 18\arcsec$ ($R_{\rm eq}\approx 14\arcsec$) to account for spiral-arm crossings.

\subsection{Milky Way}

We do not attempt to model the light profile of the Milky Way. Instead, we adopt the S{\'e}rsic profile parameters and the bulge absolute magnitude from \citet{Okamoto:2013}; the stellar spheroid mass estimate comes from \citet{Licquia:Newman:2015}.

\begin{sidewaysfigure}
\includegraphics[clip=true,trim= 11mm 1mm 3mm 5mm,width=0.249\textwidth]{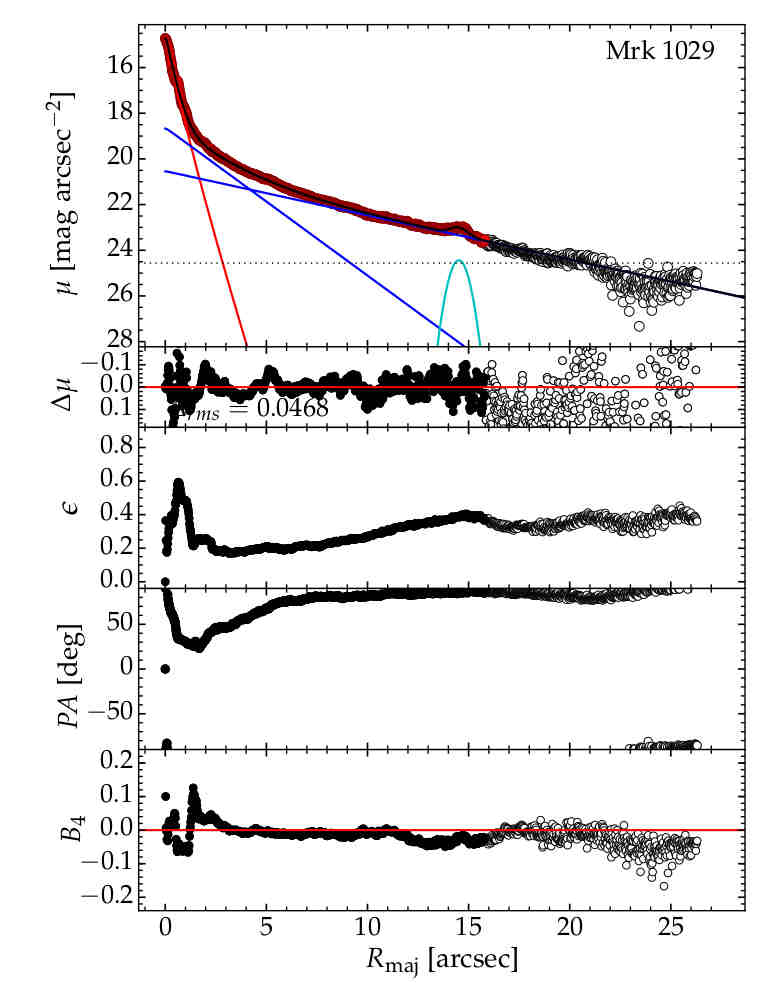}
\includegraphics[clip=true,trim= 11mm 1mm 3mm 5mm,width=0.249\textwidth]{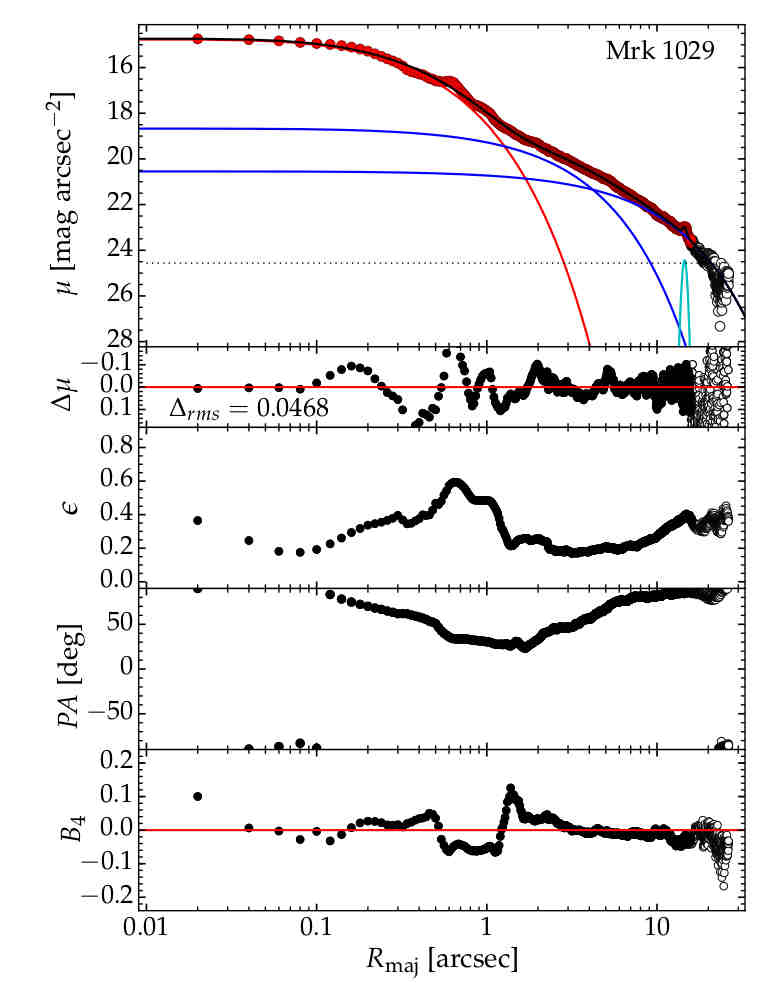}
\includegraphics[clip=true,trim= 11mm 1mm 3mm 5mm,width=0.249\textwidth]{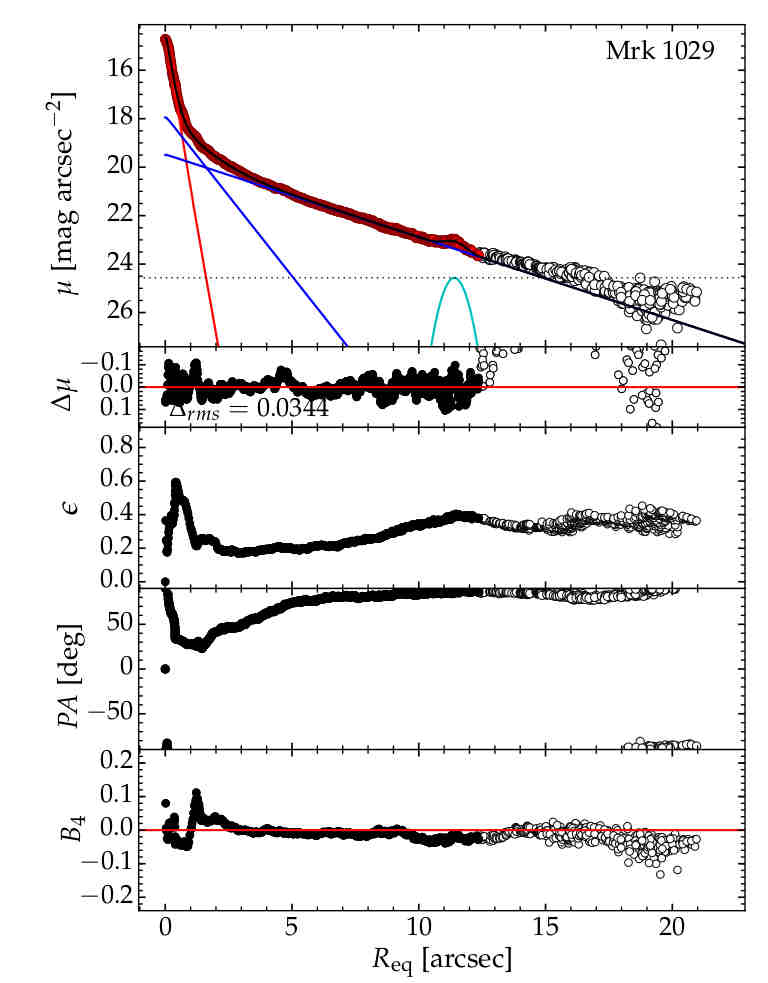}
\includegraphics[clip=true,trim= 11mm 1mm 3mm 5mm,width=0.249\textwidth]{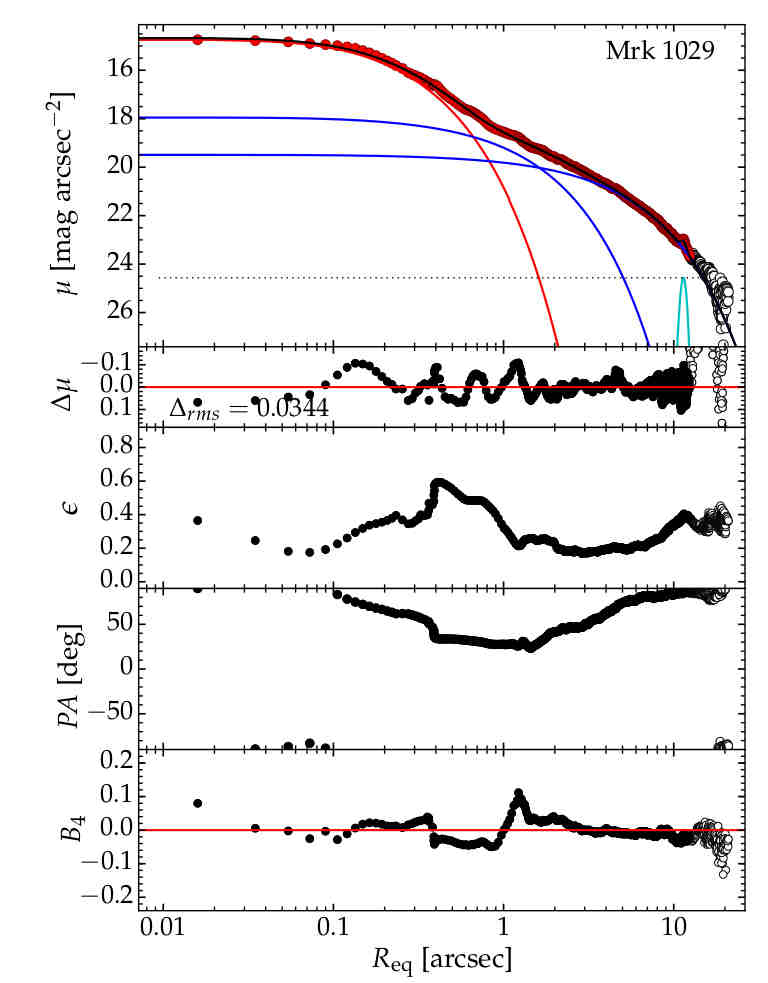}
\caption{\textit{HST} WFC3 UVIS2 F814W surface brightness profile for Mrk~1029, with a physical scale of 0.6254$\,\text{kpc}\,\text{arcsec}^{-1}$. \textbf{Left two panels}---The model represents $0\arcsec \leq R_{\rm maj} \leq 15\farcs8$ with $\Delta_{\rm rms}=0.0468\,\text{mag\,arcsec}^{-2}$ and additional data from $15\farcs8 < R_{\rm maj} \leq 26\farcs3$ is plotted, but not modeled. \underline{S{\'e}rsic Profile Parameters:} \textcolor{red}{$R_e=0\farcs47\pm0\farcs00$, $\mu_e=16.53\pm0.02\,\text{mag\,arcsec}^{-2}$, and $n=1.15\pm0.02$.} \underline{Exponential Profile Parameters:} \textcolor{blue}{$\mu_0 = 18.63\pm0.03$ \& $20.53\pm0.04\,\text{mag\,arcsec}^{-2}$ and $h = 1\farcs68\pm0\farcs04$ \& $5\farcs61\pm0\farcs10$, respectively.} \underline{Additional Parameters:} one Gaussian component added at: \textcolor{cyan}{$R_{\rm r}=14\farcs52\pm0\farcs01$, with $\mu_0 = 24.44\pm0.04$, and FWHM = $0\farcs95\pm0\farcs04$.} \textbf{Right two panels}---The model represents $0\arcsec \leq R_{\rm maj} \leq 12\farcs4$ with $\Delta_{\rm rms}=0.0344\,\text{mag\,arcsec}^{-2}$ and additional data from $12\farcs4 < R_{\rm maj} \leq 21\farcs0$ is plotted, but not modeled. \underline{S{\'e}rsic Profile Parameters:} \textcolor{red}{$R_e=0\farcs28\pm0\farcs00$, $\mu_e=16.29\pm0.02\,\text{mag\,arcsec}^{-2}$, and $n=1.07\pm0.02$.} \underline{Exponential Profile Parameters:} \textcolor{blue}{$\mu_0 = 17.88\pm0.04$ \& $19.47\pm0.01\,\text{mag\,arcsec}^{-2}$ and $h = 0\farcs82\pm0\farcs01$ \& $3\farcs18\pm0\farcs01$, respectively.} \underline{Additional Parameters:} one Gaussian component added at: \textcolor{cyan}{$R_{\rm r}=11\farcs40\pm0\farcs01$, with $\mu_0 = 24.57\pm0.03$, and FWHM = $0\farcs95\pm0\farcs03$.}}
\label{Mrk1029_plot}
\end{sidewaysfigure}

\subsection{Mrk~1029 (Figure~\ref{Mrk1029_plot})}

We have added one Gaussian component at $R_{\rm maj} \approx 15\arcsec$ ($R_{\rm eq}\approx 11\arcsec$) to account for faint spiral arms. We model Mrk~1029 with an embedded central disk plus two exponential components to the surface brightness profile to model the intermediate-scale disk.

\subsection{NGC~224 (Figure~17 from \citealt{Savorgnan:2016})}

Due to the large apparent size of NGC~224 (M31), ``the Andromeda galaxy,'' we did not model its structure. We instead refer to \citet{Savorgnan:2016}, who constructed their own mosaic with $3.6\,\micron$ imaging from the \textit{Spitzer Space Telescope}. We adopt their structural parameters, along with the subsequent spheroid stellar mass from \citet{Savorgnan:2016:II}.

\begin{sidewaysfigure}
\includegraphics[clip=true,trim= 11mm 1mm 3mm 5mm,width=0.249\textwidth]{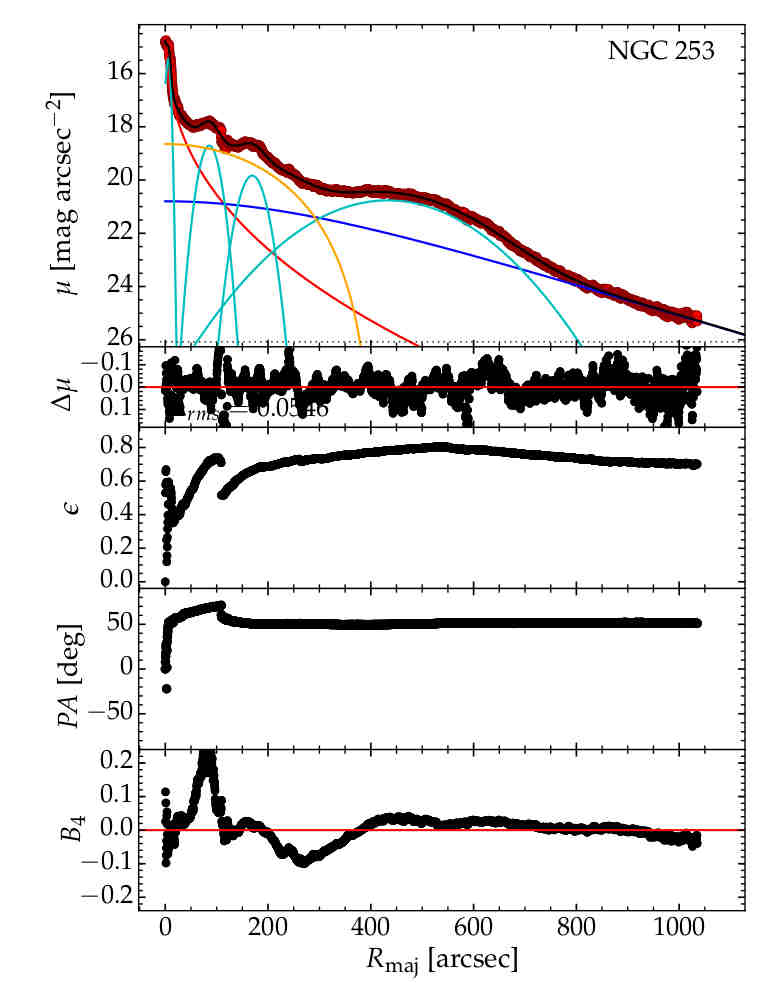}
\includegraphics[clip=true,trim= 11mm 1mm 3mm 5mm,width=0.249\textwidth]{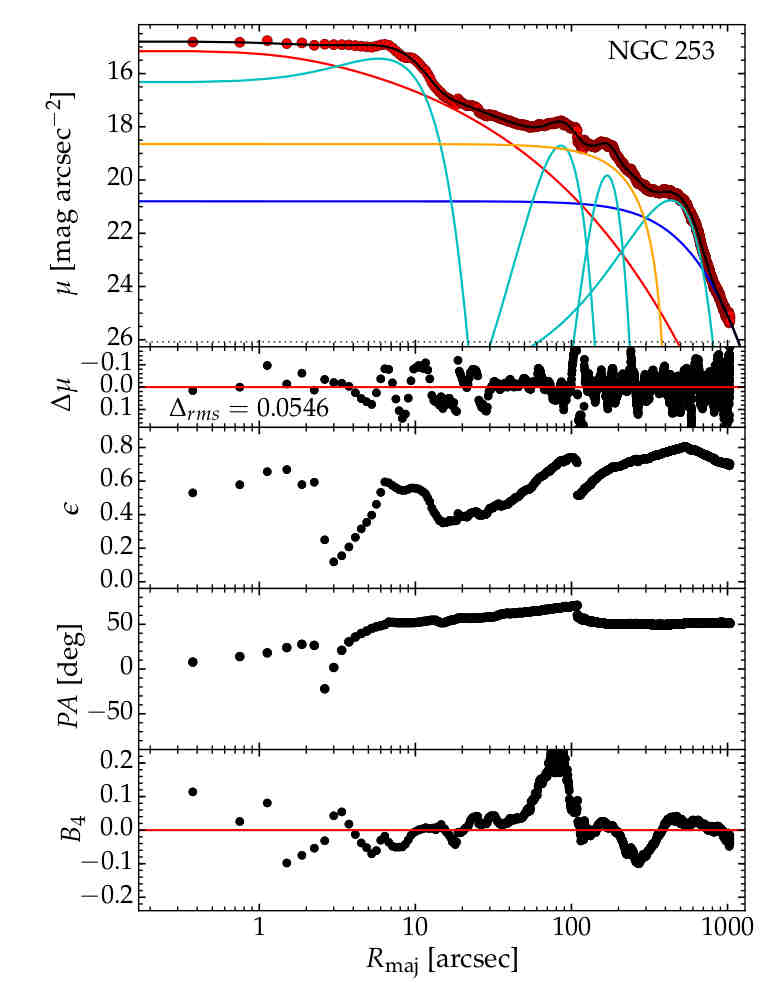}
\includegraphics[clip=true,trim= 11mm 1mm 3mm 5mm,width=0.249\textwidth]{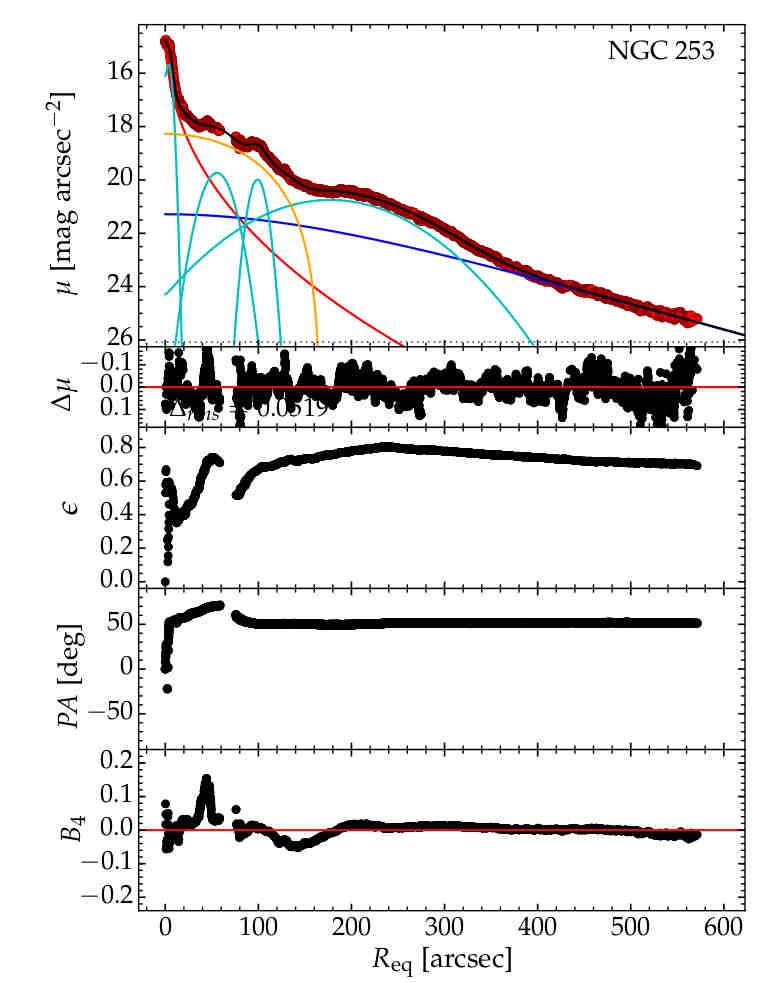}
\includegraphics[clip=true,trim= 11mm 1mm 3mm 5mm,width=0.249\textwidth]{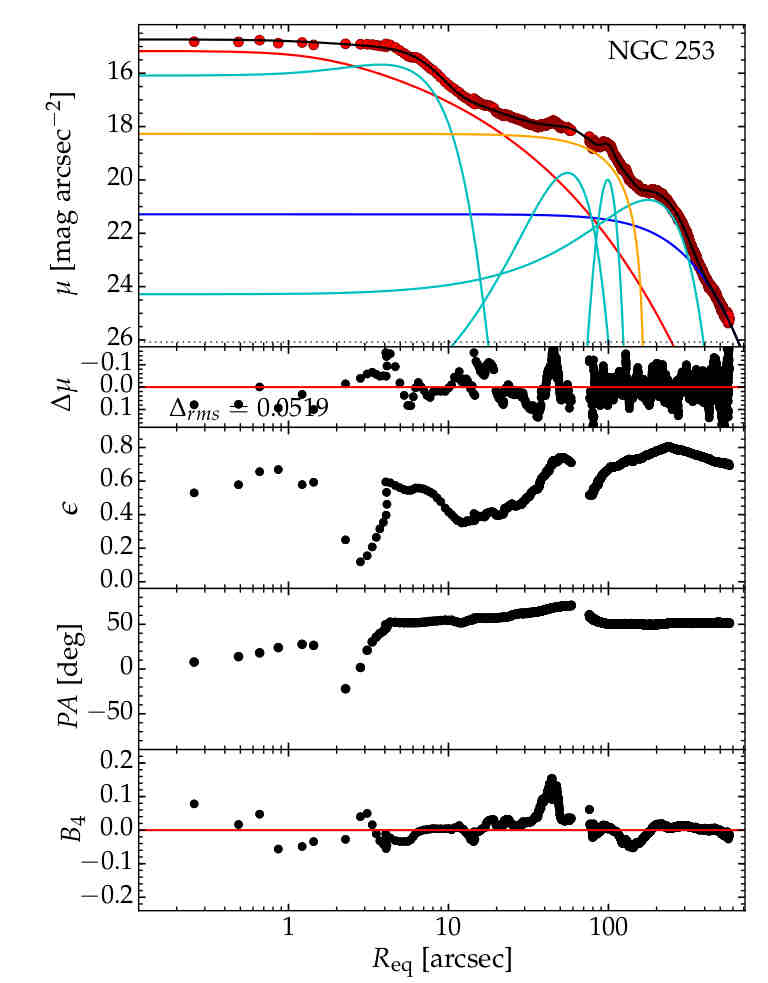}
\caption{\textit{Spitzer} $3.6\,\micron$ surface brightness profile for NGC~253, with a physical scale of 0.0168$\,\text{kpc}\,\text{arcsec}^{-1}$. \textbf{Left two panels}---The model represents $0\arcsec \leq R_{\rm maj} \leq 1035\arcsec$ with $\Delta_{\rm rms}=0.0546\,\text{mag\,arcsec}^{-2}$. \underline{S{\'e}rsic Profile Parameters:} \textcolor{red}{$R_e=55\farcs55\pm1\farcs82$, $\mu_e=18.59\pm0.06\,\text{mag\,arcsec}^{-2}$, and $n=2.53\pm0.08$}. \underline{Ferrers Profile Parameters:} \textcolor{Orange}{$\mu_0 = 18.65\pm0.02\,\text{mag\,arcsec}^{-2}$, $R_{\rm end} = 412\farcs21\pm7\farcs00$, and $\alpha = 9.16\pm0.43$.} \underline{Edge-on Disk Model Parameters:} \textcolor{blue}{$\mu_0 = 20.80\pm0.01\,\text{mag\,arcsec}^{-2}$ and $h_z = 375\farcs56\pm0\farcs80$.} \underline{Additional Parameters:} four Gaussian components added at: \textcolor{cyan}{$R_{\rm r}=5\farcs91\pm0\farcs17$, $85\farcs65\pm0\farcs24$, $169\farcs19\pm0\farcs42$, \& $433\farcs00\pm1\farcs30$; with $\mu_0 = 15.39\pm0.03$, $18.71\pm0.02$, $19.83\pm0.02$, \& $20.77\pm0.01\,\text{mag\,arcsec}^{-2}$; and FWHM = $7\farcs88\pm0\farcs28$, $35\farcs21\pm0\farcs71$, $45\farcs84\pm1\farcs06$, \& $279\farcs64\pm1\farcs67$, respectively.} \textbf{Right two panels}---The model represents $0\arcsec \leq R_{\rm eq} \leq 572\arcsec$ with $\Delta_{\rm rms}=0.0519\,\text{mag\,arcsec}^{-2}$. \underline{S{\'e}rsic Profile Parameters:} \textcolor{red}{$R_e=27\farcs89\pm0\farcs71$, $\mu_e=17.99\pm0.05\,\text{mag\,arcsec}^{-2}$, and $n=2.33\pm0.08$.} \underline{Ferrers Profile Parameters:} \textcolor{Orange}{$\mu_0 = 18.26\pm0.01\,\text{mag\,arcsec}^{-2}$, $R_{\rm end} = 167\farcs59\pm1\farcs30$, and $\alpha = 5.91\pm0.16$}. \underline{Edge-on Disk Model Parameters:} \textcolor{blue}{$\mu_0 = 21.29\pm0.01\,\text{mag\,arcsec}^{-2}$ and $h_z = 223\farcs88\pm0\farcs49$.} \underline{Additional Parameters:} four Gaussian components added at: \textcolor{cyan}{$R_{\rm r}=3\farcs85\pm0\farcs22$, $55\farcs54\pm0\farcs49$, $99\farcs04\pm0\farcs21$, \& $177\farcs09\pm0\farcs66$; with $\mu_0 = 15.59\pm0.04$, $19.77\pm0.05$, $20.03\pm0.03$, \& $20.76\pm0.01\,\text{mag\,arcsec}^{-2}$; and FWHM = $6\farcs77\pm0\farcs32$, $30\farcs57\pm1\farcs44$, $17\farcs60\pm0\farcs55$, \& $162\farcs34\pm0\farcs66$, respectively.}}
\label{NGC0253_plot}
\end{sidewaysfigure}

\subsection{NGC~253 (Figure~\ref{NGC0253_plot})}

A near edge-on galaxy, NGC~253 (the ``Sculptor Galaxy'') is  inclined by $75\fdg3\pm2\fdg0$ with respect to the plane of the sky. Because of this, we use the edge-on disk model. Four Gaussian components have been added for: the inner ring at $R_{\rm maj} \approx6\arcsec$ ($R_{\rm eq}\approx4\arcsec$) and crossings of the large $m=2$ grand design spiral arms\footnote{Here $m$ is the harmonic mode (i.e., the number of symmetric spiral arms).} along the major axis at $R_{\rm maj} \approx86\arcsec$ ($R_{\rm eq}\approx56\arcsec$), $R_{\rm maj} \approx169\arcsec$ ($R_{\rm eq}\approx99\arcsec$), and $R_{\rm maj} \approx433\arcsec$ ($R_{\rm eq}\approx177\arcsec$).

\begin{sidewaysfigure}
\includegraphics[clip=true,trim= 11mm 1mm 1mm 5mm,width=0.249\textwidth]{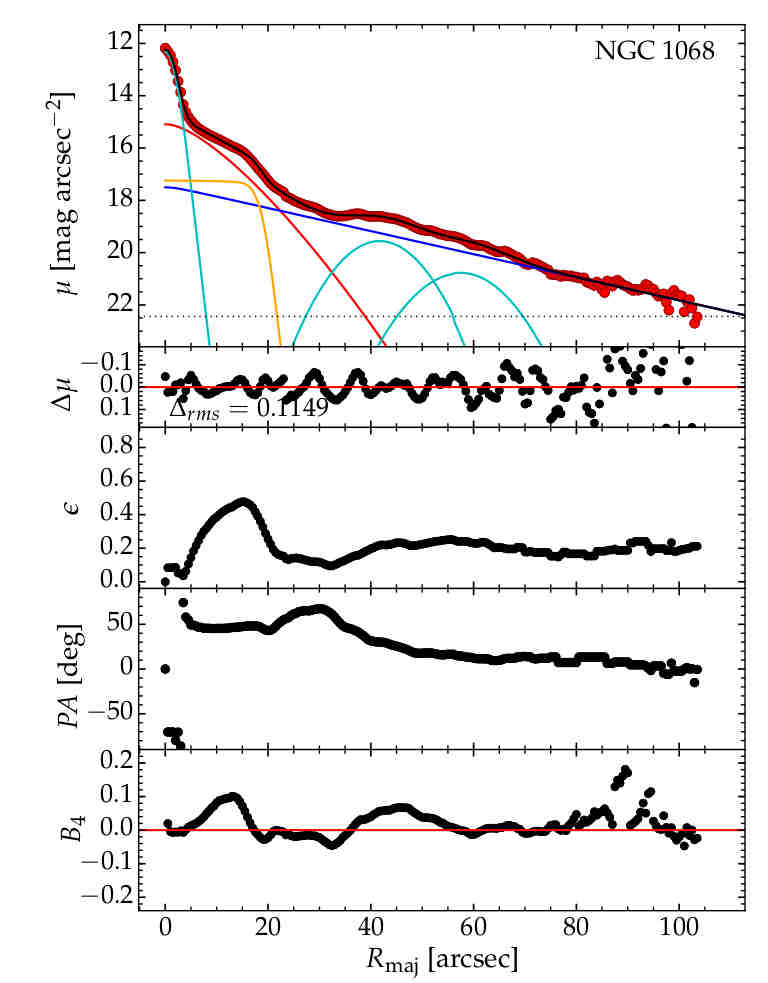}
\includegraphics[clip=true,trim= 11mm 1mm 1mm 5mm,width=0.249\textwidth]{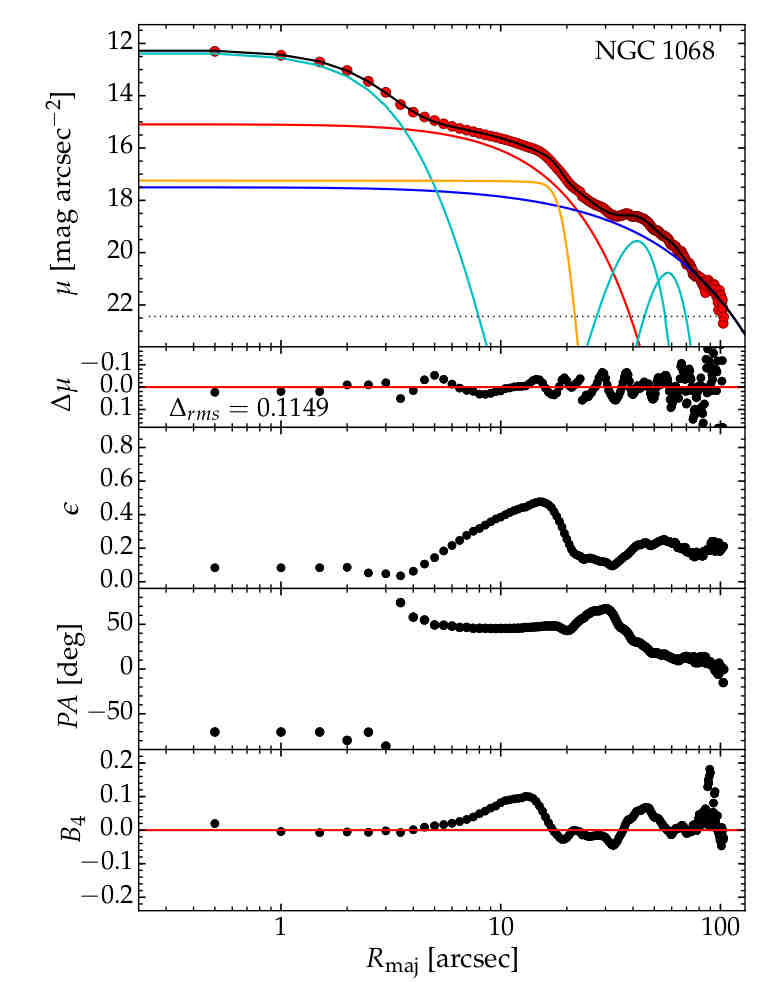}
\includegraphics[clip=true,trim= 11mm 1mm 1mm 5mm,width=0.249\textwidth]{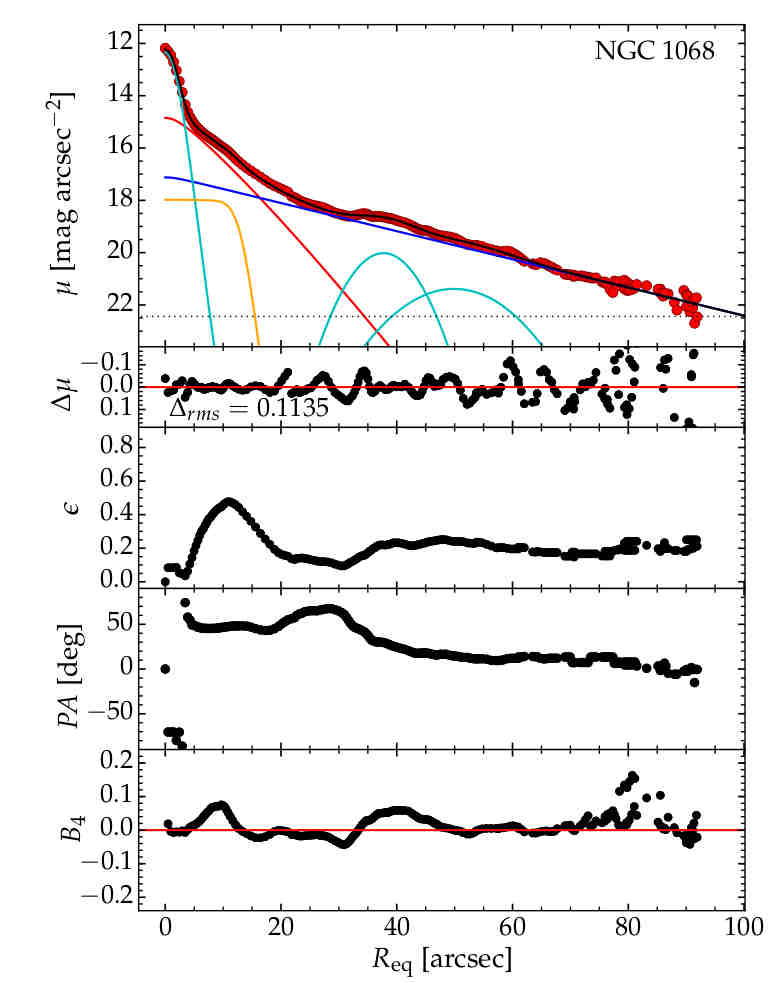}
\includegraphics[clip=true,trim= 11mm 1mm 1mm 5mm,width=0.249\textwidth]{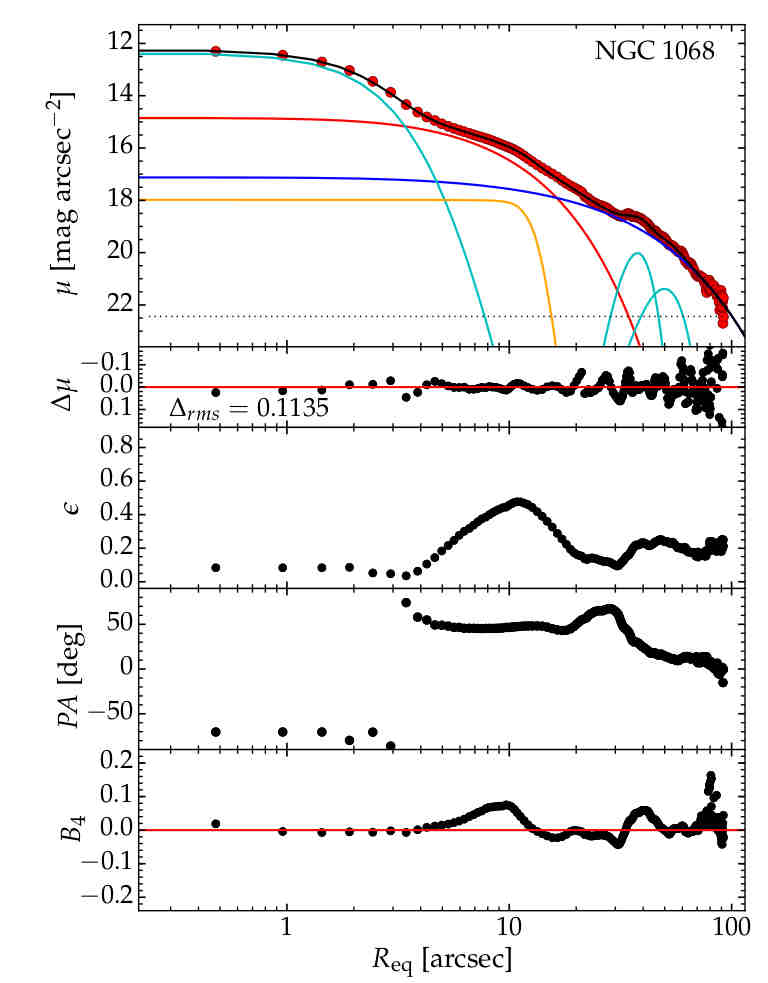}
\caption{2MASS $K_s$ filter surface brightness profile for NGC~1068, with a physical scale of 0.0489$\,\text{kpc}\,\text{arcsec}^{-1}$. \textbf{Left two panels}---The model represents $0\arcsec \leq R_{\rm maj} \leq 103\farcs5$ with $\Delta_{\rm rms}=0.1149\,\text{mag\,arcsec}^{-2}$. \underline{S{\'e}rsic Profile Parameters:} \textcolor{red}{$R_e=10\farcs52\pm0\farcs71$, $\mu_e=16.17\pm0.31\,\text{mag\,arcsec}^{-2}$, and $n=0.71\pm0.14$.} \underline{Ferrers Profile Parameters:} \textcolor{Orange}{$\mu_0 = 17.24\pm1.14\,\text{mag\,arcsec}^{-2}$, $R_{\rm end} = 18\farcs00\pm34\farcs22$, and $\alpha = 0.13\pm3.06$.} \underline{Exponential Profile Parameters:} \textcolor{blue}{$\mu_0 = 17.41\pm0.22\,\text{mag\,arcsec}^{-2}$ and $h = 24\farcs60\pm1\farcs32$.} \underline{Additional Parameters:} three Gaussian components added at: \textcolor{cyan}{$R_{\rm r}=0\arcsec$, $41\farcs72\pm1\farcs09$, \& $57\farcs45\pm3\farcs76$; with $\mu_0 = 10.53\pm0.31$, $19.53\pm0.12$, \& $20.77\pm0.33\,\text{mag\,arcsec}^{-2}$; and FWHM = $1\farcs57\pm0\farcs29$, $14\farcs27\pm2\farcs69$, \& $16\farcs34\pm4\farcs74$, respectively.} \textbf{Right two panels}---The model represents $0\arcsec \leq R_{\rm eq} \leq 92\arcsec$ with $\Delta_{\rm rms}=0.1135\,\text{mag\,arcsec}^{-2}$. \underline{S{\'e}rsic Profile Parameters:} \textcolor{red}{$R_e=8\farcs29\pm0\farcs79$, $\mu_e=16.14\pm0.35\,\text{mag\,arcsec}^{-2}$, and $n=0.87\pm0.23$.} \underline{Ferrers Profile Parameters:} \textcolor{Orange}{$\mu_0 = 17.98\pm2.45\,\text{mag\,arcsec}^{-2}$, $R_{\rm end} = 12\farcs13\pm0\farcs00$, and $\alpha = 0.07\pm3.21$.} \underline{Exponential Profile Parameters:} \textcolor{blue}{$\mu_0 = 17.03\pm0.21\,\text{mag\,arcsec}^{-2}$ and $h = 20\farcs17\pm0\farcs92$.} \underline{Additional Parameters:} three Gaussian components added at: \textcolor{cyan}{$R_{\rm r}=0\arcsec$, $37\farcs70\pm0\farcs82$, \& $49\farcs90\pm5\farcs03$; with $\mu_0 = 10.05\pm0.55$, $19.95\pm0.18$, \& $21.36\pm0.44\,\text{mag\,arcsec}^{-2}$; and FWHM = $1\farcs09\pm0\farcs31$, $9\farcs61\pm1\farcs95$, \& $17\farcs81\pm7\farcs36$, respectively.}}
\label{NGC1068_plot}
\end{sidewaysfigure}

\subsection{NGC~1068 (Figure~\ref{NGC1068_plot})}

The galaxy NGC~1068 (M77) possesses a Type 1 Seyfert AGN with broad polarized Balmer lines \citep{Veron-Cetty:2006}. \citet{Tanaka:2017} indicated that NGC~1068 underwent a minor merger several billion yr ago and speculated that spawned its nuclear activity. Three Gaussian components have been added: one central Gaussian to account for the excess nuclear emission and two Gaussians for the spiral arms at $R_{\rm maj} \approx 42\arcsec$ ($R_{\rm eq}\approx 38\arcsec$) and $R_{\rm maj} \approx 57\arcsec$ ($R_{\rm eq}\approx 50\arcsec$).

\begin{sidewaysfigure}
\includegraphics[clip=true,trim= 11mm 1mm 4mm 5mm,width=0.249\textwidth]{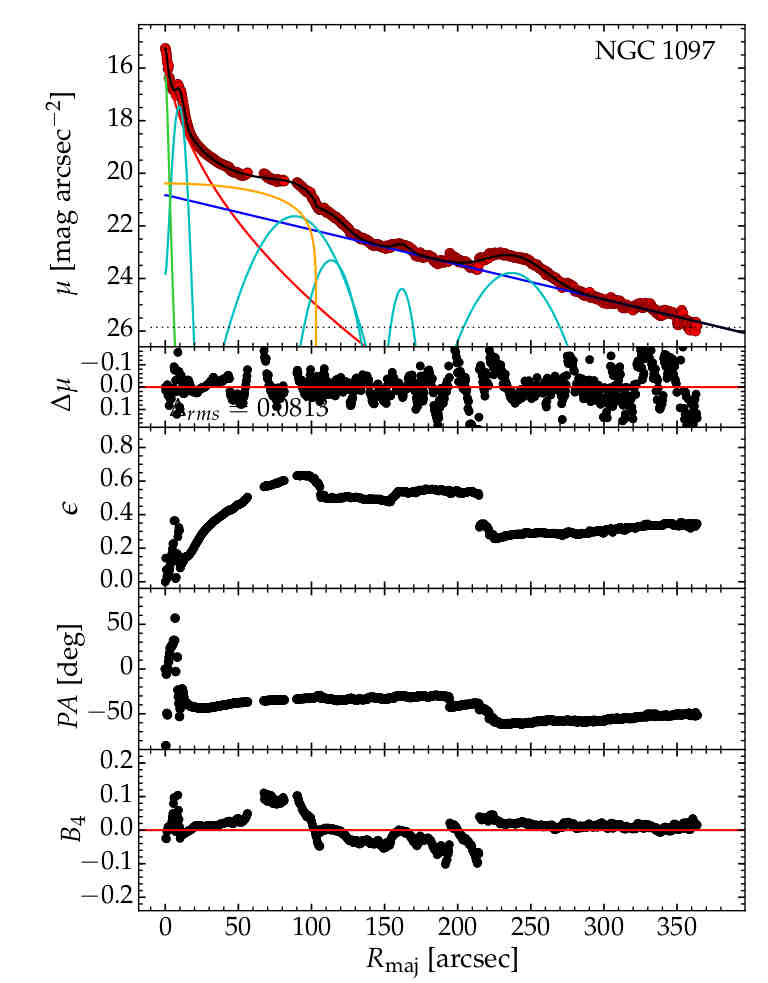}
\includegraphics[clip=true,trim= 11mm 1mm 4mm 5mm,width=0.249\textwidth]{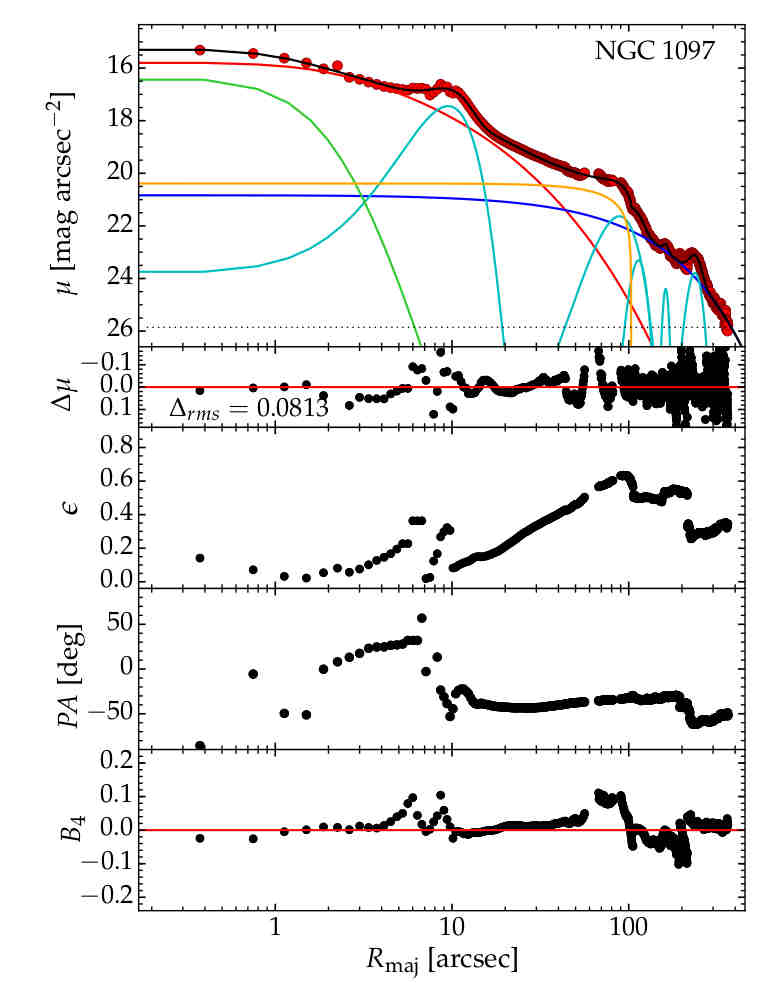}
\includegraphics[clip=true,trim= 11mm 1mm 4mm 5mm,width=0.249\textwidth]{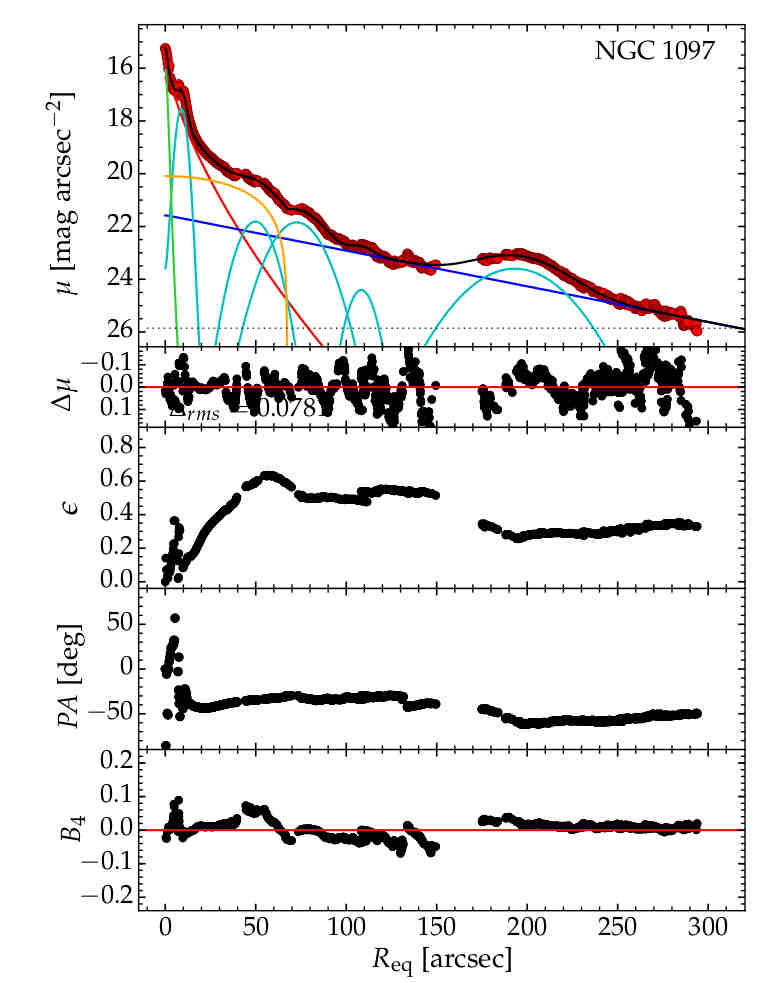}
\includegraphics[clip=true,trim= 11mm 1mm 4mm 5mm,width=0.249\textwidth]{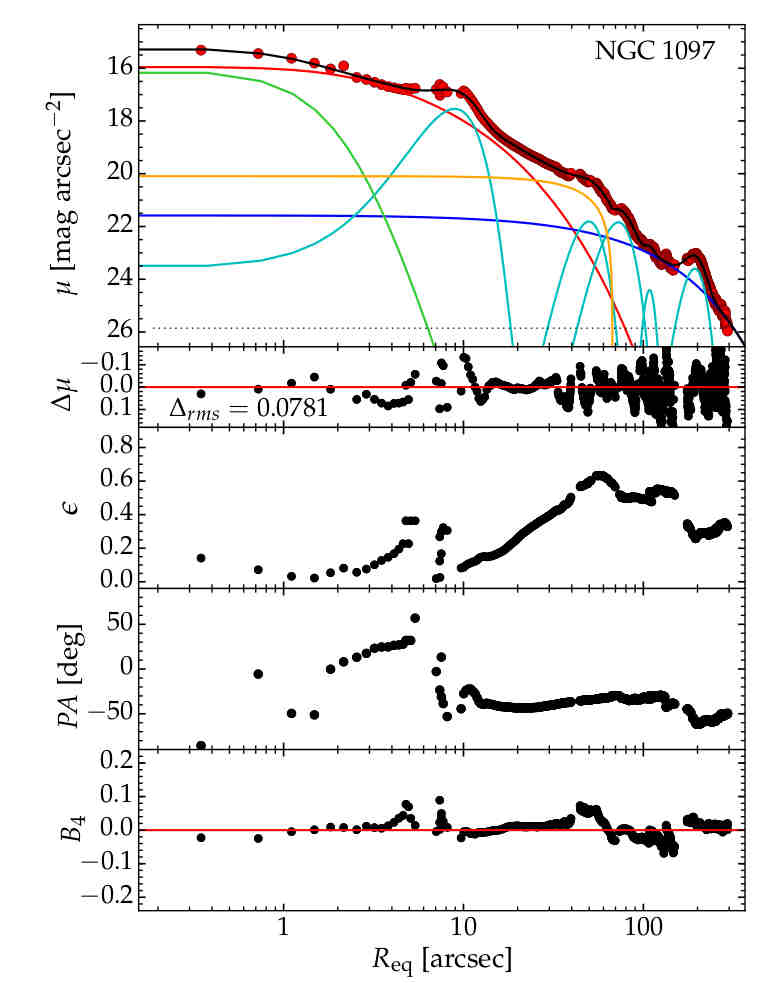}
\caption{\textit{Spitzer} $3.6\,\micron$ surface brightness profile for NGC~1097, with a physical scale of 0.1207$\,\text{kpc}\,\text{arcsec}^{-1}$. \textbf{Left two panels}---The model represents $0\arcsec \leq R_{\rm maj} \leq 364\arcsec$ with $\Delta_{\rm rms}=0.0813\,\text{mag\,arcsec}^{-2}$. \underline{Point Source:} \textcolor{LimeGreen}{$\mu_0 = 16.26\pm0.22\,\text{mag\,arcsec}^{-2}$.} \underline{S{\'e}rsic Profile Parameters:} \textcolor{red}{$R_e=15\farcs72\pm1\farcs93$, $\mu_e=18.71\pm0.24\,\text{mag\,arcsec}^{-2}$, and $n=1.95\pm0.26$.} \underline{Ferrers Profile Parameters:} \textcolor{Orange}{$\mu_0 = 20.39\pm0.12\,\text{mag\,arcsec}^{-2}$, $R_{\rm end} = 103\farcs28\pm0\farcs35$, and $\alpha = 1.61\pm0.24$.} \underline{Exponential Profile Parameters:} \textcolor{blue}{$\mu_0 = 20.83\pm0.04\,\text{mag\,arcsec}^{-2}$ and $h = 81\farcs92\pm0\farcs70$.} \underline{Additional Parameters:} five Gaussian components added at: \textcolor{cyan}{$R_{\rm r}=9\farcs51\pm0\farcs14$, $88\farcs56\pm3\farcs04$, $113\farcs47\pm3\farcs98$, $161\farcs81\pm0\farcs47$, \& $236\farcs94\pm0\farcs31$; with $\mu_0 = 17.31\pm0.06$, $21.63\pm0.17$, $23.32\pm0.98$, $24.40\pm0.10$, \& $23.79\pm0.02\,\text{mag\,arcsec}^{-2}$; and FWHM = $4\farcs92\pm0\farcs35$, $38\farcs05\pm5\farcs47$, $20\farcs29\pm7\farcs54$, $11\farcs11\pm1\farcs20$, \& $39\farcs79\pm0\farcs69$, respectively.} \textbf{Right two panels}---The model represents $0\arcsec \leq R_{\rm eq} \leq 294\arcsec$ with $\Delta_{\rm rms}=0.0781\,\text{mag\,arcsec}^{-2}$. \underline{Point Source:} \textcolor{LimeGreen}{$\mu_0 = 16.07\pm0.20\,\text{mag\,arcsec}^{-2}$.} \underline{S{\'e}rsic Profile Parameters:} \textcolor{red}{$R_e=11\farcs39\pm1\farcs88$, $\mu_e=18.27\pm0.33\,\text{mag\,arcsec}^{-2}$, and $n=1.52\pm0.33$.} \underline{Ferrers Profile Parameters:} \textcolor{Orange}{$\mu_0 = 20.10\pm0.27\,\text{mag\,arcsec}^{-2}$, $R_{\rm end} = 67\farcs40\pm0\farcs57$, and $\alpha = 2.46\pm0.40$.} \underline{Exponential Profile Parameters:} \textcolor{blue}{$\mu_0 = 21.57\pm0.03\,\text{mag\,arcsec}^{-2}$ and $h = 80\farcs29\pm0\farcs60$.} \underline{Additional Parameters:} five Gaussian components added at: \textcolor{cyan}{$R_{\rm r}=8\farcs94\pm0\farcs17$, $49\farcs82\pm1\farcs12$, $72\farcs68\pm1\farcs32$, $108\farcs32\pm0\farcs59$, \& $193\farcs10\pm0\farcs28$; with $\mu_0 = 17.40\pm0.07$, $21.82\pm0.33$, $21.85\pm0.09$, $24.41\pm0.09$, \& $23.61\pm0.01\,\text{mag\,arcsec}^{-2}$; and FWHM = $4\farcs70\pm0\farcs42$, $17\farcs68\pm3\farcs89$, $25\farcs93\pm1\farcs75$, $13\farcs99\pm1\farcs43$, \& $50\farcs38\pm0\farcs58$, respectively.} Given our focus on isolating the bulge light, we have allowed degeneracy among the components which dominate at large radii (whose parameters are therefore neither stable nor reliable) when this appears to not compromise the bulge.}
\label{NGC1097_plot}
\end{sidewaysfigure}

\subsection{NGC~1097 (Figure~\ref{NGC1097_plot})}

Imaging from \textit{HST} indicates the presence of a point source at the center of NGC~1097, and there exists a Type 3 Seyfert AGN or low-ionization nuclear emission-line region (LINER) with broad Balmer lines \citep{Veron-Cetty:2006}. Five Gaussian components have been added for the inner ring at $R_{\rm maj} \approx10\arcsec$ ($R_{\rm eq}\approx9\arcsec$), for ansae at the bar's ends at $R_{\rm maj} \approx89\arcsec$ ($R_{\rm eq}\approx50\arcsec$), for the spiral arms at $R_{\rm maj} \approx113\arcsec$ ($R_{\rm eq}\approx73\arcsec$) and $R_{\rm maj} \approx162\arcsec$ ($R_{\rm eq}\approx108\arcsec$), and to capture the northwest cloud at $R_{\rm maj} \approx 237\arcsec$ ($R_{\rm eq}\approx193\arcsec$).

\begin{sidewaysfigure}
\includegraphics[clip=true,trim= 11mm 1mm 1mm 5mm,width=0.249\textwidth]{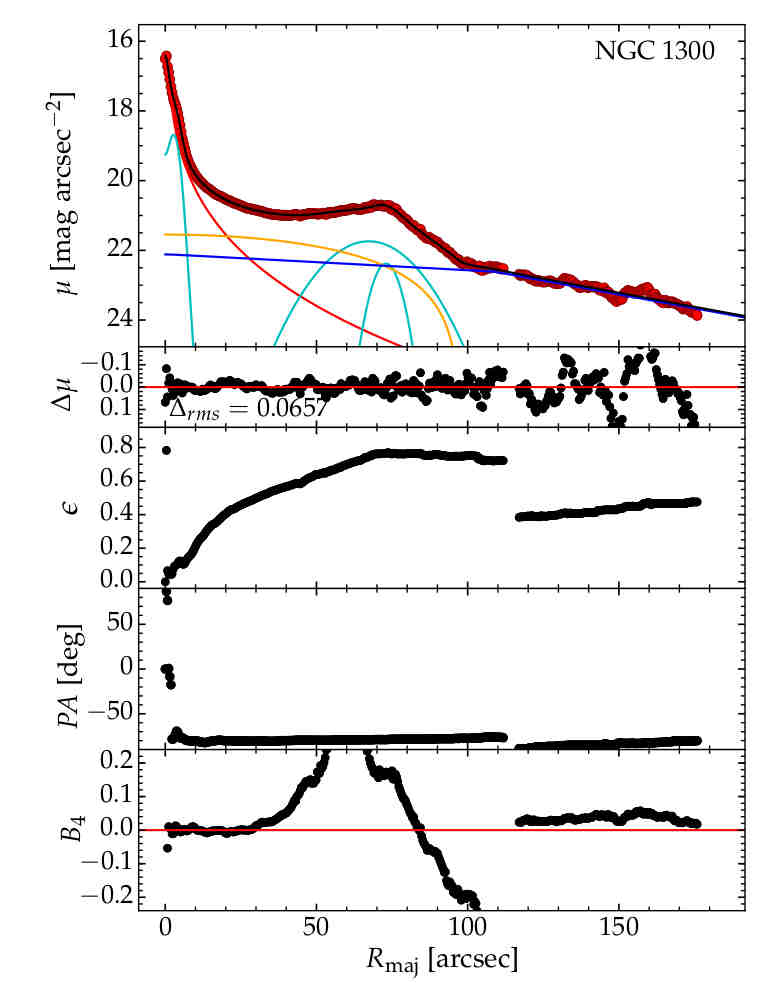}
\includegraphics[clip=true,trim= 11mm 1mm 1mm 5mm,width=0.249\textwidth]{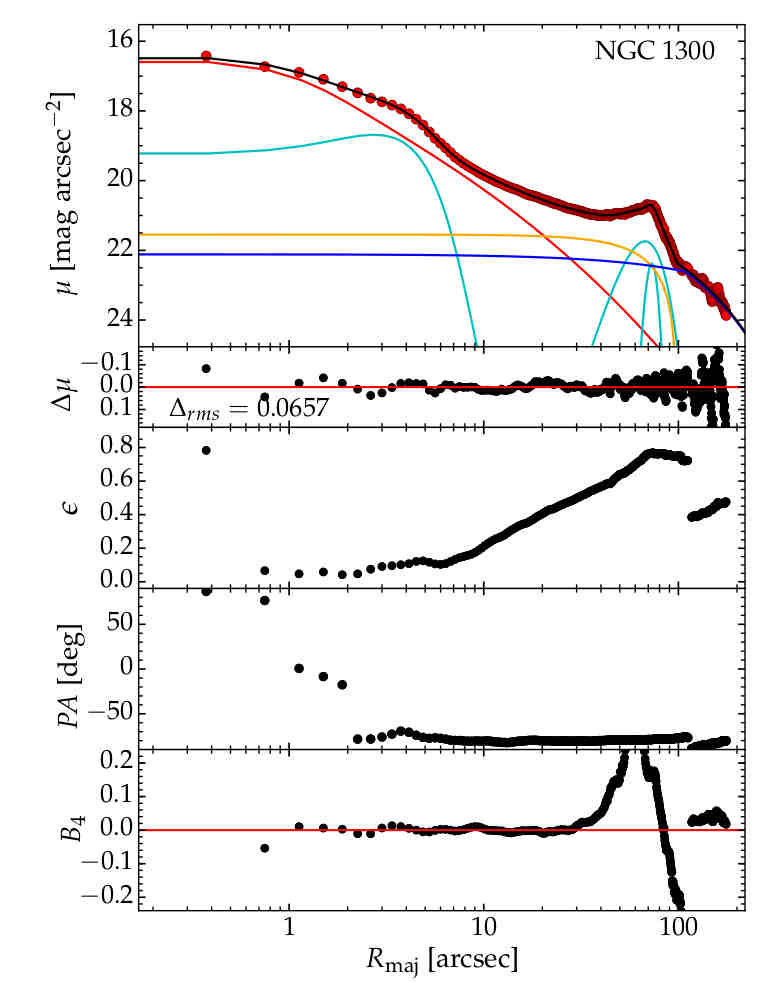}
\includegraphics[clip=true,trim= 11mm 1mm 1mm 5mm,width=0.249\textwidth]{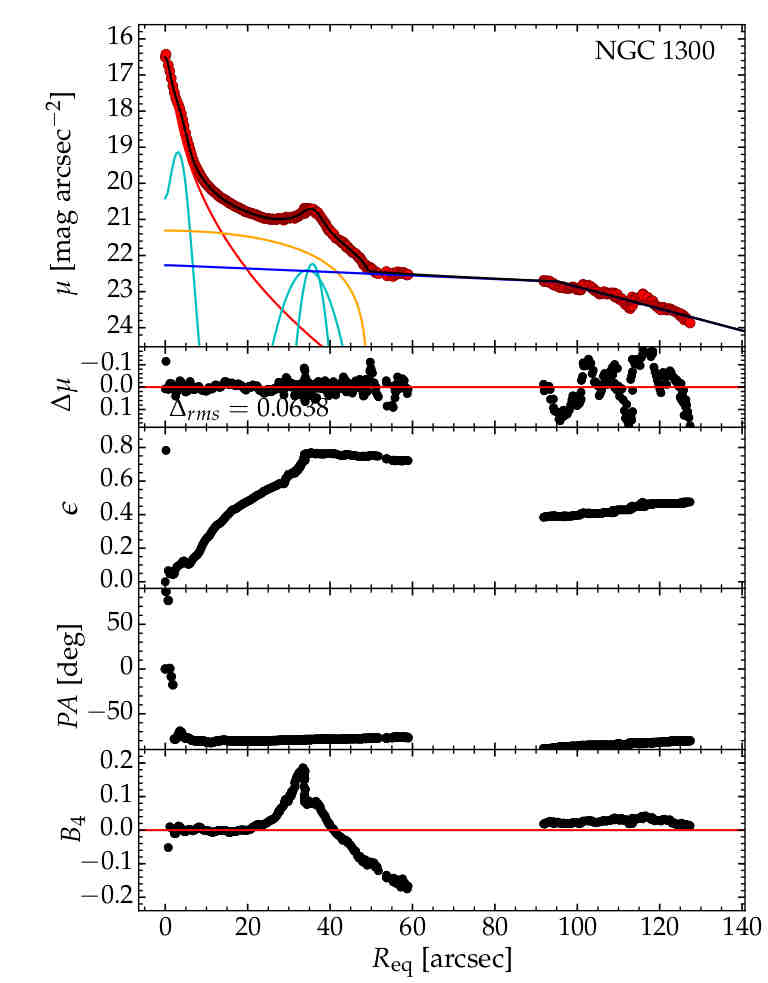}
\includegraphics[clip=true,trim= 11mm 1mm 1mm 5mm,width=0.249\textwidth]{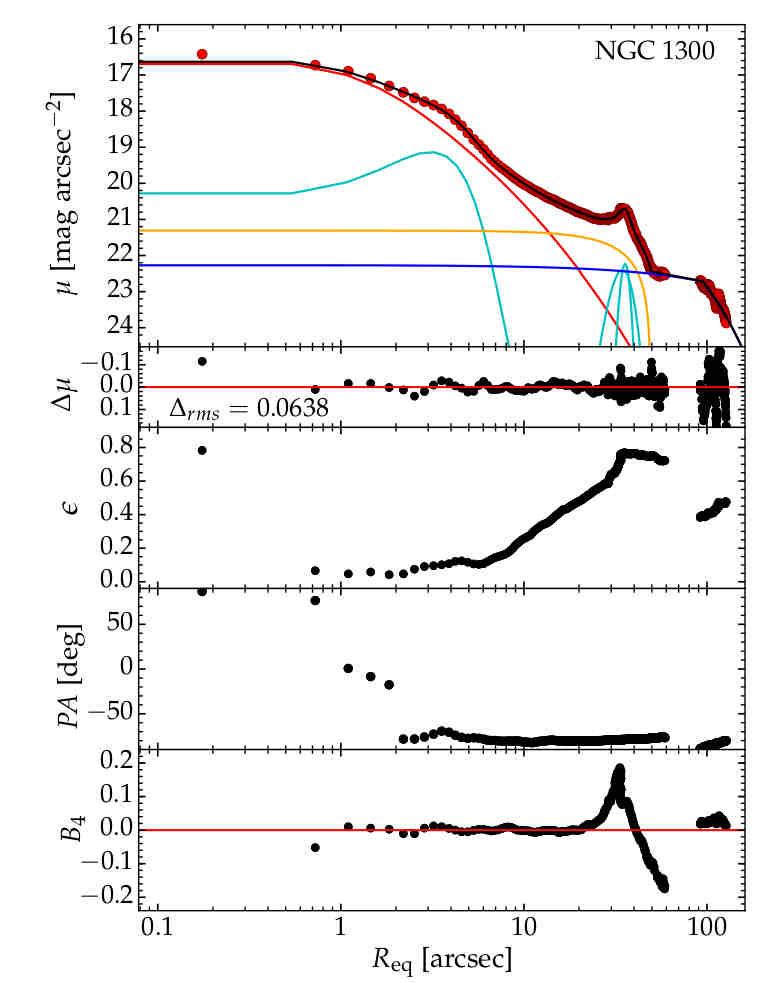}
\caption{\textit{Spitzer} $3.6\,\micron$ surface brightness profile for NGC~1300, with a physical scale of 0.0704$\,\text{kpc}\,\text{arcsec}^{-1}$. \textbf{Left two panels}---The model represents $0\arcsec \leq R_{\rm maj} \leq 176\arcsec$ with $\Delta_{\rm rms}=0.0657\,\text{mag\,arcsec}^{-2}$. \underline{S{\'e}rsic Profile Parameters:} \textcolor{red}{$R_e=24\farcs37\pm13\farcs93$, $\mu_e=21.97\pm0.94\,\text{mag\,arcsec}^{-2}$, and $n=4.20\pm0.48$.} \underline{Ferrers Profile Parameters:} \textcolor{Orange}{$\mu_0 = 21.55\pm0.30\,\text{mag\,arcsec}^{-2}$, $R_{\rm end} = 97\farcs80\pm1\farcs33$, and $\alpha = 2.58\pm0.69$.} \underline{Broken Exponential Profile Parameters:} \textcolor{blue}{$\mu_0 = 22.11\pm0.86\,\text{mag\,arcsec}^{-2}$, $R_b = 108\farcs75\pm2\farcs66$, $h_1 = 235\farcs60\pm392\farcs36$, and $h_2 = 68\farcs41\pm1\farcs39$.} \underline{Additional Parameters:} three Gaussian components added at: \textcolor{cyan}{$R_{\rm r}=2\farcs91\pm0\farcs43$, $67\farcs31\pm1\farcs07$, \& $73\farcs01\pm0\farcs42$; with $\mu_0 = 18.41\pm0.15$, $21.75\pm0.11$, \& $22.38\pm0.14\,\text{mag\,arcsec}^{-2}$; and FWHM = $3\farcs09\pm0\farcs75$, $31\farcs36\pm2\farcs19$, \& $9\farcs90\pm1\farcs27$, respectively.} \textbf{Right two panels}---The model represents $0\arcsec \leq R_{\rm eq} \leq 130\arcsec$ with $\Delta_{\rm rms}=0.0638\,\text{mag\,arcsec}^{-2}$. \underline{S{\'e}rsic Profile Parameters:} \textcolor{red}{$R_e=7\farcs39\pm2\farcs36$, $\mu_e=19.99\pm0.65\,\text{mag\,arcsec}^{-2}$, and $n=2.83\pm0.38$.} \underline{Ferrers Profile Parameters:} \textcolor{Orange}{$\mu_0 = 21.31\pm0.13\,\text{mag\,arcsec}^{-2}$, $R_{\rm end} = 49\farcs35\pm0\farcs36$, and $\alpha = 2.12\pm0.23$.} \underline{Broken Exponential Profile Parameters:} \textcolor{blue}{$\mu_0 = 22.27\pm0.10\,\text{mag\,arcsec}^{-2}$, $R_b = 95\farcs24\pm0\farcs84$, $h_1 = 223\farcs83\pm48\farcs78$, and $h_2 = 35\farcs83\pm0\farcs91$.} \underline{Additional Parameters:} three Gaussian components added at: \textcolor{cyan}{$R_{\rm r}=3\farcs17\pm0\farcs50$, $34\farcs52\pm0\farcs70$, \& $35\farcs71\pm0\farcs32$; with $\mu_0 = 18.66\pm0.23$, $22.41\pm0.45$, \& $22.23\pm0.41\,\text{mag\,arcsec}^{-2}$; and FWHM = $2\farcs15\pm1\farcs14$, $10\farcs20\pm2\farcs60$, \& $4\farcs51\pm1\farcs14$, respectively.}}
\label{NGC1300_plot}
\end{sidewaysfigure}

\subsection{NGC~1300 (Figure~\ref{NGC1300_plot})}

We use a broken exponential model to account for the redistributed disk light in the inner region of the galaxy due to the peanut-shell-shaped structure \citep{Saha:2018}. We have added three Gaussian components to account for the unresolved nuclear spiral arms that are apparent in \textit{HST} imaging at $R_{\rm maj} = R_{\rm eq} \approx 3\arcsec$ and the prominent grand design spiral arms emanating from either end of the bar at $R_{\rm maj} \approx 67\arcsec$ ($R_{\rm eq}\approx 35\arcsec$) and $R_{\rm maj} \approx 73\arcsec$ ($R_{\rm eq}\approx 36\arcsec$).

\begin{sidewaysfigure}
\includegraphics[clip=true,trim= 11mm 1mm 1mm 5mm,width=0.249\textwidth]{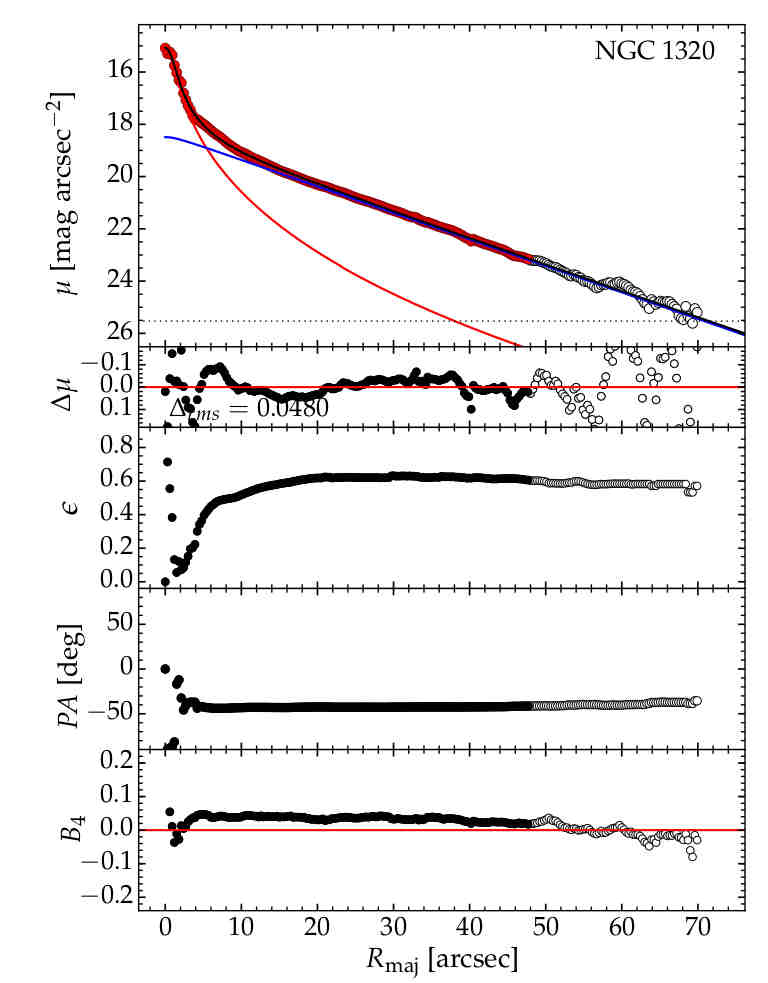}
\includegraphics[clip=true,trim= 11mm 1mm 1mm 5mm,width=0.249\textwidth]{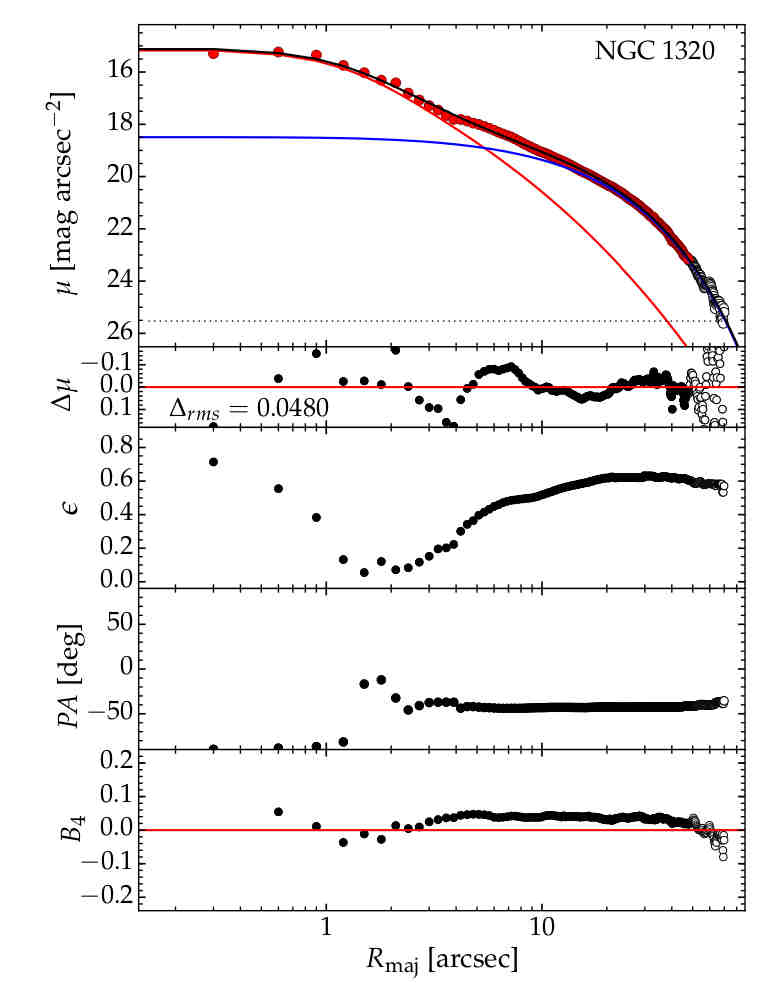}
\includegraphics[clip=true,trim= 11mm 1mm 1mm 5mm,width=0.249\textwidth]{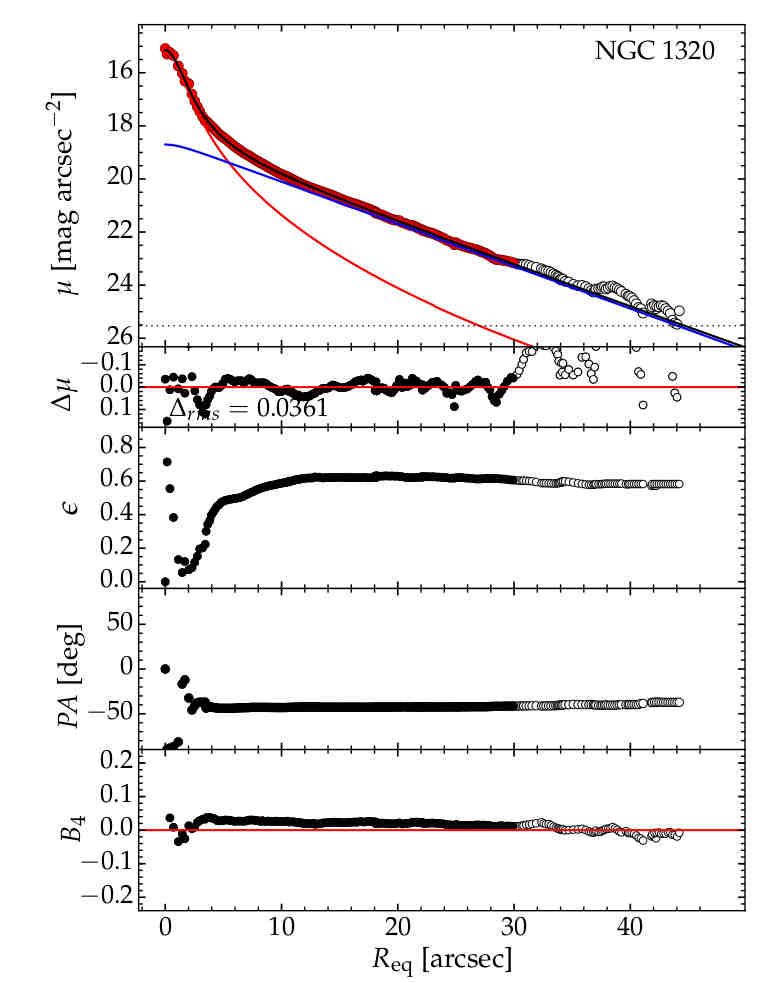}
\includegraphics[clip=true,trim= 11mm 1mm 1mm 5mm,width=0.249\textwidth]{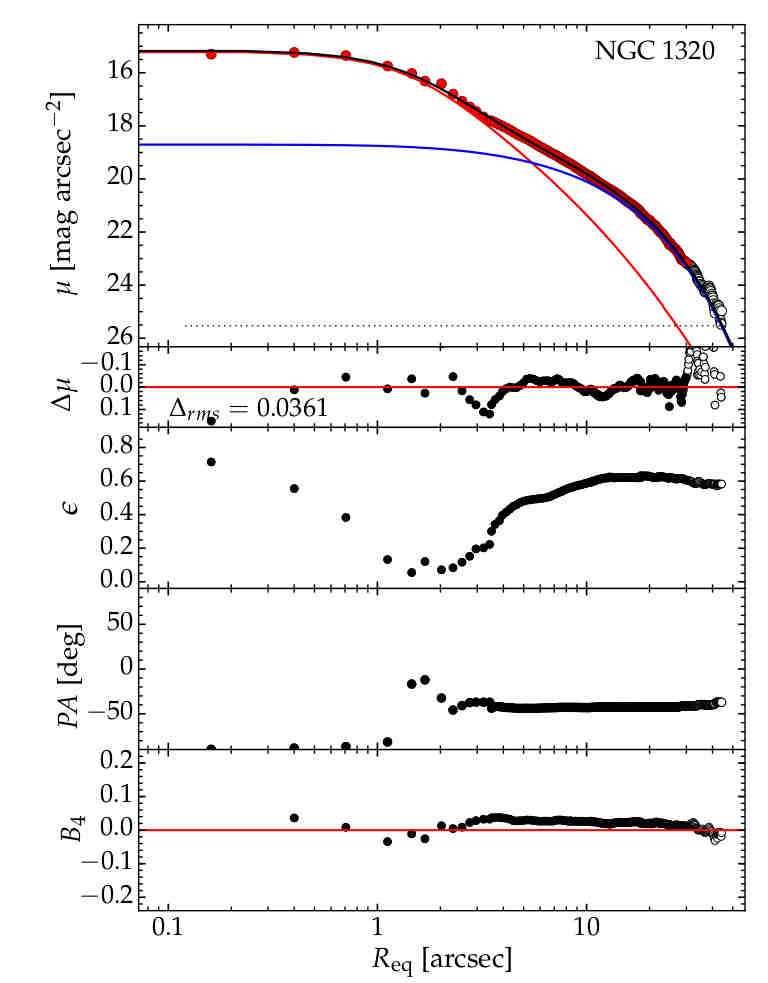}
\caption{\textit{Spitzer} $3.6\,\micron$ surface brightness profile for NGC~1320, with a physical scale of 0.1828$\,\text{kpc}\,\text{arcsec}^{-1}$. \textbf{Left two panels}---The model represents $0\arcsec \leq R_{\rm maj} \leq 48\arcsec$ with $\Delta_{\rm rms}=0.0480\,\text{mag\,arcsec}^{-2}$ and additional data from $48\arcsec < R_{\rm maj} \leq 70\arcsec$ is plotted, but not modeled. \underline{S{\'e}rsic Profile Parameters:} \textcolor{red}{$R_e=3\farcs35\pm0\farcs32$, $\mu_e=17.93\pm0.19\,\text{mag\,arcsec}^{-2}$, and $n=3.08\pm0.12$.} \underline{Exponential Profile Parameters:} \textcolor{blue}{$\mu_0 = 18.35\pm0.04\,\text{mag\,arcsec}^{-2}$ and $h = 10\farcs71\pm0\farcs06$.} \textbf{Right two panels}---The model represents $0\arcsec \leq R_{\rm eq} \leq 30\arcsec$ with $\Delta_{\rm rms}=0.0361\,\text{mag\,arcsec}^{-2}$ and additional data from $30\arcsec < R_{\rm maj} \leq 46\arcsec$ is plotted, but not modeled. \underline{S{\'e}rsic Profile Parameters:} \textcolor{red}{$R_e=2\farcs23\pm0\farcs14$, $\mu_e=17.40\pm0.13\,\text{mag\,arcsec}^{-2}$, and $n=2.87\pm0.10$.} \underline{Exponential Profile Parameters:} \textcolor{blue}{$\mu_0 = 18.52\pm0.04\,\text{mag\,arcsec}^{-2}$ and $h = 6\farcs83\pm0\farcs04$.}}
\label{NGC1320_plot}
\end{sidewaysfigure}

\subsection{NGC~1320 (Figure~\ref{NGC1320_plot})}

For NGC 1320, we only require a two-component S{\'e}rsic + exponential model to adequately model the light. We truncated the outer region of the galaxy beyond $\approx 4.3$ scale lengths, where the faint spiral arms influence the fit.

\begin{sidewaysfigure}
\includegraphics[clip=true,trim= 11mm 1mm 1mm 5mm,width=0.249\textwidth]{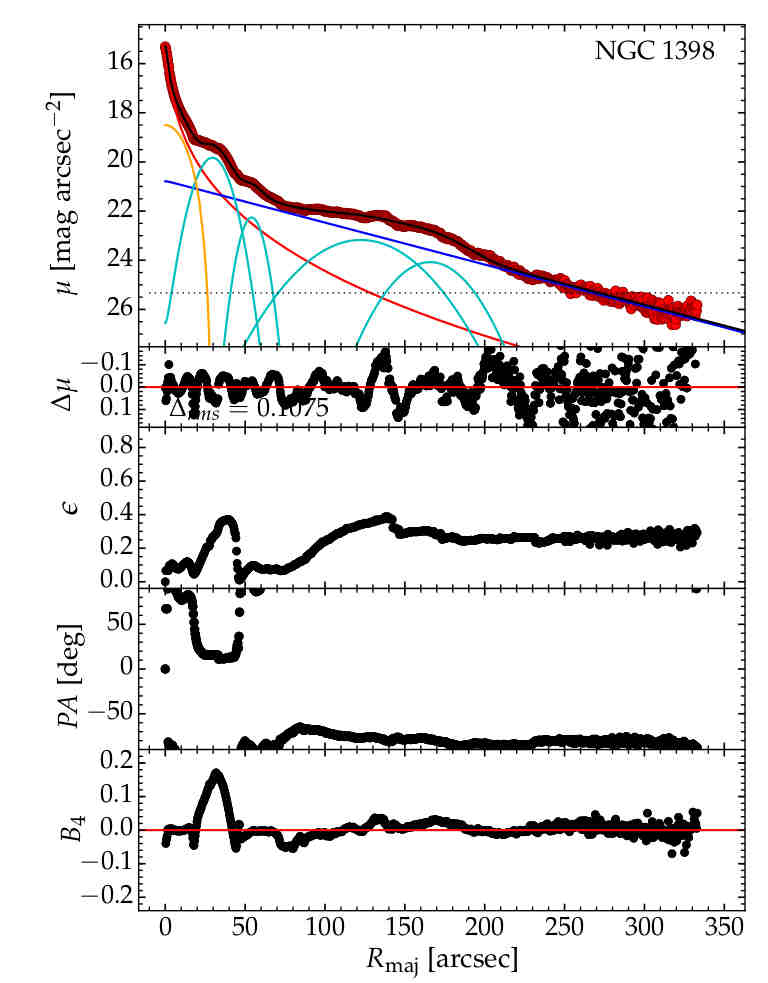}
\includegraphics[clip=true,trim= 11mm 1mm 1mm 5mm,width=0.249\textwidth]{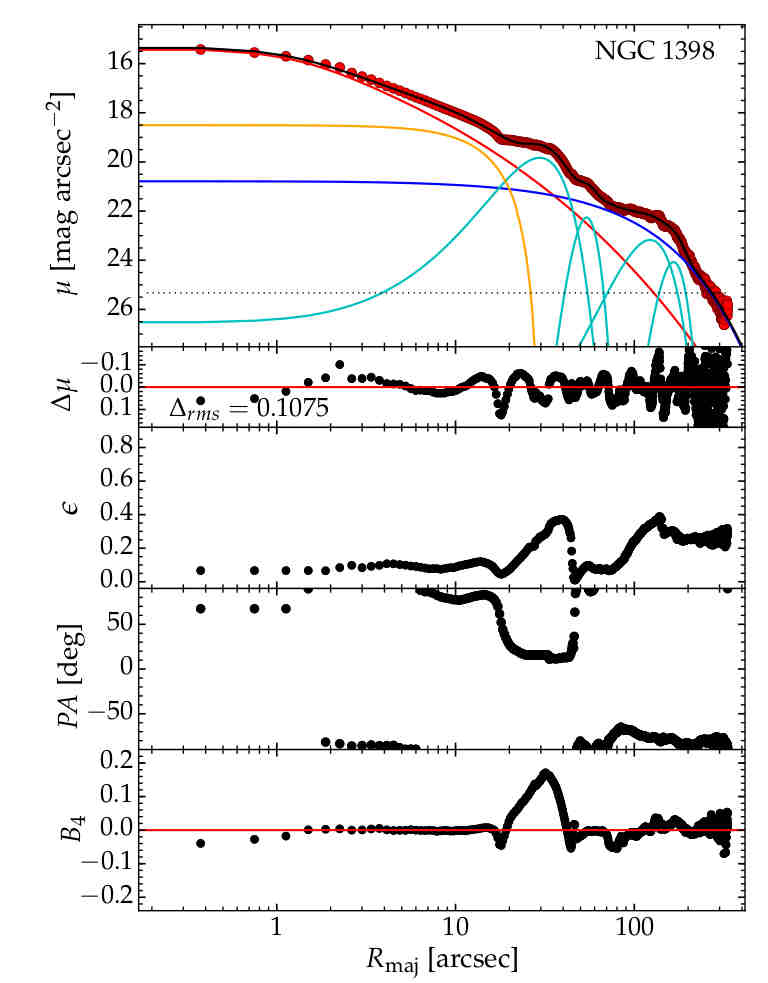}
\includegraphics[clip=true,trim= 11mm 1mm 1mm 5mm,width=0.249\textwidth]{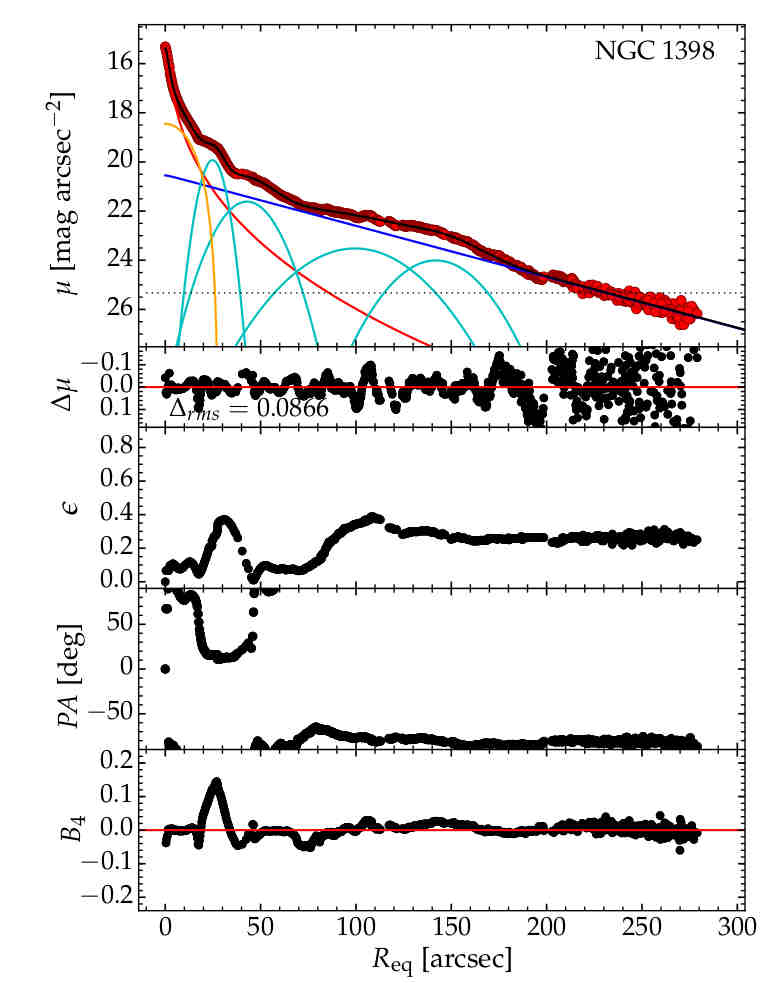}
\includegraphics[clip=true,trim= 11mm 1mm 1mm 5mm,width=0.249\textwidth]{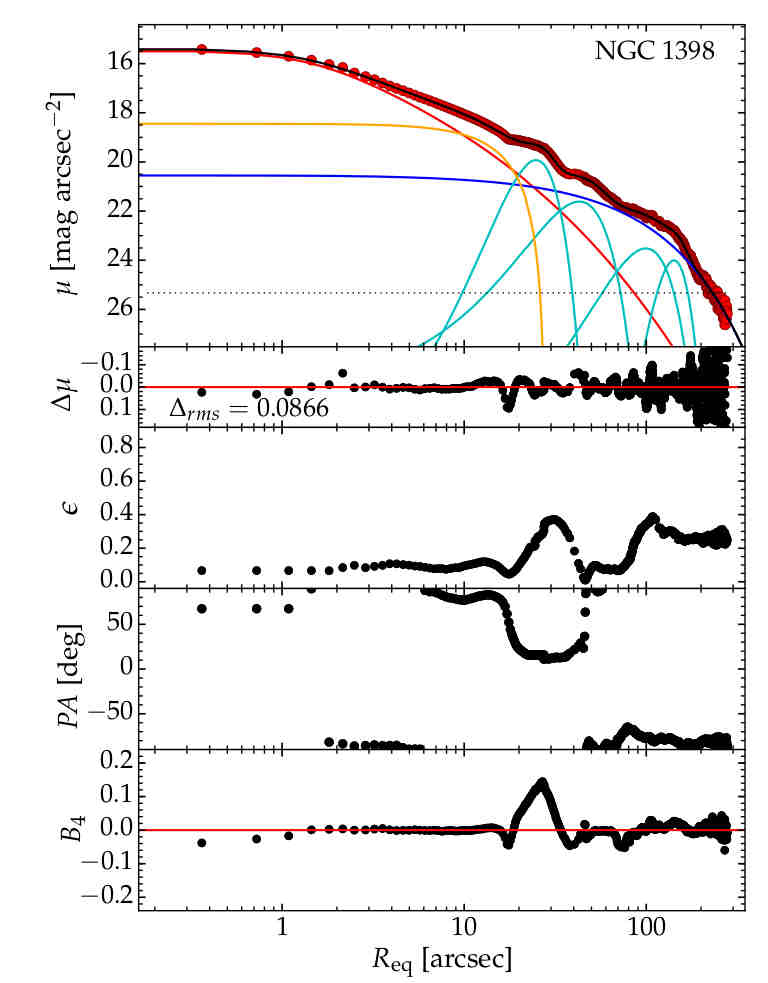}
\caption{\textit{Spitzer} $3.6\,\micron$ surface brightness profile for NGC~1398, with a physical scale of 0.1201$\,\text{kpc}\,\text{arcsec}^{-1}$. \textbf{Left two panels}---The model represents $0\arcsec \leq R_{\rm maj} \leq 333\arcsec$ with $\Delta_{\rm rms}=0.1075\,\text{mag\,arcsec}^{-2}$. \underline{S{\'e}rsic Profile Parameters:} \textcolor{red}{$R_e=17\farcs53\pm2\farcs03$, $\mu_e=19.75\pm0.26\,\text{mag\,arcsec}^{-2}$, and $n=3.44\pm0.23$.} \underline{Ferrers Profile Parameters:} \textcolor{Orange}{$\mu_0 = 18.51\pm0.31\,\text{mag\,arcsec}^{-2}$, $R_{\rm end} = 29\farcs57\pm14\farcs55$, and $\alpha = 10.00\pm12.23$.} \underline{Exponential Profile Parameters:} \textcolor{blue}{$\mu_0 = 20.76\pm0.04\,\text{mag\,arcsec}^{-2}$ and $h = 63\farcs66\pm0\farcs41$.} \underline{Additional Parameters:} four Gaussian components added at: \textcolor{cyan}{$R_{\rm r}=29\farcs53\pm0\farcs93$, $54\farcs09\pm1\farcs03$, $122\farcs51\pm5\farcs77$, \& $165\farcs82\pm6\farcs91$; with $\mu_0 = 19.84\pm0.06$, $22.30\pm0.12$, $23.23\pm0.08$, \& $24.18\pm0.69\,\text{mag\,arcsec}^{-2}$; and FWHM = $18\farcs67\pm1\farcs64$, $13\farcs20\pm1\farcs97$, $61\farcs29\pm13\farcs95$, \& $43\farcs50\pm6\farcs53$, respectively.} \textbf{Right two panels}---The model represents $0\arcsec \leq R_{\rm eq} \leq 280\arcsec$ with $\Delta_{\rm rms}=0.0866\,\text{mag\,arcsec}^{-2}$. \underline{S{\'e}rsic Profile Parameters:} \textcolor{red}{$R_e=10\farcs38\pm0\farcs84$, $\mu_e=19.04\pm0.17\,\text{mag\,arcsec}^{-2}$, and $n=3.00\pm0.17$.} \underline{Ferrers Profile Parameters:} \textcolor{Orange}{$\mu_0 = 18.44\pm0.18\,\text{mag\,arcsec}^{-2}$, $R_{\rm end} = 27\farcs81\pm13\farcs15$, and $\alpha = 7.52\pm7.61$.} \underline{Exponential Profile Parameters:} \textcolor{blue}{$\mu_0 = 20.53\pm0.03\,\text{mag\,arcsec}^{-2}$ and $h = 52\farcs25\pm0\farcs26$.} \underline{Additional Parameters:} four Gaussian components added at: \textcolor{cyan}{$R_{\rm r}=24\farcs78\pm1\farcs61$, $42\farcs88\pm1\farcs59$, $99\farcs50\pm4\farcs71$, \& $141\farcs92\pm4\farcs43$; with $\mu_0 = 19.93\pm0.22$, $21.62\pm0.09$, $23.52\pm0.08$, \& $24.01\pm0.36\,\text{mag\,arcsec}^{-2}$; and FWHM = $10\farcs81\pm1\farcs60$, $26\farcs44\pm2\farcs47$, $54\farcs54\pm12\farcs25$, \& $41\farcs07\pm3\farcs80$, respectively.}}
\label{NGC1398_plot}
\end{sidewaysfigure}

\subsection{NGC~1398 (Figure~\ref{NGC1398_plot})}

Four Gaussian components have been added to account for the ring around the end of the bar (corresponding to a large spike in the $B_4$ component) at $R_{\rm maj} \approx 30\arcsec$ ($R_{\rm eq}\approx 25\arcsec$) and for crossings of the spiral arms at $R_{\rm maj} \approx 54\arcsec$ ($R_{\rm eq}\approx 43\arcsec$), $R_{\rm maj} \approx 123\arcsec$ ($R_{\rm eq}\approx 100\arcsec$), and $R_{\rm maj} \approx 166\arcsec$ ($R_{\rm eq}\approx 142\arcsec$).

\begin{sidewaysfigure}
\includegraphics[clip=true,trim= 11mm 1mm 1mm 5mm,width=0.249\textwidth]{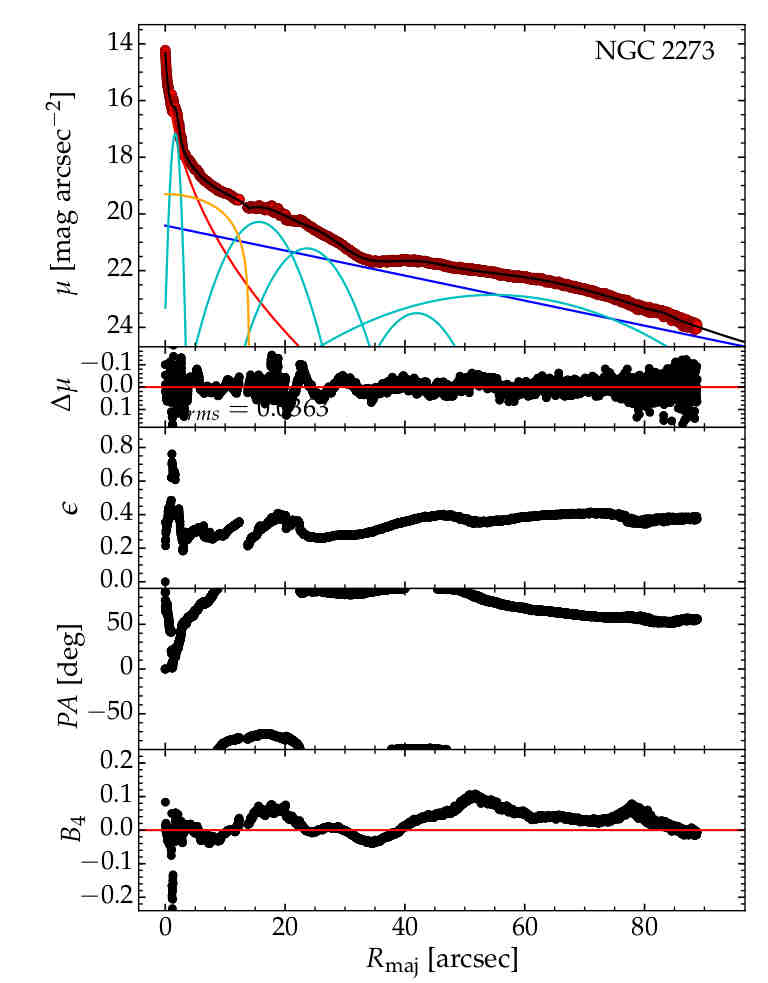}
\includegraphics[clip=true,trim= 11mm 1mm 1mm 5mm,width=0.249\textwidth]{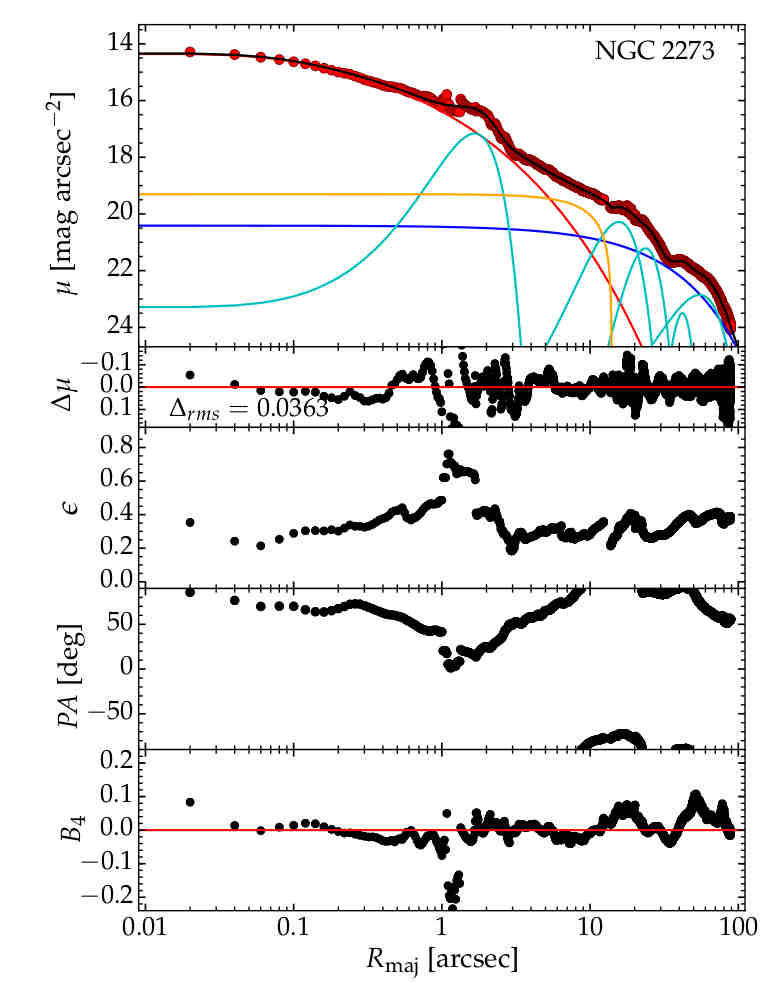}
\includegraphics[clip=true,trim= 11mm 1mm 1mm 5mm,width=0.249\textwidth]{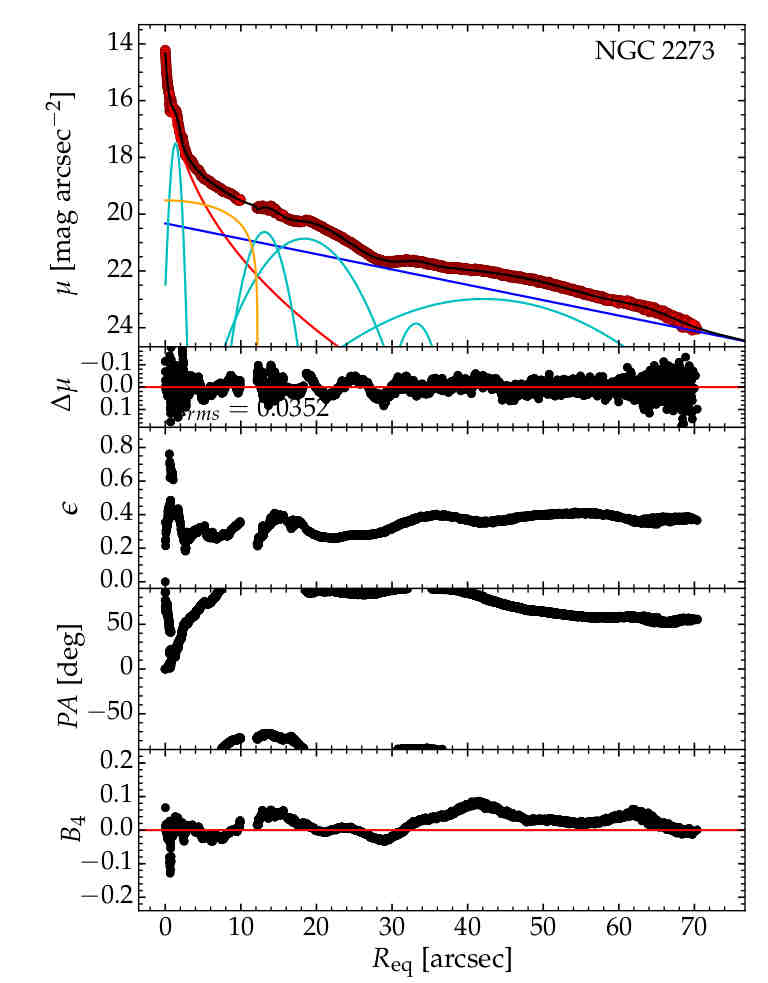}
\includegraphics[clip=true,trim= 11mm 1mm 1mm 5mm,width=0.249\textwidth]{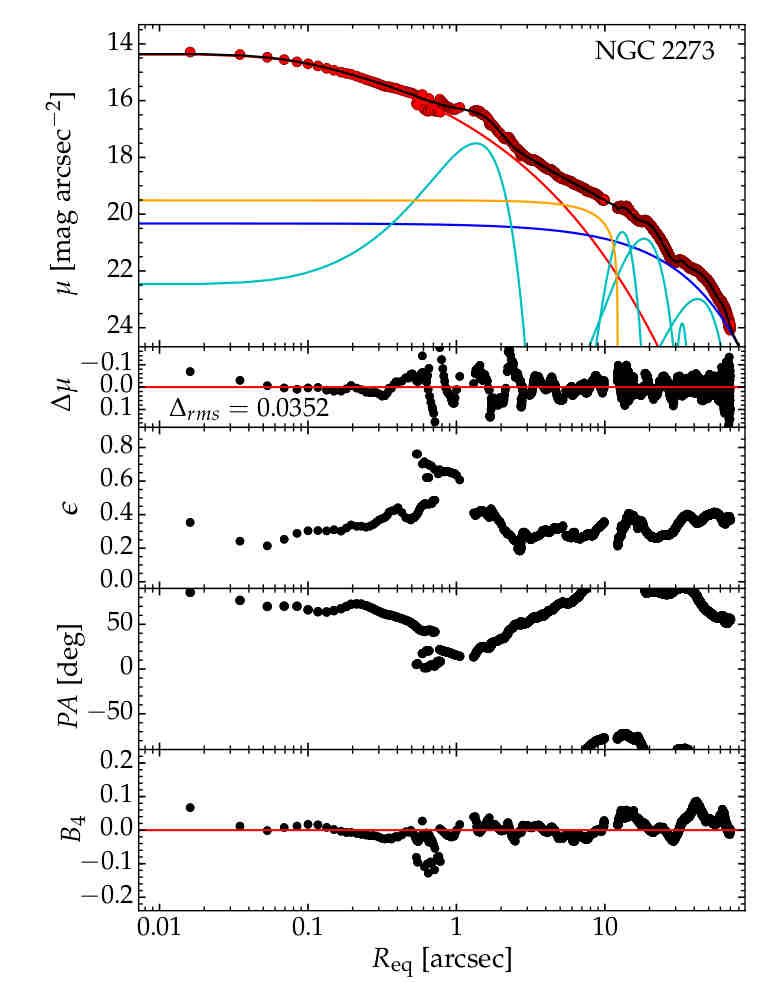}
\caption{\textit{HST} WFC3 UVIS2 F814W surface brightness profile for NGC~2273, with a physical scale of 0.1533$\,\text{kpc}\,\text{arcsec}^{-1}$. \textbf{Left two panels}---The model represents $0\arcsec \leq R_{\rm maj} \leq 88\farcs78$ with $\Delta_{\rm rms}=0.0363\,\text{mag\,arcsec}^{-2}$. \underline{S{\'e}rsic Profile Parameters:} \textcolor{red}{$R_e=2\farcs99\pm0\farcs09$, $\mu_e=18.13\pm0.05\,\text{mag\,arcsec}^{-2}$, and $n=2.24\pm0.04$}. \underline{Ferrers Profile Parameters:} \textcolor{Orange}{$\mu_0 = 19.30\pm0.03\,\text{mag\,arcsec}^{-2}$, $R_{\rm end} = 13\farcs98\pm0\farcs05$, and $\alpha = 2.36\pm0.08$.} \underline{Exponential Profile Parameters:} \textcolor{blue}{$\mu_0 = 20.41\pm0.01\,\text{mag\,arcsec}^{-2}$ and $h = 24\farcs60\pm0\farcs10$.} \underline{Additional Parameters:} six Gaussian components added at: \textcolor{cyan}{$R_{\rm r}=1\farcs66\pm0\farcs01$, $15\farcs64\pm0\farcs08$, $23\farcs68\pm0\farcs25$, $41\farcs98\pm0\farcs07$, $54\farcs78\pm0\farcs26$, \& $83\farcs08\pm0\farcs08$; with $\mu_0 = 17.16\pm0.02$, $20.28\pm0.02$, $21.22\pm0.05$, $23.50\pm0.02$, $22.86\pm0.02$, \& $26.40\pm0.06\,\text{mag\,arcsec}^{-2}$; and FWHM = $1\farcs12\pm0\farcs02$, $8\farcs78\pm0\farcs25$, $9\farcs74\pm0\farcs24$, $10\farcs18\pm0\farcs21$, $37\farcs23\pm0\farcs57$, \& $2\farcs98\pm0\farcs21$, respectively.} The outermost Gaussian is below the visible portion of the plots. \textbf{Right two panels}---The model represents $0\arcsec \leq R_{\rm eq} \leq 70\farcs37$ with $\Delta_{\rm rms}=0.0352\,\text{mag\,arcsec}^{-2}$. \underline{S{\'e}rsic Profile Parameters:} \textcolor{red}{$R_e=3\farcs15\pm0\farcs11$, $\mu_e=18.52\pm0.06\,\text{mag\,arcsec}^{-2}$, and $n=2.49\pm0.05$}. \underline{Ferrers Profile Parameters:} \textcolor{Orange}{$\mu_0 = 19.51\pm0.03\,\text{mag\,arcsec}^{-2}$, $R_{\rm end} = 12\farcs20\pm0\farcs02$, and $\alpha = 1.76\pm0.05$.} \underline{Exponential Profile Parameters:} \textcolor{blue}{$\mu_0 = 20.32\pm0.01\,\text{mag\,arcsec}^{-2}$ and $h = 20\farcs06\pm0\farcs11$.} \underline{Additional Parameters:} six Gaussian components added at: \textcolor{cyan}{$R_{\rm r}=1\farcs35\pm0\farcs01$, $13\farcs12\pm0\farcs03$, $18\farcs38\pm0\farcs05$, $33\farcs17\pm0\farcs04$, $42\farcs08\pm0\farcs10$, \& $62\farcs98\pm0\farcs04$; with $\mu_0 = 17.50\pm0.02$, $20.63\pm0.01$, $20.86\pm0.01$, $23.85\pm0.03$, $22.99\pm0.01$, \& $25.09\pm0.02\,\text{mag\,arcsec}^{-2}$; and FWHM = $1\farcs01\pm0\farcs02$, $3\farcs82\pm0\farcs07$, $9\farcs39\pm0\farcs08$, $4\farcs10\pm0\farcs12$, $24\farcs91\pm0\farcs29$, \& $8\farcs08\pm0\farcs02$, respectively.} The outermost Gaussian is below the visible portion of the plots.}
\label{NGC2273_plot}
\end{sidewaysfigure}

\subsection{NGC~2273 (Figure~\ref{NGC2273_plot})}

The galaxy NGC 2273 contains a bar encompassed by a pseudoring \citep{Comeron:2010} formed from the tight beginnings of two spiral arms. We have added six Gaussian components: $R_{\rm maj} \approx 2\arcsec$ ($R_{\rm eq}\approx 1\arcsec$) for a nuclear ring, $R_{\rm maj} \approx 16\arcsec$ ($R_{\rm eq}\approx 13\arcsec$) for bar anase, $R_{\rm maj} \approx 24\arcsec$ ($R_{\rm eq}\approx 18\arcsec$) for the pseudoring, and, at $R_{\rm maj} \approx 42\arcsec$ ($R_{\rm eq}\approx 33\arcsec$), $R_{\rm maj} \approx 55\arcsec$ ($R_{\rm eq}\approx 42\arcsec$), and $R_{\rm maj} \approx 83\arcsec$ ($R_{\rm eq}\approx 63\arcsec$), for the spiral arms.

\begin{sidewaysfigure}
\includegraphics[clip=true,trim= 11mm 1mm 1mm 6mm,width=0.249\textwidth]{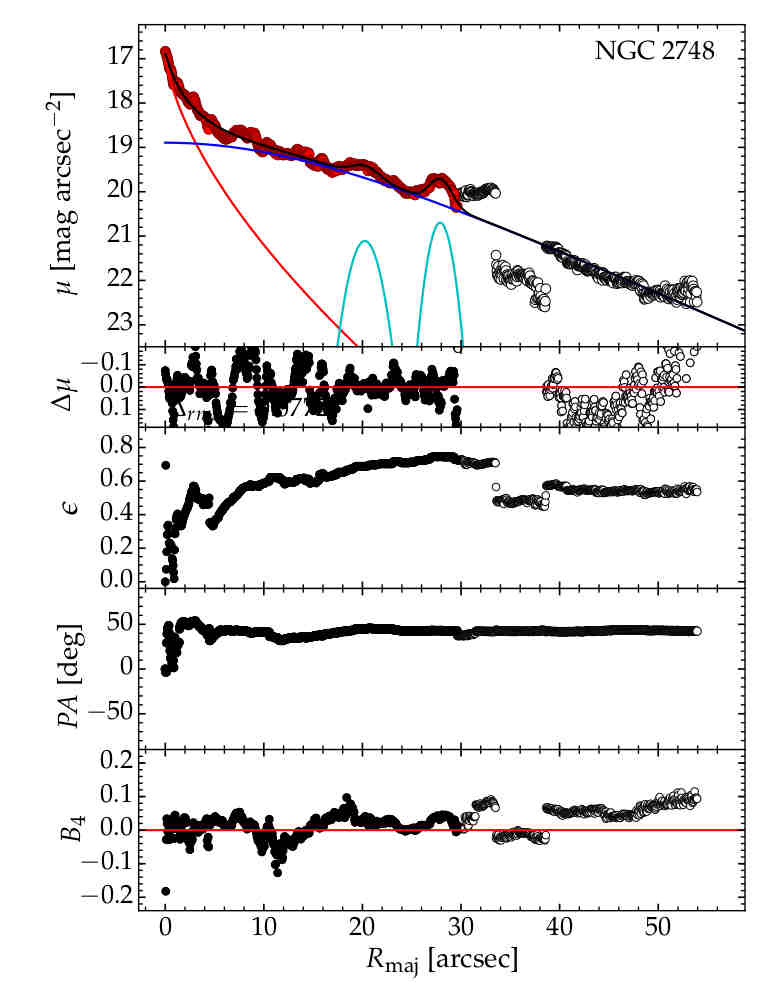}
\includegraphics[clip=true,trim= 11mm 1mm 1mm 6mm,width=0.249\textwidth]{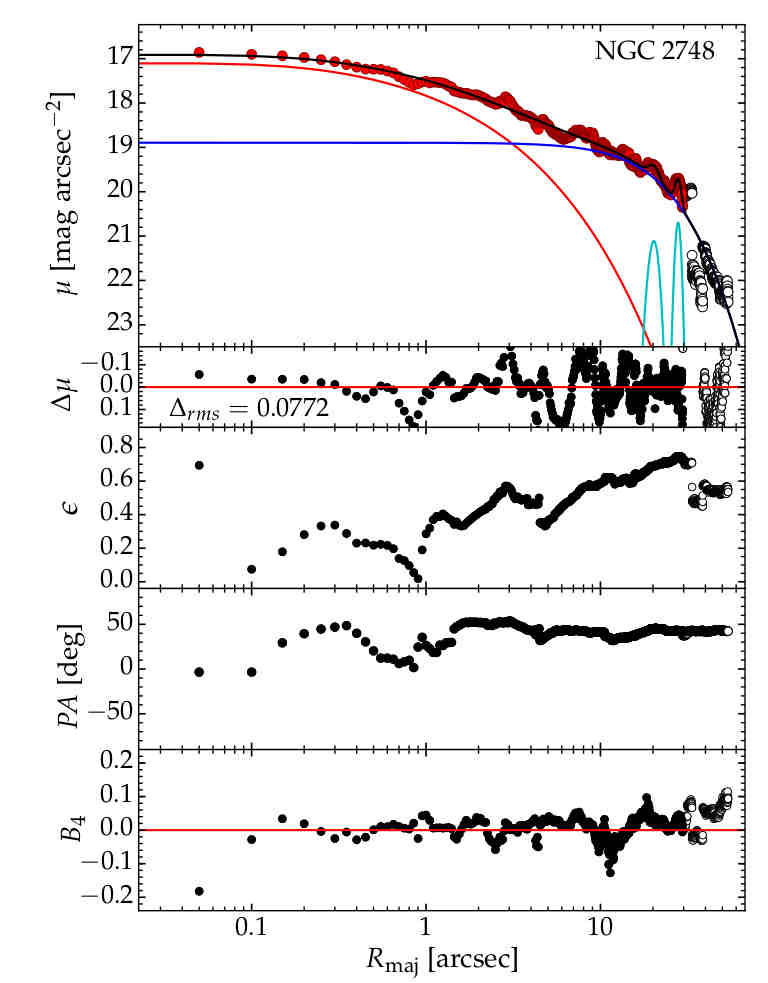}
\includegraphics[clip=true,trim= 11mm 1mm 1mm 6mm,width=0.249\textwidth]{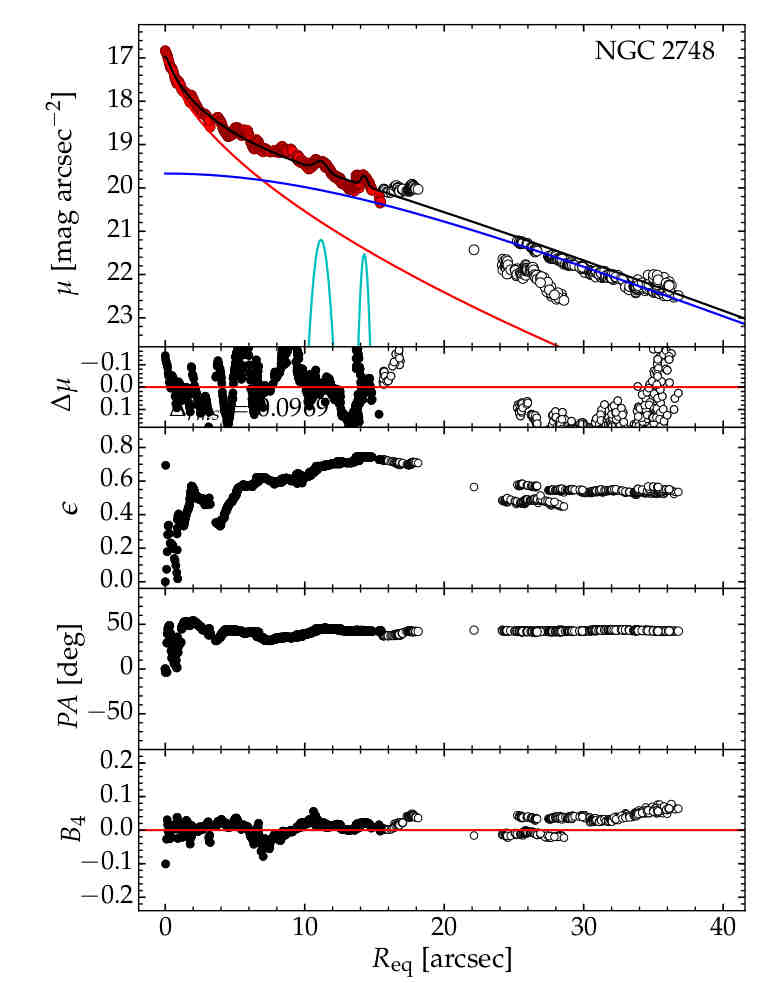}
\includegraphics[clip=true,trim= 11mm 1mm 1mm 6mm,width=0.249\textwidth]{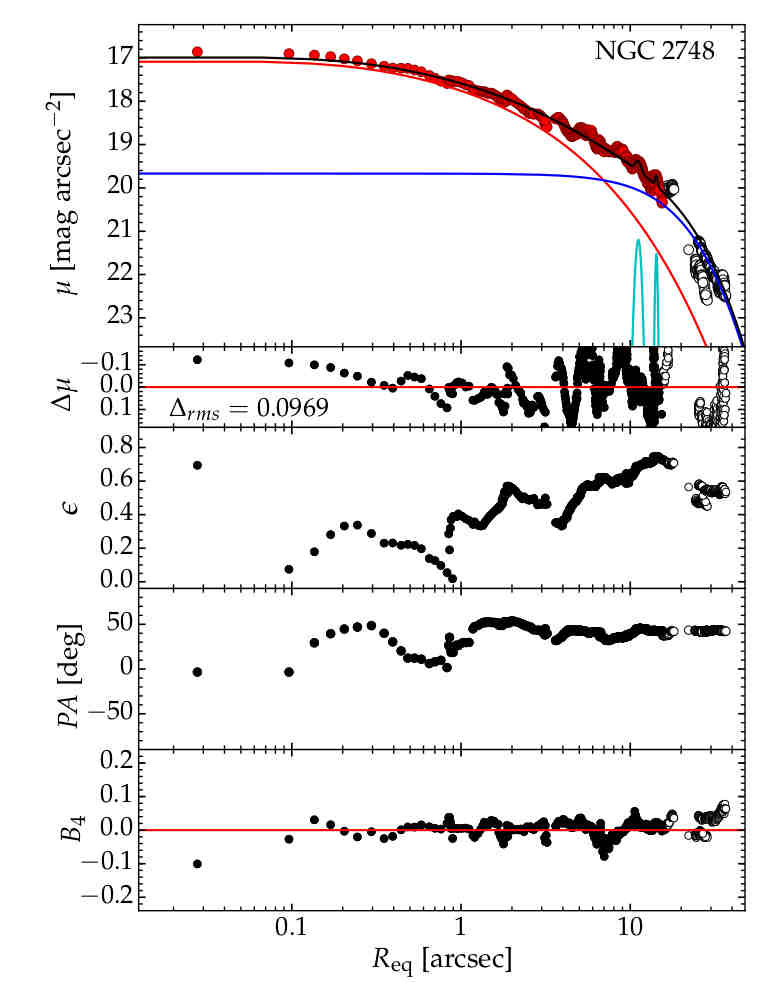}
\caption{\textit{HST} WFPC2/WFC F814W surface brightness profile for NGC~2748, with a physical scale of 0.0882$\,\text{kpc}\,\text{arcsec}^{-1}$. \textbf{Left two panels}---The model represents $0\arcsec \leq R_{\rm maj} \leq 29\farcs64$ with $\Delta_{\rm rms}=0.0772\,\text{mag\,arcsec}^{-2}$ and additional data from $29\farcs64 < R_{\rm maj} \leq 53\farcs96$ is plotted, but not modeled. \underline{S{\'e}rsic Profile Parameters:} \textcolor{red}{$R_e=5\farcs74\pm0\farcs79$, $\mu_e=19.91\pm0.17\,\text{mag\,arcsec}^{-2}$, and $n=1.59\pm0.11$.} \underline{Edge-on Disk Model Parameters:} \textcolor{blue}{$\mu_0 = 18.89\pm0.04\,\text{mag\,arcsec}^{-2}$ and $h_z = 22\farcs25\pm0\farcs48$.} \underline{Additional Parameters:} two Gaussian components added at: \textcolor{cyan}{$R_{\rm r}=20\farcs24\pm0\farcs09$ \& $27\farcs88\pm0\farcs04$; with $\mu_0 = 21.11\pm0.07$ \& $20.70\pm0.04\,\text{mag\,arcsec}^{-2}$; and FWHM = $3\farcs13\pm0\farcs24$ \& $2\farcs42\pm0\farcs12$, respectively.} \textbf{Right two panels}---The model represents $0\arcsec \leq R_{\rm maj} \leq 15\farcs60$ with $\Delta_{\rm rms}=0.0969\,\text{mag\,arcsec}^{-2}$ and additional data from $15\farcs60 < R_{\rm maj} \leq 38\farcs26$ is plotted, but not modeled. \underline{S{\'e}rsic Profile Parameters:} \textcolor{red}{$R_e=8\farcs29\pm0\farcs38$, $\mu_e=20.15\pm0.08\,\text{mag\,arcsec}^{-2}$, and $n=1.71\pm0.06$.} \underline{Edge-on Disk Model Parameters:} \textcolor{blue}{$\mu_0 = 19.67\pm0.05\,\text{mag\,arcsec}^{-2}$ and $h_z = 18\farcs20\pm0\farcs25$.} \underline{Additional Parameters:} two Gaussian components added at: \textcolor{cyan}{$R_{\rm r}=11\farcs16\pm0\farcs08$ \& $14\farcs27\pm0\farcs06$; with $\mu_0 = 21.20\pm0.19$ \& $21.53\pm0.27\,\text{mag\,arcsec}^{-2}$; and FWHM = $0\farcs97\pm0\farcs19$ \& $0\farcs51\pm0\farcs14$, respectively.}}
\label{NGC2748_plot}
\end{sidewaysfigure}

\subsection{NGC~2748 (Figure~\ref{NGC2748_plot})}

A near edge-on galaxy, NGC~2748 is inclined by $62\fdg4\pm10\fdg7$ with respect to the plane of the sky. Because of this, we use the edge-on disk model. We add two Gaussian components for spiral-arm contributions to the surface brightness profile at $R_{\rm maj} \approx 20\arcsec$ ($R_{\rm eq}\approx 11\arcsec$) and $R_{\rm maj} \approx 28\arcsec$ ($R_{\rm eq}\approx 14\arcsec$). The inner S{\'e}rsic model encapsulates a likely nuclear (disk) component, not a spheroid.

\begin{sidewaysfigure}
\includegraphics[clip=true,trim= 11mm 1mm 1mm 5mm,width=0.249\textwidth]{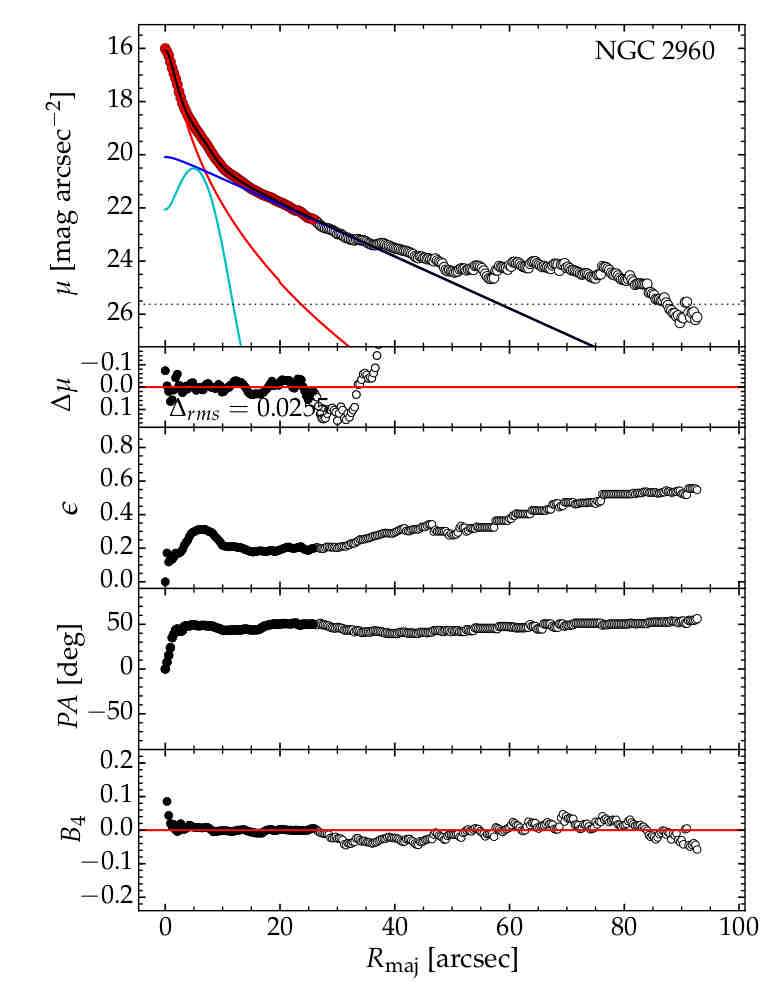}
\includegraphics[clip=true,trim= 11mm 1mm 1mm 5mm,width=0.249\textwidth]{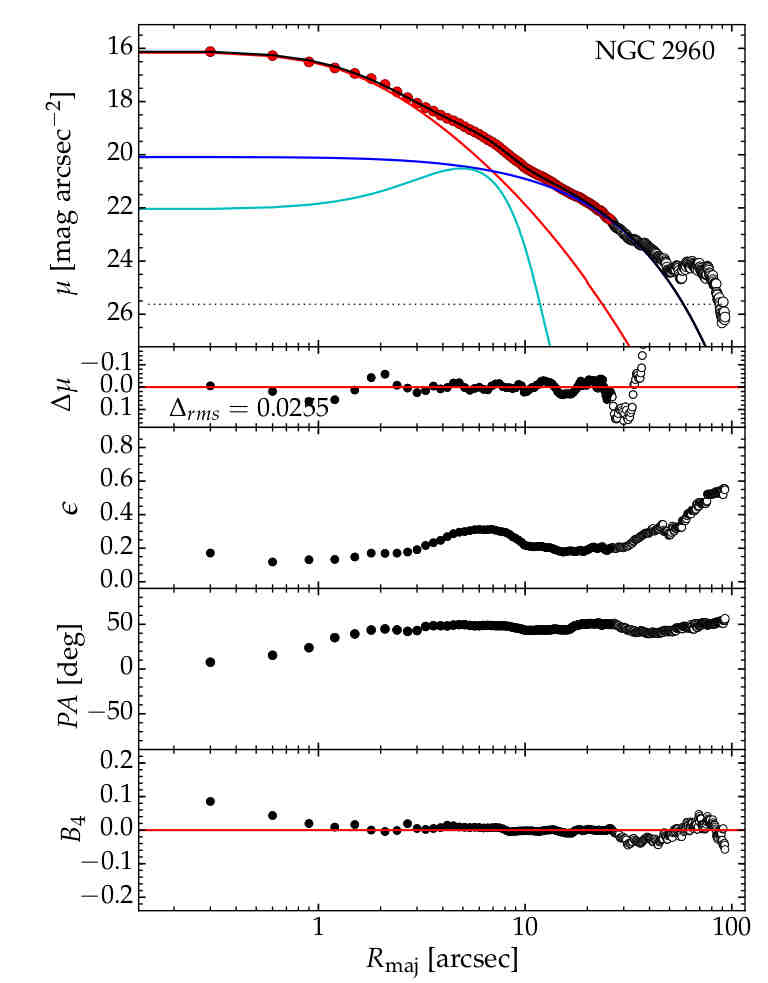}
\includegraphics[clip=true,trim= 11mm 1mm 1mm 5mm,width=0.249\textwidth]{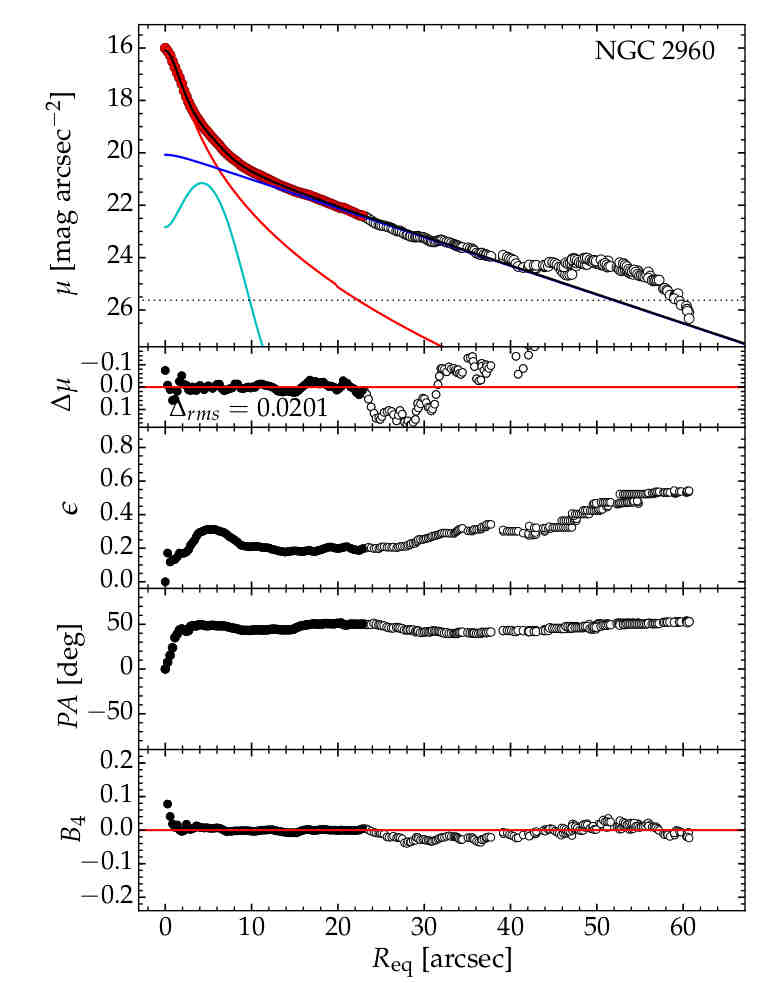}
\includegraphics[clip=true,trim= 11mm 1mm 1mm 5mm,width=0.249\textwidth]{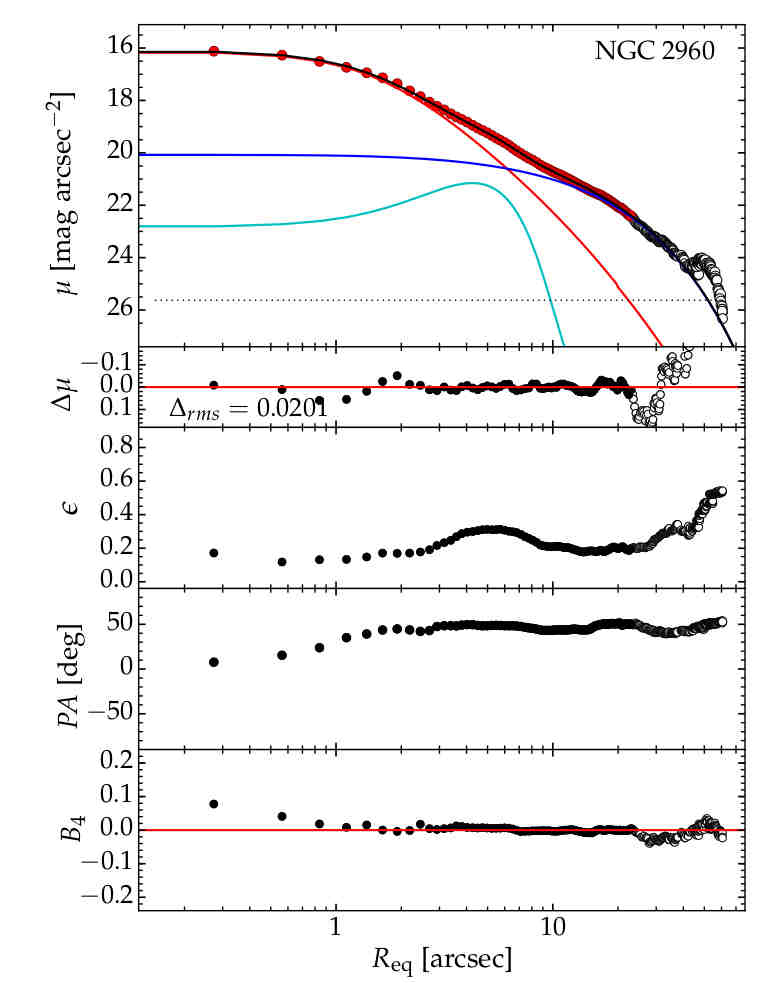}
\caption{\textit{Spitzer} $3.6\,\micron$ surface brightness profile for NGC~2960, with a physical scale of 0.3447$\,\text{kpc}\,\text{arcsec}^{-1}$. \textbf{Left two panels}---The model represents $0\arcsec \leq R_{\rm maj} \leq 26\arcsec$ with $\Delta_{\rm rms}=0.0255\,\text{mag\,arcsec}^{-2}$ and additional data from $26\arcsec < R_{\rm maj} \leq 93\arcsec$ is plotted, but not modeled. \underline{S{\'e}rsic Profile Parameters:} \textcolor{red}{$R_e=2\farcs35\pm0\farcs27$, $\mu_e=18.04\pm0.20\,\text{mag\,arcsec}^{-2}$, and $n=2.59\pm0.12$.} \underline{Exponential Profile Parameters:} \textcolor{blue}{$\mu_0 = 19.93\pm0.11\,\text{mag\,arcsec}^{-2}$ and $h = 11\farcs12\pm0\farcs43$.} \underline{Additional Parameters:} one Gaussian component added at: \textcolor{cyan}{$R_{\rm r}=5\farcs07\pm0\farcs36$ with $\mu_0 = 20.32\pm0.15\,\text{mag\,arcsec}^{-2}$ and FWHM = $4\farcs15\pm0\farcs59$.} \textbf{Right two panels}---The model represents $0\arcsec \leq R_{\rm eq} \leq 23\arcsec$ with $\Delta_{\rm rms}=0.0201\,\text{mag\,arcsec}^{-2}$ and additional data from $23\arcsec < R_{\rm maj} \leq 62\arcsec$ is plotted, but not modeled. \underline{S{\'e}rsic Profile Parameters:} \textcolor{red}{$R_e=2\farcs19\pm0\farcs24$, $\mu_e=18.30\pm0.19\,\text{mag\,arcsec}^{-2}$, and $n=2.86\pm0.11$.} \underline{Exponential Profile Parameters:} \textcolor{blue}{$\mu_0 = 19.93\pm0.09\,\text{mag\,arcsec}^{-2}$ and $h = 9\farcs89\pm0\farcs30$.} \underline{Additional Parameters:} one Gaussian component added at: \textcolor{cyan}{$R_{\rm r}=4\farcs32\pm0\farcs46$ with $\mu_0 = 20.88\pm0.19\,\text{mag\,arcsec}^{-2}$ and FWHM = $3\farcs28\pm0\farcs79$.}}
\label{NGC2960_plot}
\end{sidewaysfigure}

\subsection{NGC~2960 (Figure~\ref{NGC2960_plot})}

We have added a Gaussian component at $R_{\rm maj} \approx 5\arcsec$ ($R_{\rm eq}\approx4\arcsec$) to account for increased light from the spiral arms in the disk.

\begin{sidewaysfigure}
\includegraphics[clip=true,trim= 11mm 1mm 1mm 5mm,width=0.249\textwidth]{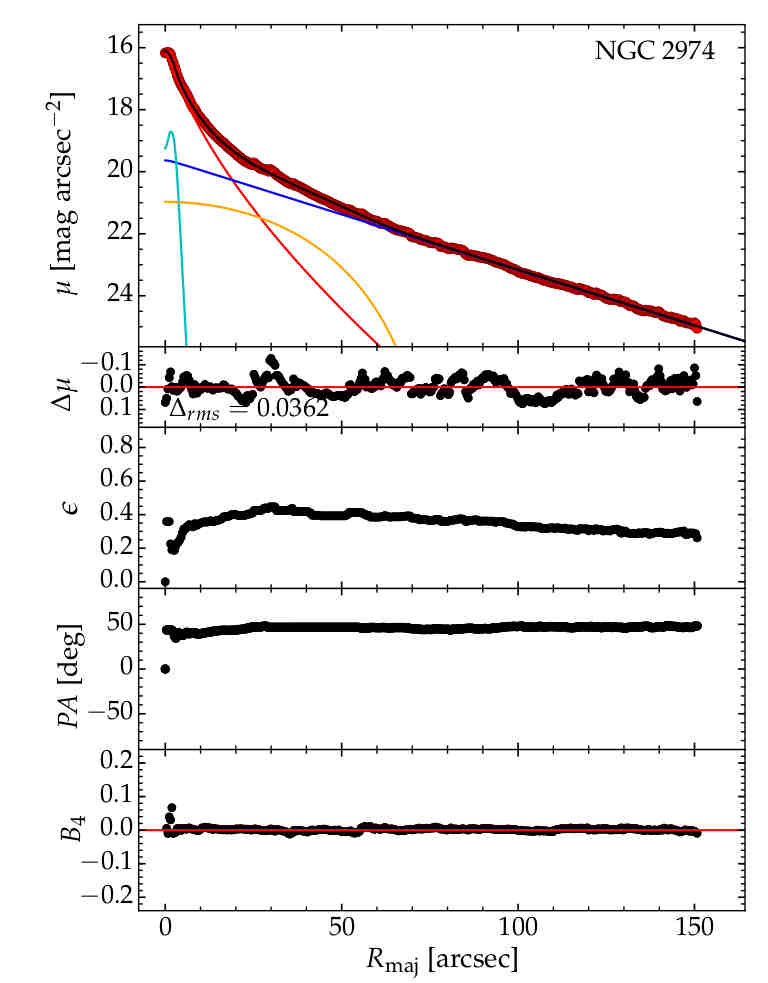}
\includegraphics[clip=true,trim= 11mm 1mm 1mm 5mm,width=0.249\textwidth]{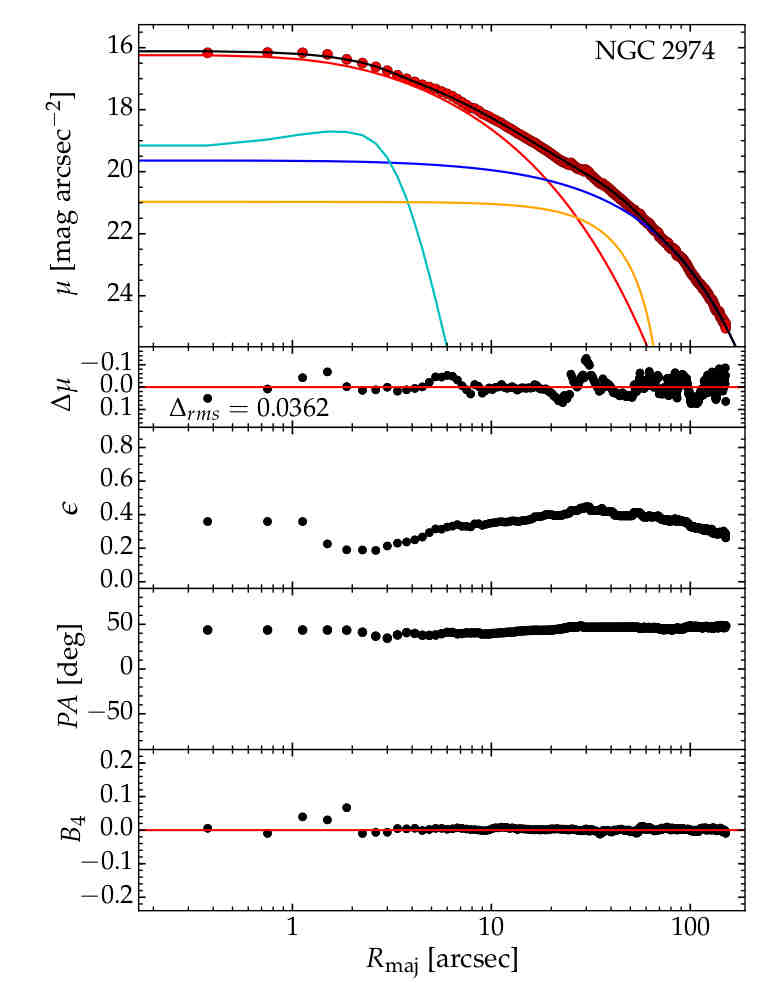}
\includegraphics[clip=true,trim= 11mm 1mm 1mm 5mm,width=0.249\textwidth]{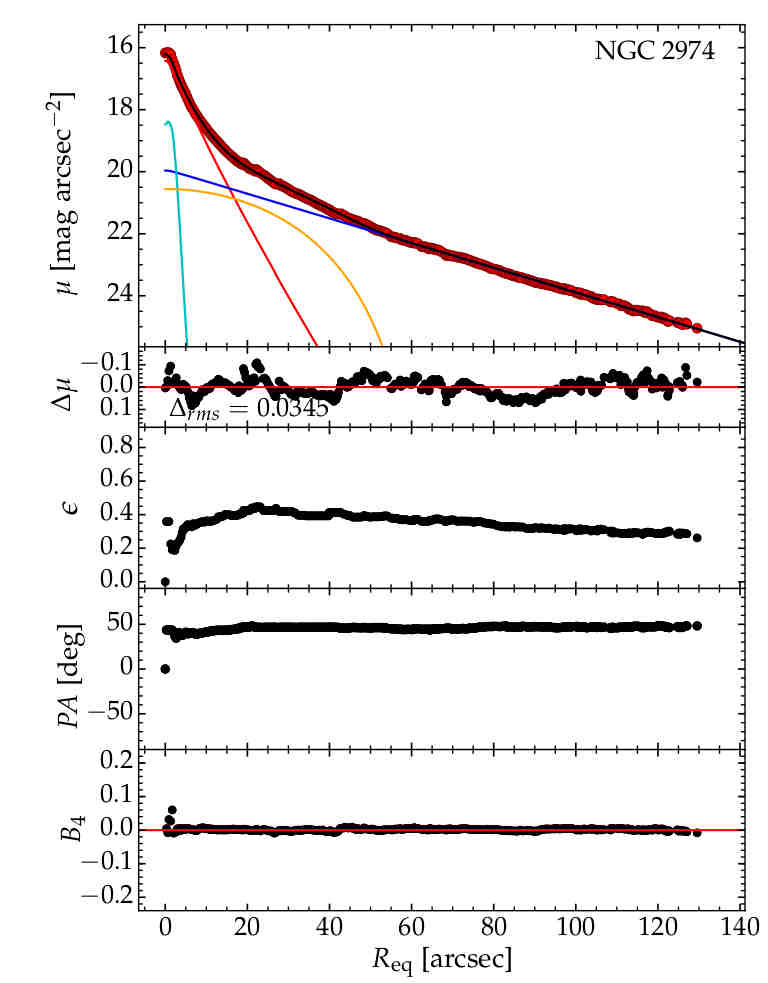}
\includegraphics[clip=true,trim= 11mm 1mm 1mm 5mm,width=0.249\textwidth]{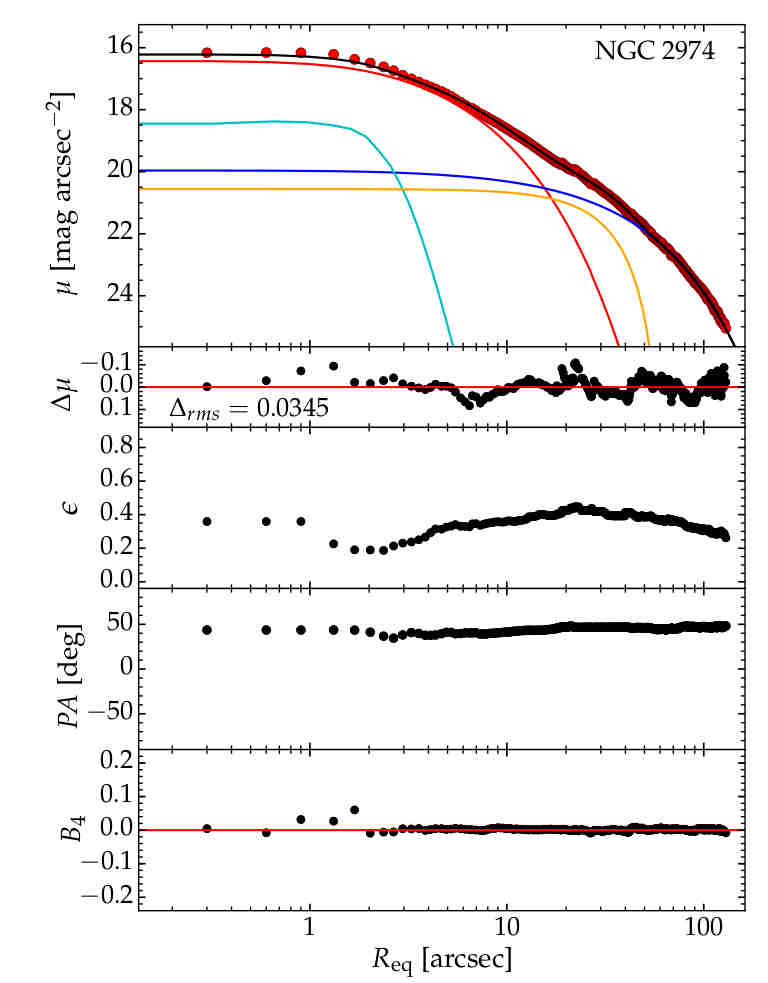}
\caption{\textit{Spitzer} $3.6\,\micron$ surface brightness profile for NGC~2974, with a physical scale of 0.1042$\,\text{kpc}\,\text{arcsec}^{-1}$. \textbf{Left two panels}---The model represents $0\arcsec \leq R_{\rm maj} \leq 151\arcsec$ with $\Delta_{\rm rms}=0.0362\,\text{mag\,arcsec}^{-2}$. \underline{S{\'e}rsic Profile Parameters:} \textcolor{red}{$R_e=9\farcs21\pm0\farcs45$, $\mu_e=18.49\pm0.08\,\text{mag\,arcsec}^{-2}$, and $n=1.56\pm0.09$}. \underline{Ferrers Profile Parameters:} \textcolor{Orange}{$\mu_0 = 20.97\pm0.19\,\text{mag\,arcsec}^{-2}$, $R_{\rm end} = 80\farcs60\pm7\farcs32$, and $\alpha = 10.00\pm2.30$.} \underline{Exponential Profile Parameters:} \textcolor{blue}{$\mu_0 = 19.60\pm0.02\,\text{mag\,arcsec}^{-2}$ and $h = 30\farcs35\pm0\farcs11$.} \underline{Additional Parameters:} one Gaussian component added at: \textcolor{cyan}{$R_{\rm r}=2\farcs04\pm999\arcsec$ with $\mu_0 = 15.36\pm999\,\text{mag\,arcsec}^{-2}$ and FWHM = $0\farcs22\pm999\arcsec$.} \textbf{Right two panels}---The model represents $0\arcsec \leq R_{\rm eq} \leq 130\arcsec$ with $\Delta_{\rm rms}=0.0345\,\text{mag\,arcsec}^{-2}$. \underline{S{\'e}rsic Profile Parameters:} \textcolor{red}{$R_e=6\farcs53\pm0\farcs14$, $\mu_e=18.12\pm0.04\,\text{mag\,arcsec}^{-2}$, and $n=1.17\pm0.06$}. \underline{Ferrers Profile Parameters:} \textcolor{Orange}{$\mu_0 = 20.57\pm0.11\,\text{mag\,arcsec}^{-2}$, $R_{\rm end} = 63\farcs73\pm3\farcs43$, and $\alpha = 10.00\pm1.77$}. \underline{Exponential Profile Parameters:} \textcolor{blue}{$\mu_0 = 19.92\pm0.01\,\text{mag\,arcsec}^{-2}$ and $h = 27\farcs29\pm0\farcs09$.} \underline{Additional Parameters:} one Gaussian component added at: \textcolor{cyan}{$R_{\rm r}=1\farcs34\pm999\arcsec$ with $\mu_0 = 16.04\pm999\,\text{mag\,arcsec}^{-2}$ and FWHM = $0\farcs16\pm999\arcsec$.}}
\label{NGC2974_plot}
\end{sidewaysfigure}

\subsection{NGC~2974 (Figure~\ref{NGC2974_plot})}

Until recently, NGC~2974 had been classified as an elliptical galaxy; it was \citet{Savorgnan:2016} who first identified it as a barred spiral galaxy by removing the obscuration caused by a bright foreground star. It hosts a Type 2 Seyfert AGN \citep{Veron-Cetty:2006} with filamentary dust in its nucleus \citep{Tran:2001}. We model this nuclear component with a Gaussian at $R_{\rm maj} \approx 2\arcsec$ ($R_{\rm eq}\approx1\arcsec$).

\begin{sidewaysfigure}
\includegraphics[clip=true,trim= 11mm 1mm 0mm 6mm,width=0.249\textwidth]{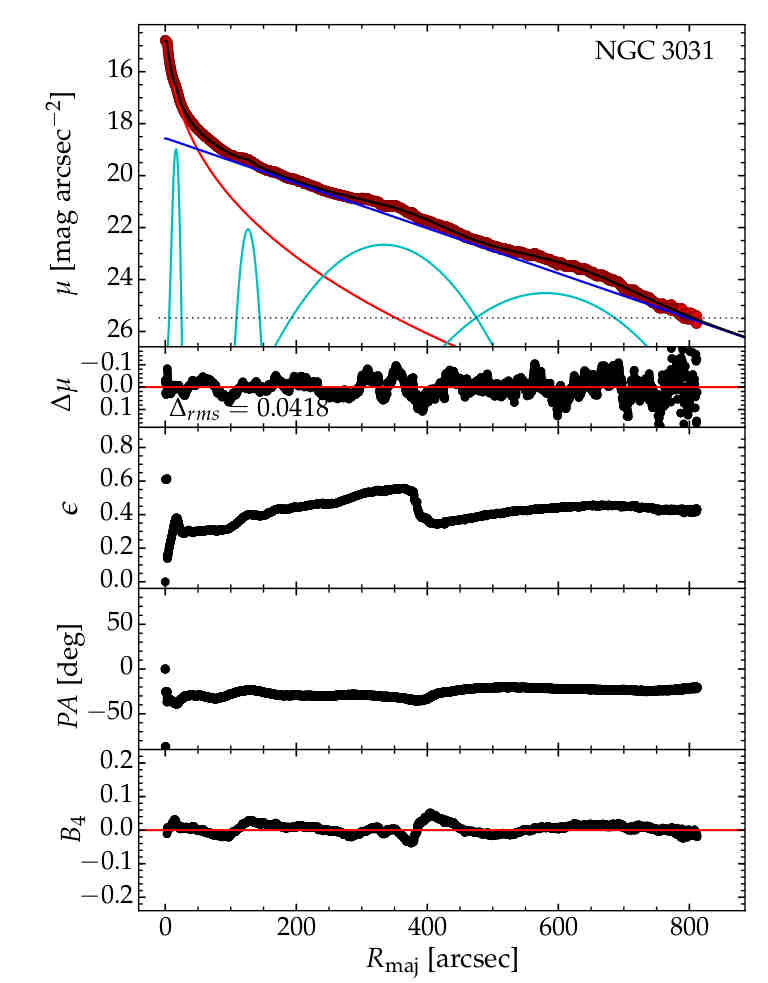}
\includegraphics[clip=true,trim= 11mm 1mm 0mm 6mm,width=0.249\textwidth]{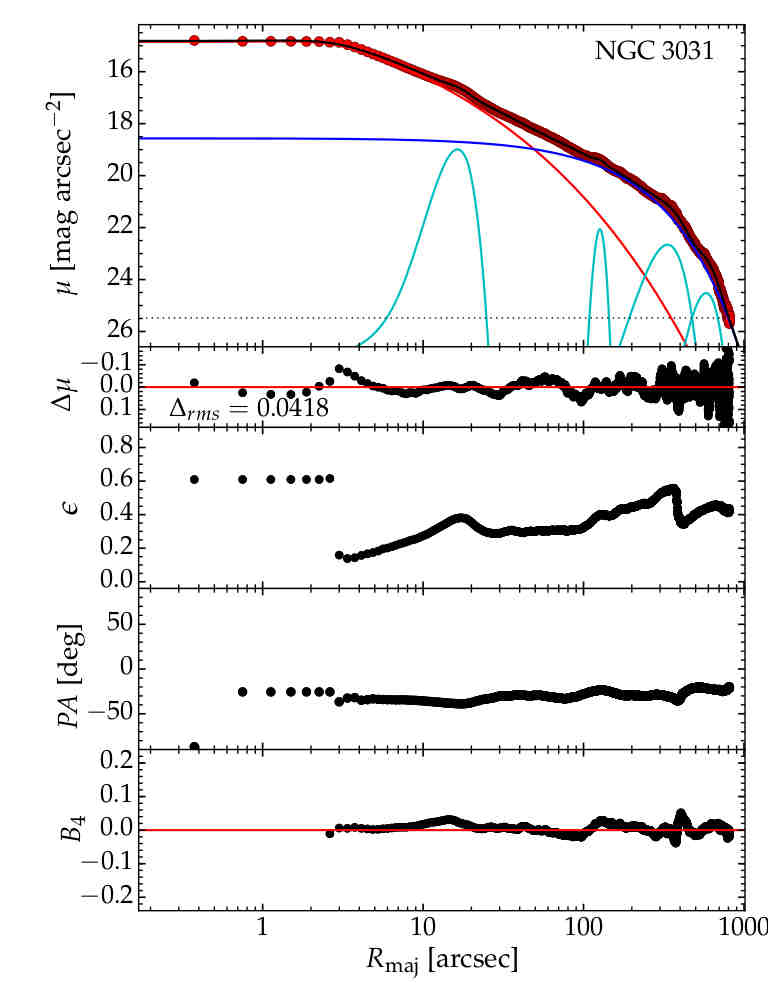}
\includegraphics[clip=true,trim= 11mm 1mm 0mm 6mm,width=0.249\textwidth]{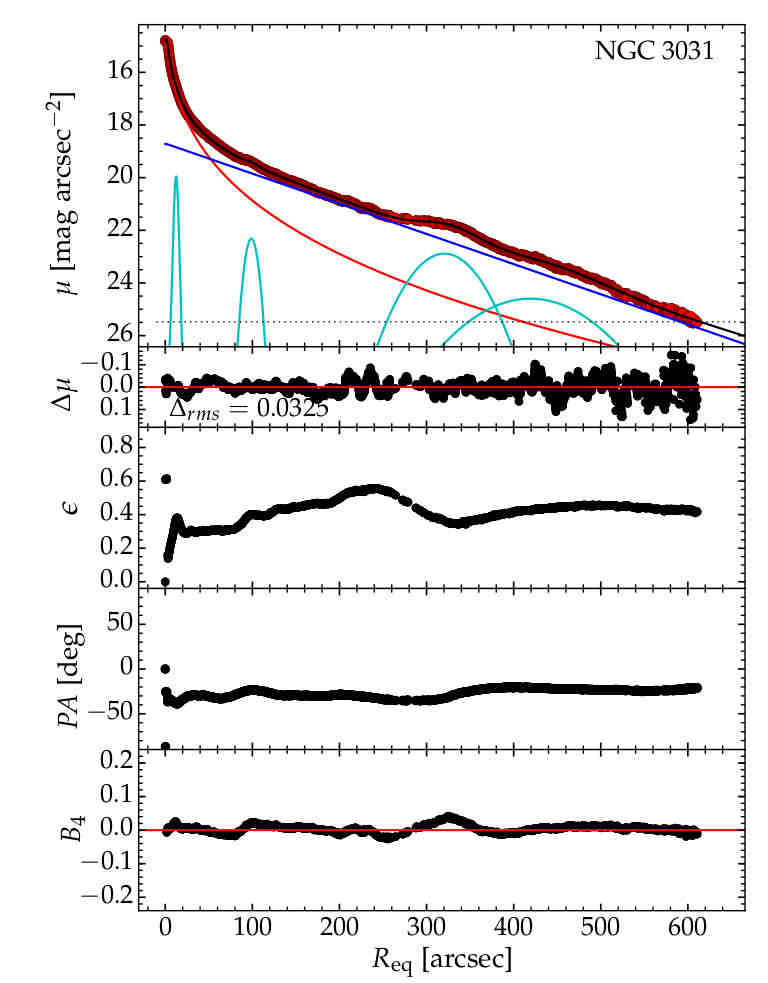}
\includegraphics[clip=true,trim= 11mm 1mm 0mm 6mm,width=0.249\textwidth]{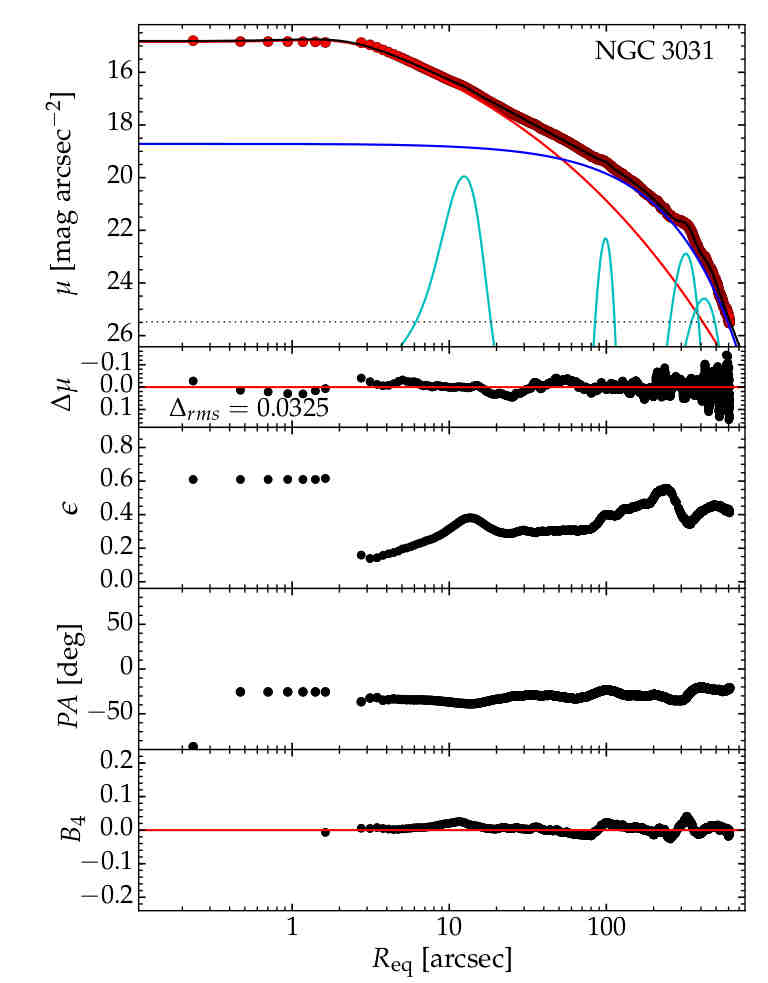}
\caption{\textit{Spitzer} $3.6\,\micron$ surface brightness profile for NGC~3031, with a physical scale of 0.0169$\,\text{kpc}\,\text{arcsec}^{-1}$. \textbf{Left two panels}---The model represents $0\arcsec \leq R_{\rm maj} \leq 812\arcsec$ with $\Delta_{\rm rms}=0.0418\,\text{mag\,arcsec}^{-2}$. \underline{Core-S{\'e}rsic Profile Parameters:} \textcolor{red}{$\gamma=-0.16\pm0.15$, $R_b=2\farcs14\pm0\farcs33$, $R_e=36\farcs19\pm1\farcs43$, $\mu_0'=12.58\pm0.16\,\text{mag\,arcsec}^{-2}$, and $n=2.81\pm0.11$.} \underline{Exponential Profile Parameters:} \textcolor{blue}{$\mu_0 = 18.26\pm0.01\,\text{mag\,arcsec}^{-2}$ and $h = 125\farcs10\pm0\farcs25$.} \underline{Additional Parameters:} four Gaussian components added at: \textcolor{cyan}{$R_{\rm r}=16\farcs37\pm0\farcs41$, $126\farcs42\pm0\farcs77$, $333\farcs88\pm1\farcs32$, \& $580\farcs25\pm1\farcs48$; with $\mu_0 = 18.56\pm0.01$, $22.07\pm0.11$, $22.66\pm0.02$, \& $24.53\pm0.02\,\text{mag\,arcsec}^{-2}$; and FWHM = $5\farcs71\pm0\farcs99$, $16\farcs68\pm1\farcs88$, $146\farcs11\pm2\farcs51$, \& $188\farcs02\pm2\farcs96$, respectively.} \textbf{Right two panels}---The model represents $0\arcsec \leq R_{\rm eq} \leq 613\arcsec$ with $\Delta_{\rm rms}=0.0325\,\text{mag\,arcsec}^{-2}$. \underline{Core-S{\'e}rsic Profile Parameters:} \textcolor{red}{$\gamma=-0.79\pm0.33$, $R_b=1\farcs39\pm0\farcs16$, $R_e=42\farcs98\pm0\farcs74$, $\mu_0'=11.75\pm0.11\,\text{mag\,arcsec}^{-2}$, and $n=3.46\pm0.06$}. \underline{Exponential Profile Parameters:} \textcolor{blue}{$\mu_0 = 18.71\pm0.00\,\text{mag\,arcsec}^{-2}$ and $h = 94\farcs78\pm0\farcs34$.} \underline{Additional Parameters:} four Gaussian components added at: \textcolor{cyan}{$R_{\rm r}=12\farcs44\pm0\farcs46$, $98\farcs84\pm0\farcs55$, $320\farcs52\pm0\farcs46$, \& $419\farcs08\pm1\farcs57$; with $\mu_0 = 19.61\pm0.45$, $22.32\pm0.10$, $22.89\pm0.01$, \& $24.60\pm0.02\,\text{mag\,arcsec}^{-2}$; and FWHM = $2\farcs77\pm1\farcs58$, $13\farcs71\pm1\farcs39$, $71\farcs54\pm0\farcs97$, \& $131\farcs27\pm2\farcs34$, respectively.}}
\label{NGC3031_plot}
\end{sidewaysfigure}

\subsection{NGC~3031 (Figure~\ref{NGC3031_plot})}

Previous studies of NGC~3031 (M81, ``Bode's Galaxy'') have identified a nuclear bar at $R_{\rm maj} \lesssim 17\arcsec$ and a large-scale bar at $R_{\rm maj} \lesssim 130\arcsec$ \citep{Elmegreen:1995,Gutierrez:2011,Erwin:2013}. In the $3.6\,\micron$ imaging, we find that evidence of such bars is extremely faint and only contributes two small bumps in the surface brightness profile at $R_{\rm maj} \approx 16\arcsec$ ($R_{\rm eq}\approx12\arcsec$) and $R_{\rm maj} \approx 126\arcsec$ ($R_{\rm eq}\approx99\arcsec$). We model these minor contributions with Gaussians rather than Ferrers profiles. We have added two additional Gaussians at $R_{\rm maj} \approx 334\arcsec$ ($R_{\rm eq}\approx321\arcsec$) and $R_{\rm maj} \approx 580\arcsec$ ($R_{\rm eq}\approx419\arcsec$) to represent the crossings of the large $m=2$ grand design spiral arms along the major axis profile of the galaxy. The \textit{Spitzer} image displays diminished light in the nucleus, and we thus model it with a core-S{\'e}rsic model.

\begin{sidewaysfigure}
\includegraphics[clip=true,trim= 11mm 1mm 4mm 6mm,width=0.249\textwidth]{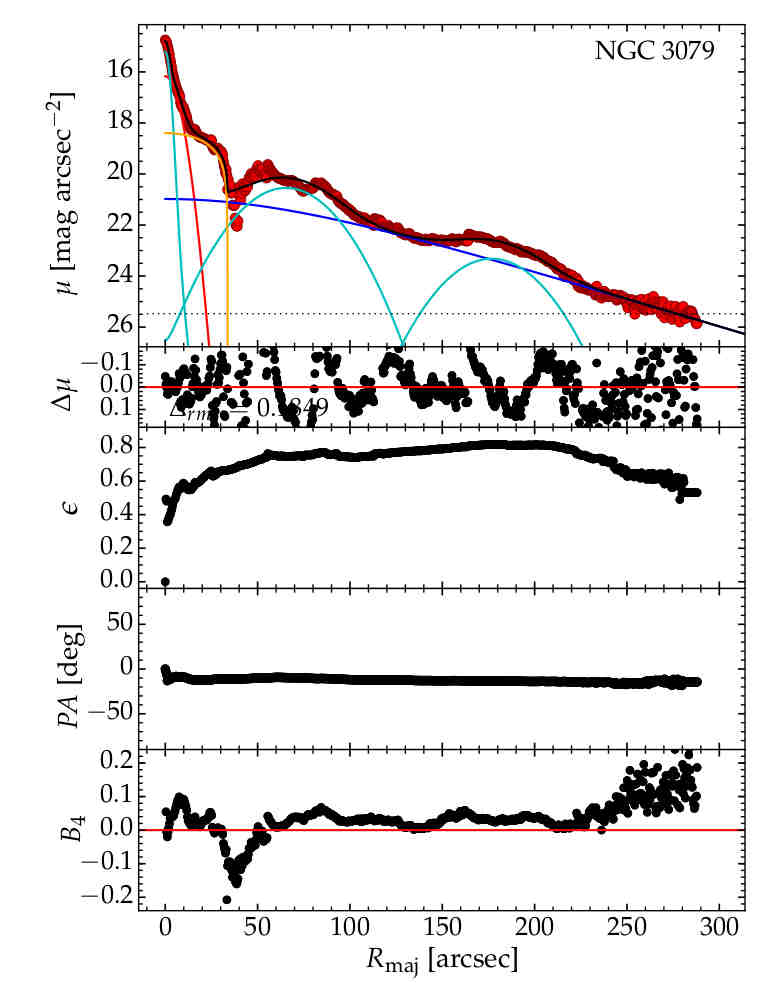}
\includegraphics[clip=true,trim= 11mm 1mm 4mm 6mm,width=0.249\textwidth]{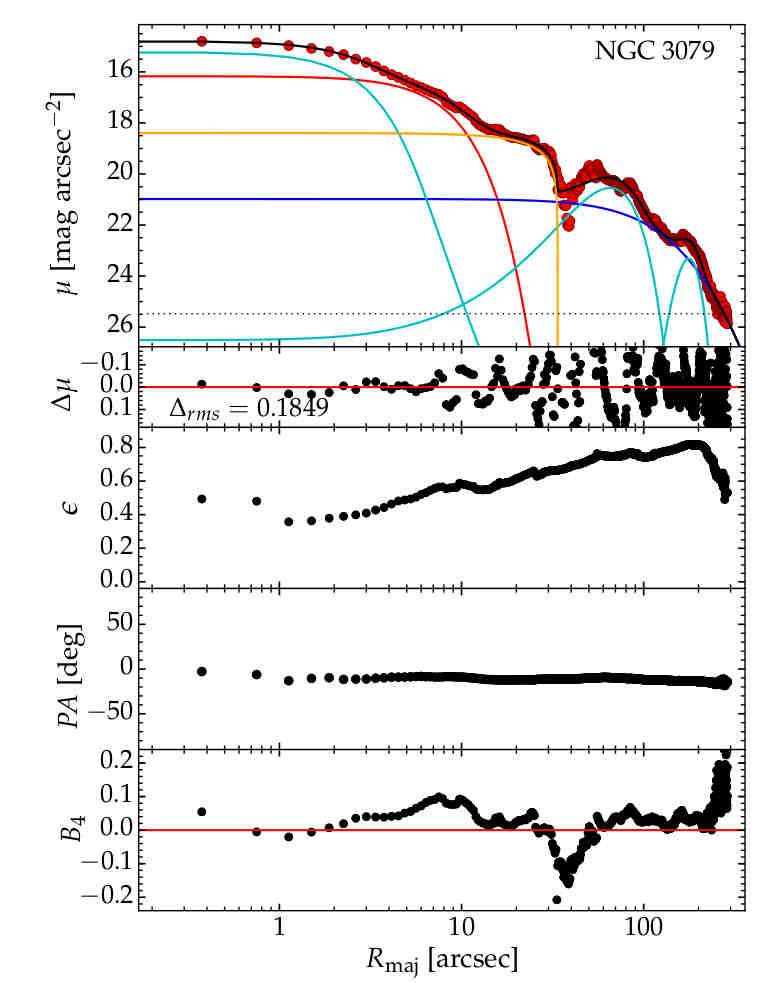}
\includegraphics[clip=true,trim= 11mm 1mm 4mm 6mm,width=0.249\textwidth]{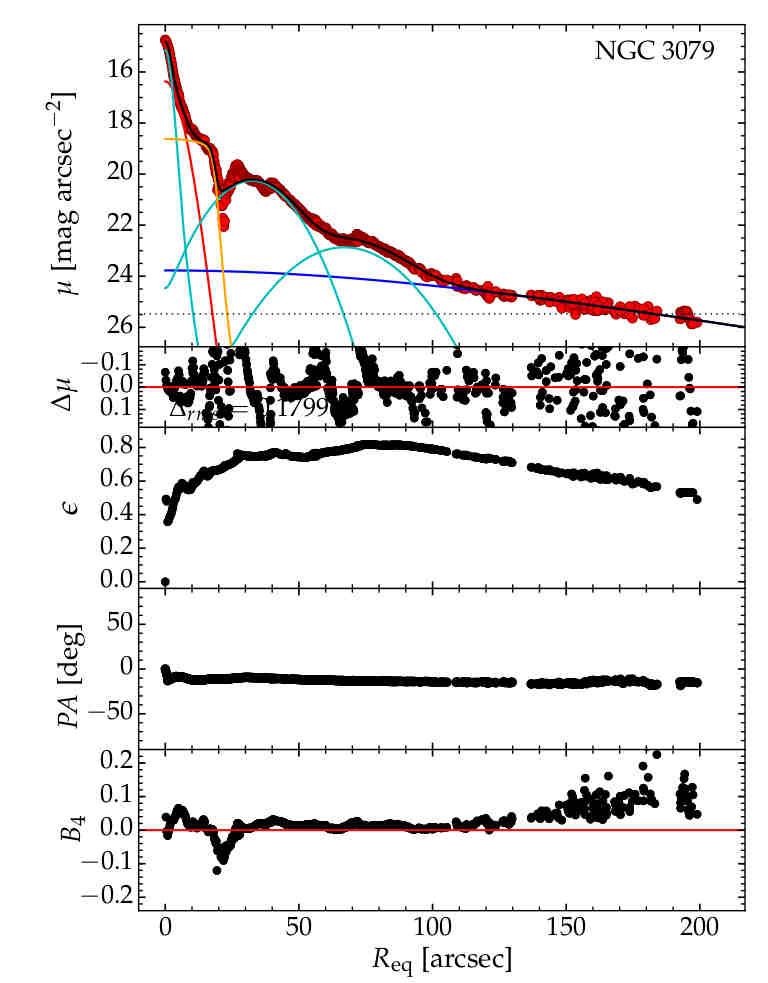}
\includegraphics[clip=true,trim= 11mm 1mm 4mm 6mm,width=0.249\textwidth]{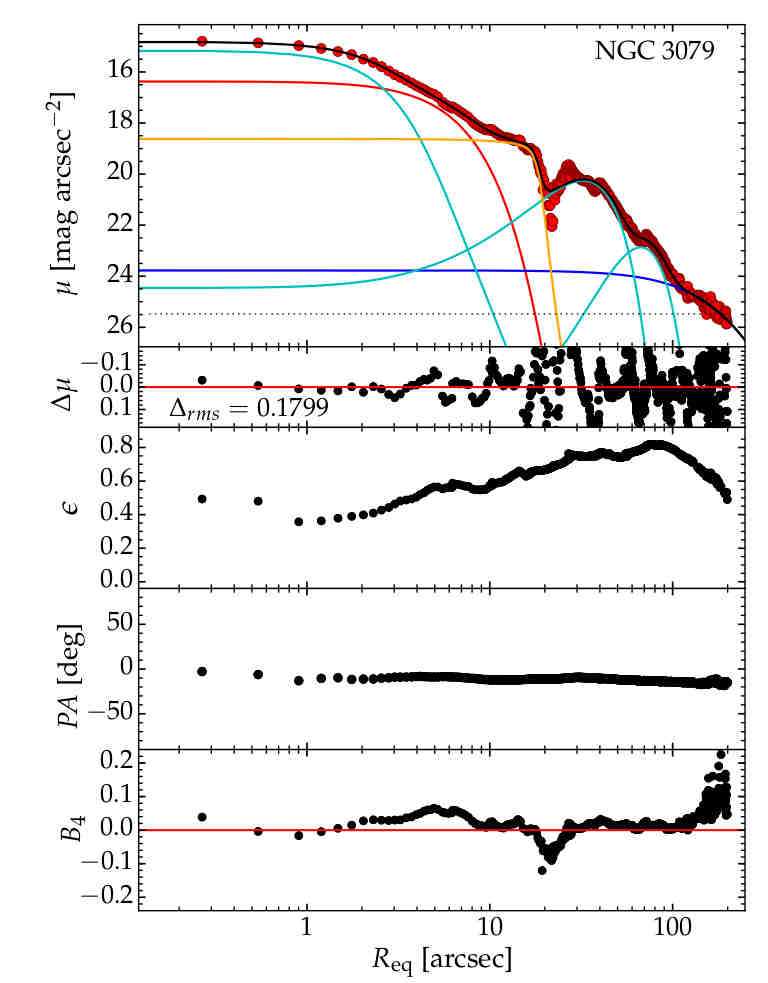}
\caption{\textit{Spitzer} $3.6\,\micron$ surface brightness profile for NGC~3079, with a physical scale of 0.0801$\,\text{kpc}\,\text{arcsec}^{-1}$. \textbf{Left two panels}---The model represents $0\arcsec \leq R_{\rm maj} \leq 288\arcsec$ with $\Delta_{\rm rms}=0.1849\,\text{mag\,arcsec}^{-2}$. \underline{S{\'e}rsic Profile Parameters:} \textcolor{red}{$R_e=5\farcs91\pm0\farcs57$, $\mu_e=16.79\pm0.25\,\text{mag\,arcsec}^{-2}$, and $n=0.52\pm0.21$.} \underline{Ferrers Profile Parameters:} \textcolor{Orange}{$\mu_0 = 18.39\pm0.07\,\text{mag\,arcsec}^{-2}$, $R_{\rm end} = 33\farcs77\pm0\farcs06$, and $\alpha = 1.64\pm0.15$.} \underline{Edge-on Disk Model Parameters:} \textcolor{blue}{$\mu_0 = 20.98\pm0.06\,\text{mag\,arcsec}^{-2}$ and $h_z = 100\farcs29\pm1\farcs13$.} \underline{Additional Parameters:} three Gaussian components added at: \textcolor{cyan}{$R_{\rm r}=0\arcsec$, $65\farcs30\pm0\farcs57$, \& $176\farcs75\pm1\farcs06$; with $\mu_0 = 14.22\pm0.27$, $20.55\pm0.03$, \& $23.33\pm0.04\,\text{mag\,arcsec}^{-2}$; and FWHM = $3\farcs23\pm0\farcs81$, $44\farcs40\pm1\farcs52$, \& $45\farcs99\pm1\farcs98$, respectively.} \textbf{Right two panels}---The model represents $0\arcsec \leq R_{\rm eq} \leq 199\arcsec$ with $\Delta_{\rm rms}=0.1799\,\text{mag\,arcsec}^{-2}$. \underline{S{\'e}rsic Profile Parameters:} \textcolor{red}{$R_e=4\farcs35\pm0\farcs54$, $\mu_e=17.13\pm0.35\,\text{mag\,arcsec}^{-2}$, and $n=0.58\pm0.47$.} \underline{Ferrers Profile Parameters:} \textcolor{Orange}{$\mu_0 = 18.62\pm0.40\,\text{mag\,arcsec}^{-2}$, $R_{\rm end} = 18\farcs39\pm2\farcs03$, and $\alpha = 0.50\pm0.92$.} \underline{Edge-on Disk Model Parameters:} \textcolor{blue}{$\mu_0 = 23.78\pm0.04\,\text{mag\,arcsec}^{-2}$ and $h_z = 128\farcs76\pm2\farcs13$.} \underline{Additional Parameters:} three Gaussian components added at: \textcolor{cyan}{$R_{\rm r}=0\arcsec$, $32\farcs43\pm0\farcs24$, \& $66\farcs98\pm1\farcs30$; with $\mu_0 = 14.31\pm0.20$, $20.26\pm0.02$, \& $22.87\pm0.04\,\text{mag\,arcsec}^{-2}$; and FWHM = $2\farcs45\pm0\farcs62$, $26\farcs00\pm0\farcs59$, \& $36\farcs91\pm1\farcs60$, respectively.} Given our focus on isolating the bulge light, we have allowed degeneracy among the components which dominate at large radii (whose parameters are therefore neither stable nor reliable) when this appears to not compromise the bulge.}
\label{NGC3079_plot}
\end{sidewaysfigure}

\subsection{NGC~3079 (Figure~\ref{NGC3079_plot})}

A near edge-on galaxy, NGC~3079 is inclined by $75\fdg0\pm3\fdg9$ with respect to the plane of the sky. We use the edge-on disk model to describe it. It possess a Type 2 Seyfert AGN \citep{Veron-Cetty:2006}, which we model with a central Gaussian component. Two additional Gaussian components have been added at $R_{\rm maj} \approx 65\arcsec$ ($R_{\rm eq}\approx32\arcsec$) and $R_{\rm maj} \approx 177\arcsec$ ($R_{\rm eq}\approx67\arcsec$) to account for the multiple crossings of spiral arms along the radius of the galaxy.
 
\begin{sidewaysfigure}
\includegraphics[clip=true,trim= 11mm 1mm 4mm 6mm,width=0.249\textwidth]{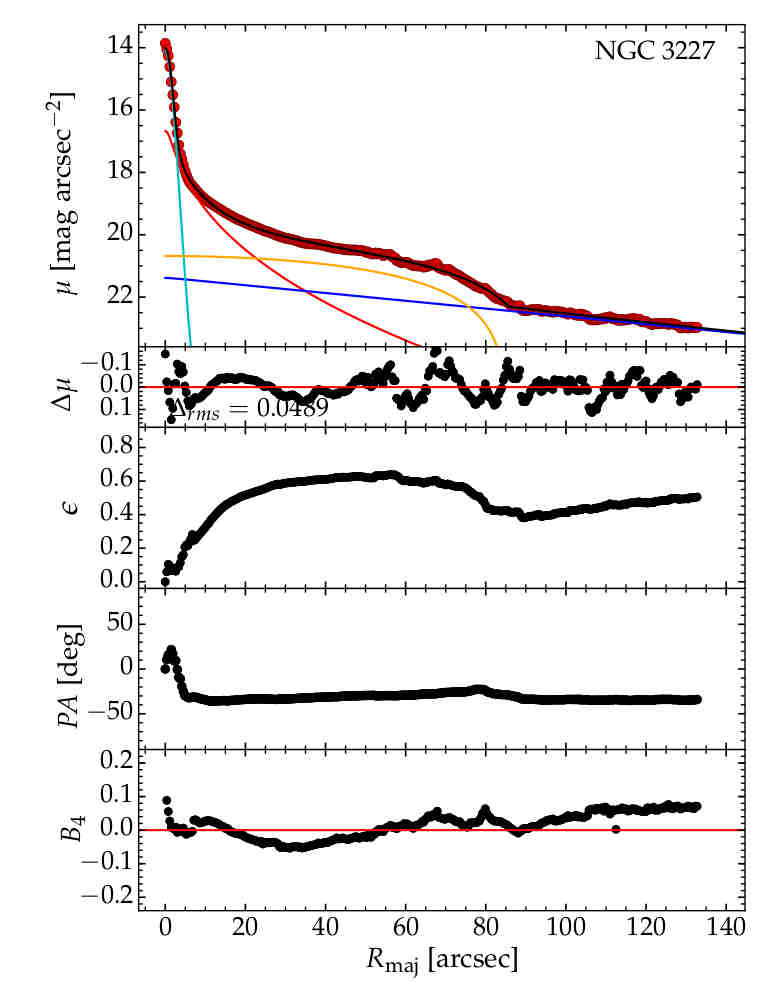}
\includegraphics[clip=true,trim= 11mm 1mm 4mm 6mm,width=0.249\textwidth]{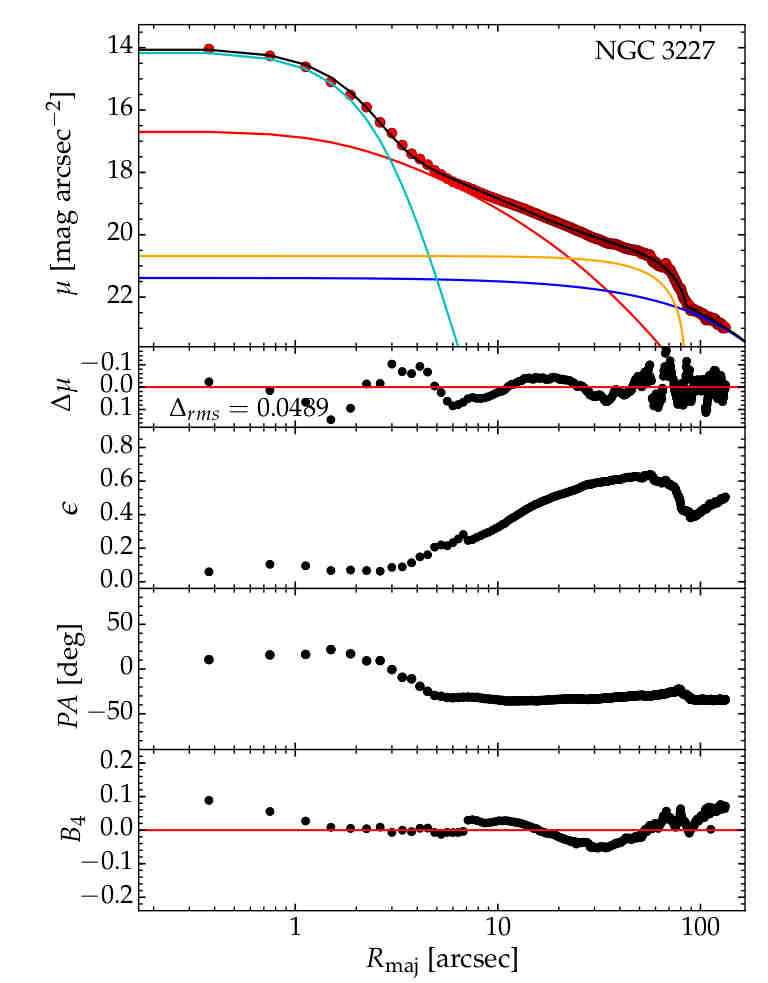}
\includegraphics[clip=true,trim= 11mm 1mm 3mm 6mm,width=0.249\textwidth]{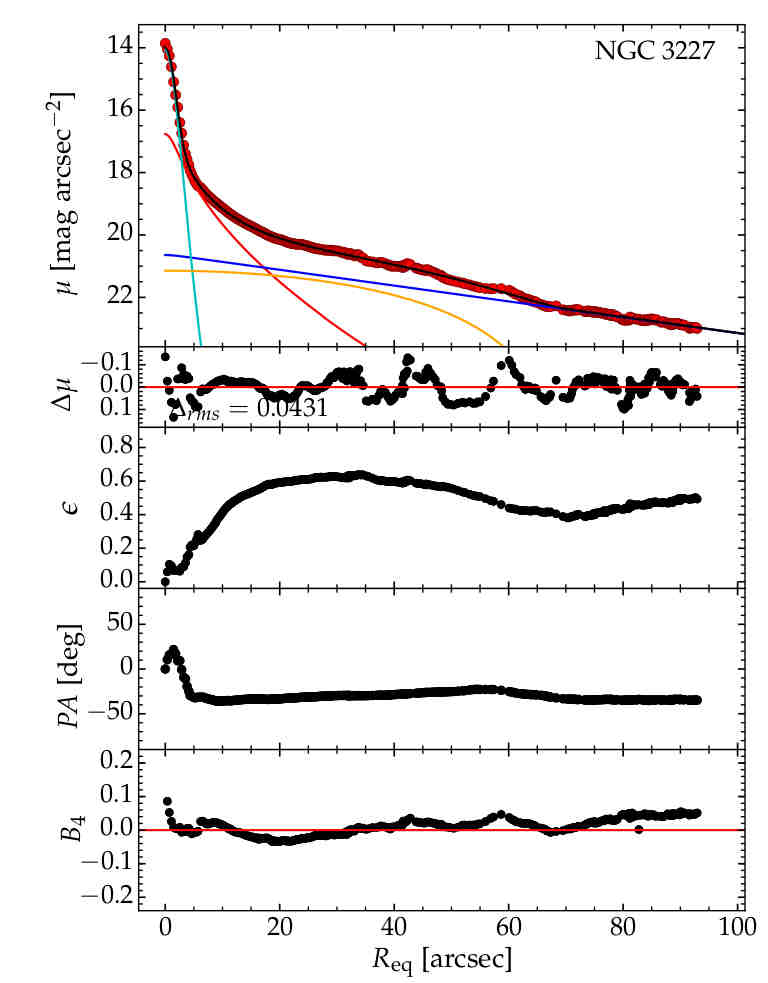}
\includegraphics[clip=true,trim= 11mm 1mm 4mm 6mm,width=0.249\textwidth]{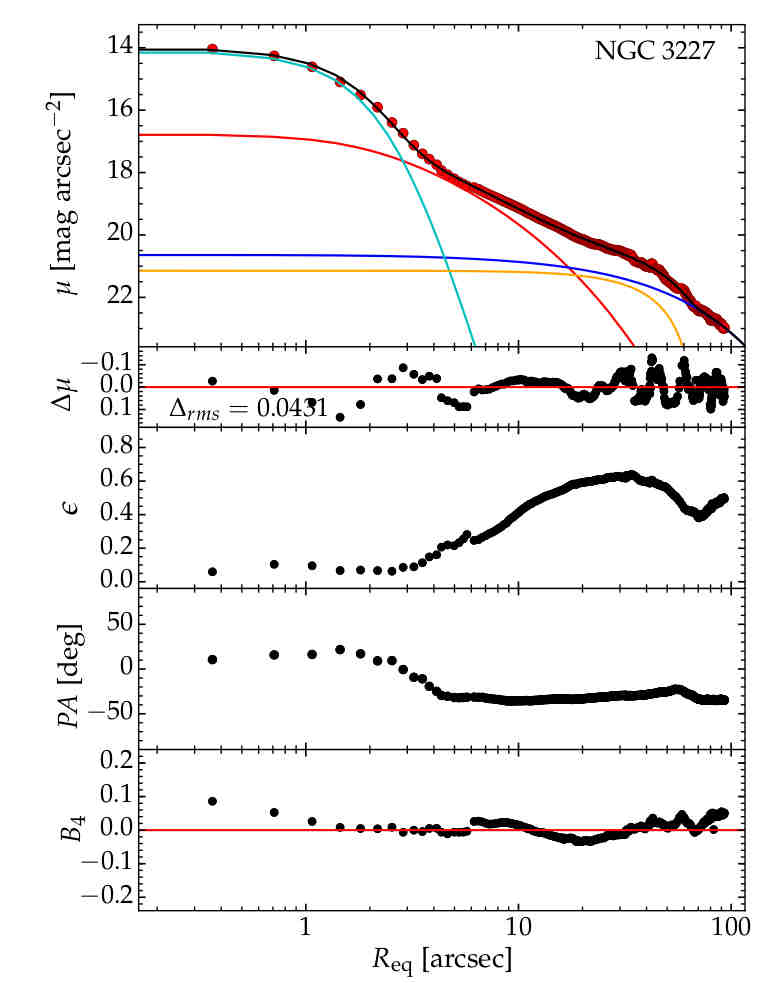}
\caption{\textit{Spitzer} $3.6\,\micron$ surface brightness profile for NGC~3227, with a physical scale of 0.1022$\,\text{kpc}\,\text{arcsec}^{-1}$. \textbf{Left two panels}---The model represents $0\arcsec \leq R_{\rm maj} \leq 133\arcsec$ with $\Delta_{\rm rms}=0.0489\,\text{mag\,arcsec}^{-2}$. \underline{S{\'e}rsic Profile Parameters:} \textcolor{red}{$R_e=17\farcs91\pm3\farcs31$, $\mu_e=20.26\pm0.32\,\text{mag\,arcsec}^{-2}$, and $n=2.60\pm0.44$.} \underline{Ferrers Profile Parameters:} \textcolor{Orange}{$\mu_0 = 20.69\pm0.04\,\text{mag\,arcsec}^{-2}$, $R_{\rm end} = 85\farcs72\pm0\farcs28$, and $\alpha = 2.47\pm0.09$.} \underline{Exponential Profile Parameters:} \textcolor{blue}{$\mu_0 = 21.37\pm0.14\,\text{mag\,arcsec}^{-2}$ and $h = 86\farcs86\pm5\farcs84$.} \underline{Additional Parameters:} one Gaussian component added at: \textcolor{cyan}{$R_{\rm r}=0\arcsec$, with $\mu_0 = 12.53\pm0.09\,\text{mag\,arcsec}^{-2}$, and FWHM = $1\farcs23\pm0\farcs06$.} \textbf{Right two panels}---The model represents $0\arcsec \leq R_{\rm eq} \leq 93\arcsec$ with $\Delta_{\rm rms}=0.0431\,\text{mag\,arcsec}^{-2}$. \underline{S{\'e}rsic Profile Parameters:} \textcolor{red}{$R_e=8\farcs34\pm0\farcs60$, $\mu_e=19.32\pm0.13\,\text{mag\,arcsec}^{-2}$, and $n=1.90\pm0.28$.} \underline{Ferrers Profile Parameters:} \textcolor{Orange}{$\mu_0 = 21.15\pm0.10\,\text{mag\,arcsec}^{-2}$, $R_{\rm end} = 70\farcs72\pm1\farcs67$, and $\alpha = 4.67\pm0.51$.} \underline{Exponential Profile Parameters:} \textcolor{blue}{$\mu_0 = 20.61\pm0.08\,\text{mag\,arcsec}^{-2}$ and $h = 42\farcs84\pm1\farcs66$.} \underline{Additional Parameters:} one Gaussian component added at: \textcolor{cyan}{$R_{\rm r}=0\arcsec$, with $\mu_0 = 12.23\pm0.14\,\text{mag\,arcsec}^{-2}$, and FWHM = $0\farcs98\pm0\farcs07$.} Given our focus on isolating the bulge light, we have allowed degeneracy among the components which dominate at large radii (whose parameters are therefore neither stable nor reliable) when this appears to not compromise the bulge.}
\label{NGC3227_plot}
\end{sidewaysfigure}

\subsection{NGC~3227 (Figure~\ref{NGC3227_plot})}\label{NGC3227}

We have modeled NGC~3227 with a central Gaussian component to account for the Type 1.5 intermediate Seyfert AGN \citep{Khachikian:Weedman:1974,Veron-Cetty:2006}.

\begin{sidewaysfigure}
\includegraphics[clip=true,trim= 11mm 1mm 1mm 5mm,width=0.249\textwidth]{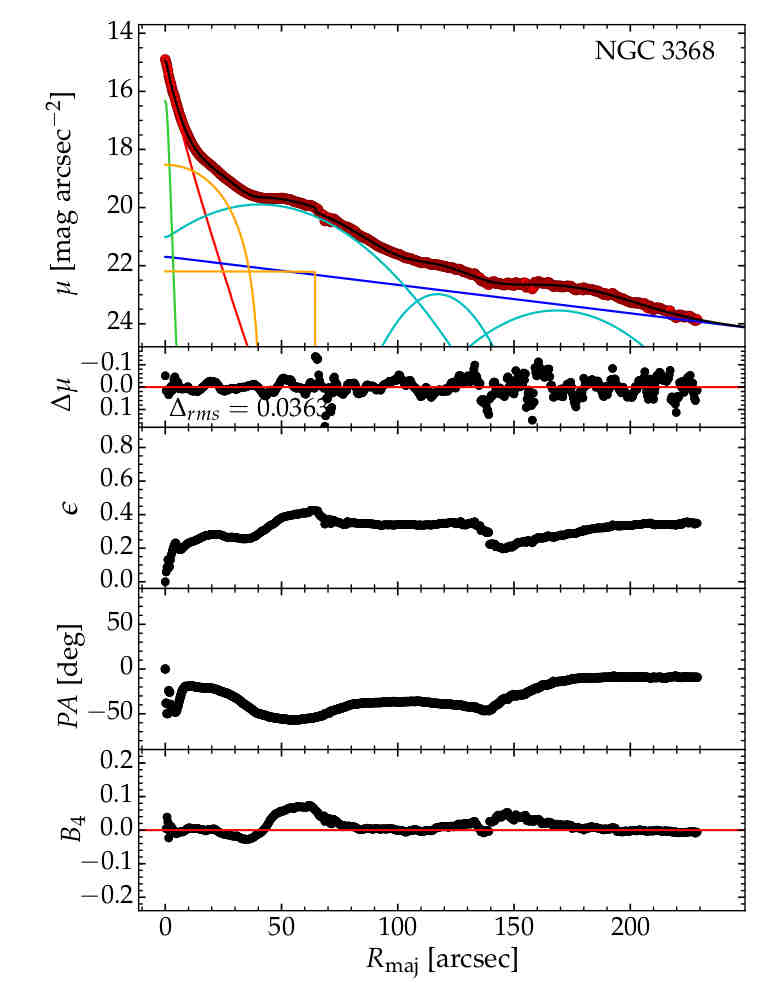}
\includegraphics[clip=true,trim= 11mm 1mm 1mm 5mm,width=0.249\textwidth]{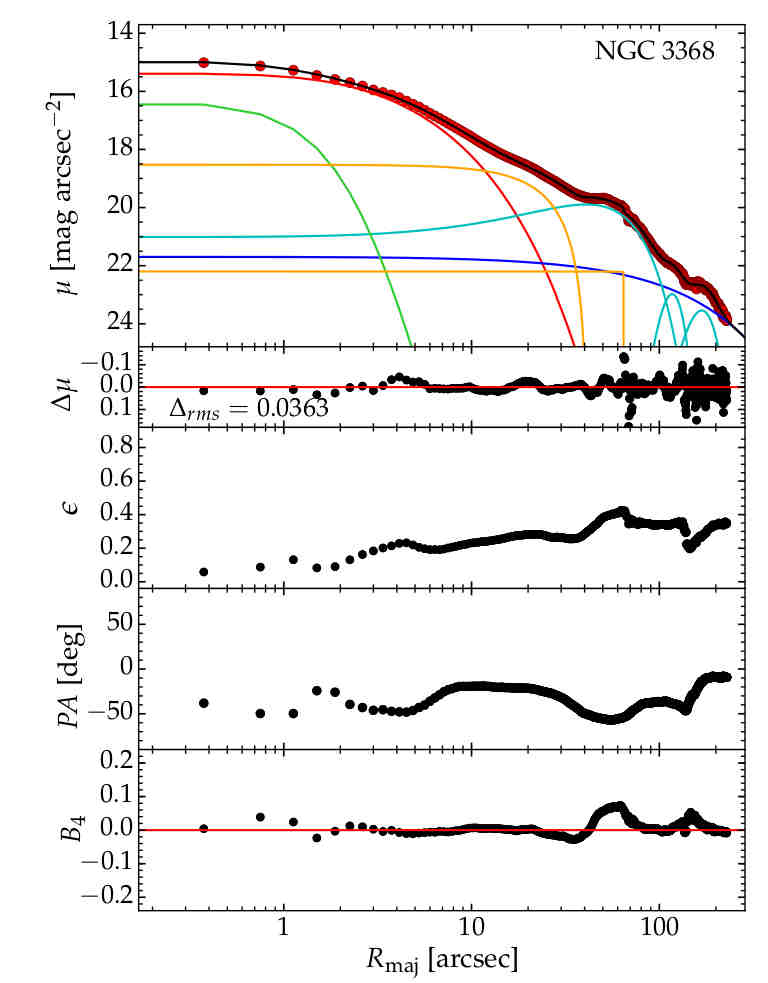}
\includegraphics[clip=true,trim= 11mm 1mm 1mm 5mm,width=0.249\textwidth]{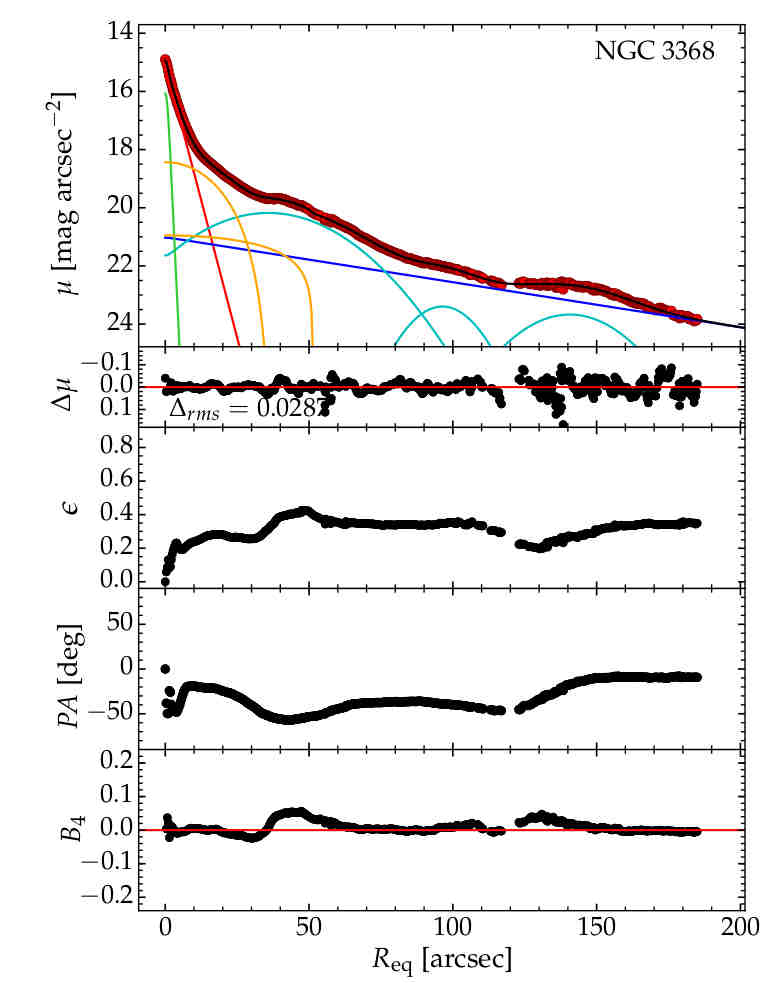}
\includegraphics[clip=true,trim= 11mm 1mm 1mm 5mm,width=0.249\textwidth]{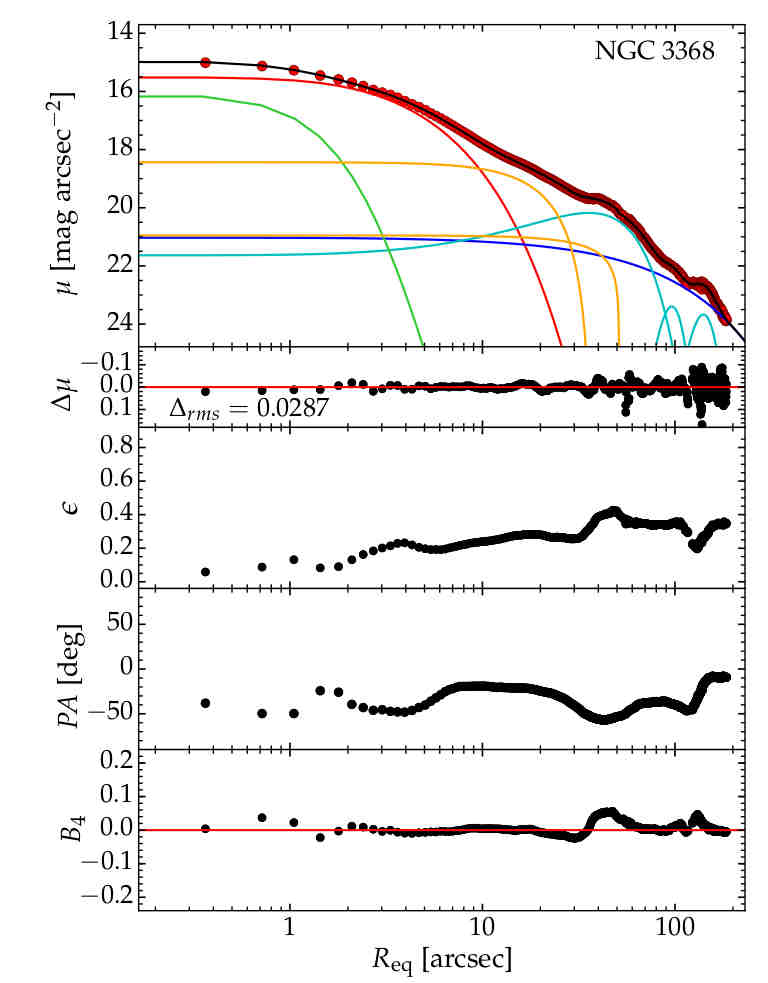}
\caption{\textit{Spitzer} $3.6\,\micron$ surface brightness profile for NGC~3368, with a physical scale of 0.0519$\,\text{kpc}\,\text{arcsec}^{-1}$. \textbf{Left two panels}---The model represents $0\arcsec \leq R_{\rm maj} \leq 229\arcsec$ with $\Delta_{\rm rms}=0.0363\,\text{mag\,arcsec}^{-2}$. \underline{Point Source:} \textcolor{LimeGreen}{$\mu_0 = 16.34\pm0.17\,\text{mag\,arcsec}^{-2}$.} \underline{S{\'e}rsic Profile Parameters:} \textcolor{red}{$R_e=5\farcs98\pm0\farcs31$, $\mu_e=17.07\pm0.08\,\text{mag\,arcsec}^{-2}$, and $n=1.19\pm0.09$.} \underline{Ferrers Profile Parameters:} \textcolor{Orange}{$\mu_0 = 18.53\pm0.11$ \& $22.23\pm0.25\,\text{mag\,arcsec}^{-2}$; $R_{\rm end} = 41\farcs30\pm1\farcs54$ \& $64\farcs50\pm0\farcs00$; and $\alpha = 5.81\pm0.85$ \& $0.01\pm0.19$.} \underline{Exponential Profile Parameters:} \textcolor{blue}{$\mu_0 = 21.69\pm1.09\,\text{mag\,arcsec}^{-2}$ and $h = 111\farcs10\pm51\farcs01$.} \underline{Additional Parameters:} three Gaussian components added at: \textcolor{cyan}{$R_{\rm r}=40\farcs64\pm1\farcs93$, $117\farcs13\pm0\farcs72$, \& $168\farcs27\pm5\farcs20$; with $\mu_0 = 19.90\pm0.12$, $22.98\pm0.34$, \& $23.54\pm0.41\,\text{mag\,arcsec}^{-2}$; and FWHM = $64\farcs49\pm4\farcs77$, $30\farcs84\pm3\farcs37$, \& $58\farcs29\pm9\farcs35$, respectively.} \textbf{Right two panels}---The model represents $0\arcsec \leq R_{\rm eq} \leq 185\arcsec$ with $\Delta_{\rm rms}=0.0287\,\text{mag\,arcsec}^{-2}$. \underline{Point Source:} \textcolor{LimeGreen}{$\mu_0 = 16.08\pm0.09\,\text{mag\,arcsec}^{-2}$.} \underline{S{\'e}rsic Profile Parameters:} \textcolor{red}{$R_e=4\farcs83\pm0\farcs14$, $\mu_e=16.92\pm0.04\,\text{mag\,arcsec}^{-2}$, and $n=1.00\pm0.05$.} \underline{Ferrers Profile Parameters:} \textcolor{Orange}{$\mu_0 = 18.43\pm0.08$ \& $21.95\pm0.23\,\text{mag\,arcsec}^{-2}$; $R_{\rm end} = 36\farcs94\pm1\farcs45$ \& $51\farcs31\pm0\farcs24$; and $\alpha = 7.20\pm0.95$ \& $1.50\pm0.22$.} \underline{Exponential Profile Parameters:} \textcolor{blue}{$\mu_0 = 21.02\pm0.29\,\text{mag\,arcsec}^{-2}$ and $h = 70\farcs06\pm6\farcs79$.} \underline{Additional Parameters:} three Gaussian components added at: \textcolor{cyan}{$R_{\rm r}=35\farcs79\pm2\farcs11$, $96\farcs34\pm0\farcs48$, \& $140\farcs64\pm1\farcs22$; with $\mu_0 = 20.18\pm0.09$, $23.40\pm0.22$, \& $23.68\pm0.12\,\text{mag\,arcsec}^{-2}$; and FWHM = $49\farcs66\pm2\farcs54$, $24\farcs85\pm2\farcs28$, \& $38\farcs03\pm2\farcs33$, respectively.} Given our focus on isolating the bulge light, we have allowed degeneracy among the components which dominate at large radii (whose parameters are therefore neither stable nor reliable) when this appears to not compromise the bulge.}
\label{NGC3368_plot}
\end{sidewaysfigure}

\subsection{NGC~3368 (Figure~\ref{NGC3368_plot})}

The galaxy NGC~3368 (M96) possesses two bars \citep{Erwin:2004,Nowak:2010}, and we fit them with Ferrers profiles. Three Gaussian components have been added at $R_{\rm maj} \approx 41\arcsec$ ($R_{\rm eq}\approx 36\arcsec$), $R_{\rm maj} \approx 117\arcsec$ ($R_{\rm eq}\approx 96\arcsec$), and $R_{\rm maj} \approx 168\arcsec$ ($R_{\rm eq}\approx 141\arcsec$) to account for the multiple crossings of spiral arms along the major axis profile of the galaxy.

\begin{sidewaysfigure}
\includegraphics[clip=true,trim= 11mm 1mm 1mm 5mm,width=0.249\textwidth]{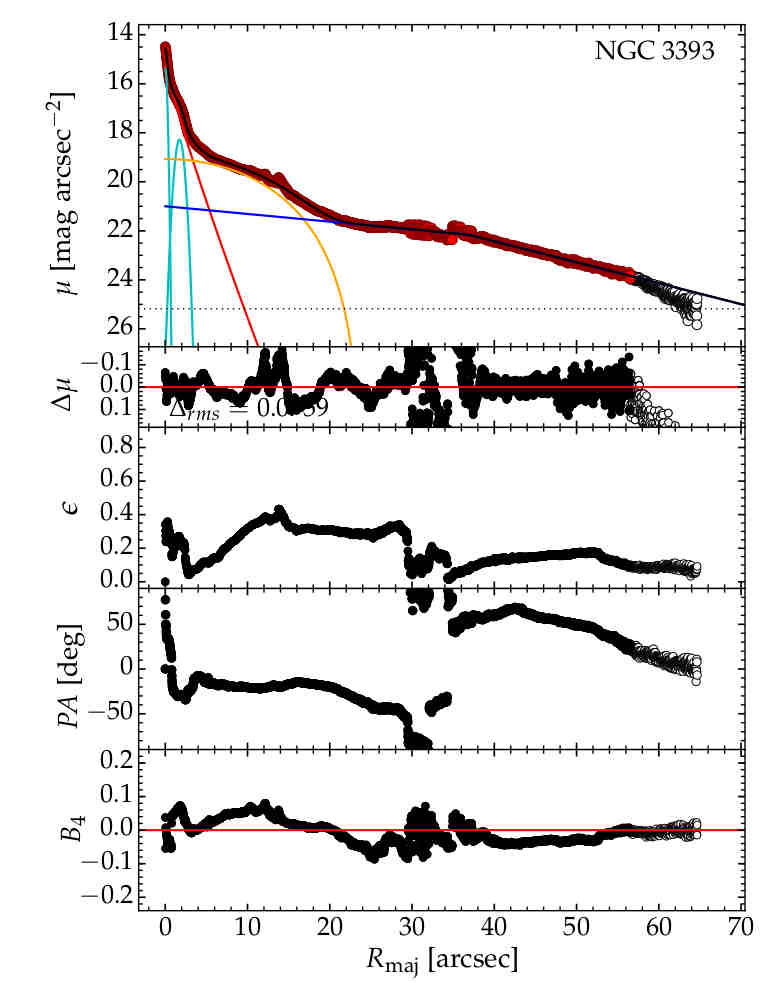}
\includegraphics[clip=true,trim= 11mm 1mm 1mm 5mm,width=0.249\textwidth]{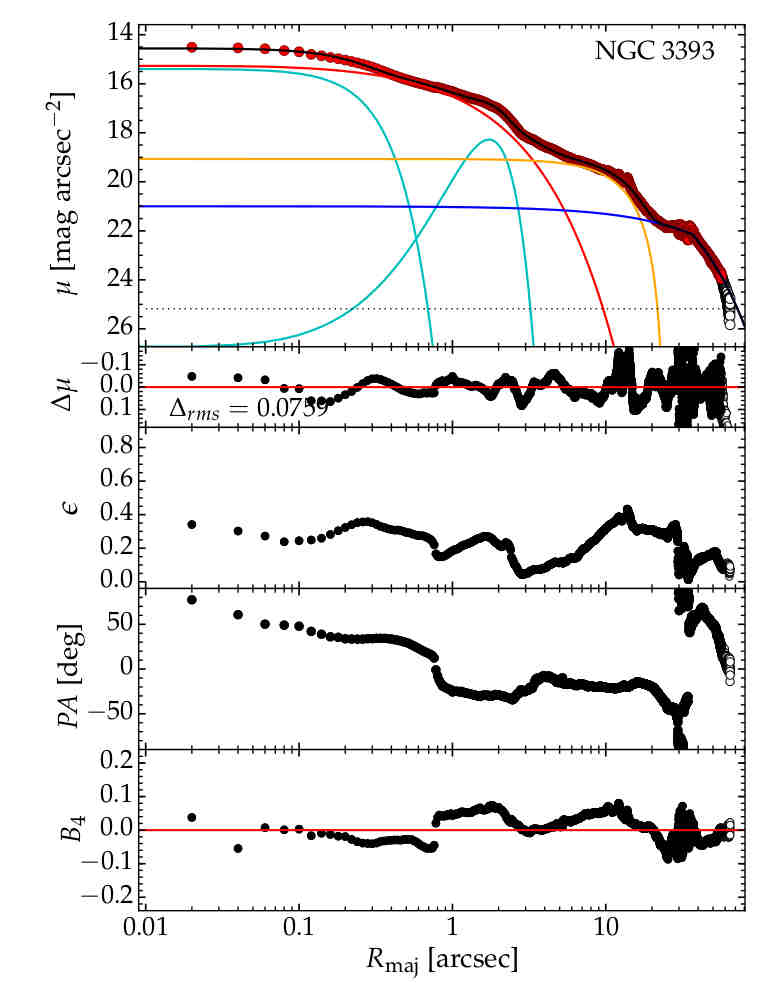}
\includegraphics[clip=true,trim= 11mm 1mm 1mm 5mm,width=0.249\textwidth]{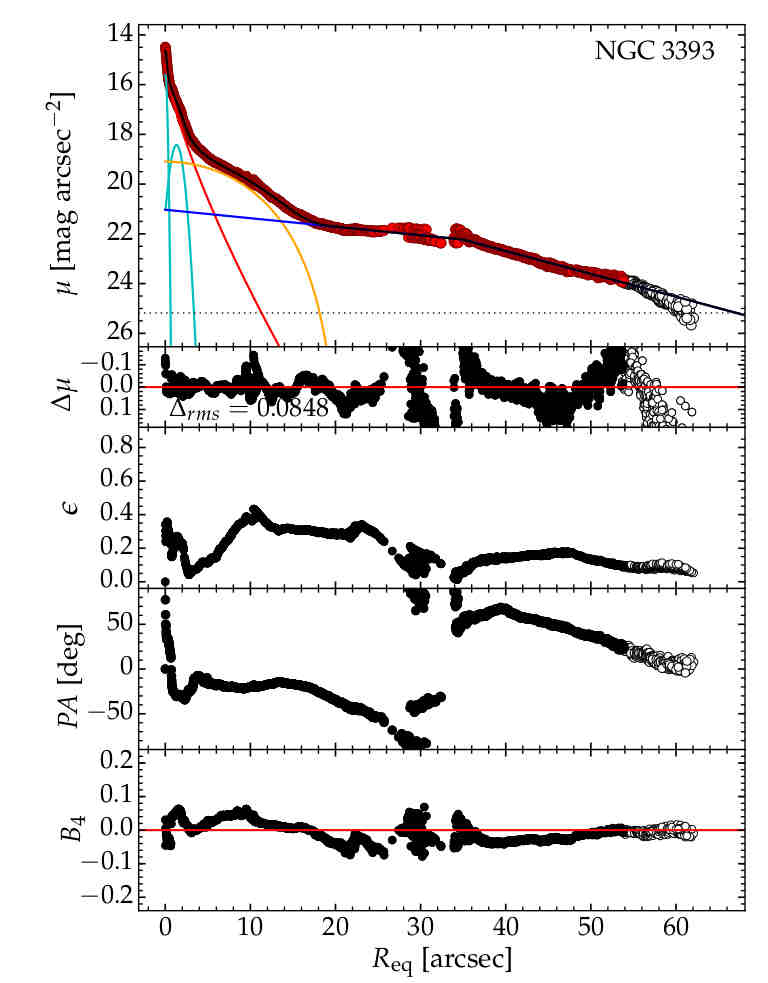}
\includegraphics[clip=true,trim= 11mm 1mm 1mm 5mm,width=0.249\textwidth]{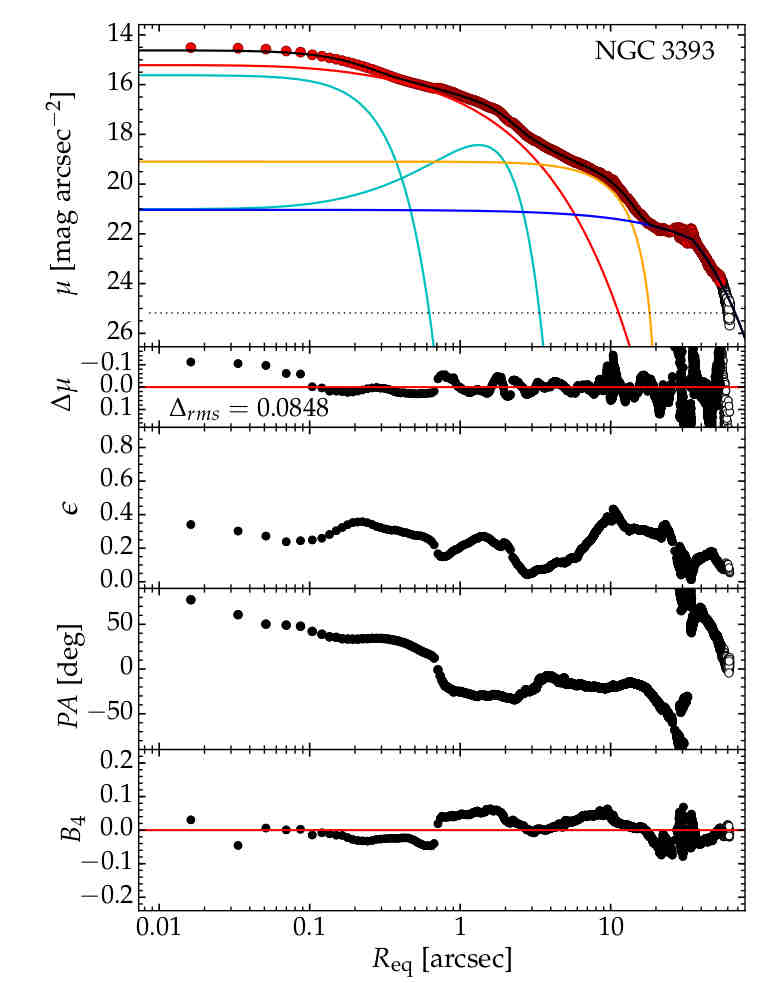}
\caption{\textit{HST} WFC3 UVIS2 F814W surface brightness profile for NGC~3393, with a physical scale of 0.2640$\,\text{kpc}\,\text{arcsec}^{-1}$. \textbf{Left two panels}---The model represents $0\arcsec \leq R_{\rm maj} \leq 56\farcs56$ with $\Delta_{\rm rms}=0.0759\,\text{mag\,arcsec}^{-2}$ and additional data from $56\farcs56 < R_{\rm maj} \leq 64\farcs68$ is plotted, but not modeled. \underline{S{\'e}rsic Profile Parameters:} \textcolor{red}{$R_e=1\farcs64\pm0\farcs02$, $\mu_e=17.27\pm0.05\,\text{mag\,arcsec}^{-2}$, and $n=1.14\pm0.07$.} \underline{Ferrers Profile Parameters:} \textcolor{Orange}{$\mu_0 = 19.07\pm0.02\,\text{mag\,arcsec}^{-2}$, $R_{\rm end} = 24\farcs20\pm0\farcs30$, and $\alpha = 8.55\pm0.34$.} \underline{Broken Exponential Profile Parameters:} \textcolor{blue}{$\mu_0 = 21.00\pm0.02\,\text{mag\,arcsec}^{-2}$, $R_b = 36\farcs79\pm0\farcs14$, $h_1 = 34\farcs54\pm0\farcs77$, and $h_2 = 12\farcs70\pm0\farcs06$.} \underline{Additional Parameters:} two Gaussian components added at: \textcolor{cyan}{$R_{\rm r}=0\arcsec$ \& $1\farcs73\pm0\farcs04$; with $\mu_0 = 15.26\pm0.10$ \& $18.28\pm0.09\,\text{mag\,arcsec}^{-2}$; and FWHM = $0\farcs36\pm0\farcs02$ \& $0\farcs99\pm0\farcs09$, respectively.} \textbf{Right two panels}---The model represents $0\arcsec \leq R_{\rm eq} \leq 53\farcs92$ with $\Delta_{\rm rms}=0.0848\,\text{mag\,arcsec}^{-2}$ and additional data from $53\farcs92 < R_{\rm maj} \leq 62\farcs51$ is plotted, but not modeled. \underline{S{\'e}rsic Profile Parameters:} \textcolor{red}{$R_e=1\farcs77\pm0\farcs09$, $\mu_e=17.63\pm0.15\,\text{mag\,arcsec}^{-2}$, and $n=1.36\pm0.13$.} \underline{Ferrers Profile Parameters:} \textcolor{Orange}{$\mu_0 = 19.09\pm0.05\,\text{mag\,arcsec}^{-2}$, $R_{\rm end} = 20\farcs91\pm0\farcs49$, and $\alpha = 10.00\pm0.77$.} \underline{Broken Exponential Profile Parameters:} \textcolor{blue}{$\mu_0 = 21.03\pm0.02\,\text{mag\,arcsec}^{-2}$, $R_b = 34\farcs82\pm0\farcs13$, $h_1 = 31\farcs86\pm0\farcs58$, and $h_2 = 11\farcs83\pm0\farcs06$.} \underline{Additional Parameters:} two Gaussian components added at: \textcolor{cyan}{$R_{\rm r}=0\arcsec$ \& $1\farcs32\pm0\farcs13$; with $\mu_0 = 15.51\pm0.16$ \& $18.43\pm0.25\,\text{mag\,arcsec}^{-2}$; and FWHM = $0\farcs33\pm0\farcs03$ \& $1\farcs37\pm0\farcs22$, respectively.}}
\label{NGC3393_plot}
\end{sidewaysfigure}

\subsection{NGC~3393 (Figure~\ref{NGC3393_plot})}

The galaxy NGC~3393 possesses a Type 2 Seyfert AGN \citep{Veron-Cetty:2006} with circumnuclear dust \citep{Martini:2003}. We have added a central Gaussian at $R_{\rm maj} \approx 2\arcsec$ ($R_{\rm eq}\approx 1\arcsec$) and a noncentral Gaussian at $R_{\rm maj} \approx 50\arcsec$ ($R_{\rm eq}\approx52\arcsec$) to account for enhanced nuclear light and nuclear spiral arms, respectively.

\begin{sidewaysfigure}
\includegraphics[clip=true,trim= 11mm 1mm 4mm 6mm,width=0.249\textwidth]{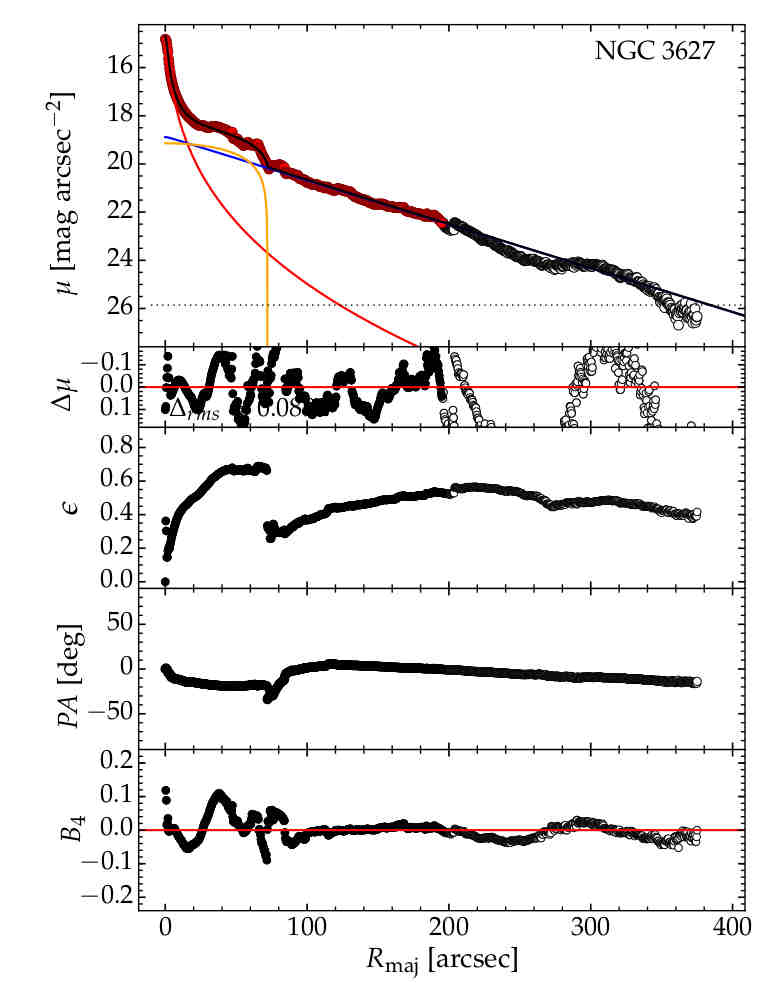}
\includegraphics[clip=true,trim= 11mm 1mm 4mm 6mm,width=0.249\textwidth]{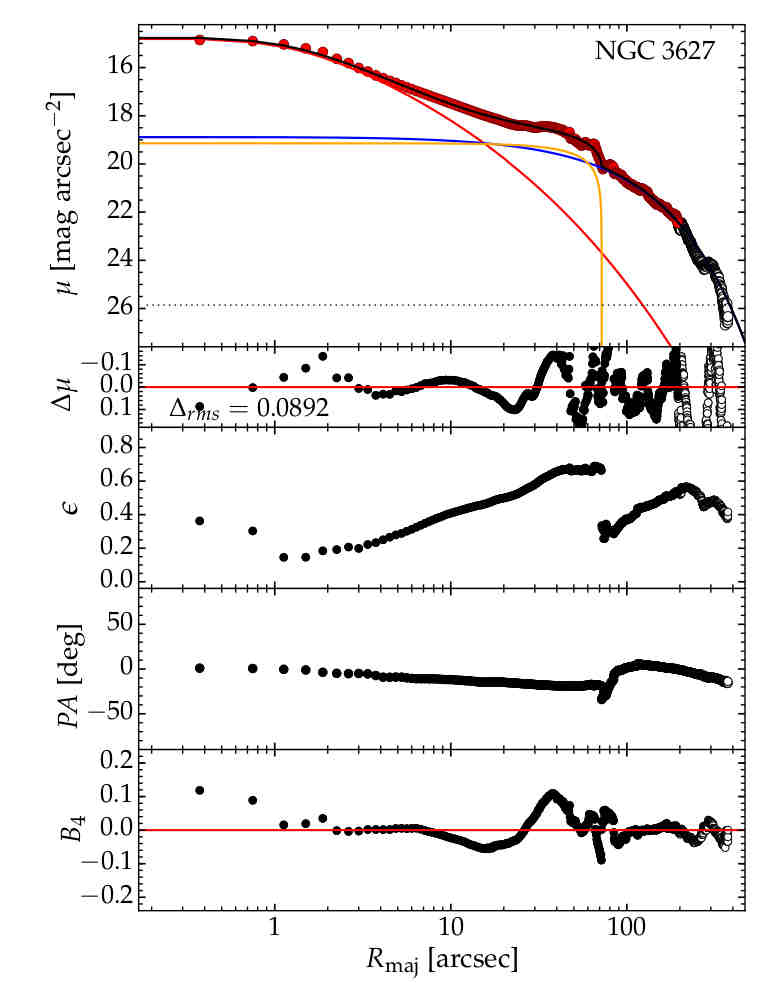}
\includegraphics[clip=true,trim= 11mm 1mm 4mm 6mm,width=0.249\textwidth]{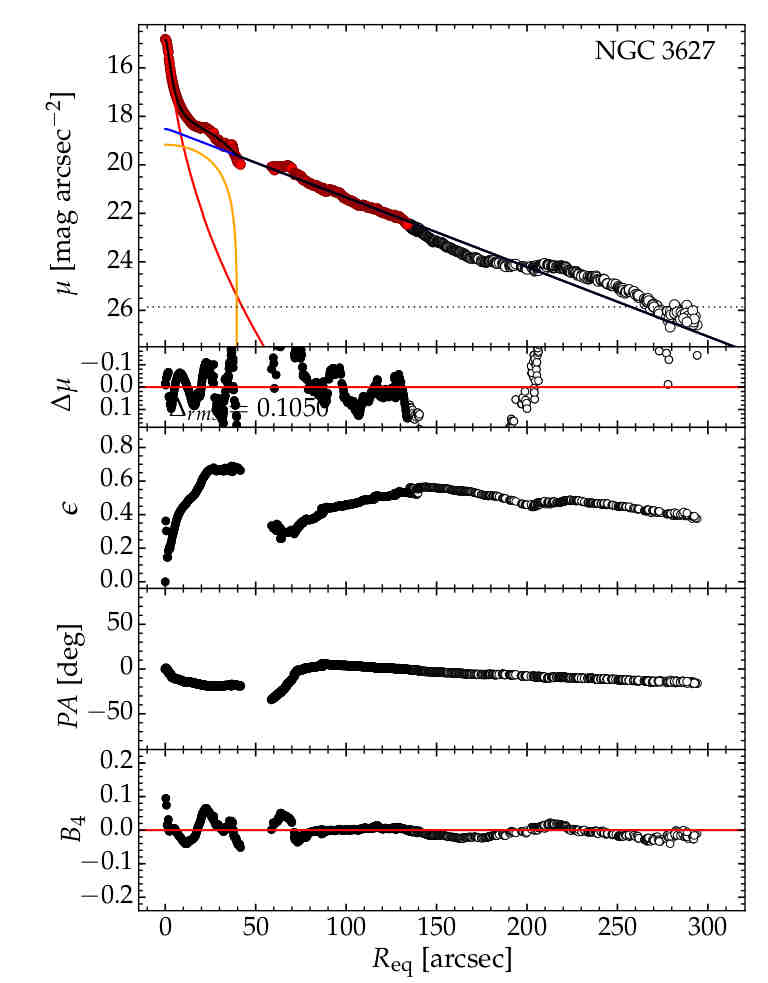}
\includegraphics[clip=true,trim= 11mm 1mm 4mm 6mm,width=0.249\textwidth]{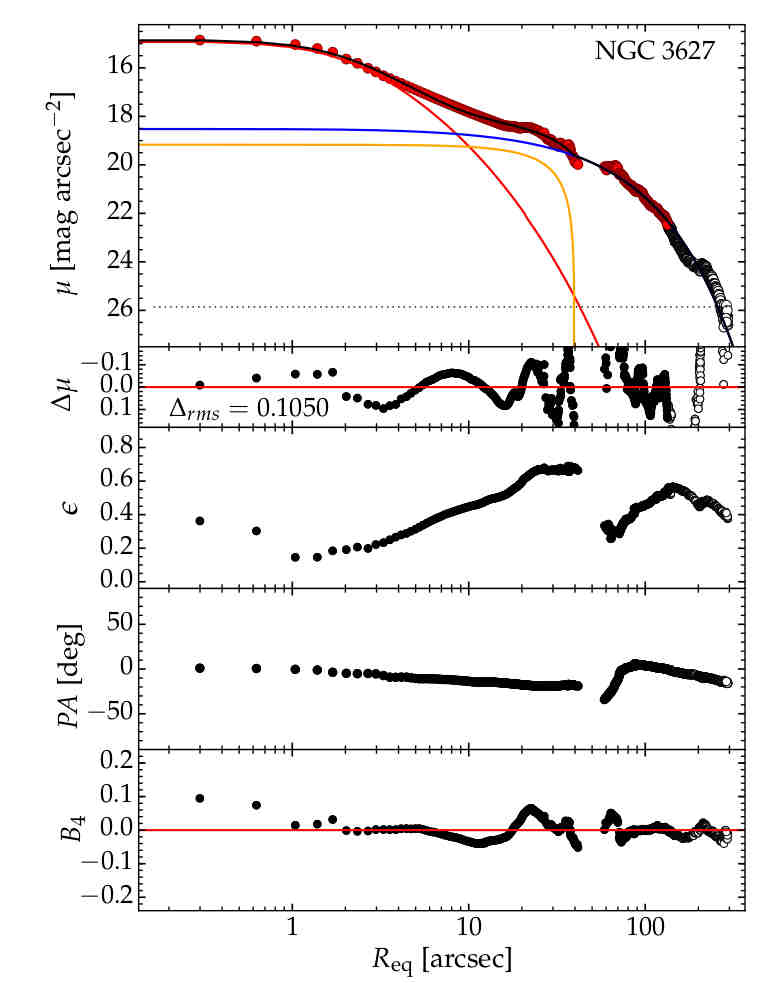}
\caption{\textit{Spitzer} $3.6\,\micron$ surface brightness profile for NGC~3627, with a physical scale of 0.0512$\,\text{kpc}\,\text{arcsec}^{-1}$. \textbf{Left two panels}---The model represents $0\arcsec \leq R_{\rm maj} \leq 195\arcsec$ with $\Delta_{\rm rms}=0.0892\,\text{mag\,arcsec}^{-2}$ and additional data from $195\arcsec < R_{\rm maj} \leq 375\arcsec$ is plotted, but not modeled. \underline{S{\'e}rsic Profile Parameters:} \textcolor{red}{$R_e=11\farcs07\pm1\farcs54$, $\mu_e=18.44\pm0.24\,\text{mag\,arcsec}^{-2}$, and $n=3.17\pm0.19$.} \underline{Ferrers Profile Parameters:} \textcolor{Orange}{$\mu_0 = 19.14\pm0.04\,\text{mag\,arcsec}^{-2}$, $R_{\rm end} = 72\farcs02\pm0\farcs12$, and $\alpha = 1.59\pm0.08$.} \underline{Exponential Profile Parameters:} \textcolor{blue}{$\mu_0 = 18.86\pm0.03\,\text{mag\,arcsec}^{-2}$ and $h = 59\farcs55\pm0\farcs53$.} \textbf{Right two panels}---The model represents $0\arcsec \leq R_{\rm eq} \leq 134\arcsec$ with $\Delta_{\rm rms}=0.1050\,\text{mag\,arcsec}^{-2}$ and additional data from $134\arcsec < R_{\rm eq} \leq 295\arcsec$ is plotted, but not modeled. \underline{S{\'e}rsic Profile Parameters:} \textcolor{red}{$R_e=3\farcs92\pm0\farcs57$, $\mu_e=16.35\pm0.29\,\text{mag\,arcsec}^{-2}$, and $n=2.10\pm0.31$.} \underline{Ferrers Profile Parameters:} \textcolor{Orange}{$\mu_0 = 19.17\pm0.16\,\text{mag\,arcsec}^{-2}$, $R_{\rm end} = 39\farcs78\pm1\farcs65$, and $\alpha = 3.08\pm0.79$.} \underline{Exponential Profile Parameters:} \textcolor{blue}{$\mu_0 = 18.49\pm0.02\,\text{mag\,arcsec}^{-2}$ and $h = 37\farcs89\pm0\farcs35$.}}
\label{NGC3627_plot}
\end{sidewaysfigure}

\subsection{NGC~3627 (Figure~\ref{NGC3627_plot})}

For NGC~3627 (M66), the model was fit to the range $0\arcsec \leq R_{\rm maj} \leq 195\arcsec$ ($0\arcsec \leq R_{\rm eq} \leq 134\arcsec$); this extends just beyond the visible outer spiral arms. The outer data from $195\arcsec < R_{\rm maj} \leq 375\arcsec$ ($134\arcsec < R_{\rm eq} \leq 295\arcsec$) are plotted but omitted from the fit. The exponential model continues to follow the profile (at least for the major axis profile) fairly well, but we do not include this outer range due to potential contamination northwest of the galaxy from the aggressive masking of the two bright foreground stars and a potential remnant tidal stream further out.

\begin{sidewaysfigure}
\includegraphics[clip=true,trim= 11mm 1mm 3mm 5mm,width=0.249\textwidth]{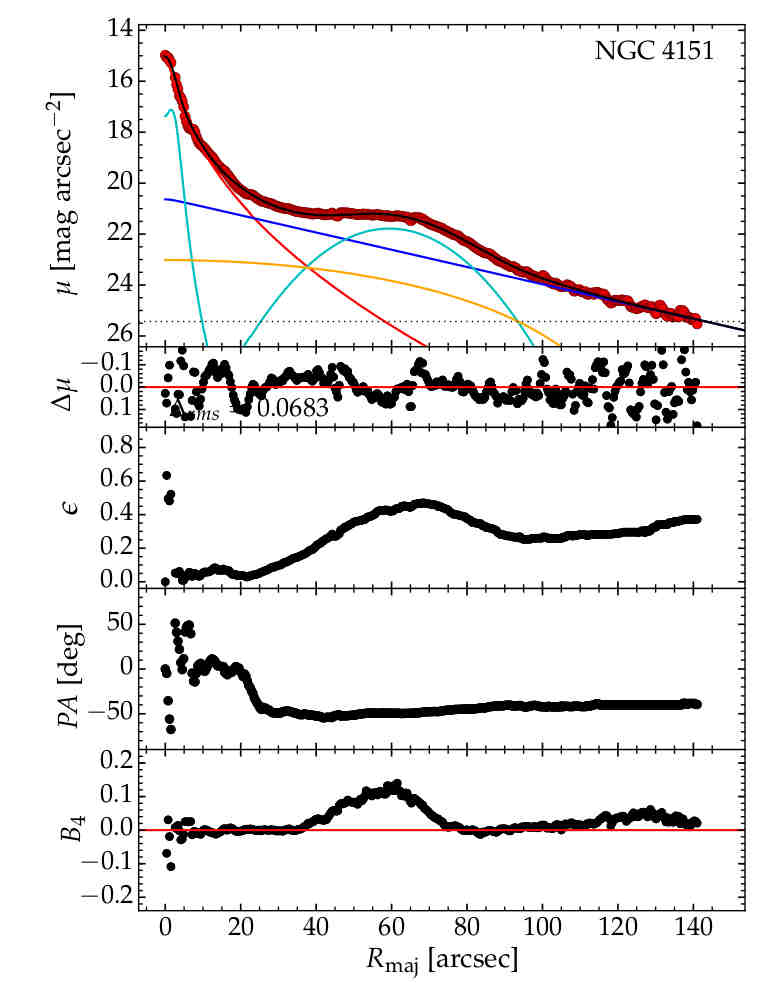}
\includegraphics[clip=true,trim= 11mm 1mm 3mm 5mm,width=0.249\textwidth]{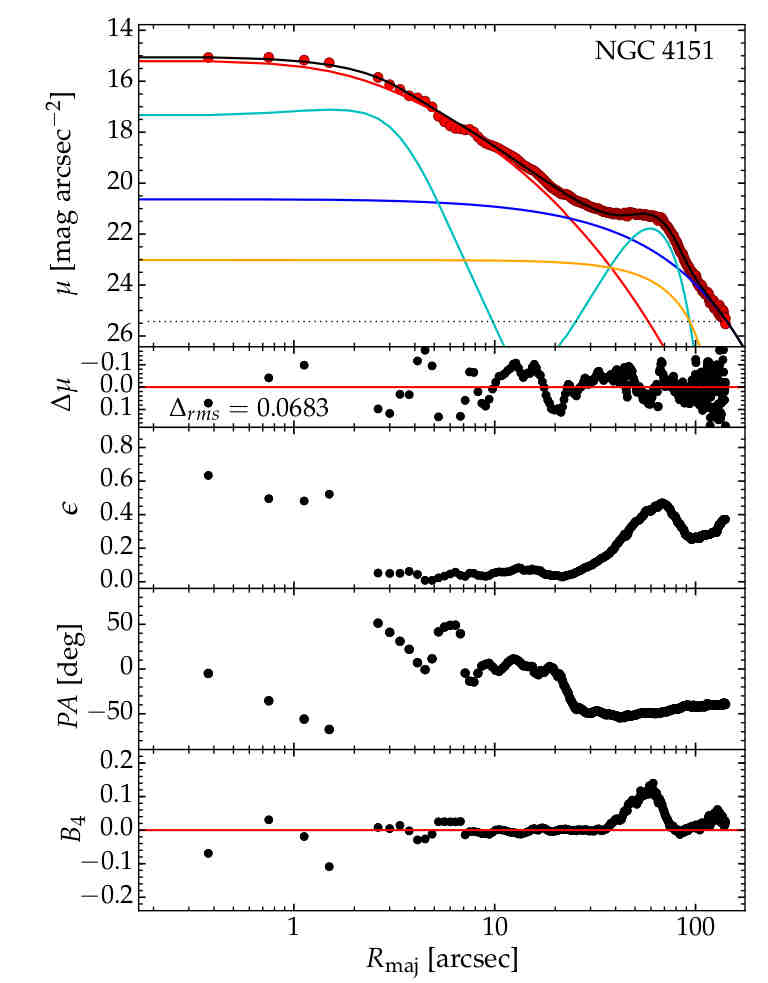}
\includegraphics[clip=true,trim= 11mm 1mm 3mm 5mm,width=0.249\textwidth]{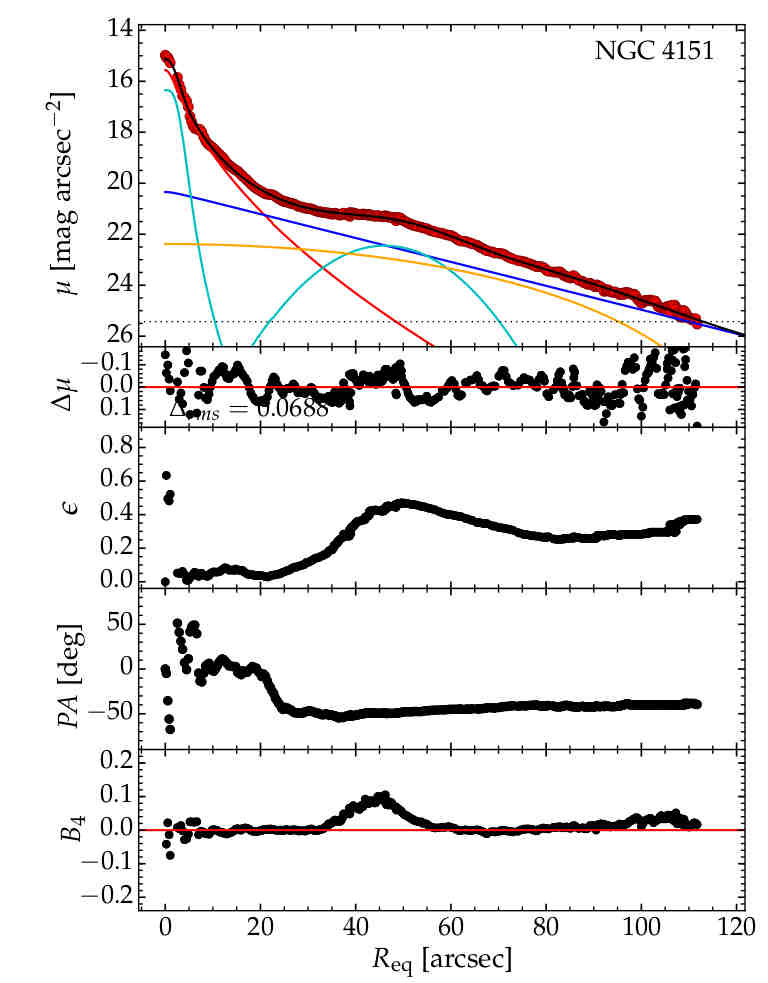}
\includegraphics[clip=true,trim= 11mm 1mm 3mm 5mm,width=0.249\textwidth]{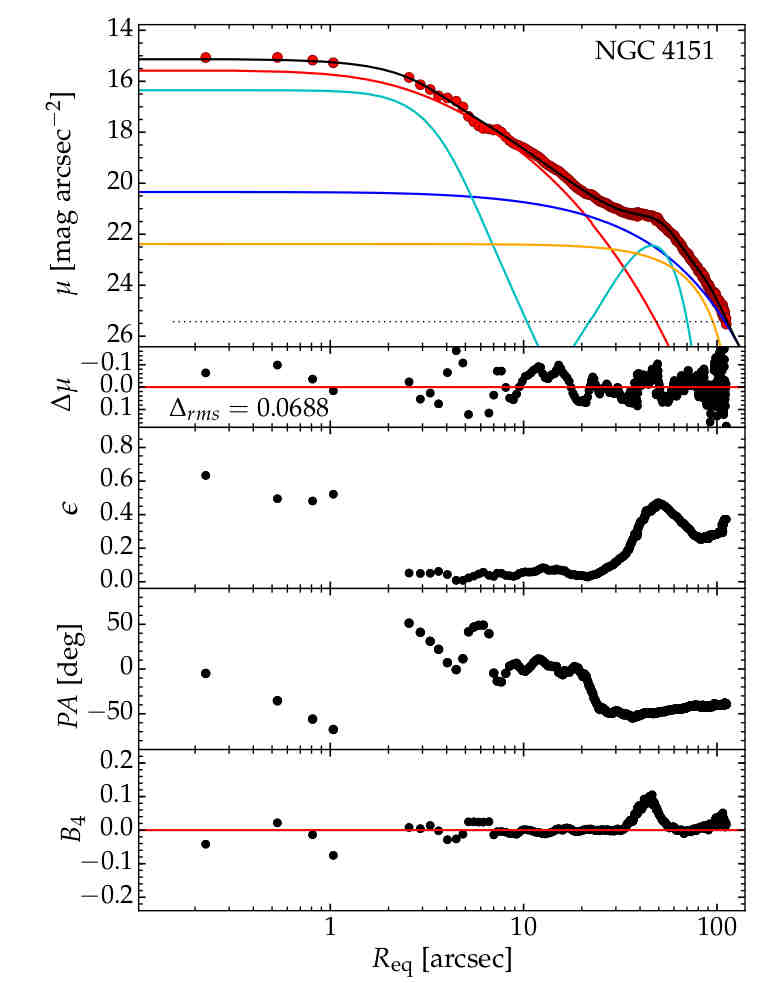}
\caption{\textit{Spitzer} $3.6\,\micron$ surface brightness profile for NGC~4151, with a physical scale of 0.0921$\,\text{kpc}\,\text{arcsec}^{-1}$. \textbf{Left two panels}---The model represents $0\arcsec \leq R_{\rm maj} \leq 141\arcsec$ with $\Delta_{\rm rms}=0.0683\,\text{mag\,arcsec}^{-2}$. \underline{S{\'e}rsic Profile Parameters:} \textcolor{red}{$R_e=6\farcs23\pm0\farcs35$, $\mu_e=17.75\pm0.12\,\text{mag\,arcsec}^{-2}$, and $n=2.24\pm0.33$.} \underline{Ferrers Profile Parameters:} \textcolor{Orange}{$\mu_0 = 22.82\pm0.93\,\text{mag\,arcsec}^{-2}$, $R_{\rm end} = 136\farcs01\pm30\farcs28$, and $\alpha = 8.64\pm7.24$.} \underline{Exponential Profile Parameters:} \textcolor{blue}{$\mu_0 = 20.58\pm0.36\,\text{mag\,arcsec}^{-2}$ and $h = 32\farcs08\pm2\farcs50$.} \underline{Additional Parameters:} two Gaussian components added at \textcolor{cyan}{$R_{\rm r}=1\farcs87\pm1\farcs30$ \& $59\farcs37\pm0\farcs50$; with $\mu_0 = 15.91\pm7.32$ \& $21.80\pm0.05\,\text{mag\,arcsec}^{-2}$; and FWHM = $1\farcs06\pm3\farcs89$ \& $30\farcs95\pm0\farcs97$, respectively.} \textbf{Right two panels}---The model represents $0\arcsec \leq R_{\rm eq} \leq 112\arcsec$ with $\Delta_{\rm rms}=0.0688\,\text{mag\,arcsec}^{-2}$. \underline{S{\'e}rsic Profile Parameters:} \textcolor{red}{$R_e=6\farcs00\pm0\farcs34$, $\mu_e=17.77\pm0.05\,\text{mag\,arcsec}^{-2}$, and $n=1.85\pm0.27$.} \underline{Ferrers Profile Parameters:} \textcolor{Orange}{$\mu_0 = 22.43\pm0.98\,\text{mag\,arcsec}^{-2}$, $R_{\rm end} = 134\farcs33\pm58\farcs06$, and $\alpha = 10.00\pm12.96$.} \underline{Exponential Profile Parameters:} \textcolor{blue}{$\mu_0 = 20.25\pm0.35\,\text{mag\,arcsec}^{-2}$ and $h = 23\farcs15\pm1\farcs80$.} \underline{Additional Parameters:} two Gaussian components added at \textcolor{cyan}{$R_{\rm r}=1\farcs37\pm0\farcs95$ \& $46\farcs11\pm0\farcs74$; with $\mu_0 = 15.91\pm7.32$ \& $22.48\pm0.13\,\text{mag\,arcsec}^{-2}$; and FWHM = $1\farcs81\pm6\farcs69$ \& $24\farcs19\pm1\farcs84$, respectively.}}
\label{NGC4151_plot}
\end{sidewaysfigure}

\subsection{NGC~4151 (Figure~\ref{NGC4151_plot})}

Two Gaussian components have been added: one at $R_{\rm maj} \approx 2\arcsec$ ($R_{\rm eq}\approx1\arcsec$), to account for a ring around the Type 1.5 intermediate Seyfert AGN \citep{Veron-Cetty:2006}, and one at $R_{\rm maj} \approx 59\arcsec$ ($R_{\rm eq}\approx46\arcsec$), coinciding with the confluence of the bar and the beginning of the spiral arms.

\begin{sidewaysfigure}
\includegraphics[clip=true,trim= 11mm 1mm 4mm 6mm,width=0.249\textwidth]{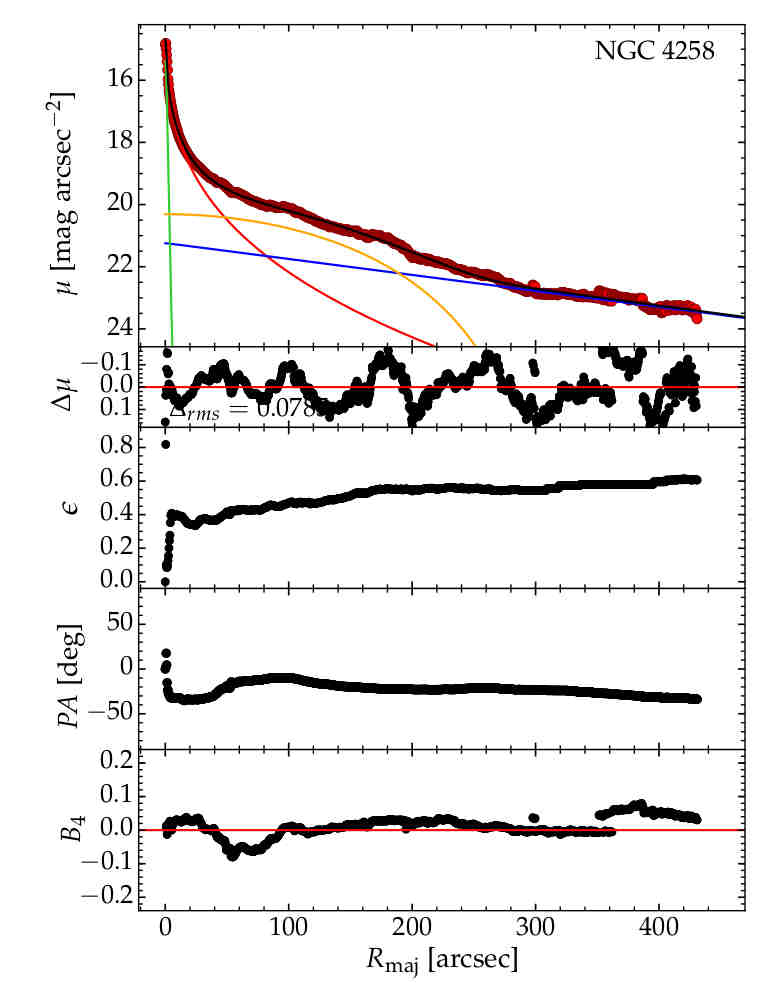}
\includegraphics[clip=true,trim= 11mm 1mm 4mm 6mm,width=0.249\textwidth]{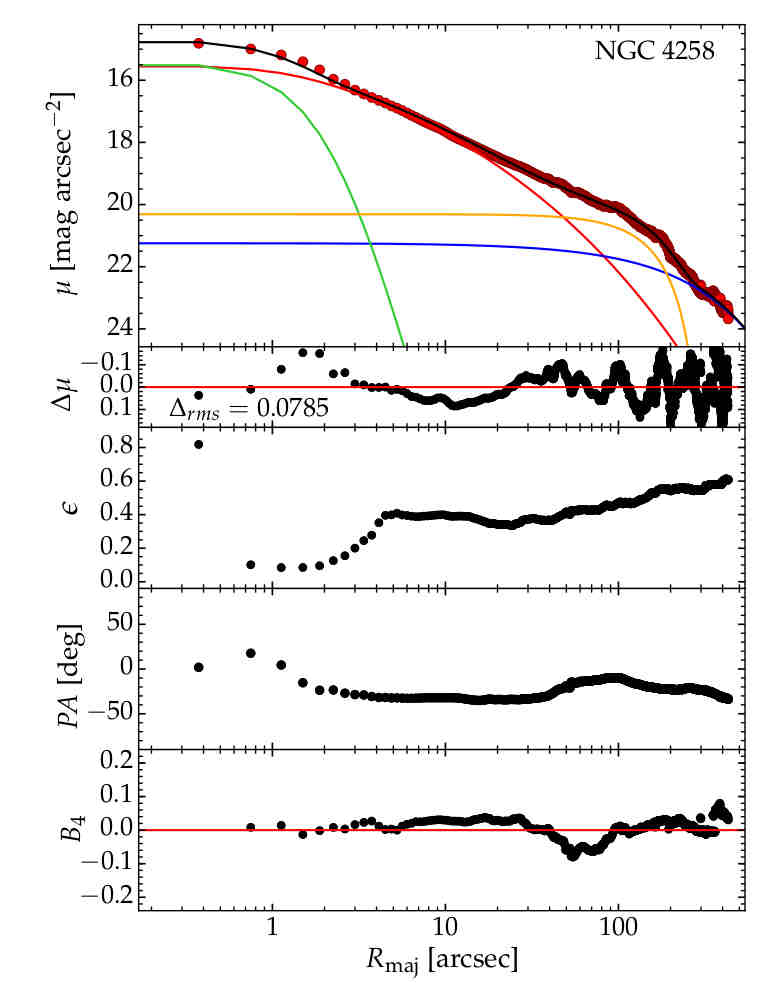}
\includegraphics[clip=true,trim= 11mm 1mm 4mm 6mm,width=0.249\textwidth]{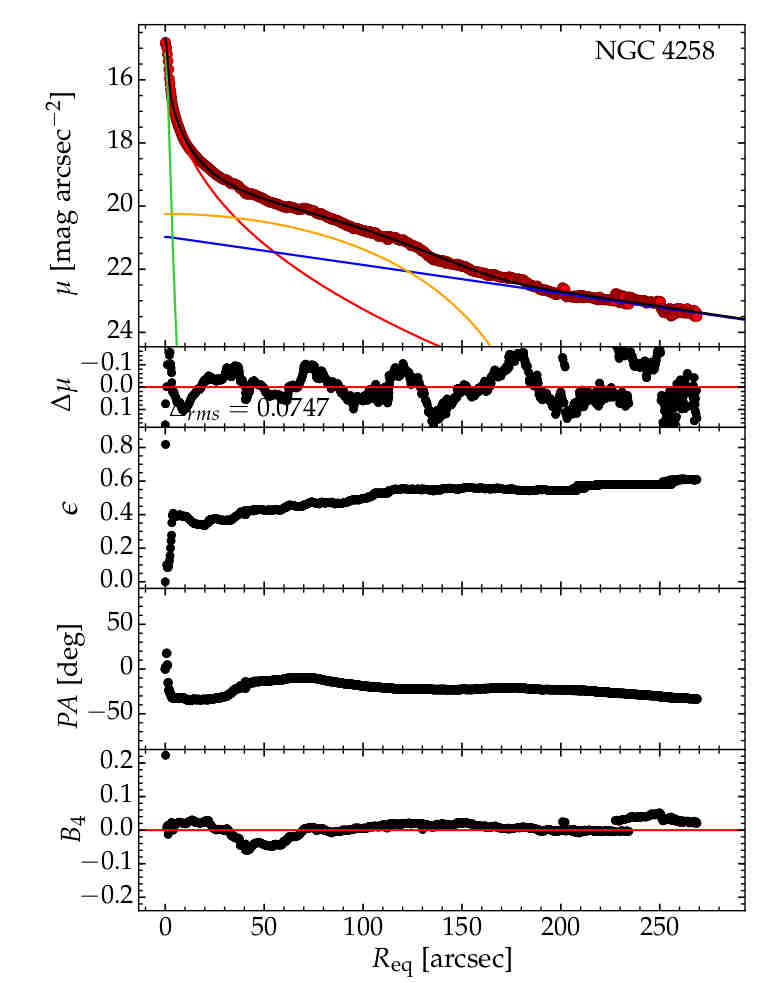}
\includegraphics[clip=true,trim= 11mm 1mm 4mm 6mm,width=0.249\textwidth]{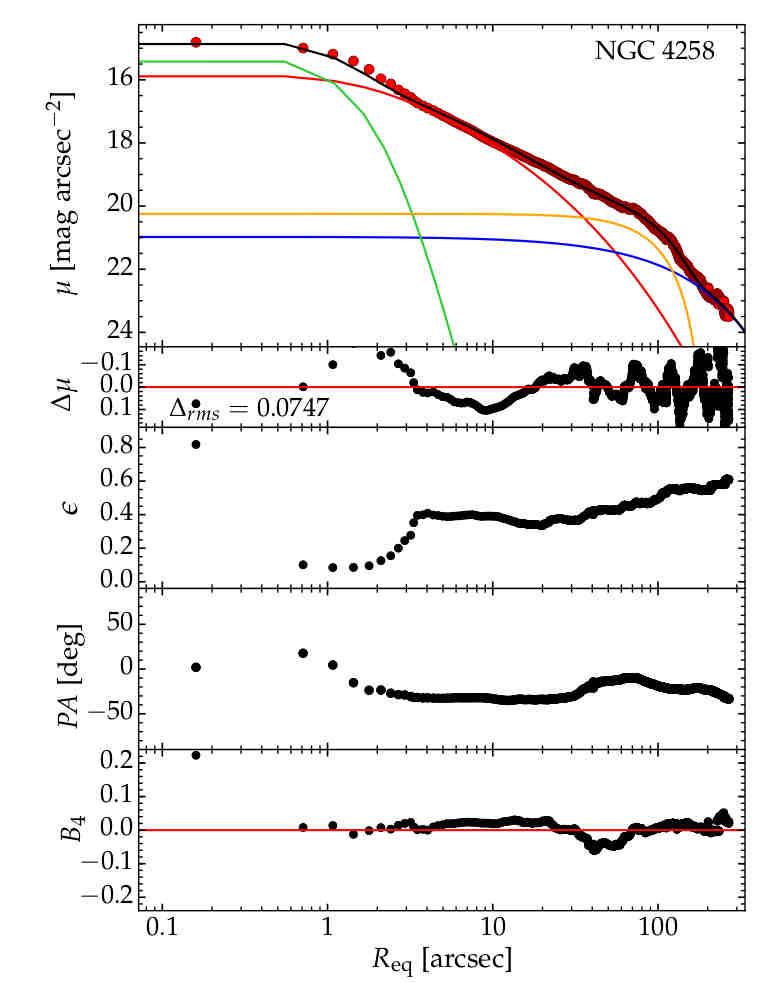}
\caption{\textit{Spitzer} $3.6\,\micron$ surface brightness profile for NGC~4258, with a physical scale of 0.0368$\,\text{kpc}\,\text{arcsec}^{-1}$. \textbf{Left two panels}---The model represents $0\arcsec \leq R_{\rm maj} \leq 431\arcsec$ with $\Delta_{\rm rms}=0.0785\,\text{mag\,arcsec}^{-2}$. \underline{Point Source:} \textcolor{LimeGreen}{$\mu_0 = 15.37\pm0.15\,\text{mag\,arcsec}^{-2}$.} \underline{S{\'e}rsic Profile Parameters:} \textcolor{red}{$R_e=41\farcs80\pm6\farcs71$, $\mu_e=20.14\pm0.27\,\text{mag\,arcsec}^{-2}$, and $n=3.21\pm0.31$.} \underline{Ferrers Profile Parameters:} \textcolor{Orange}{$\mu_0 = 20.28\pm0.05\,\text{mag\,arcsec}^{-2}$, $R_{\rm end} = 318\farcs48\pm10\farcs20$, and $\alpha = 9.99\pm0.94$.} \underline{Exponential Profile Parameters:} \textcolor{blue}{$\mu_0 = 21.25\pm0.07\,\text{mag\,arcsec}^{-2}$ and $h = 211\farcs37\pm5\farcs38$.} \textbf{Right two panels}---The model represents $0\arcsec \leq R_{\rm eq} \leq 270\arcsec$ with $\Delta_{\rm rms}=0.0747\,\text{mag\,arcsec}^{-2}$. \underline{Point Source:} \textcolor{LimeGreen}{$\mu_0 = 15.13\pm0.12\,\text{mag\,arcsec}^{-2}$.} \underline{S{\'e}rsic Profile Parameters:} \textcolor{red}{$R_e=26\farcs40\pm3\farcs90$, $\mu_e=19.73\pm0.25\,\text{mag\,arcsec}^{-2}$, and $n=2.60\pm0.28$.} \underline{Ferrers Profile Parameters:} \textcolor{Orange}{$\mu_0 = 20.23\pm0.04\,\text{mag\,arcsec}^{-2}$, $R_{\rm end} = 206\farcs70\pm1\farcs73$, and $\alpha = 9.60\pm0.00$.} \underline{Exponential Profile Parameters:} \textcolor{blue}{$\mu_0 = 20.93\pm0.10\,\text{mag\,arcsec}^{-2}$ and $h = 120\farcs70\pm4\farcs68$.}}
\label{NGC4258_plot}
\end{sidewaysfigure}

\subsection{NGC~4258 (Figure~\ref{NGC4258_plot})}\label{NGC4258}

We have included a point source in our model of NGC~4258 (M106) to account for light from the Type 2 Seyfert AGN \citep{Veron-Cetty:2006}. Our decomposition of the surface brightness profile differs significantly from Figure~29 of \citet{Savorgnan:2016}, whose small radial range likely prevented fitting a bar and disk.

\begin{sidewaysfigure}
\includegraphics[clip=true,trim= 11mm 1mm 5mm 5mm,width=0.249\textwidth]{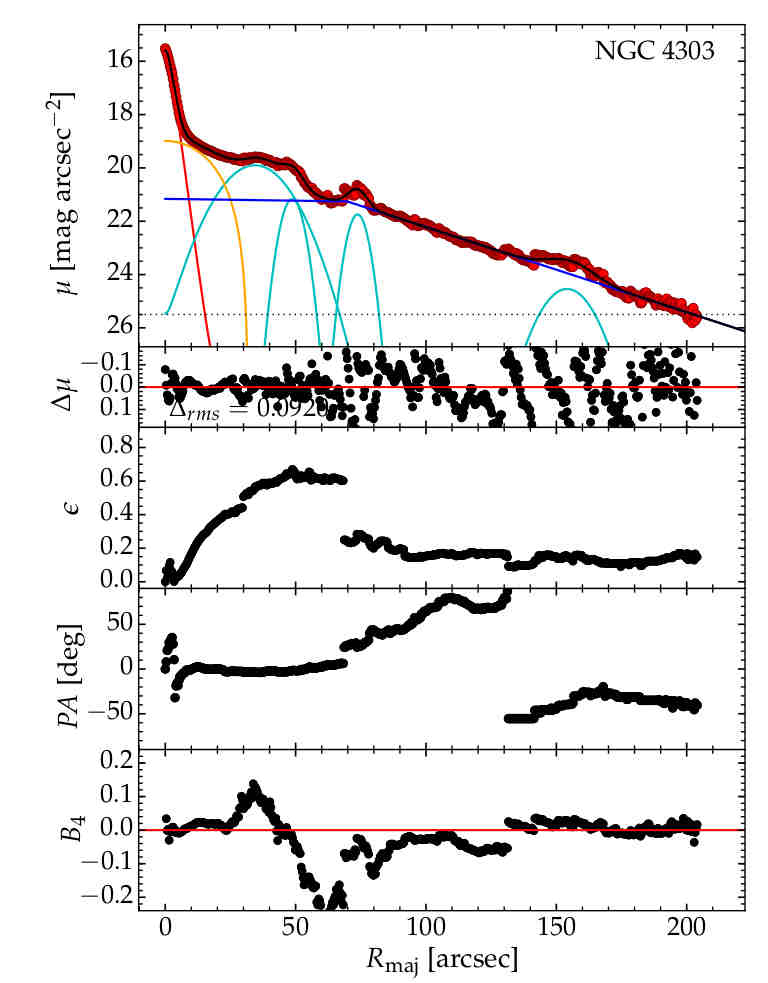}
\includegraphics[clip=true,trim= 11mm 1mm 5mm 5mm,width=0.249\textwidth]{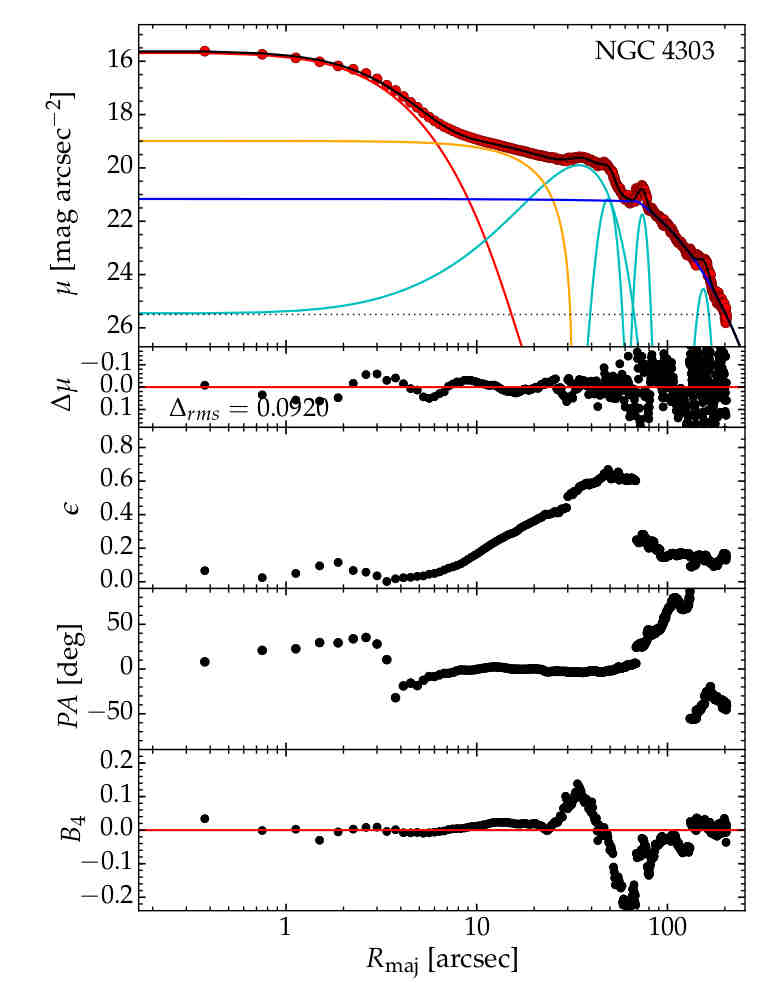}
\includegraphics[clip=true,trim= 11mm 1mm 5mm 5mm,width=0.249\textwidth]{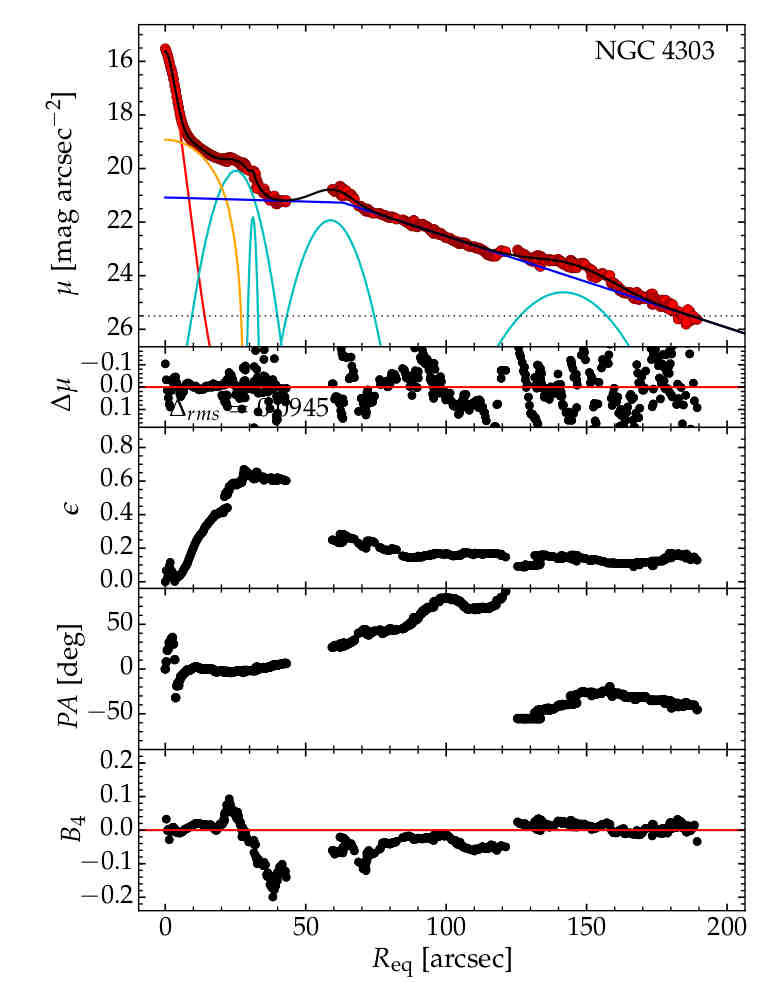}
\includegraphics[clip=true,trim= 11mm 1mm 5mm 5mm,width=0.249\textwidth]{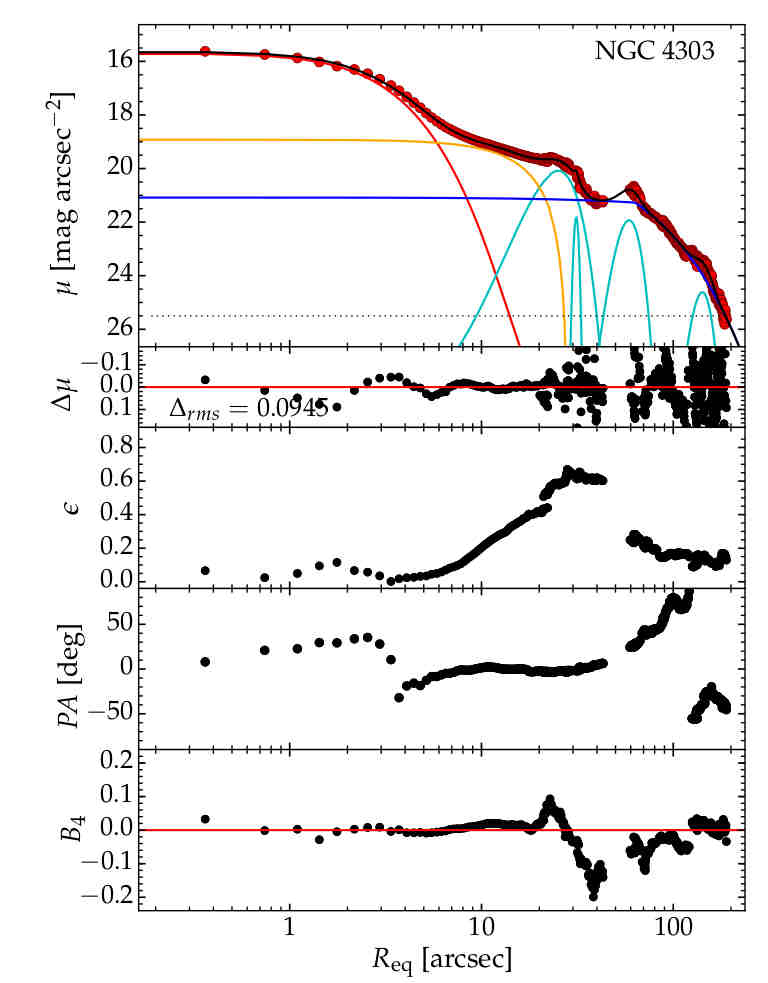}
\caption{\textit{Spitzer} $3.6\,\micron$ surface brightness profile for NGC~4303, with a physical scale of 0.0597$\,\text{kpc}\,\text{arcsec}^{-1}$. \textbf{Left two panels}---The model represents $0\arcsec \leq R_{\rm maj} \leq 204\arcsec$ with $\Delta_{\rm rms}=0.0920\,\text{mag\,arcsec}^{-2}$. \underline{S{\'e}rsic Profile Parameters:} \textcolor{red}{$R_e=2\farcs28\pm0\farcs09$, $\mu_e=15.88\pm0.10\,\text{mag\,arcsec}^{-2}$, and $n=1.02\pm0.13$.} \underline{Ferrers Profile Parameters:} \textcolor{Orange}{$\mu_0 = 18.99\pm0.28\,\text{mag\,arcsec}^{-2}$, $R_{\rm end} = 31\farcs91\pm3\farcs32$, and $\alpha = 5.33\pm1.60$.} \underline{Broken Exponential Profile Parameters:} \textcolor{blue}{$\mu_0 = 21.17\pm2.13\,\text{mag\,arcsec}^{-2}$, $R_b = 69\farcs75\pm2\farcs33$, $h_1 = 800\farcs34\pm18500\arcsec$, and $h_2 = 34\farcs01\pm0\farcs17$.} \underline{Additional Parameters:} four Gaussian components added at \textcolor{cyan}{$R_{\rm r}=34\farcs38\pm1\farcs53$, $48\farcs68\pm0\farcs39$, $73\farcs75\pm0\farcs28$, \& $154\farcs11\pm0\farcs38$; with $\mu_0 = 19.91\pm0.31$, $21.17\pm0.28$, $21.75\pm0.07$, \& $24.54\pm0.05\,\text{mag\,arcsec}^{-2}$; and FWHM = $23\farcs83\pm6\farcs48$, $7\farcs82\pm1\farcs48$, $7\farcs39\pm0\farcs61$, \& $18\farcs92\pm0\farcs84$, respectively.} \textbf{Right two panels}---The model represents $0\arcsec \leq R_{\rm eq} \leq 190\arcsec$ with $\Delta_{\rm rms}=0.0945\,\text{mag\,arcsec}^{-2}$. \underline{S{\'e}rsic Profile Parameters:} \textcolor{red}{$R_e=2\farcs16\pm0\farcs09$, $\mu_e=15.15\pm0.11\,\text{mag\,arcsec}^{-2}$, and $n=0.90\pm0.13$.} \underline{Ferrers Profile Parameters:} \textcolor{Orange}{$\mu_0 = 18.92\pm0.13\,\text{mag\,arcsec}^{-2}$, $R_{\rm end} = 28\farcs06\pm5\farcs80$, and $\alpha = 5.83\pm2.44$.} \underline{Broken Exponential Profile Parameters:} \textcolor{blue}{$\mu_0 = 21.08\pm0.17\,\text{mag\,arcsec}^{-2}$, $R_b = 63\farcs23\pm2\farcs49$, $h_1 = 355\farcs69\pm445\farcs92$, and $h_2 = 31\farcs80\pm0\farcs17$.} \underline{Additional Parameters:} four Gaussian components added at \textcolor{cyan}{$R_{\rm r}=25\farcs10\pm1\farcs13$, $31\farcs20\pm0\farcs20$, $58\farcs83\pm0\farcs42$, \& $141\farcs74\pm0\farcs52$; with $\mu_0 = 20.09\pm0.17$, $21.81\pm0.29$, $21.94\pm0.12$, \& $24.62\pm0.04\,\text{mag\,arcsec}^{-2}$; and FWHM = $11\farcs05\pm1\farcs03$, $1\farcs64\pm0\farcs50$, $14\farcs30\pm0\farcs78$, \& $28\farcs57\pm1\farcs12$, respectively.}}
\label{NGC4303_plot}
\end{sidewaysfigure}

\subsection{NGC~4303 (Figure~\ref{NGC4303_plot})}

For NGC~4303 (M61), we use a broken exponential model to account for redistributed disk light in the inner region of the galaxy \citep{Saha:2018}. We add four Gaussian components at $R_{\rm maj} \approx34\arcsec$ ($R_{\rm eq}\approx25\arcsec$), $R_{\rm maj} \approx49\arcsec$ ($R_{\rm eq}\approx31\arcsec$), $R_{\rm maj} \approx74\arcsec$ ($R_{\rm eq}\approx59\arcsec$), and $R_{\rm maj} \approx154\arcsec$ ($R_{\rm eq}\approx142\arcsec$) to account for the end of the bar, the staggered beginnings of two spiral arms, and the confluence of the two spiral outer arms, respectively.

\begin{sidewaysfigure}
\includegraphics[clip=true,trim= 11mm 1mm 5mm 5mm,width=0.249\textwidth]{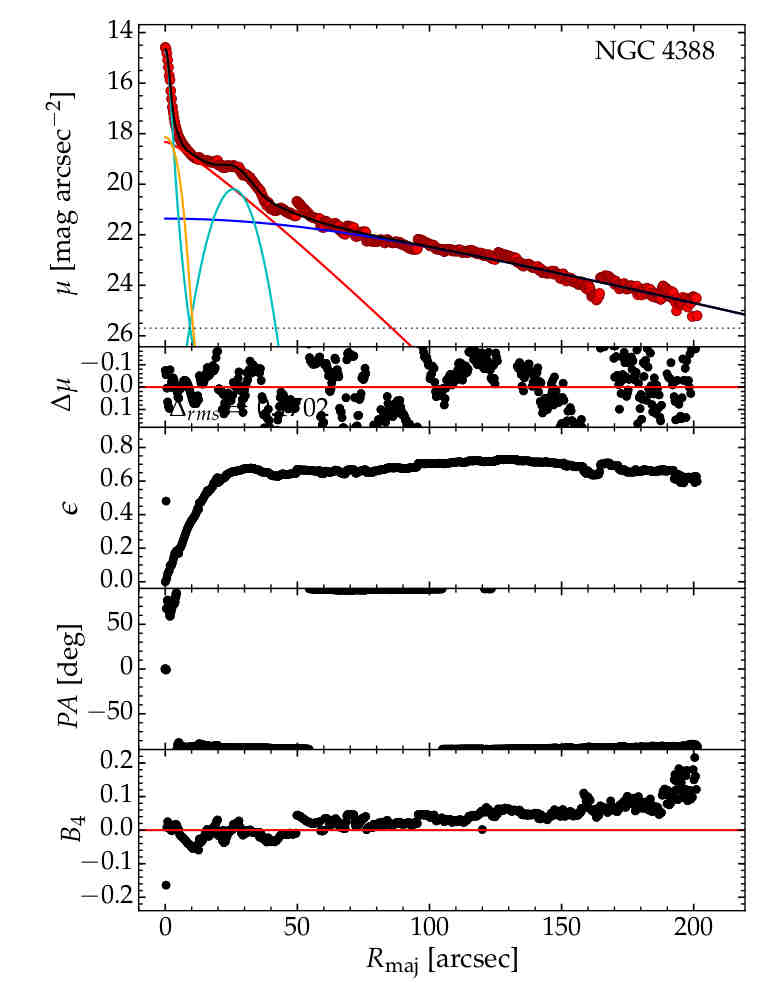}
\includegraphics[clip=true,trim= 11mm 1mm 5mm 5mm,width=0.249\textwidth]{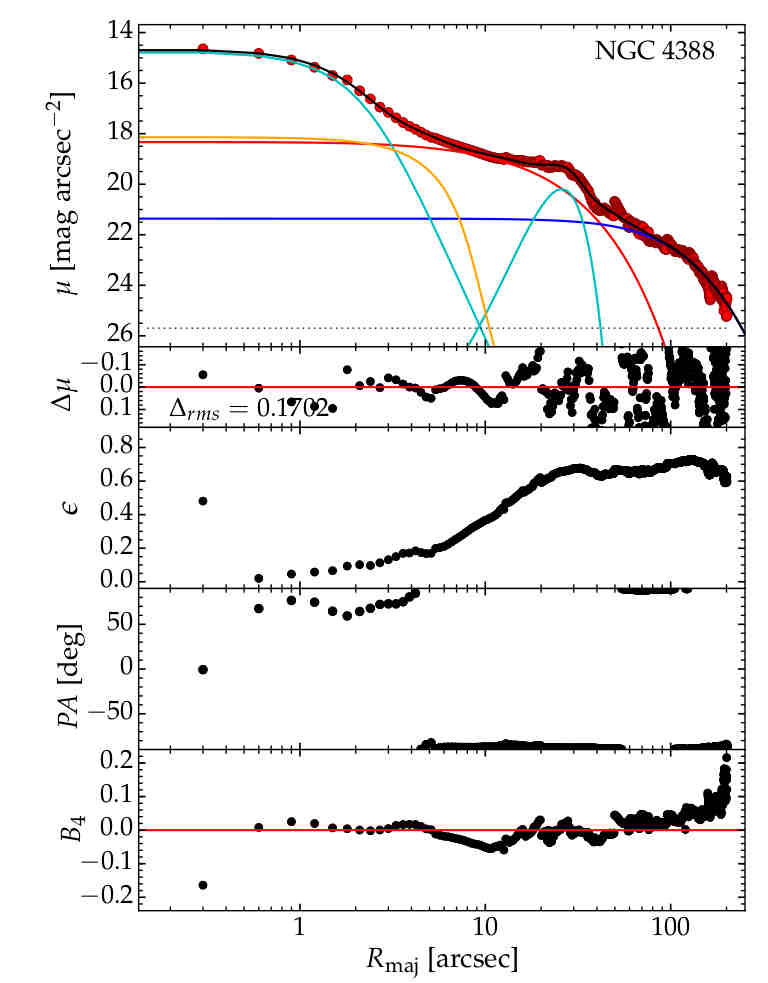}
\includegraphics[clip=true,trim= 11mm 1mm 5mm 5mm,width=0.249\textwidth]{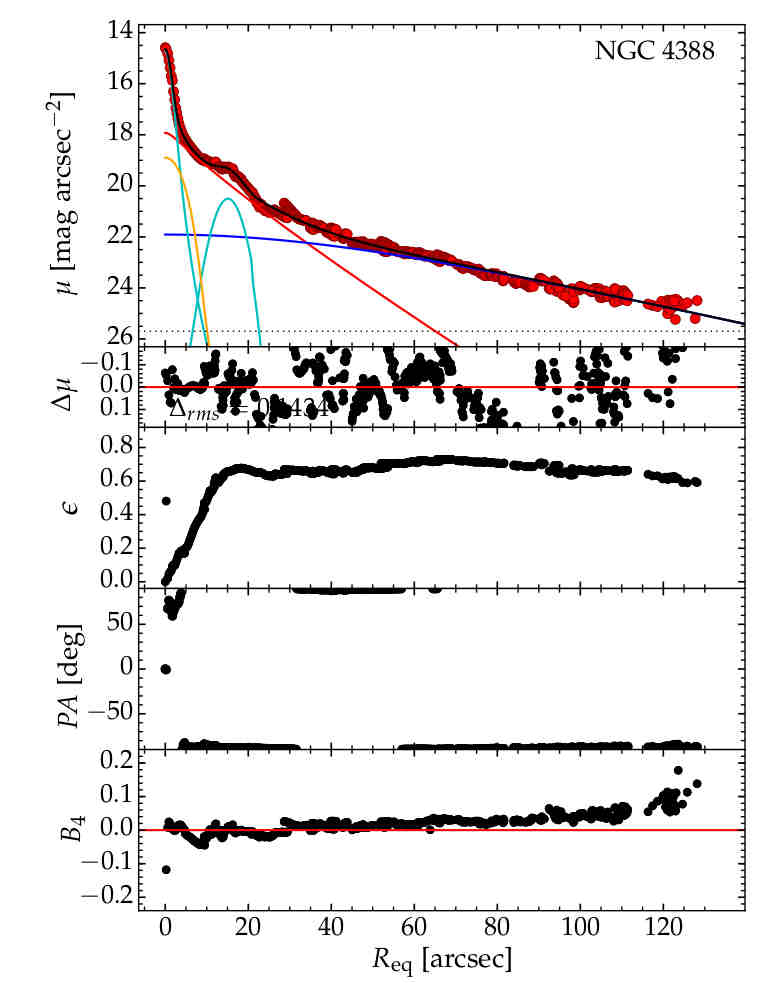}
\includegraphics[clip=true,trim= 11mm 1mm 5mm 5mm,width=0.249\textwidth]{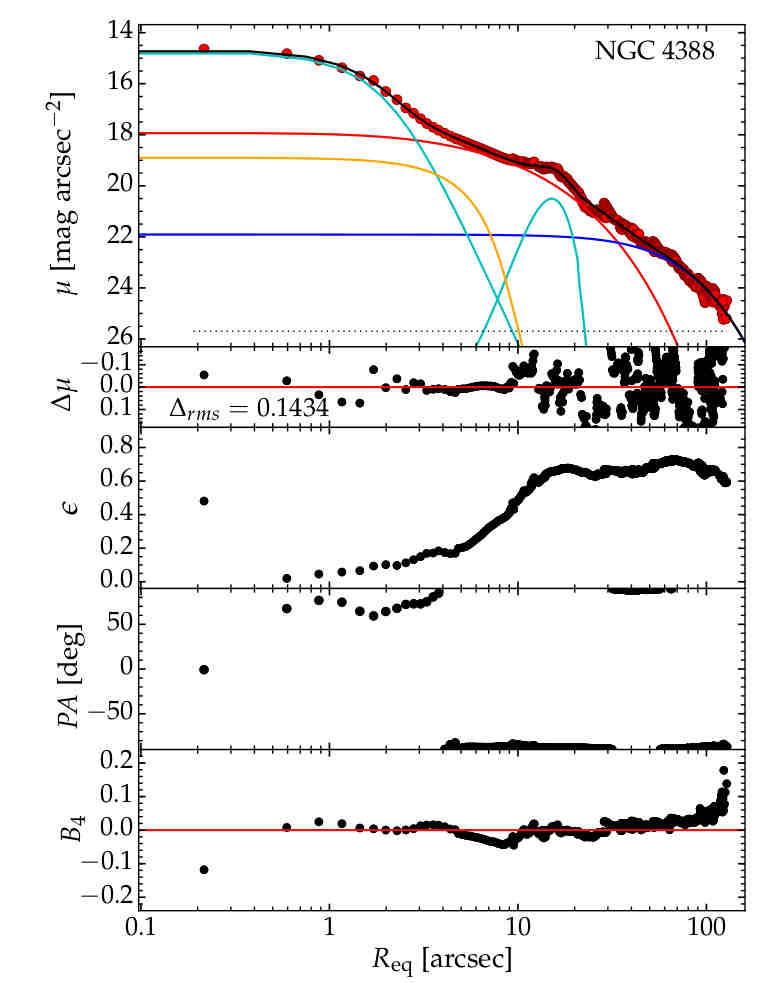}
\caption{\textit{Spitzer} $3.6\,\micron$ surface brightness profile for NGC~4388, with a physical scale of 0.0862$\,\text{kpc}\,\text{arcsec}^{-1}$. \textbf{Left two panels}---The model represents $0\arcsec \leq R_{\rm maj} \leq 202\arcsec$ with $\Delta_{\rm rms}=0.1702\,\text{mag\,arcsec}^{-2}$. \underline{S{\'e}rsic Profile Parameters:} \textcolor{red}{$R_e=21\farcs68\pm0\farcs54$, $\mu_e=19.83\pm0.08\,\text{mag\,arcsec}^{-2}$, and $n=0.89\pm0.13$.} \underline{Ferrers Profile Parameters:} \textcolor{Orange}{$\mu_0 = 17.99\pm1.26\,\text{mag\,arcsec}^{-2}$, $R_{\rm end} = 9\farcs17\pm15\farcs99$, and $\alpha = 10.00\pm47.60$.} \underline{Edge-on Disk Model Parameters:} \textcolor{blue}{$\mu_0 = 21.36\pm0.04\,\text{mag\,arcsec}^{-2}$ and $h_z = 90\farcs08\pm1\farcs10$.} \underline{Additional Parameters:} two Gaussian components added at: \textcolor{cyan}{$R_{\rm r}=0\arcsec$ \& $25\farcs79\pm0\farcs55$; with $\mu_0 = 13.21\pm0.38$ \& $20.20\pm0.11\,\text{mag\,arcsec}^{-2}$; and FWHM = $1\farcs28\pm0\farcs30$ \& $11\farcs90\pm1\farcs32$, respectively.} \textbf{Right two panels}---The model represents $0\arcsec \leq R_{\rm eq} \leq 129\arcsec$ with $\Delta_{\rm rms}=0.1434\,\text{mag\,arcsec}^{-2}$. \underline{S{\'e}rsic Profile Parameters:} \textcolor{red}{$R_e=14\farcs30\pm0\farcs55$, $\mu_e=19.82\pm0.10\,\text{mag\,arcsec}^{-2}$, and $n=1.15\pm0.09$.} \underline{Ferrers Profile Parameters:} \textcolor{Orange}{$\mu_0 = 18.78\pm1.85\,\text{mag\,arcsec}^{-2}$, $R_{\rm end} = 9\farcs30\pm25\farcs47$, and $\alpha = 10.00\pm78.67$.} \underline{Edge-on Disk Model Parameters:} \textcolor{blue}{$\mu_0 = 21.91\pm0.04\,\text{mag\,arcsec}^{-2}$ and $h_z = 60\farcs78\pm0\farcs76$.} \underline{Additional Parameters:} two Gaussian components added at: \textcolor{cyan}{$R_{\rm r}=0\arcsec$ \& $15\farcs12\pm0\farcs44$; with $\mu_0 = 12.98\pm0.46$ \& $20.39\pm0.16\,\text{mag\,arcsec}^{-2}$; and FWHM = $1\farcs04\pm0\farcs26$ \& $5\farcs64\pm1\farcs08$, respectively.}}
\label{NGC4388_plot}
\end{sidewaysfigure}

\subsection{NGC~4388 (Figure~\ref{NGC4388_plot})}\label{NGC4388}

The galaxy NGC~4388 possesses a Type 1 Seyfert AGN with broad polarized Balmer lines \citep{Veron-Cetty:2006}, which we modeled with a central Gaussian. It is a near edge-on galaxy, inclined by $71\fdg6\pm1\fdg9$ with respect to the plane of the sky. Because of this, we use the edge-on disk model. We add another Gaussian component at $R_{\rm maj} \approx 26\arcsec$ ($R_{\rm eq}\approx 15\arcsec$) to account for the spiral arms.

\begin{sidewaysfigure}
\includegraphics[clip=true,trim= 11mm 1mm 5mm 6mm,width=0.249\textwidth]{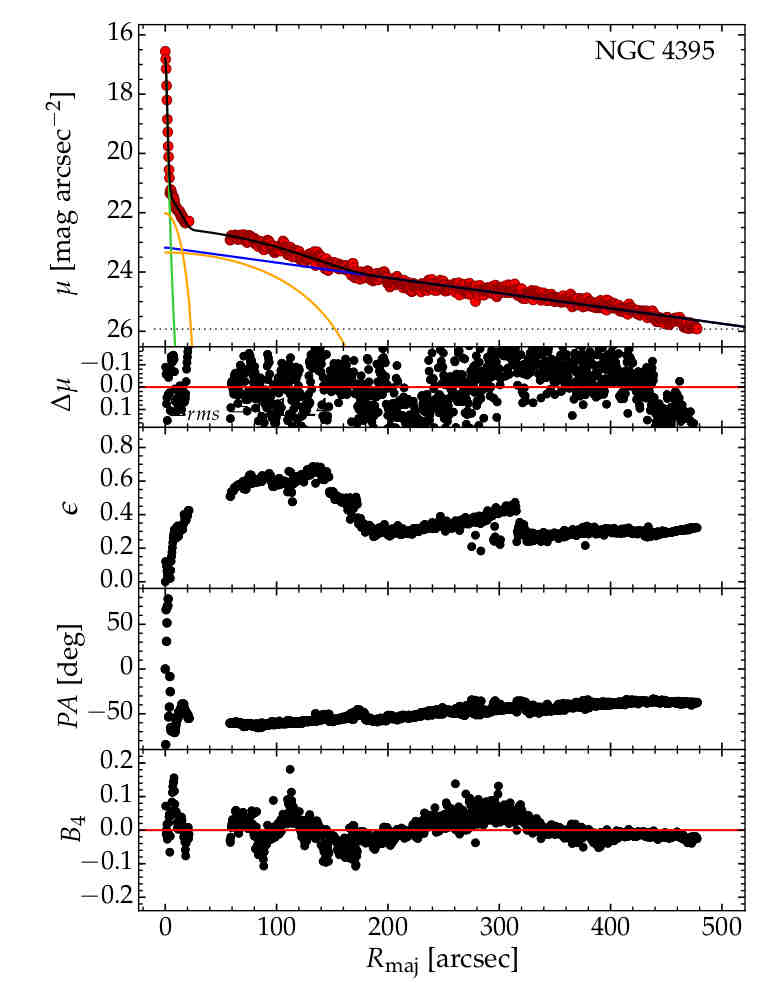}
\includegraphics[clip=true,trim= 11mm 1mm 5mm 6mm,width=0.249\textwidth]{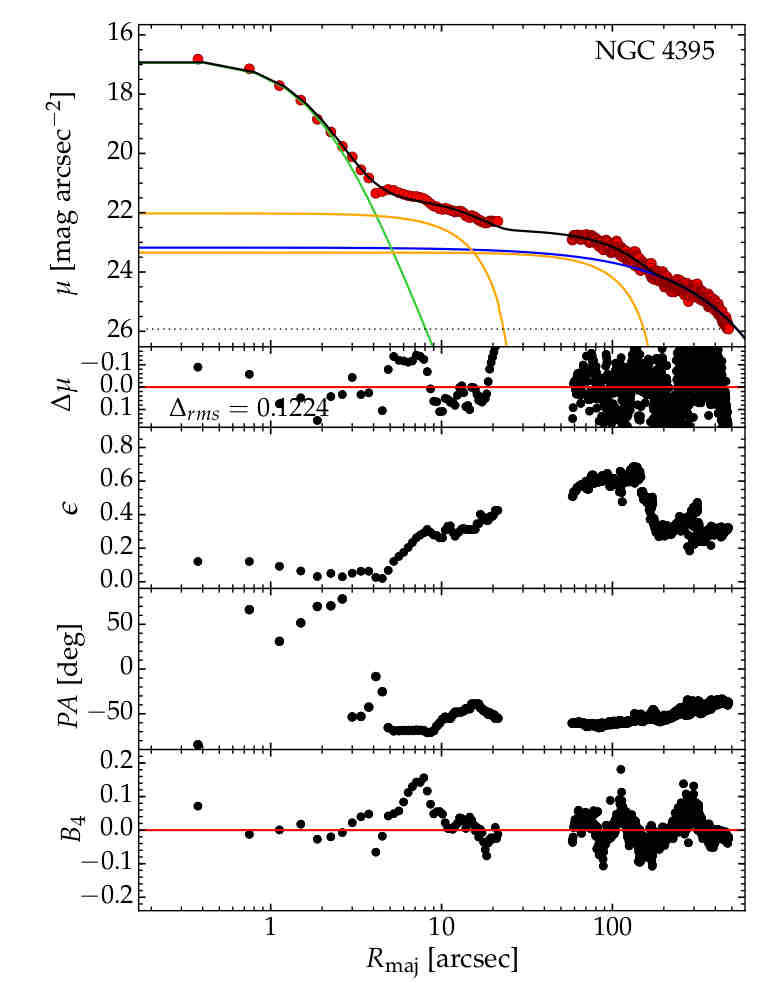}
\includegraphics[clip=true,trim= 11mm 1mm 5mm 6mm,width=0.249\textwidth]{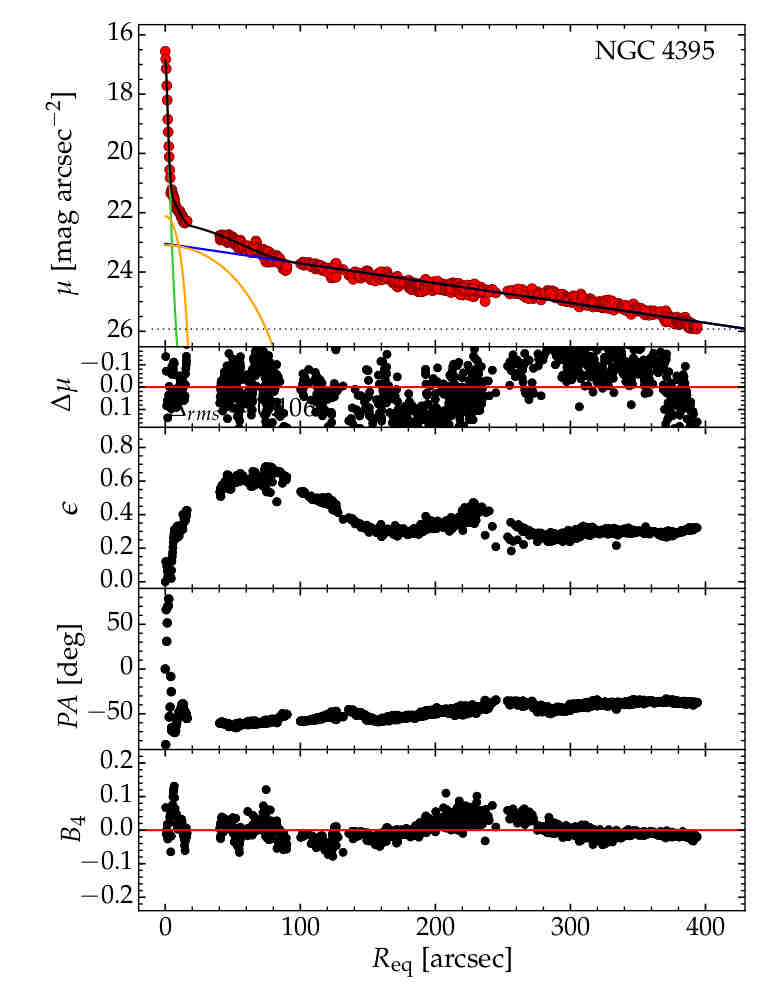}
\includegraphics[clip=true,trim= 11mm 1mm 5mm 6mm,width=0.249\textwidth]{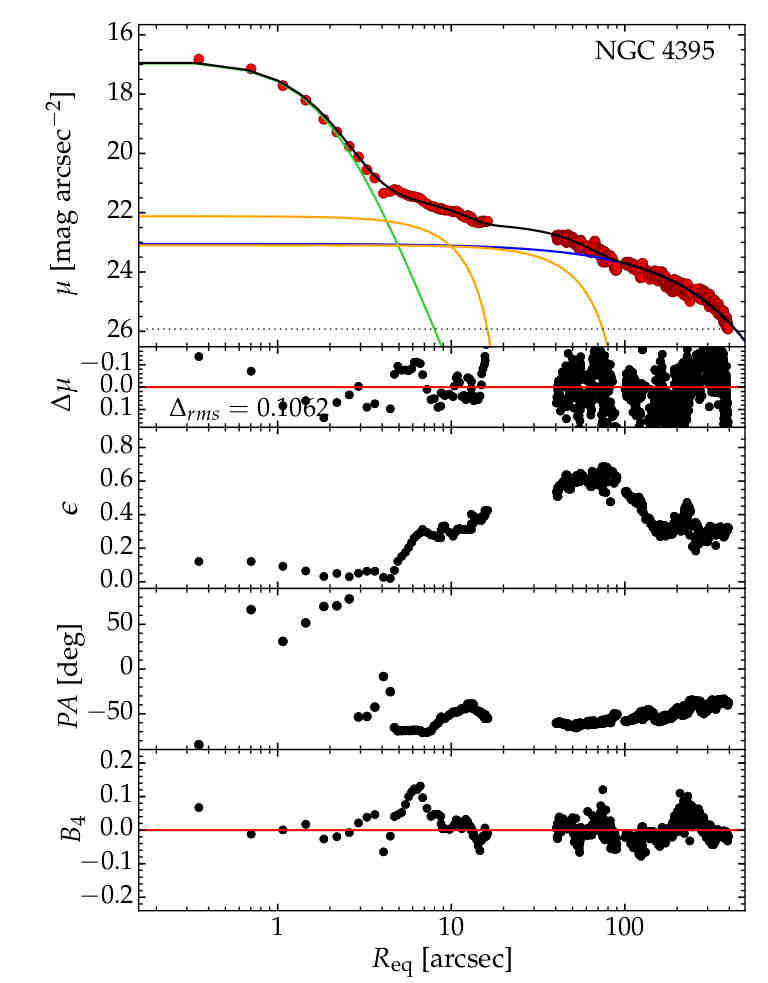}
\caption{\textit{Spitzer} $3.6\,\micron$ surface brightness profile for NGC~4395, with a physical scale of 0.0231$\,\text{kpc}\,\text{arcsec}^{-1}$. \textbf{Left two panels}---The model represents $0\arcsec \leq R_{\rm maj} \leq 21\arcsec$ \& $58\arcsec \leq R_{\rm maj} \leq 478\arcsec$ with $\Delta_{\rm rms}=0.1224\,\text{mag\,arcsec}^{-2}$. \underline{Point Source:} \textcolor{LimeGreen}{$\mu_0 = 16.82\pm0.04\,\text{mag\,arcsec}^{-2}$.} \underline{Ferrers Profile Parameters:} \textcolor{Orange}{$\mu_0 = 22.01\pm0.11$ \& $23.35\pm0.06\,\text{mag\,arcsec}^{-2}$; $R_{\rm end} = 29\farcs95\pm13\farcs61$ \& $196\farcs87\pm11\farcs63$; with $\alpha = 9.99\pm13.73$ \& $6.47\pm1.24$, respectively.} \underline{Exponential Profile Parameters:} \textcolor{blue}{$\mu_0 = 23.17\pm0.02\,\text{mag\,arcsec}^{-2}$ and $h = 210\farcs84\pm2\farcs17$.} \textbf{Right two panels}---The model represents $0\arcsec \leq R_{\rm maj} \leq 16\arcsec$ \& $41\arcsec \leq R_{\rm maj} \leq 394\arcsec$ with $\Delta_{\rm rms}=0.1062\,\text{mag\,arcsec}^{-2}$. \underline{Point Source:} \textcolor{LimeGreen}{$\mu_0 = 16.89\pm0.05\,\text{mag\,arcsec}^{-2}$.} \underline{Ferrers Profile Parameters:} \textcolor{Orange}{$\mu_0 = 22.09\pm0.16$ \& $23.10\pm0.09\,\text{mag\,arcsec}^{-2}$; $R_{\rm end} = 18\farcs51\pm5\farcs67$ \& $108\farcs11\pm7\farcs51$; with $\alpha = 7.01\pm6.15$ \& $10.00\pm2.02$, respectively.} \underline{Exponential Profile Parameters:} \textcolor{blue}{$\mu_0 = 23.05\pm0.01\,\text{mag\,arcsec}^{-2}$ and $h = 162\farcs53\pm1\farcs24$.}}
\label{NGC4395_plot}
\end{sidewaysfigure}

\subsection{NGC~4395 (Figure~\ref{NGC4395_plot})}

The galaxy NGC~4395 is bulgeless \citep{Sandage:1981,Brok:2015} with a Type 1.8 intermediate AGN \citep{Veron-Cetty:2006}. From \textit{HST} observations, \citet{Filippenko:Ho:2003} revealed the presence of a nuclear star cluster. Our isophote fitting failed to converge over the range $21\arcsec < R_{\rm maj} < 58\arcsec$ ($16\arcsec < R_{\rm eq} < 41\arcsec$); thus, these regions were omitted from our surface brightness profiles. We model the nuclear region without a S{\'e}rsic profile, only a PSF. Additionally, we model a bar plus an off-nuclear, possibly H$\alpha$, gas cloud.

\begin{sidewaysfigure}
\includegraphics[clip=true,trim= 11mm 1mm 4mm 5mm,width=0.249\textwidth]{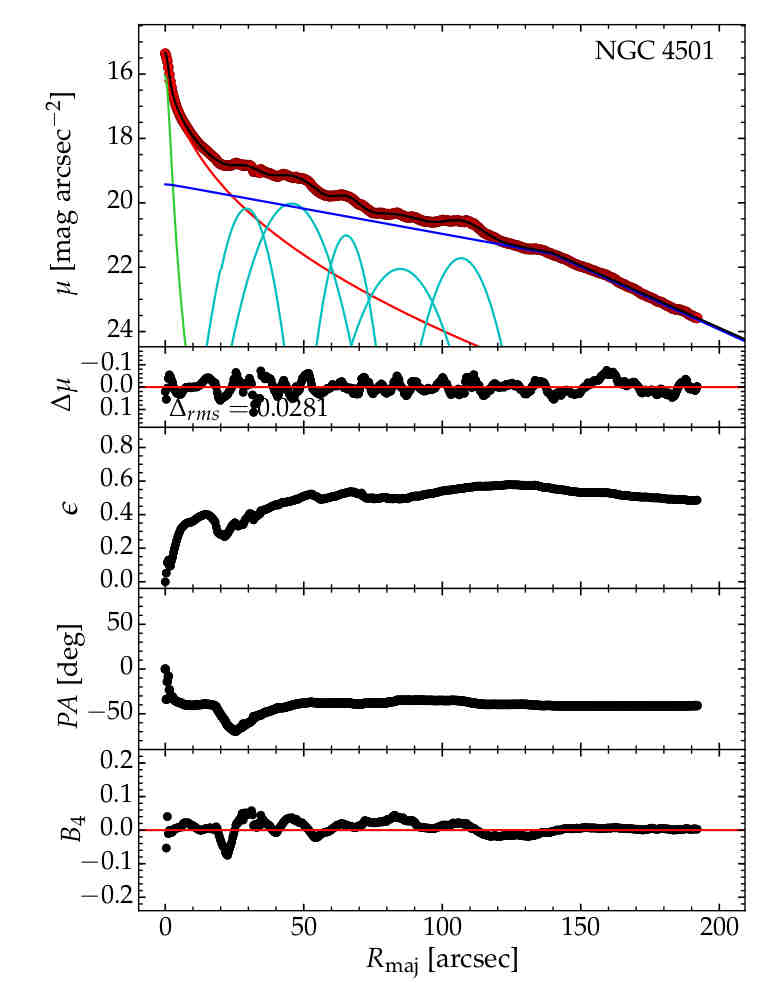}
\includegraphics[clip=true,trim= 11mm 1mm 4mm 5mm,width=0.249\textwidth]{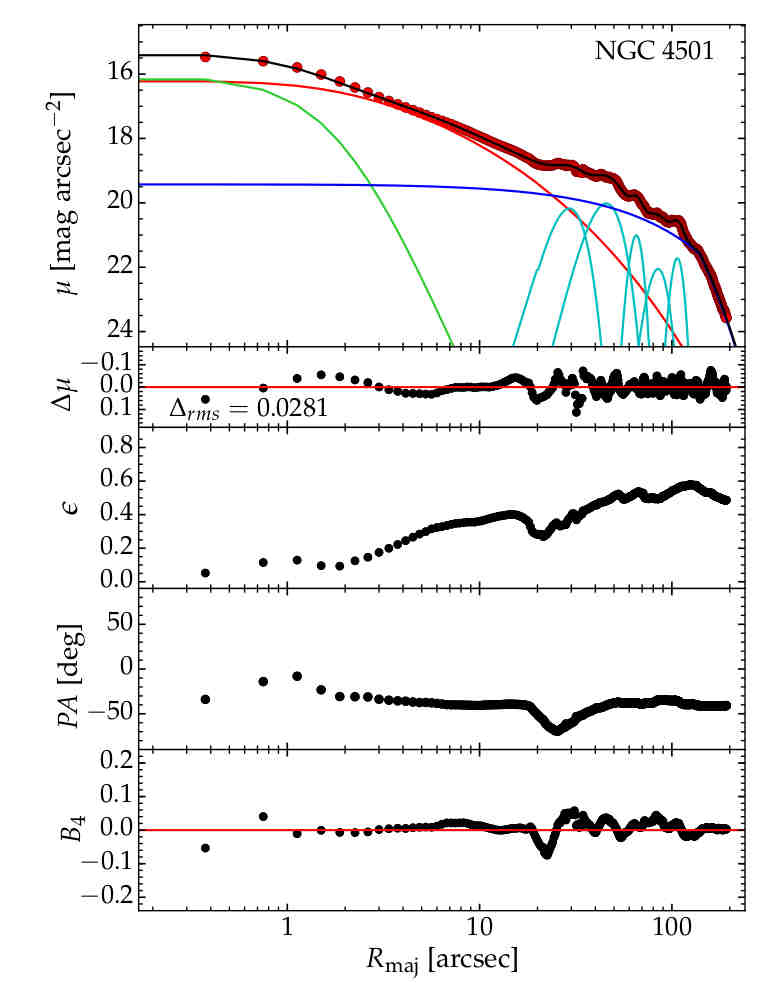}
\includegraphics[clip=true,trim= 11mm 1mm 4mm 5mm,width=0.249\textwidth]{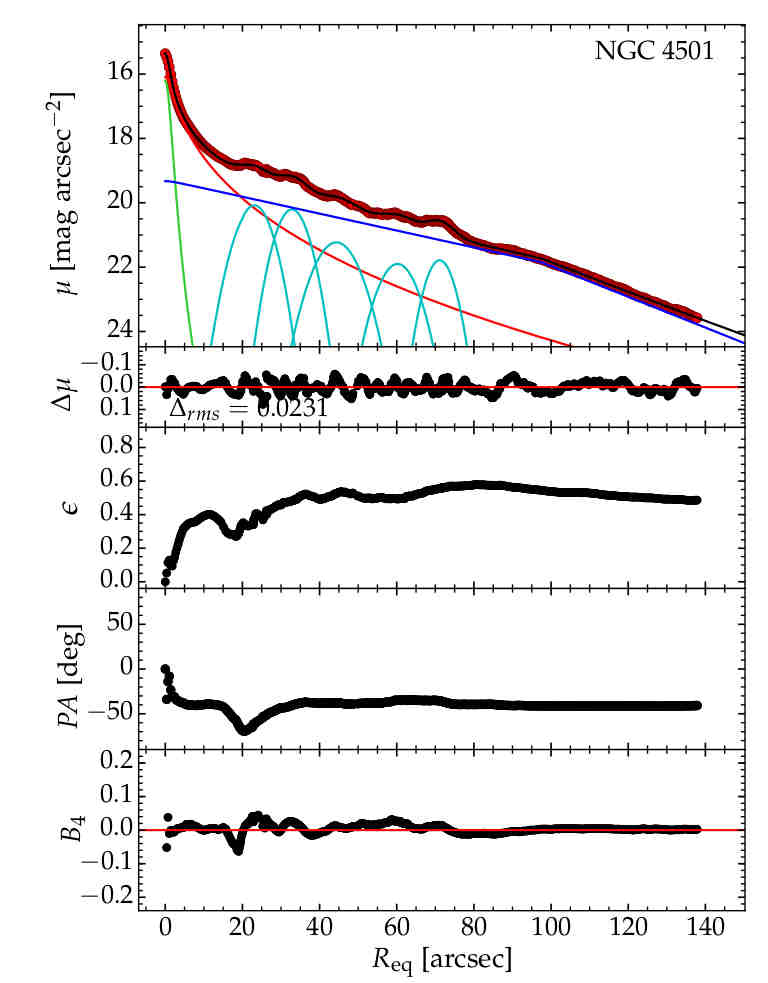}
\includegraphics[clip=true,trim= 11mm 1mm 4mm 5mm,width=0.249\textwidth]{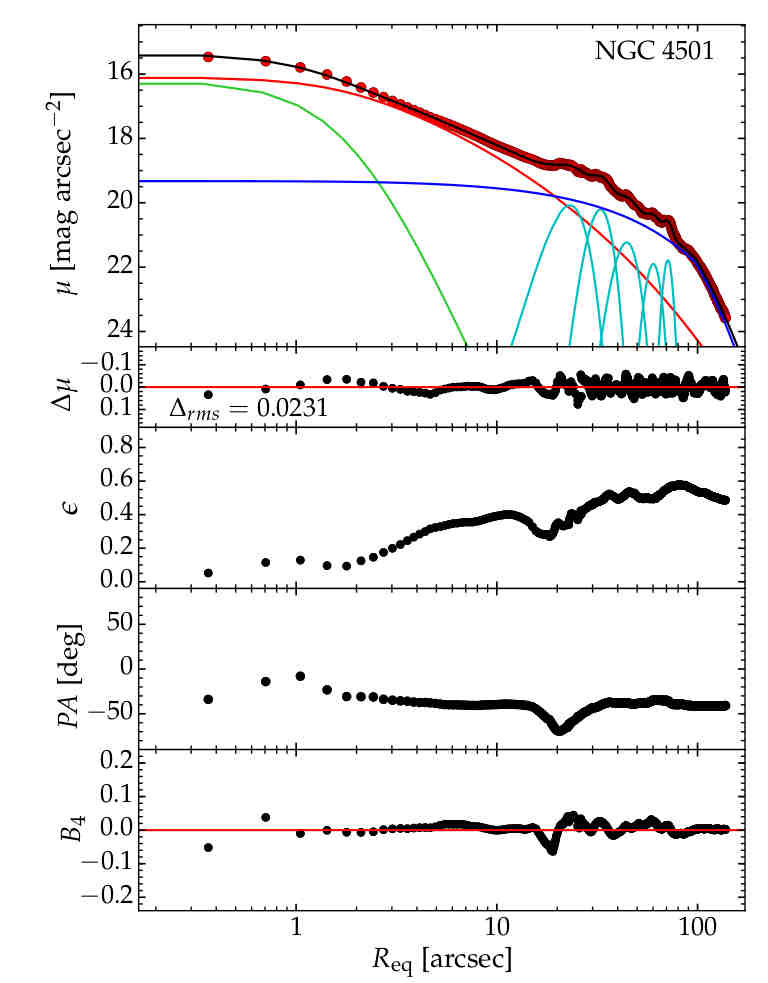}
\caption{\textit{Spitzer} $3.6\,\micron$ surface brightness profile for NGC~4501, with a physical scale of 0.0541$\,\text{kpc}\,\text{arcsec}^{-1}$. \textbf{Left two panels}---The model represents $0\arcsec \leq R_{\rm maj} \leq 192\arcsec$ with $\Delta_{\rm rms}=0.0281\,\text{mag\,arcsec}^{-2}$. \underline{Point Source:} \textcolor{LimeGreen}{$\mu_0 = 15.42\pm0.07\,\text{mag\,arcsec}^{-2}$.} \underline{S{\'e}rsic Profile Parameters:} \textcolor{red}{$R_e=21\farcs22\pm3\farcs93$, $\mu_e=19.53\pm0.26\,\text{mag\,arcsec}^{-2}$, and $n=2.33\pm0.23$.} \underline{Broken Exponential Profile Parameters:} \textcolor{blue}{$\mu_0 = 19.40\pm0.18\,\text{mag\,arcsec}^{-2}$, $R_b = 139\farcs69\pm0\farcs40$, $h_1 = 69\farcs39\pm5\farcs32$, and $h_2 = 27\farcs81\pm0\farcs36$.} \underline{Additional Parameters:} five Gaussian components added at: \textcolor{cyan}{$R_{\rm r}=29\farcs30\pm0\farcs24$, $45\farcs56\pm0\farcs25$, $65\farcs31\pm0\farcs17$, $84\farcs80\pm0\farcs56$, and $106\farcs80\pm0\farcs26$; with $\mu_0 = 20.18\pm0.06$, $20.03\pm0.05$, $21.02\pm0.06$, $22.06\pm0.14$, and $21.73\pm0.04\,\text{mag\,arcsec}^{-2}$; and FWHM = $11\farcs47\pm0\farcs63$, $17\farcs83\pm0\farcs74$, $10\farcs04\pm0\farcs47$, $19\farcs87\pm1\farcs79$, and $15\farcs41\pm0\farcs55$.} \textbf{Right two panels}---The model represents $0\arcsec \leq R_{\rm eq} \leq 138\arcsec$ with $\Delta_{\rm rms}=0.0231\,\text{mag\,arcsec}^{-2}$. \underline{Point Source:} \textcolor{LimeGreen}{$\mu_0 = 16.21\pm0.07\,\text{mag\,arcsec}^{-2}$.} \underline{S{\'e}rsic Profile Parameters:} \textcolor{red}{$R_e=20\farcs35\pm2\farcs79$, $\mu_e=19.91\pm0.21\,\text{mag\,arcsec}^{-2}$, and $n=2.83\pm0.20$.} \underline{Broken Exponential Profile Parameters:} \textcolor{blue}{$\mu_0 = 19.29\pm0.09\,\text{mag\,arcsec}^{-2}$, $R_b = 97\farcs54\pm0\farcs39$, $h_1 = 41\farcs29\pm1\farcs14$, and $h_2 = 22\farcs73\pm0\farcs64$.} \underline{Additional Parameters:} five Gaussian components added at: \textcolor{cyan}{$R_{\rm r}=23\farcs10\pm0\farcs18$, $32\farcs82\pm0\farcs17$, $44\farcs36\pm0\farcs30$, $60\farcs21\pm0\farcs19$, and $71\farcs12\pm0\farcs14$; with $\mu_0 = 20.08\pm0.04$, $20.21\pm0.03$, $21.24\pm0.05$, $21.91\pm0.05$, and $21.79\pm0.03\,\text{mag\,arcsec}^{-2}$; and FWHM = $8\farcs68\pm0\farcs50$, $8\farcs27\pm0\farcs46$, $11\farcs31\pm0\farcs81$, $10\farcs04\pm0\farcs68$, and $7\farcs52\pm0\farcs26$.}}
\label{NGC4501_plot}
\end{sidewaysfigure}

\subsection{NGC~4501 (Figure~\ref{NGC4501_plot})}\label{NGC4501}

The galaxy NGC~4501 (M88) contains a Type 2 Seyfert AGN \citep{Veron-Cetty:2006}, for which we have added a point source. It is a very flocculent spiral galaxy. We have added five Gaussian components to account for numerous spiral-arm crossings at $R_{\rm maj} \approx29\arcsec$ ($R_{\rm eq}\approx23\arcsec$), $R_{\rm maj} \approx46\arcsec$ ($R_{\rm eq}\approx33\arcsec$), $R_{\rm maj} \approx65\arcsec$ ($R_{\rm eq}\approx44\arcsec$), $R_{\rm maj} \approx85\arcsec$ ($R_{\rm eq}\approx60\arcsec$), and $R_{\rm maj} \approx107\arcsec$ ($R_{\rm eq}\approx71\arcsec$). We modeled NGC~4501 with a truncated exponential profile.

\begin{sidewaysfigure}
\includegraphics[clip=true,trim= 11mm 1mm 4mm 5mm,width=0.249\textwidth]{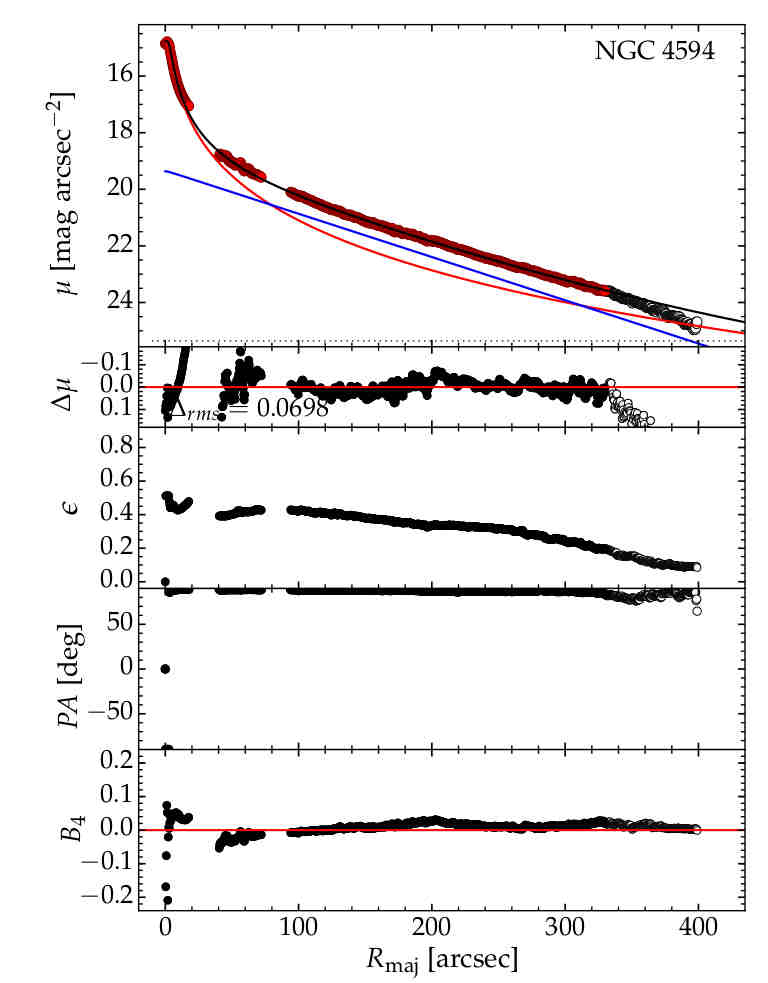}
\includegraphics[clip=true,trim= 11mm 1mm 4mm 5mm,width=0.249\textwidth]{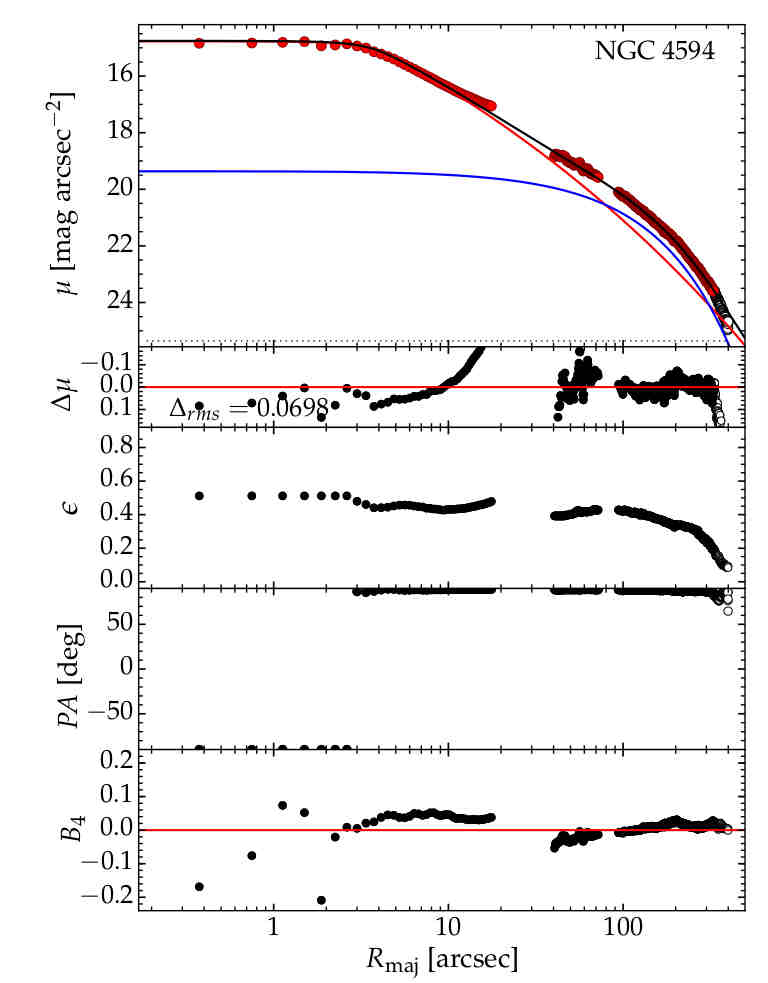}
\includegraphics[clip=true,trim= 11mm 1mm 4mm 5mm,width=0.249\textwidth]{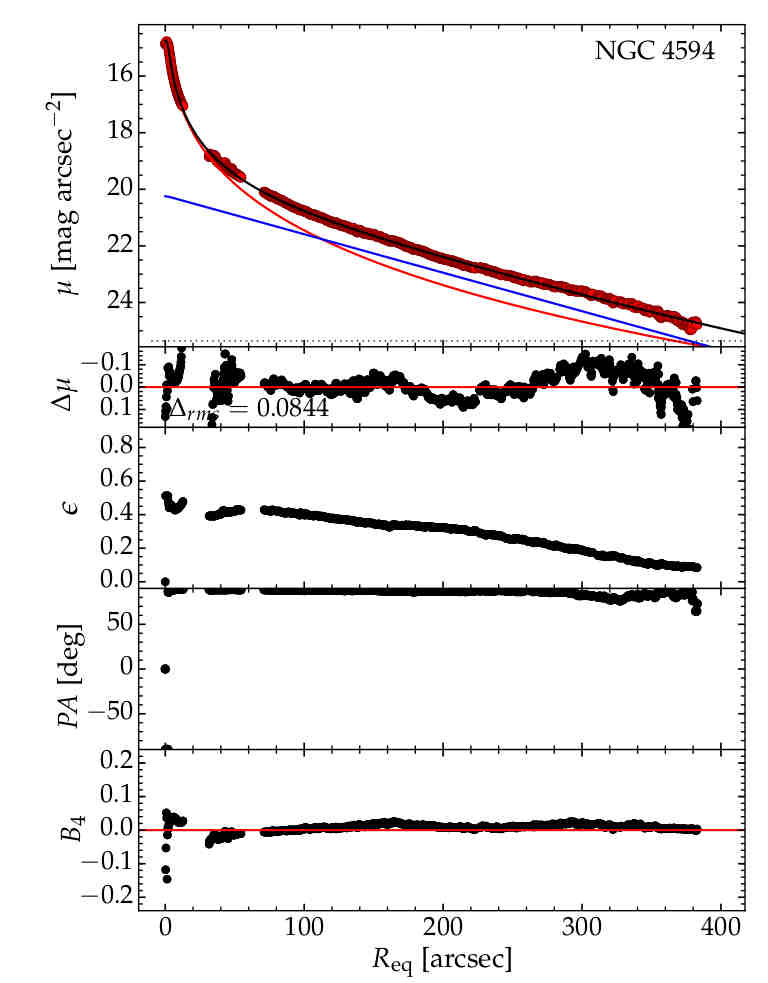}
\includegraphics[clip=true,trim= 11mm 1mm 4mm 5mm,width=0.249\textwidth]{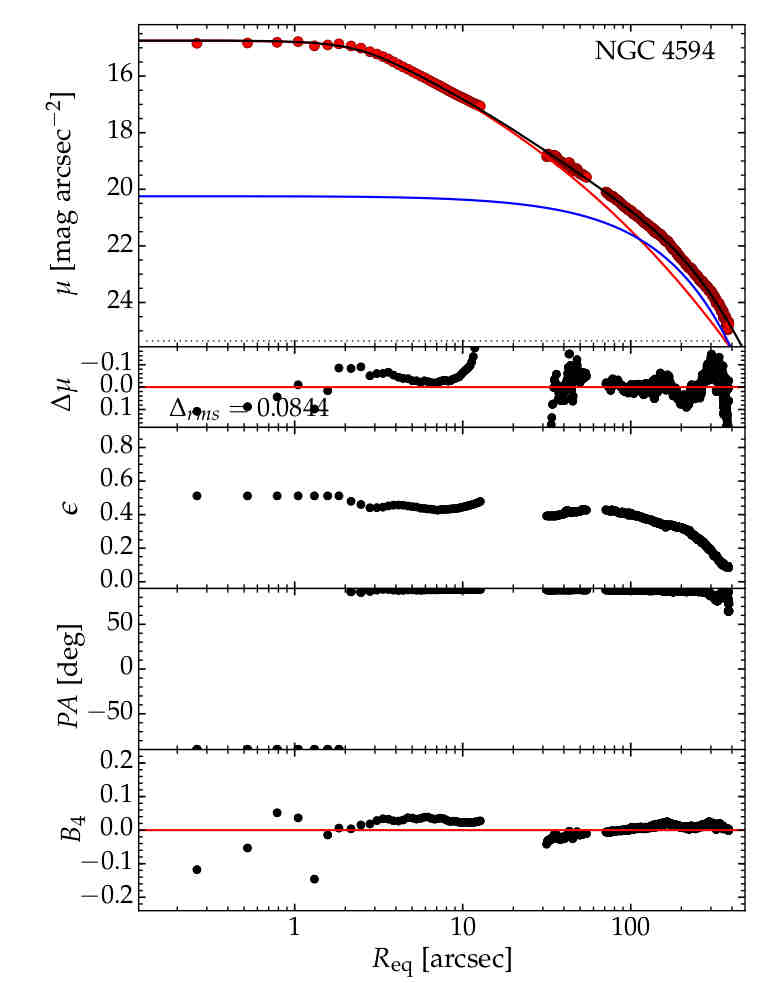}
\caption{\textit{Spitzer} $3.6\,\micron$ surface brightness profile for NGC~4594, with a physical scale of 0.0463$\,\text{kpc}\,\text{arcsec}^{-1}$. \textbf{Left two panels}---The model represents $0\arcsec \leq R_{\rm maj} \leq 330\arcsec$ with $\Delta_{\rm rms}=0.0698\,\text{mag\,arcsec}^{-2}$ and additional data from $330\arcsec < R_{\rm maj} \leq 399\arcsec$ is plotted, but not modeled. \underline{Core-S{\'e}rsic Profile Parameters:} \textcolor{red}{$\gamma=0.0082\pm0.0076$, $R_b=3\farcs14\pm0\farcs15$, $R_e=44\farcs94\pm2\farcs88$, $\mu_0'=6.39\pm1.05\,\text{mag\,arcsec}^{-2}$, and $n=6.14\pm0.54$.} \underline{Exponential Profile Parameters:} \textcolor{blue}{$\mu_0 = 19.29\pm0.05\,\text{mag\,arcsec}^{-2}$ and $h = 71\farcs01\pm1\farcs54$.} \textbf{Right two panels}---The model represents $0\arcsec \leq R_{\rm eq} \leq 383\arcsec$ with $\Delta_{\rm rms}=0.0844\,\text{mag\,arcsec}^{-2}$. \underline{Core-S{\'e}rsic Profile Parameters:} \textcolor{red}{$\gamma=0.0038\pm0.0137$, $R_b=1\farcs60\pm0\farcs14$, $R_e=41\farcs36\pm1\farcs94$, $\mu_0'=10.59\pm0.36\,\text{mag\,arcsec}^{-2}$, and $n=4.24\pm0.20$.} \underline{Exponential Profile Parameters:} \textcolor{blue}{$\mu_0 = 20.21\pm0.03\,\text{mag\,arcsec}^{-2}$ and $h = 79\farcs84\pm1\farcs65$.} We note that the best-fitting solution to the major- and equivalent axis profiles yields a different central surface brightness for the disk, however we remain uncertain as to which is correct.}
\label{NGC4594_plot}
\end{sidewaysfigure}

\subsection{NGC~4594 (Figure~\ref{NGC4594_plot})}

A fascinating example of a galaxy with dual morphology, NGC~4594 (M104, the ``Sombrero Galaxy'') appears to be simultaneously elliptical and spiral \citep{Gadotti:2012}. It also possesses a Type 1.9 intermediate Seyfert AGN \citep{Veron-Cetty:2006}. Our photometric decomposition of NGC~4594 models the elliptical component of the galaxy, leaving a mostly intact residual inner disk. Our isophote fitting failed to converge over the ranges $23<R_{\rm maj}<72$ ($17<R_{\rm eq}<56$) and $128<R_{\rm maj}<167$ ($97<R_{\rm eq}<126$); thus, these regions were omitted from the surface brightness profiles. The \textit{Spitzer} image displays diminished light in the nucleus; thus, we model it with a core-S{\'e}rsic model, in agreement with \citet{Jardel:2011}. Our decomposition of the surface brightness profile differs significantly from Figure~27 of \citet{Savorgnan:2016}, which may have been due to their restrictive FoV for this galaxy's image. However, we do find nice agreement with Figure~2 from \cite{Gadotti:2012}.

\begin{sidewaysfigure}
\includegraphics[clip=true,trim= 11mm 1mm 4mm 5mm,width=0.249\textwidth]{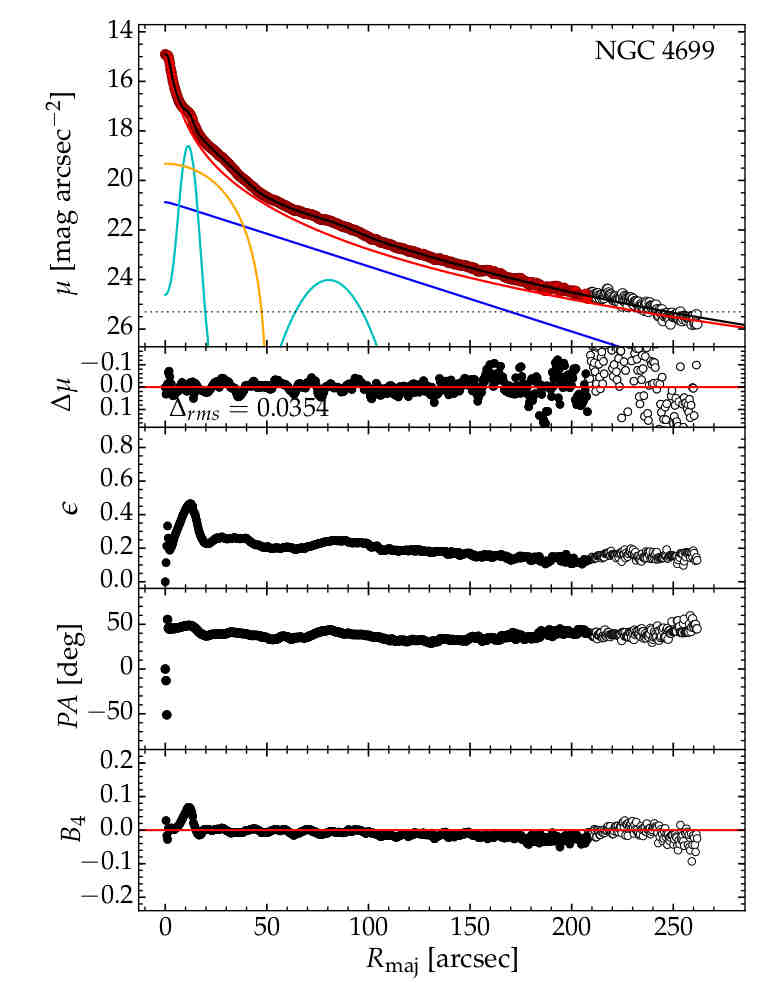}
\includegraphics[clip=true,trim= 11mm 1mm 4mm 5mm,width=0.249\textwidth]{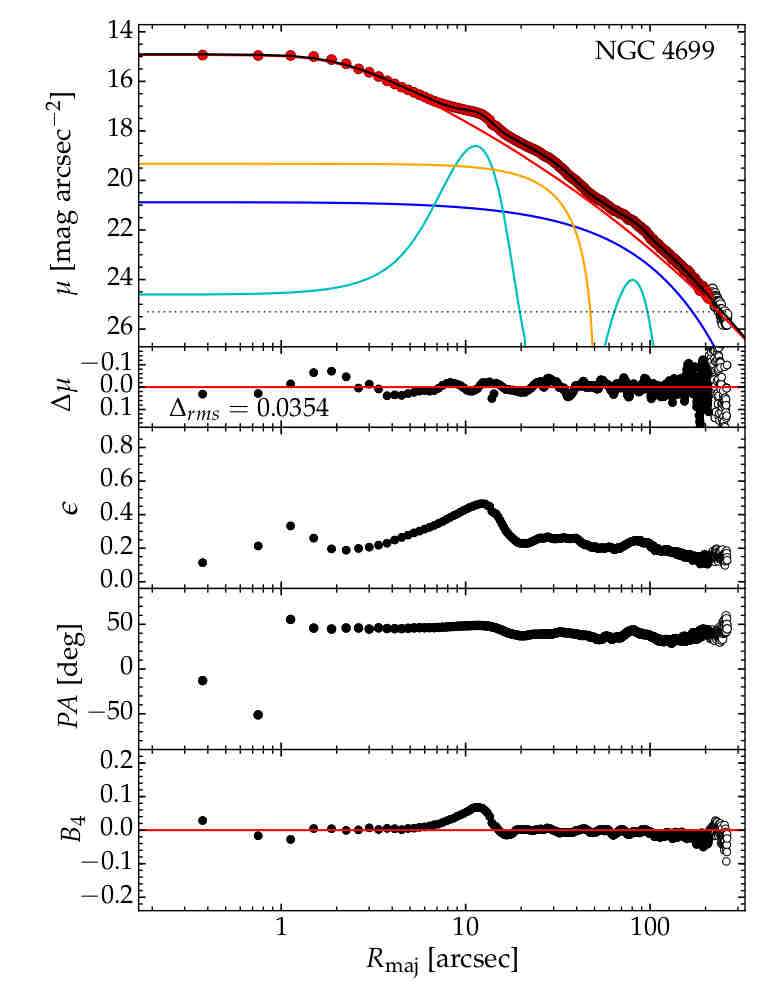}
\includegraphics[clip=true,trim= 11mm 1mm 4mm 5mm,width=0.249\textwidth]{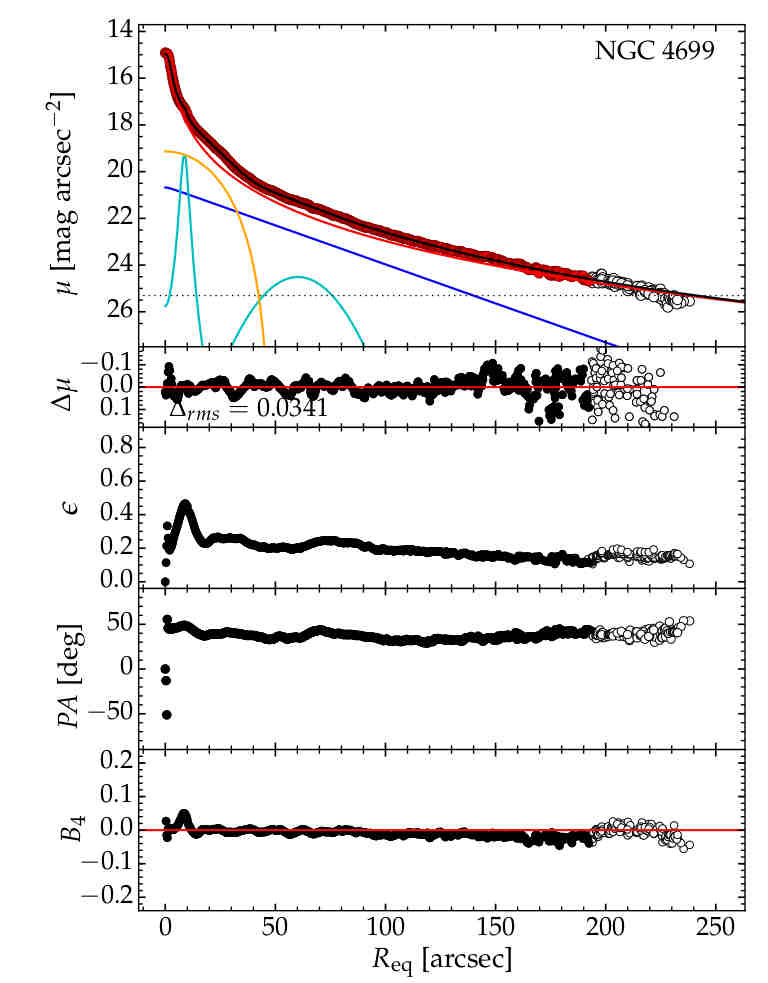}
\includegraphics[clip=true,trim= 11mm 1mm 4mm 5mm,width=0.249\textwidth]{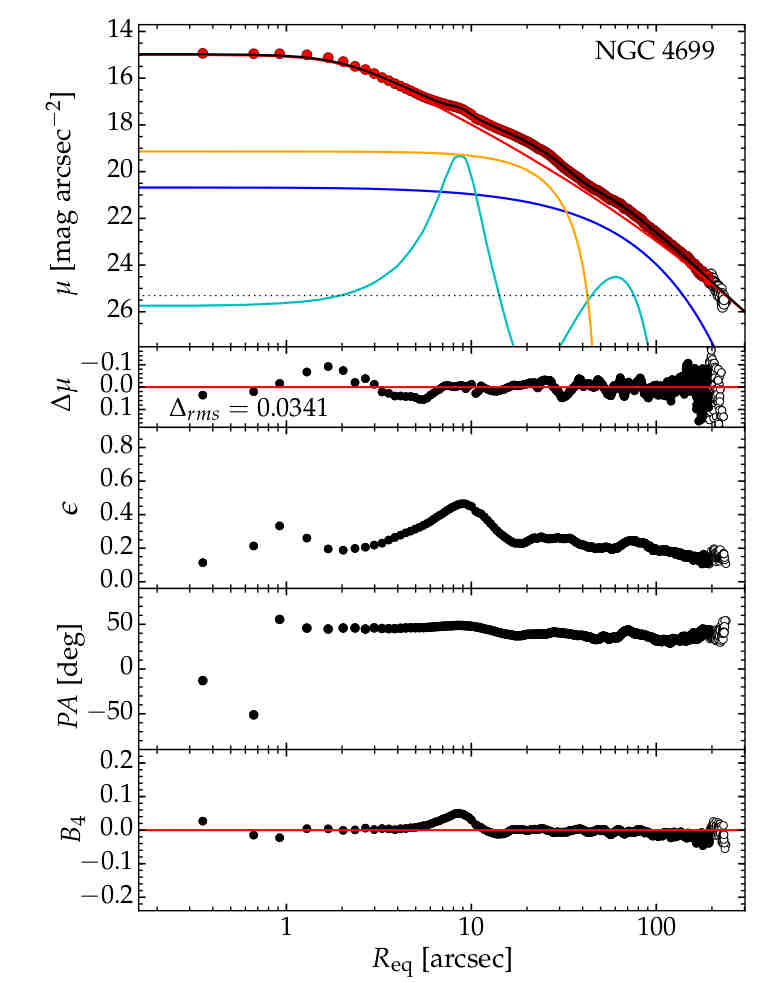}
\caption{\textit{Spitzer} $3.6\,\micron$ surface brightness profile for NGC~4699, with a physical scale of 0.1147$\,\text{kpc}\,\text{arcsec}^{-1}$. \textbf{Left two panels}---The model represents $0\arcsec \leq R_{\rm maj} \leq 208\arcsec$ with $\Delta_{\rm rms}=0.0354\,\text{mag\,arcsec}^{-2}$, and additional unfit data are plotted for $208\arcsec < R_{\rm maj} \leq 262\arcsec$. \underline{Core-S{\'e}rsic Profile Parameters:} \textcolor{red}{$\gamma=-6.95\pm99.99$, $R_b=0\farcs55\pm9\farcs99$, $R_e=24\farcs44\pm0\farcs46$, $\mu_0'=8.13\pm0.56\,\text{mag\,arcsec}^{-2}$, and $n=5.35\pm0.28$.} \underline{Ferrers Profile Parameters:} \textcolor{Orange}{$\mu_0 = 19.32\pm0.09\,\text{mag\,arcsec}^{-2}$, $R_{\rm end} = 50\farcs96\pm2\farcs03$, and $\alpha = 6.85\pm0.97$.} \underline{Exponential Profile Parameters:} \textcolor{blue}{$\mu_0 = 20.84\pm0.14\,\text{mag\,arcsec}^{-2}$ and $h = 41\farcs24\pm1\farcs67$.} \underline{Additional Parameters:} two Gaussian components added at: \textcolor{cyan}{$R_{\rm r}=11\farcs35\pm0\farcs12$ \& $80\farcs47\pm0\farcs87$; with $\mu_0 = 18.38\pm0.06$ \& $23.02\pm0.07\,\text{mag\,arcsec}^{-2}$; and FWHM = $3\farcs98\pm0\farcs35$ \& $25\farcs13\pm1\farcs99$, respectively.} \textbf{Right two panels}---The model represents $0\arcsec \leq R_{\rm eq} \leq 194\arcsec$ with $\Delta_{\rm rms}=0.0341\,\text{mag\,arcsec}^{-2}$, and additional unfit data are plotted for $194\arcsec < R_{\rm maj} \leq 242\arcsec$. \underline{Core-S{\'e}rsic Profile Parameters:} \textcolor{red}{$\gamma=-10.00\pm869.97$, $R_b=0\farcs62\pm1\farcs40$, $R_e=29\farcs75\pm0\farcs22$, $\mu_0'=5.80\pm0.18\,\text{mag\,arcsec}^{-2}$, and $n=6.77\pm0.08$.} \underline{Ferrers Profile Parameters:} \textcolor{Orange}{$\mu_0 = 19.14\pm0.04\,\text{mag\,arcsec}^{-2}$, $R_{\rm end} = 48\farcs72\pm2\farcs21$, and $\alpha = 10.00\pm1.28$.} \underline{Exponential Profile Parameters:} \textcolor{blue}{$\mu_0 = 20.67\pm0.06\,\text{mag\,arcsec}^{-2}$ and $h = 32\farcs51\pm0\farcs41$.} \underline{Additional Parameters:} two Gaussian components added at: \textcolor{cyan}{$R_{\rm r}=8\farcs63\pm20\farcs85$ \& $60\farcs23\pm2\farcs41$; with $\mu_0 = 16.73\pm618.78$ \& $24.44\pm0.18\,\text{mag\,arcsec}^{-2}$; and FWHM = $0\farcs24\pm61\farcs47$ \& $30\farcs34\pm4\farcs82$, respectively.}}
\label{NGC4699_plot}
\end{sidewaysfigure}

\subsection{NGC~4699 (Figure~\ref{NGC4699_plot})}\label{NGC4699}

Previous studies of NGC~4699 have remarked on the presence of a small classical bulge embedded within a larger pseudobulge. However, \citet{Weinzirl:2009} only fit the inner region and did not fit the outer disk. Additionally, \citet{Erwin:2015} did not account for the presence of a bar. The \textit{Spitzer} image displays diminished light in the nucleus; thus, we model it with a core-S{\'e}rsic model. The bar displays prominent ansae, which we model with a Gaussian centered at $R_{\rm maj} \approx 11\arcsec$ ($R_{\rm eq}\approx9\arcsec$). We model the strong/broad influence of superposed spiral arms in the disk with a Gaussian centered at $R_{\rm maj} \approx 80\arcsec$ ($R_{\rm eq}\approx 60\arcsec$).

\begin{sidewaysfigure}
\includegraphics[clip=true,trim= 11mm 1mm 4mm 5mm,width=0.249\textwidth]{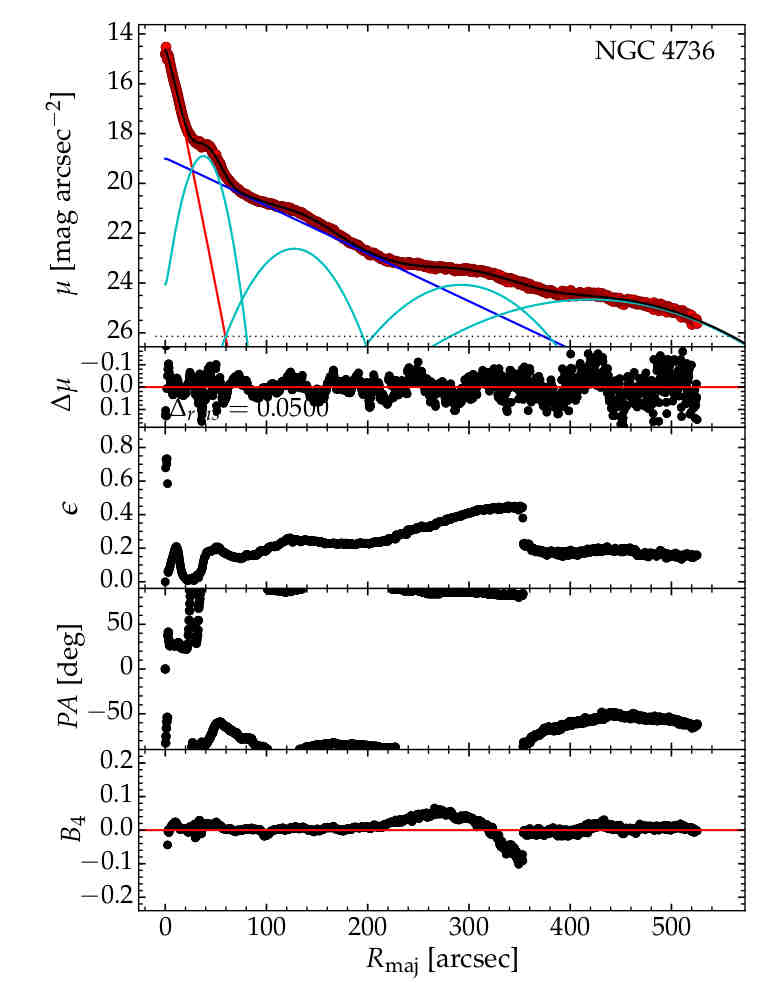}
\includegraphics[clip=true,trim= 11mm 1mm 4mm 5mm,width=0.249\textwidth]{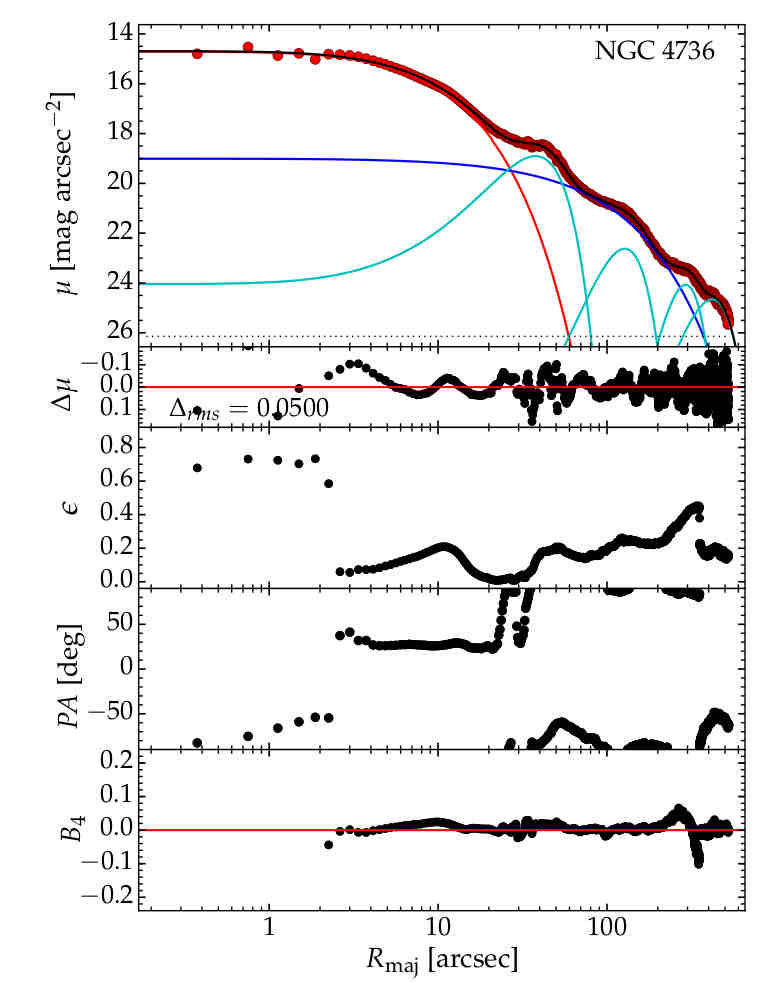}
\includegraphics[clip=true,trim= 11mm 1mm 4mm 5mm,width=0.249\textwidth]{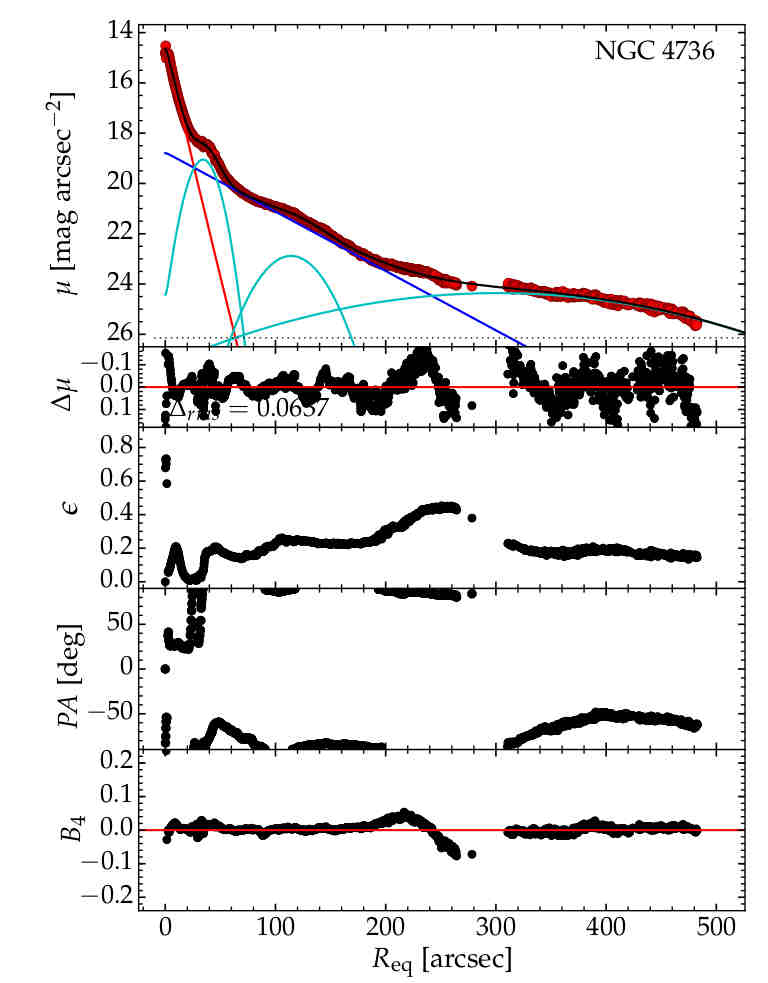}
\includegraphics[clip=true,trim= 11mm 1mm 4mm 5mm,width=0.249\textwidth]{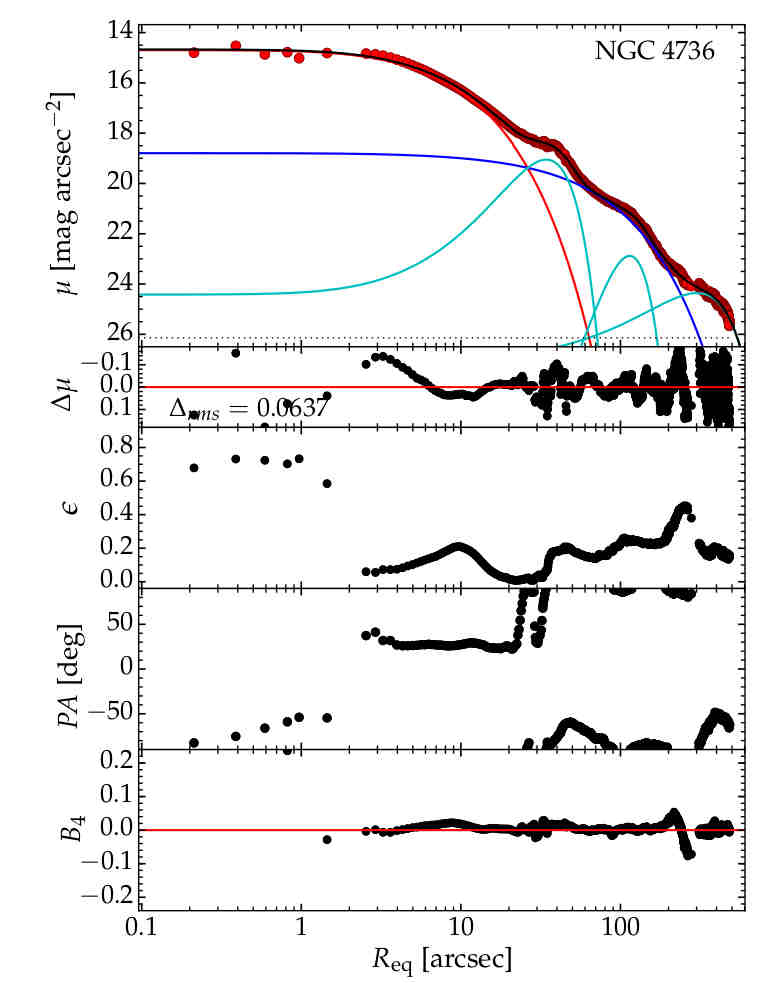}
\caption{\textit{Spitzer} $3.6\,\micron$ surface brightness profile for NGC~4736, with a physical scale of 0.0214$\,\text{kpc}\,\text{arcsec}^{-1}$. \textbf{Left two panels}---The model represents $0\arcsec \leq R_{\rm maj} \leq 526\arcsec$ with $\Delta_{\rm rms}=0.0500\,\text{mag\,arcsec}^{-2}$. \underline{S{\'e}rsic Profile Parameters:} \textcolor{red}{$R_e=9\farcs79\pm0\farcs10$, $\mu_e=16.17\pm0.03\,\text{mag\,arcsec}^{-2}$, and $n=0.93\pm0.02$}. \underline{Exponential Profile Parameters:} \textcolor{blue}{$\mu_0 = 18.99\pm0.02\,\text{mag\,arcsec}^{-2}$ and $h = 56\farcs63\pm0\farcs35$.} \underline{Additional Parameters:} four Gaussian components added at: \textcolor{cyan}{$R_{\rm r}=37\farcs68\pm0\farcs38$, $127\farcs43\pm1\farcs24$, $292\farcs12\pm0\farcs76$, \& $419\farcs85\pm2\farcs25$; with $\mu_0 = 18.99\pm0.02$, $22.63\pm0.04$, $24.08\pm0.02$, \& $24.66\pm0.01\,\text{mag\,arcsec}^{-2}$; and FWHM = $27\farcs48\pm0\farcs55$, $62\farcs64\pm2\farcs32$, $104\farcs76\pm2\farcs21$, \& $197\farcs06\pm4\farcs15$, respectively.} \textbf{Right two panels}---The model represents $0\arcsec \leq R_{\rm eq} \leq 483\arcsec$ with $\Delta_{\rm rms}=0.0637\,\text{mag\,arcsec}^{-2}$. \underline{S{\'e}rsic Profile Parameters:} \textcolor{red}{$R_e=9\farcs65\pm0\farcs13$, $\mu_e=16.31\pm0.03\,\text{mag\,arcsec}^{-2}$, and $n=1.03\pm0.02$.} \underline{Exponential Profile Parameters:} \textcolor{blue}{$\mu_0 = 18.77\pm0.02\,\text{mag\,arcsec}^{-2}$ and $h = 45\farcs99\pm0\farcs24$.} \underline{Additional Parameters:} three Gaussian components added at: \textcolor{cyan}{$R_{\rm r}=34\farcs25\pm0\farcs39$, $114\farcs15\pm1\farcs04$, \& $301\farcs35\pm2\farcs32$; with $\mu_0 = 19.06\pm0.03$, $22.89\pm0.04$, \& $24.36\pm0.01\,\text{mag\,arcsec}^{-2}$; and FWHM = $24\farcs32\pm0\farcs54$, $52\farcs64\pm1\farcs81$, \& $309\farcs51\pm3\farcs53$, respectively.}}
\label{NGC4736_plot}
\end{sidewaysfigure}

\subsection{NGC~4736 (Figure~\ref{NGC4736_plot})}

The galaxy NGC~4736 (M94) possesses a faint nuclear bar at $R_{\rm maj} = R_{\rm eq} \lesssim 23\arcsec$; however, it has little effect on the surface brightness profile, and we do not model it. We have added four (three for the equivalent axis) Gaussian components at $R_{\rm maj} \approx 38\arcsec$ ($R_{\rm eq}\approx34\arcsec$), $R_{\rm maj} \approx 127\arcsec$ ($R_{\rm eq}\approx114\arcsec$), $R_{\rm maj} \approx 292\arcsec$, and $R_{\rm maj} \approx 420\arcsec$ ($R_{\rm eq}\approx301\arcsec$) to account for the inner ring, outer ring, and the two outer spiral arms, respectively.

\begin{sidewaysfigure}
\includegraphics[clip=true,trim= 11mm 1mm 4mm 5mm,width=0.249\textwidth]{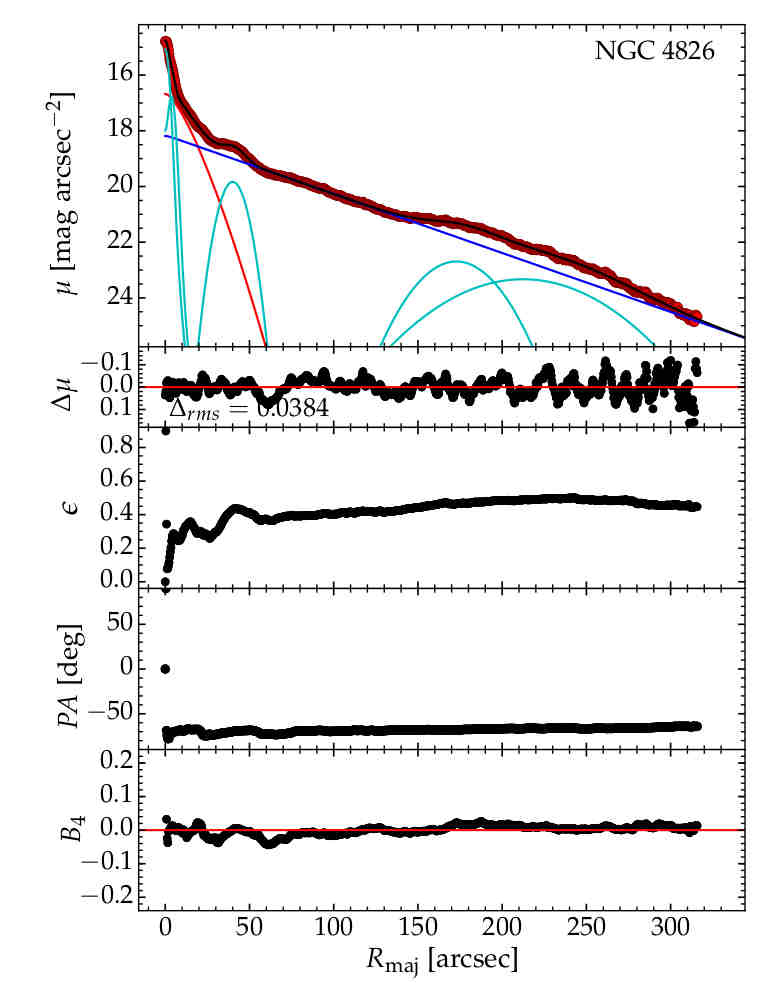}
\includegraphics[clip=true,trim= 11mm 1mm 4mm 5mm,width=0.249\textwidth]{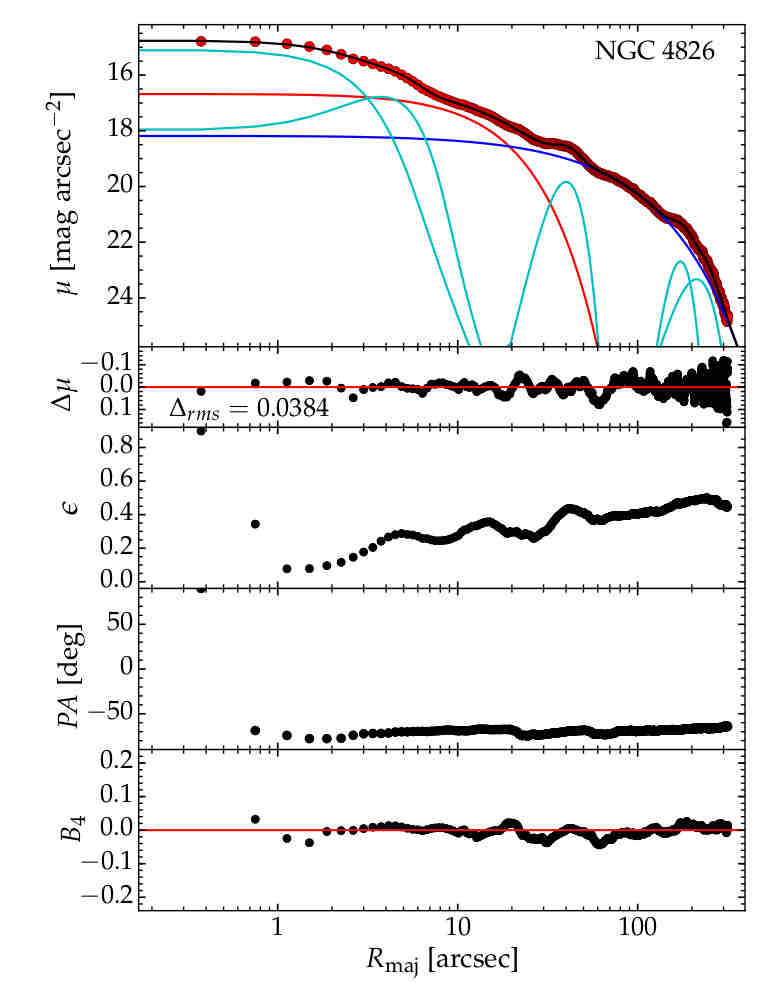}
\includegraphics[clip=true,trim= 11mm 1mm 4mm 5mm,width=0.249\textwidth]{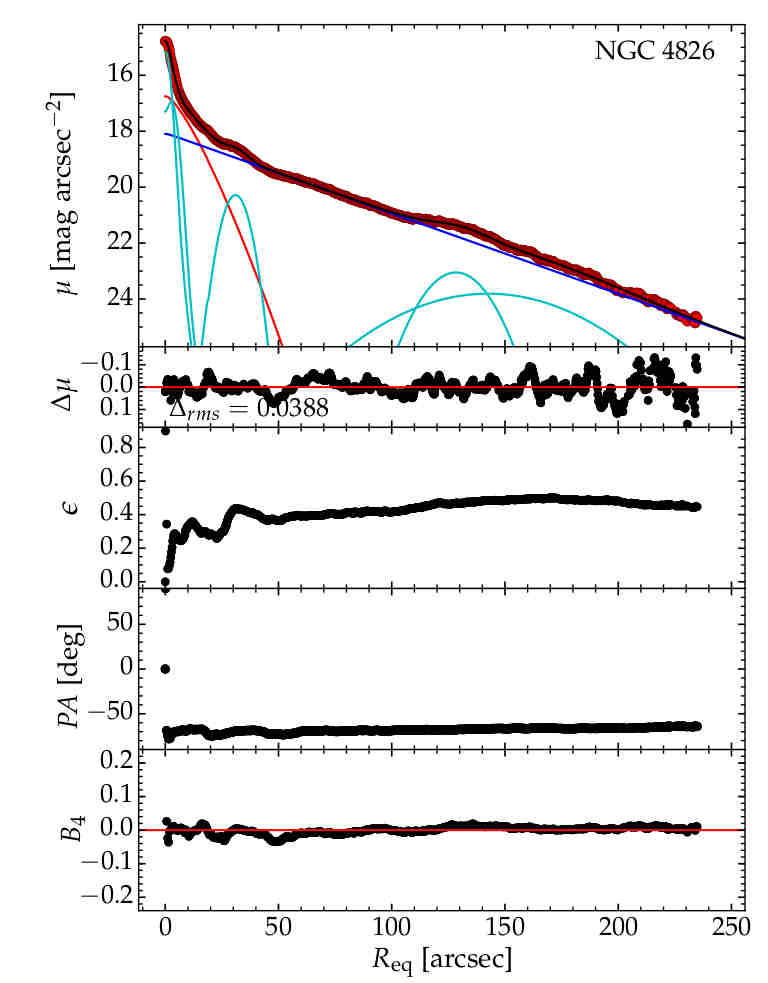}
\includegraphics[clip=true,trim= 11mm 1mm 4mm 5mm,width=0.249\textwidth]{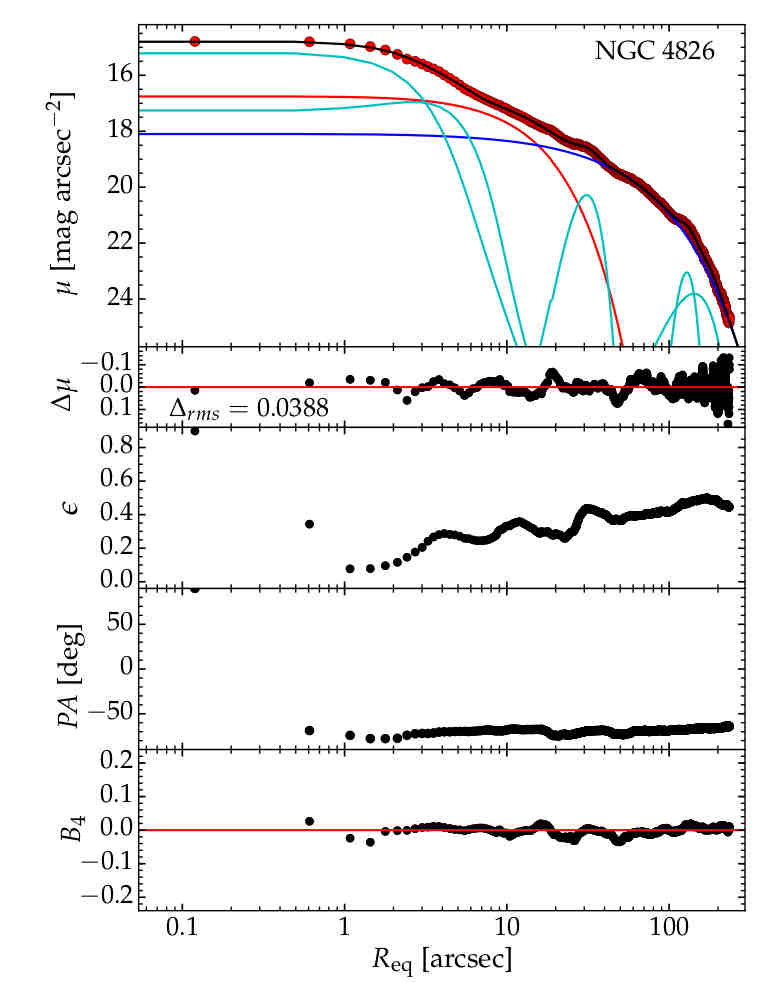}
\caption{\textit{Spitzer} $3.6\,\micron$ surface brightness profile for NGC~4826, with a physical scale of 0.0269$\,\text{kpc}\,\text{arcsec}^{-1}$. \textbf{Left two panels}---The model represents $0\arcsec \leq R_{\rm maj} \leq 316\arcsec$ with $\Delta_{\rm rms}=0.0384\,\text{mag\,arcsec}^{-2}$. \underline{S{\'e}rsic Profile Parameters:} \textcolor{red}{$R_e=13\farcs89\pm0\farcs19$, $\mu_e=17.86\pm0.03\,\text{mag\,arcsec}^{-2}$, and $n=0.73\pm0.07$}. \underline{Exponential Profile Parameters:} \textcolor{blue}{$\mu_0 = 18.16\pm0.01\,\text{mag\,arcsec}^{-2}$ and $h = 51\farcs25\pm0\farcs23$.} \underline{Additional Parameters:} five Gaussian components added at: \textcolor{cyan}{$R_{\rm r}=0\arcsec$, $3\farcs87\pm0\farcs84$, $40\farcs03\pm0\farcs30$, $172\farcs82\pm0\farcs45$, \& $212\farcs36\pm3\farcs57$; with $\mu_0 = 14.52\pm0.09$, $16.48\pm0.32$, $19.84\pm0.04$, $22.70\pm0.06$, \& $23.34\pm0.07\,\text{mag\,arcsec}^{-2}$; and FWHM = $3\farcs27\pm0\farcs71$, $3\farcs05\pm1\farcs01$, $14\farcs68\pm0\farcs63$, $44\farcs87\pm1\farcs54$, \& $87\farcs92\pm4\farcs09$, respectively.} \textbf{Right two panels}---The model represents $0\arcsec \leq R_{\rm eq} \leq 235\arcsec$ with $\Delta_{\rm rms}=0.0388\,\text{mag\,arcsec}^{-2}$. \underline{S{\'e}rsic Profile Parameters:} \textcolor{red}{$R_e=11\farcs93\pm0\farcs29$, $\mu_e=17.98\pm0.05\,\text{mag\,arcsec}^{-2}$, and $n=0.76\pm0.05$.} \underline{Exponential Profile Parameters:} \textcolor{blue}{$\mu_0 = 18.06\pm0.01\,\text{mag\,arcsec}^{-2}$ and $h = 37\farcs73\pm0\farcs20$.} \underline{Additional Parameters:} five Gaussian components added at: \textcolor{cyan}{$R_{\rm r}=0\arcsec$, $2\farcs80\pm0\farcs61$, $31\farcs02\pm0\farcs40$, $128\farcs32\pm0\farcs63$, \& $142\farcs59\pm5\farcs78$; with $\mu_0 = 14.68\pm0.09$, $16.79\pm0.33$, $20.29\pm0.09$, $23.06\pm0.08$, \& $23.81\pm0.15\,\text{mag\,arcsec}^{-2}$; and FWHM = $3\farcs11\pm0\farcs68$, $3\farcs99\pm1\farcs32$, $10\farcs73\pm0\farcs96$, $27\farcs25\pm1\farcs78$, \& $77\farcs36\pm6\farcs64$, respectively.}}
\label{NGC4826_plot}
\end{sidewaysfigure}

\subsection{NGC~4826 (Figure~\ref{NGC4826_plot})}

The galaxy NGC~4826 (M64) contains an unclassified Seyfert AGN \citep{Veron-Cetty:2006} and unresolved nuclear spiral arms (evident in \textit{HST} imaging). We account for the AGN and inner nuclear spiral arms with a central Gaussian and the outer nuclear spiral arms with another Gaussian at $R_{\rm maj} \approx 4\arcsec$ ($R_{\rm eq}\approx3\arcsec$). Additionally, we account for light contribution from a ring with a Gaussian at $R_{\rm maj} \approx 40\arcsec$ ($R_{\rm eq}\approx31\arcsec$) and the two outer spiral arms with two Gaussian components at $R_{\rm maj} \approx 173\arcsec$ ($R_{\rm eq}\approx128\arcsec$) and $R_{\rm maj} \approx 212\arcsec$ ($R_{\rm eq}\approx143\arcsec$).

\begin{sidewaysfigure}
\includegraphics[clip=true,trim= 11mm 1mm 1mm 5mm,width=0.249\textwidth]{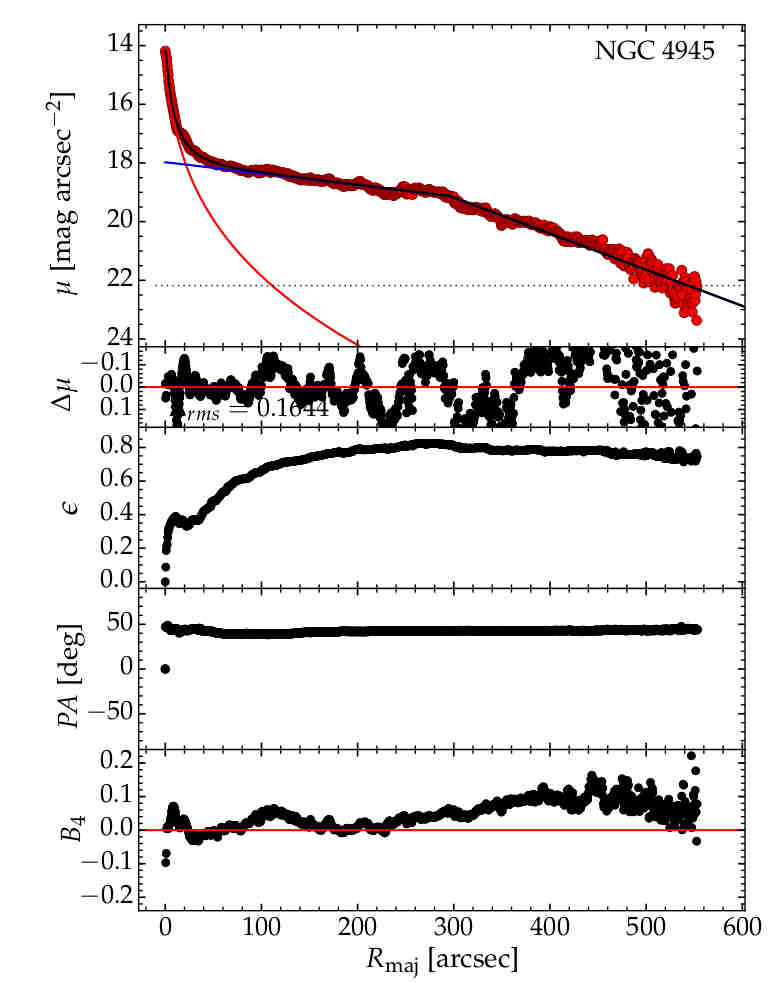}
\includegraphics[clip=true,trim= 11mm 1mm 1mm 5mm,width=0.249\textwidth]{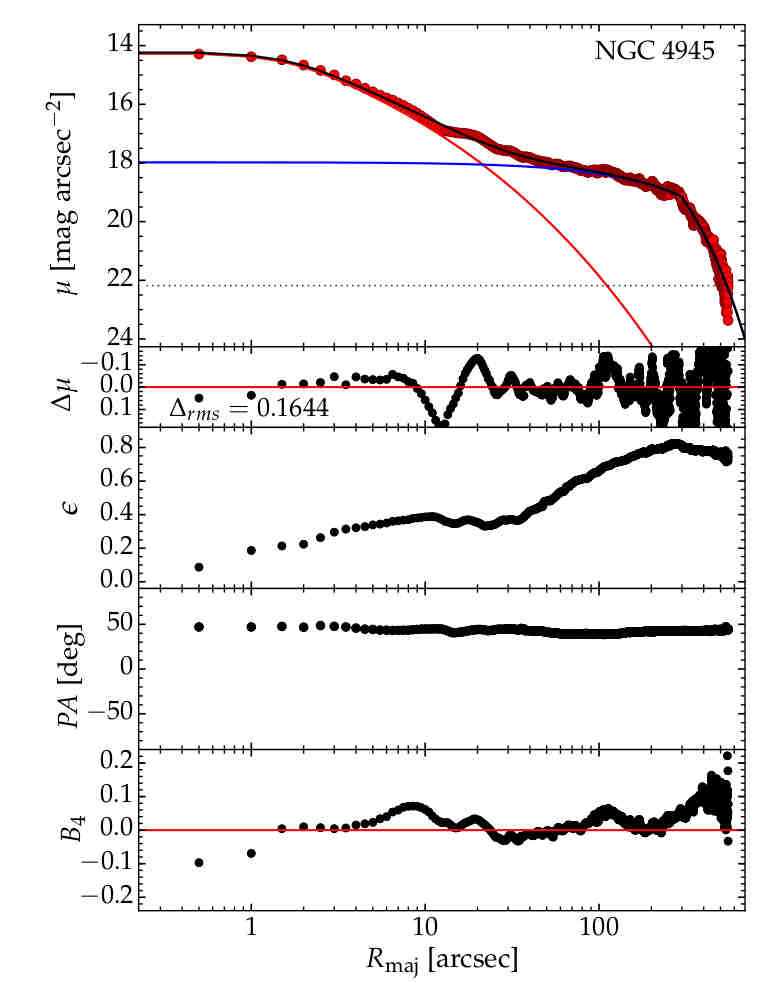}
\includegraphics[clip=true,trim= 11mm 1mm 1mm 5mm,width=0.249\textwidth]{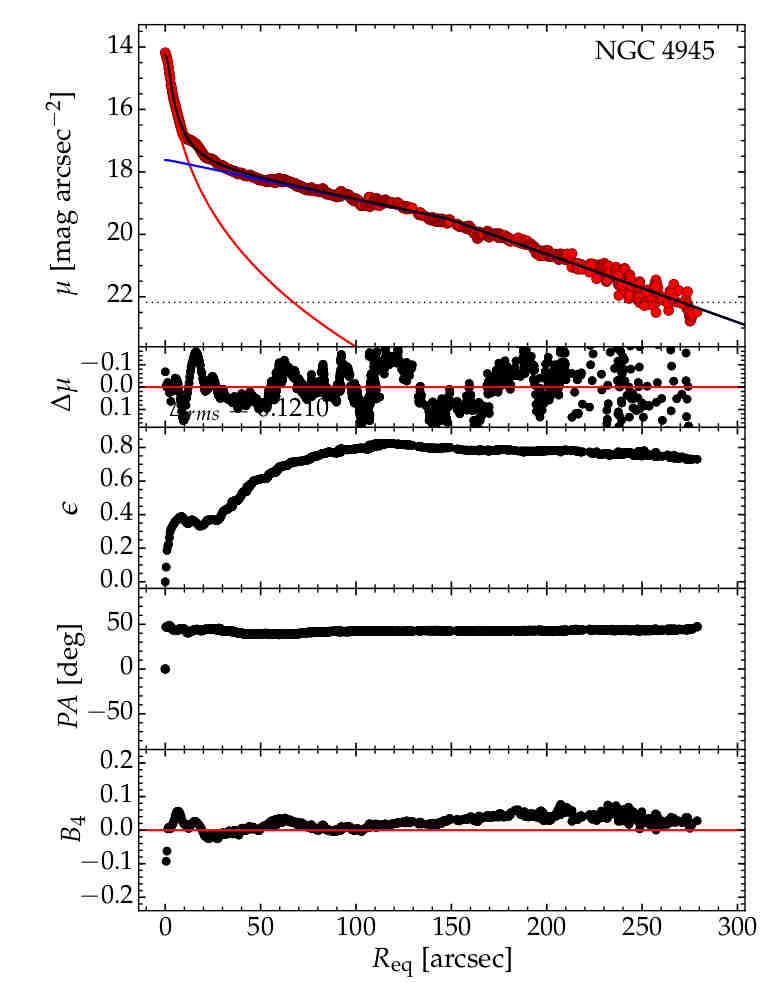}
\includegraphics[clip=true,trim= 11mm 1mm 1mm 5mm,width=0.249\textwidth]{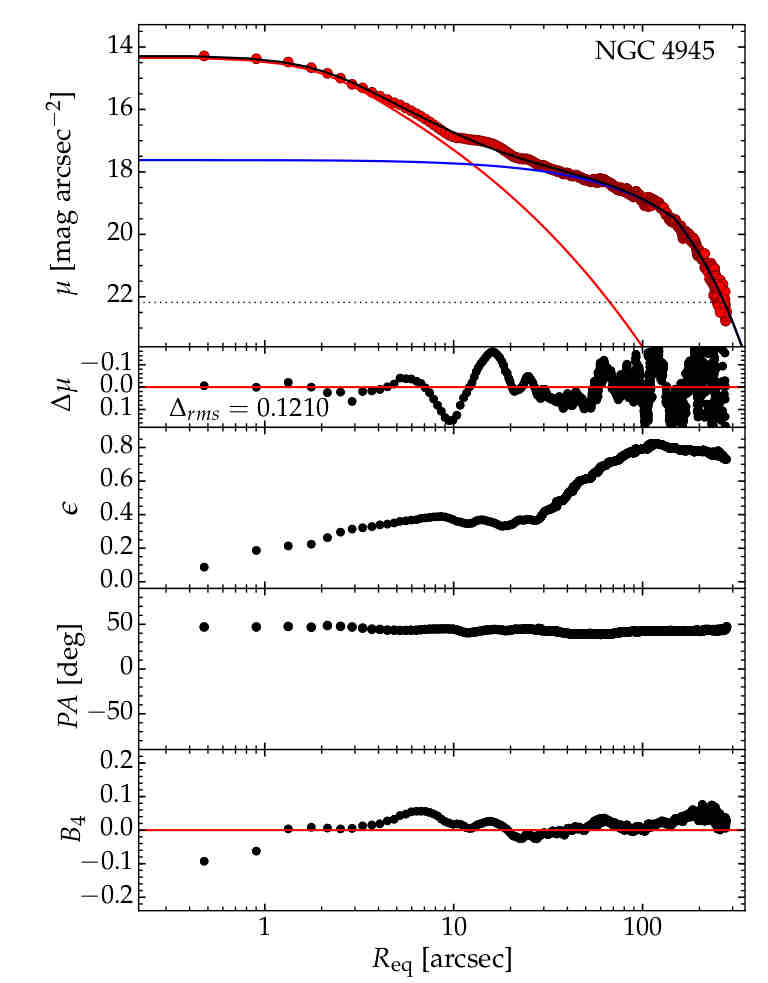}
\caption{2MASS $K_s$ filter surface brightness profile for NGC~4945, with a physical scale of 0.0180$\,\text{kpc}\,\text{arcsec}^{-1}$. \textbf{Left two panels}---The model represents $0\arcsec \leq R_{\rm maj} \leq 553\arcsec$ with $\Delta_{\rm rms}=0.1644\,\text{mag\,arcsec}^{-2}$. \underline{S{\'e}rsic Profile Parameters:} \textcolor{red}{$R_e=26\farcs33\pm5\farcs17$, $\mu_e=18.48\pm0.34\,\text{mag\,arcsec}^{-2}$, and $n=3.40\pm0.28$.} \underline{Broken Exponential Profile Parameters:} \textcolor{blue}{$\mu_0 = 17.97\pm0.03\,\text{mag\,arcsec}^{-2}$, $R_b = 295\farcs96\pm2\farcs59$, $h_1 = 277\farcs59\pm10\farcs93$, and $h_2 = 88\farcs08\pm0\farcs70$.} \textbf{Right two panels}---The model represents $0\arcsec \leq R_{\rm eq} \leq 279\arcsec$ with $\Delta_{\rm rms}=0.1210\,\text{mag\,arcsec}^{-2}$. \underline{S{\'e}rsic Profile Parameters:} \textcolor{red}{$R_e=13\farcs93\pm2\farcs81$, $\mu_e=17.99\pm0.35\,\text{mag\,arcsec}^{-2}$, and $n=3.19\pm0.29$.} \underline{Broken Exponential Profile Parameters:} \textcolor{blue}{$\mu_0 = 17.60\pm0.04\,\text{mag\,arcsec}^{-2}$, $R_b = 147\farcs82\pm2\farcs41$, $h_1 = 84\farcs60\pm2\farcs34$, and $h_2 = 49\farcs74\pm0\farcs43$.}}
\label{NGC4945_plot}
\end{sidewaysfigure}

\subsection{NGC~4945 (Figure~\ref{NGC4945_plot})}

A near edge-on galaxy, NGC~4945 is inclined by $77\fdg0\pm1\fdg7$ with respect to the plane of the sky. We model this with a broken exponential. The surface brightness profile displays a slight hint of a nuclear bar at $R_{\rm maj}\lesssim20\arcsec$. However, we do not model this component.

\begin{sidewaysfigure}
\includegraphics[clip=true,trim= 11mm 1mm 4mm 5mm,width=0.249\textwidth]{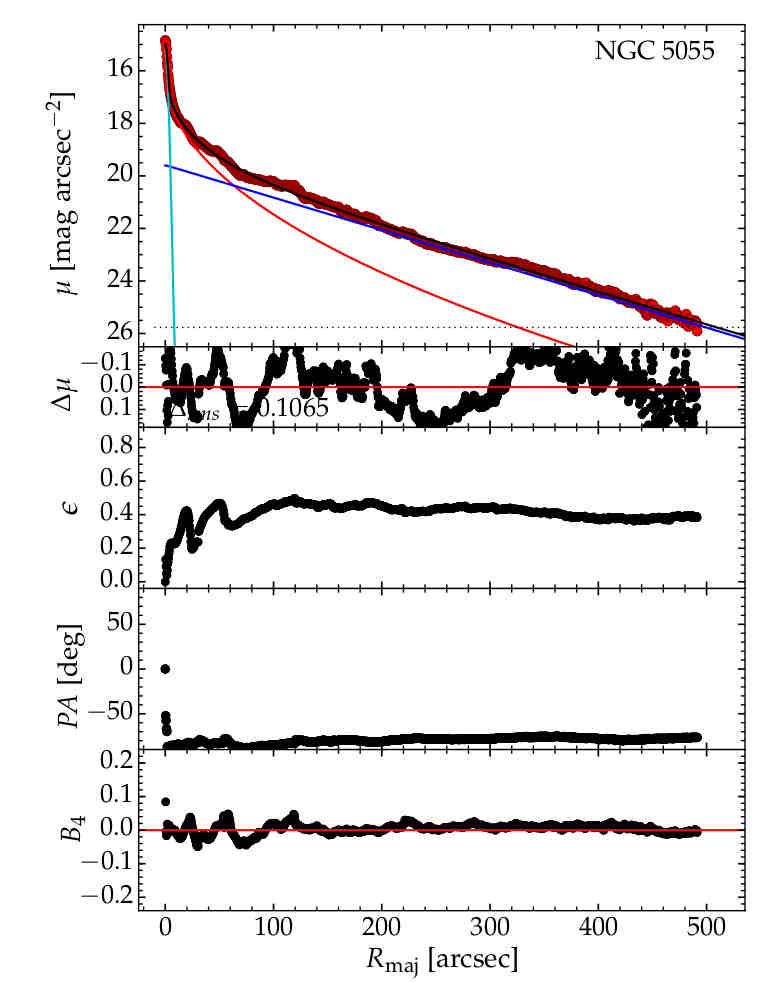}
\includegraphics[clip=true,trim= 11mm 1mm 4mm 5mm,width=0.249\textwidth]{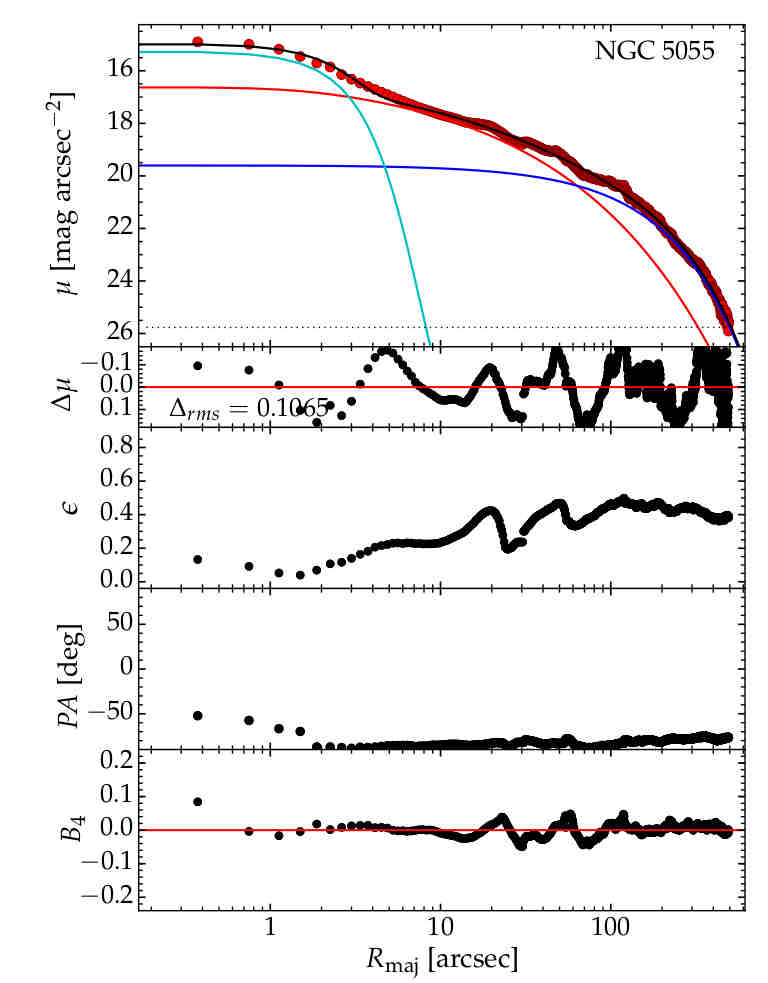}
\includegraphics[clip=true,trim= 11mm 1mm 4mm 5mm,width=0.249\textwidth]{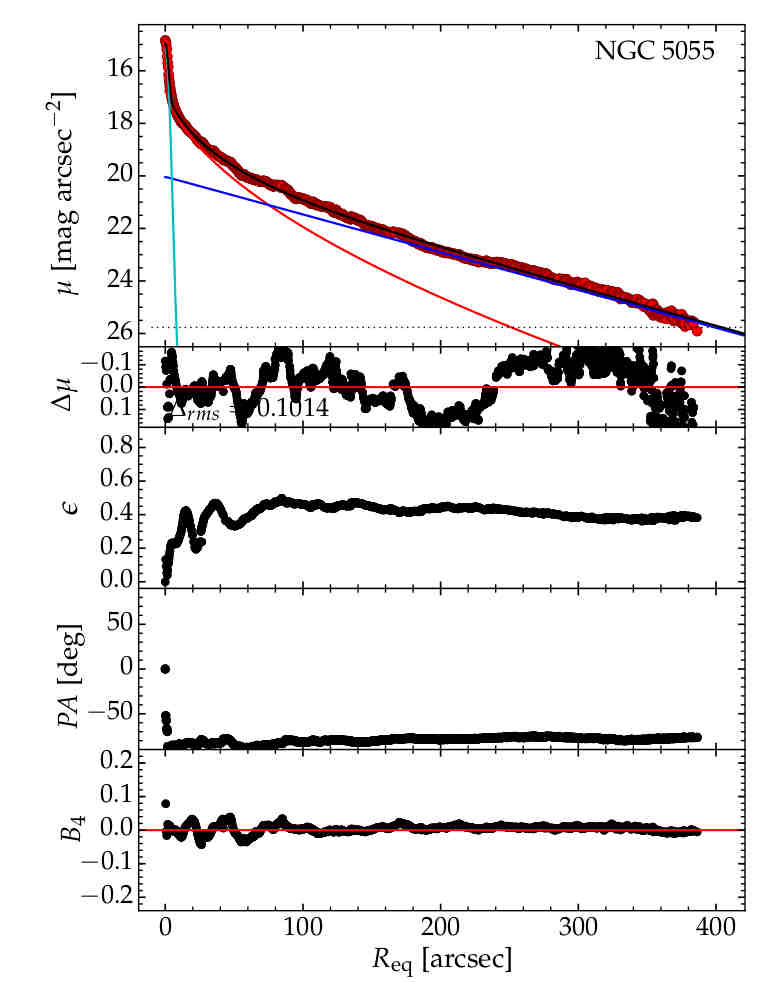}
\includegraphics[clip=true,trim= 11mm 1mm 4mm 5mm,width=0.249\textwidth]{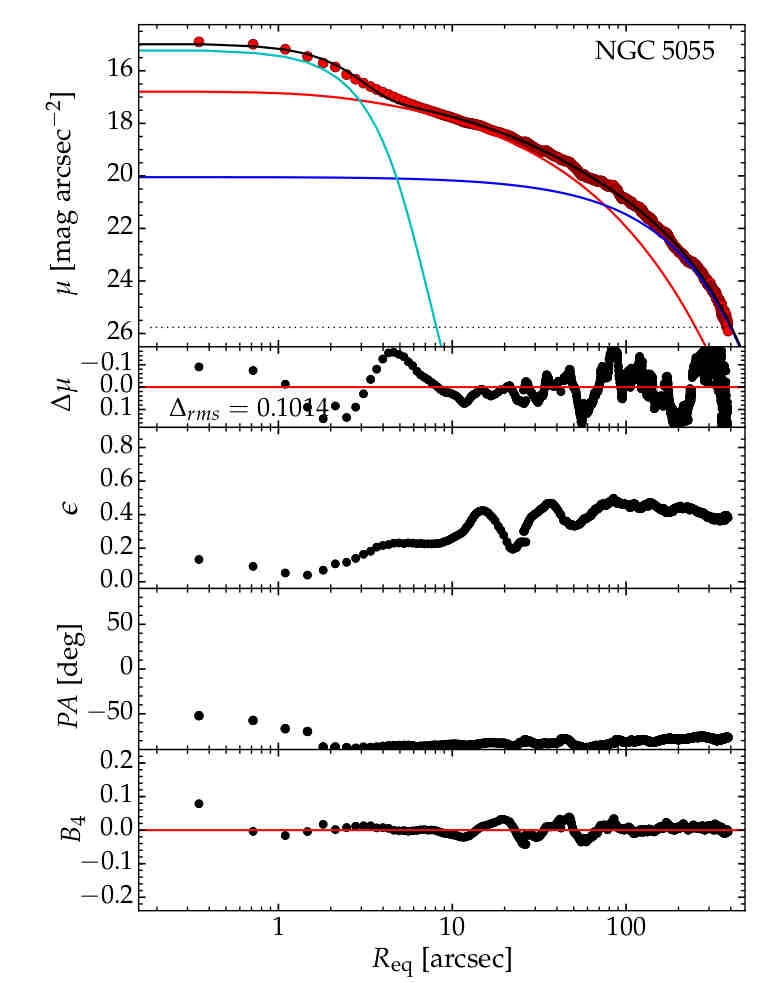}
\caption{\textit{Spitzer} $3.6\,\micron$ surface brightness profile for NGC~5055, with a physical scale of 0.0438$\,\text{kpc}\,\text{arcsec}^{-1}$. \textbf{Left two panels}---The model represents $0\arcsec \leq R_{\rm maj} \leq 491\arcsec$ with $\Delta_{\rm rms}=0.1065\,\text{mag\,arcsec}^{-2}$. \underline{S{\'e}rsic Profile Parameters:} \textcolor{red}{$R_e=55\farcs12\pm4\farcs56$, $\mu_e=20.09\pm0.12\,\text{mag\,arcsec}^{-2}$, and $n=2.02\pm0.13$.} \underline{Exponential Profile Parameters:} \textcolor{blue}{$\mu_0 = 19.59\pm0.05\,\text{mag\,arcsec}^{-2}$ and $h = 87\farcs74\pm0\farcs53$.} \underline{Additional Parameters:} one Gaussian component added at: \textcolor{cyan}{$R_{\rm r}=0\arcsec$, with $\mu_0 = 14.75\pm0.10$, and FWHM = $3\farcs00\pm0\farcs18$.} \textbf{Right two panels}---The model represents $0\arcsec \leq R_{\rm eq} \leq 387\arcsec$ with $\Delta_{\rm rms}=0.1014\,\text{mag\,arcsec}^{-2}$. \underline{S{\'e}rsic Profile Parameters:} \textcolor{red}{$R_e=43\farcs52\pm3\farcs15$, $\mu_e=19.85\pm0.10\,\text{mag\,arcsec}^{-2}$, and $n=1.76\pm0.11$.} \underline{Exponential Profile Parameters:} \textcolor{blue}{$\mu_0 = 20.04\pm0.10\,\text{mag\,arcsec}^{-2}$ and $h = 75\farcs52\pm0\farcs80$.} \underline{Additional Parameters:} one Gaussian component added at: \textcolor{cyan}{$R_{\rm r}=0\arcsec$, with $\mu_0 = 14.70\pm0.09$, and FWHM = $2\farcs78\pm0\farcs16$.}}
\label{NGC5055_plot}
\end{sidewaysfigure}

\subsection{NGC~5055 (Figure~\ref{NGC5055_plot})}

The galaxy NGC~5055 (M63, the ``Sunflower Galaxy'') is classified as possessing an AGN with intermediate emission-line ratios between the LINER and H \textsc{ii} regions \citep{Ho:1997}. We have added a central Gaussian to account for this excess nuclear light.

\begin{sidewaysfigure}
\includegraphics[clip=true,trim= 11mm 1mm 3mm 5mm,width=0.249\textwidth]{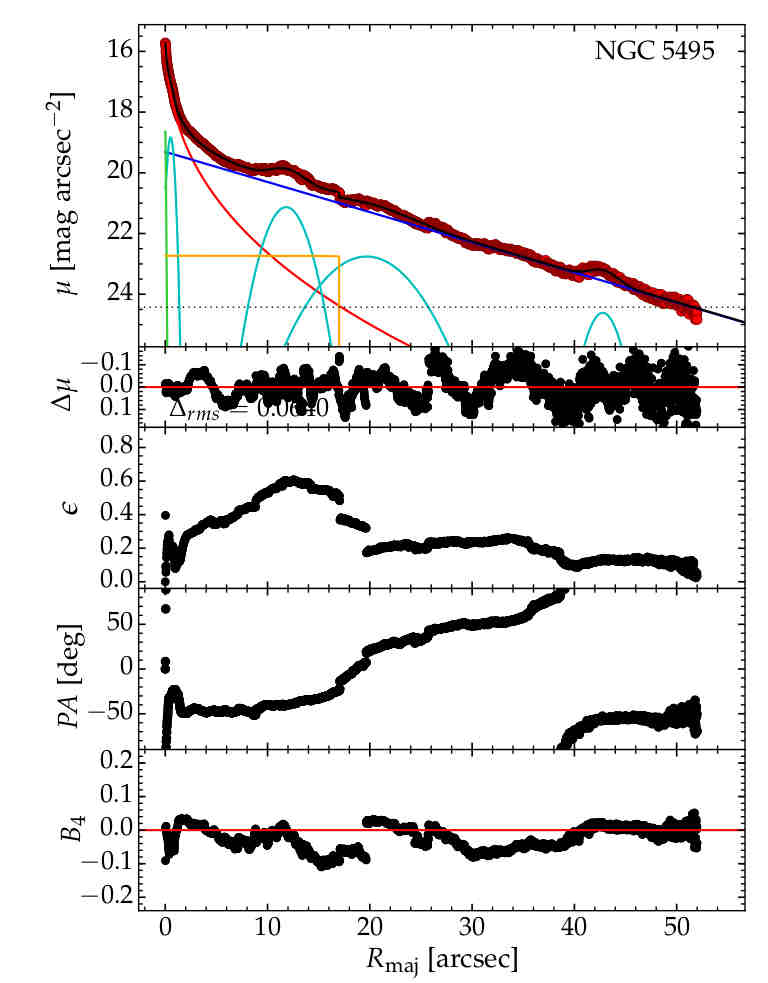}
\includegraphics[clip=true,trim= 11mm 1mm 3mm 5mm,width=0.249\textwidth]{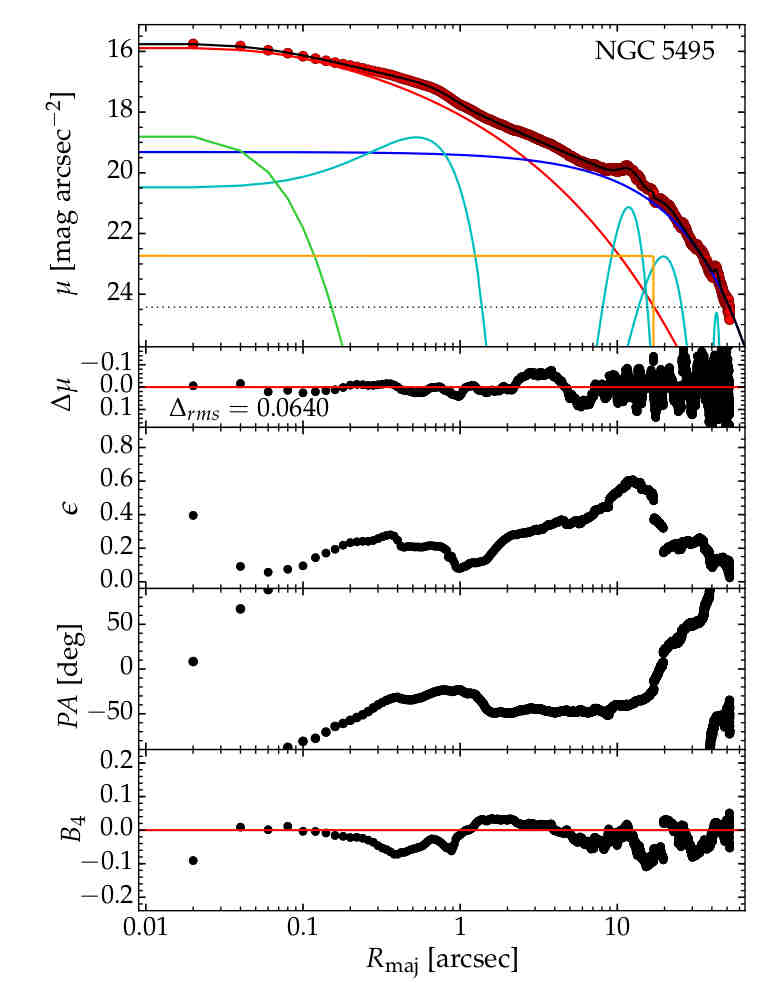}
\includegraphics[clip=true,trim= 11mm 1mm 3mm 5mm,width=0.249\textwidth]{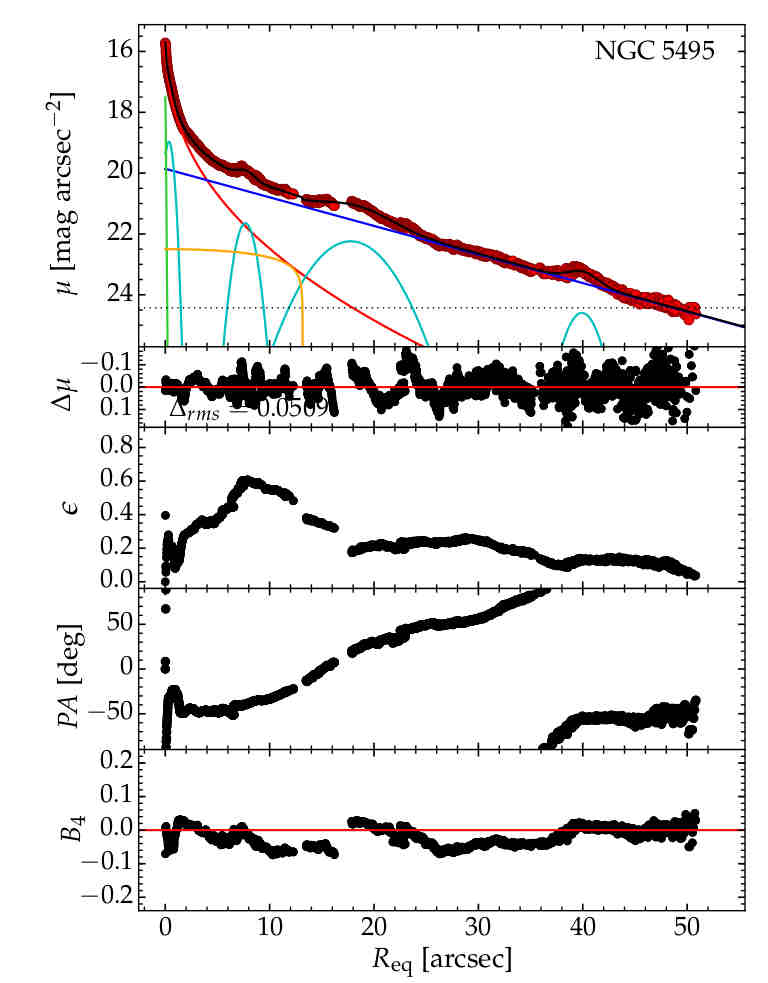}
\includegraphics[clip=true,trim= 11mm 1mm 3mm 5mm,width=0.249\textwidth]{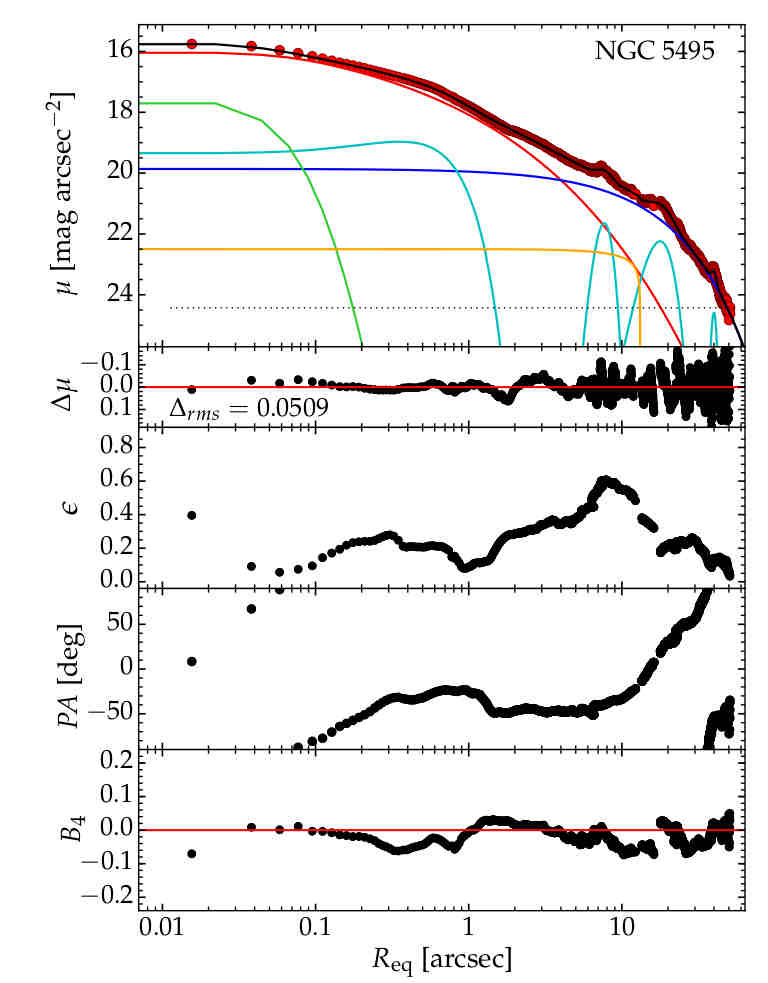}
\caption{\textit{HST} WFC3 UVIS2 F814W surface brightness profile for NGC~5495, with a physical scale of 0.4686$\,\text{kpc}\,\text{arcsec}^{-1}$. \textbf{Left two panels}---The model represents $0\arcsec \leq R_{\rm maj} \leq 52\arcsec$ with $\Delta_{\rm rms}=0.0640\,\text{mag\,arcsec}^{-2}$. \underline{Point Source:} \textcolor{LimeGreen}{$\mu_0 = 18.65\pm1.12\,\text{mag\,arcsec}^{-2}$.} \underline{S{\'e}rsic Profile Parameters:} \textcolor{red}{$R_e=3\farcs75\pm0\farcs34$, $\mu_e=20.21\pm0.14\,\text{mag\,arcsec}^{-2}$, and $n=2.60\pm0.12$.} \underline{Ferrers Profile Parameters:} \textcolor{Orange}{$\mu_0 = 22.73\pm0.18\,\text{mag\,arcsec}^{-2}$, $R_{\rm end} = 17\farcs00\pm0\farcs00$, and $\alpha = 0.01\pm0.13$.} \underline{Exponential Profile Parameters:} \textcolor{blue}{$\mu_0 = 19.31\pm0.02\,\text{mag\,arcsec}^{-2}$ and $h = 10\farcs93\pm0\farcs04$.} \underline{Additional Parameters:} four Gaussian components added at: \textcolor{cyan}{$R_{\rm r}=0\farcs53\pm0\farcs05$, $11\farcs81\pm0\farcs03$, $19\farcs69\pm0\farcs15$, \& $42\farcs78\pm0\farcs03$; with $\mu_0 = 18.82\pm0.13$, $21.13\pm0.03$, $22.76\pm0.03$, \& $24.61\pm0.03\,\text{mag\,arcsec}^{-2}$; and FWHM = $0\farcs62\pm0\farcs10$, $3\farcs58\pm0\farcs12$, $8\farcs11\pm0\farcs28$, \& $3\farcs01\pm0\farcs08$, respectively.} \textbf{Right two panels}---The model represents $0\arcsec \leq R_{\rm maj} \leq 51\arcsec$ with $\Delta_{\rm rms}=0.0509\,\text{mag\,arcsec}^{-2}$. \underline{Point Source:} \textcolor{LimeGreen}{$\mu_0 = 17.51\pm0.35\,\text{mag\,arcsec}^{-2}$.} \underline{S{\'e}rsic Profile Parameters:} \textcolor{red}{$R_e=3\farcs99\pm0\farcs27$, $\mu_e=20.23\pm0.11\,\text{mag\,arcsec}^{-2}$, and $n=2.46\pm0.12$.} \underline{Ferrers Profile Parameters:} \textcolor{Orange}{$\mu_0 = 22.50\pm0.16\,\text{mag\,arcsec}^{-2}$, $R_{\rm end} = 13\farcs16\pm0\farcs06$, and $\alpha = 0.73\pm0.18$.} \underline{Exponential Profile Parameters:} \textcolor{blue}{$\mu_0 = 19.86\pm0.02\,\text{mag\,arcsec}^{-2}$ and $h = 11\farcs56\pm0\farcs03$.} \underline{Additional Parameters:} four Gaussian components added at: \textcolor{cyan}{$R_{\rm r}=0\farcs36\pm0\farcs14$, $7\farcs68\pm0\farcs03$, $17\farcs76\pm0\farcs05$, \& $39\farcs92\pm0\farcs03$; with $\mu_0 = 18.96\pm0.28$, $21.65\pm0.04$, $22.24\pm0.02$, \& $24.59\pm0.02\,\text{mag\,arcsec}^{-2}$; and FWHM = $0\farcs84\pm0\farcs18$, $1\farcs82\pm0\farcs08$, $6\farcs95\pm0\farcs11$, \& $3\farcs17\pm0\farcs06$, respectively.}}
\label{NGC5495_plot}
\end{sidewaysfigure}

\subsection{NGC~5495 (Figure~\ref{NGC5495_plot})}

We have added a central point source and four Gaussian components at $R_{\rm maj} \approx 1\arcsec$ ($R_{\rm eq}\approx0\arcsec$), $R_{\rm maj} \approx 12\arcsec$ ($R_{\rm eq}\approx 8\arcsec$), $R_{\rm maj} \approx 20\arcsec$ ($R_{\rm eq}\approx 18\arcsec$), and $R_{\rm maj} \approx 43\arcsec$ ($R_{\rm eq}\approx 40\arcsec$). This accounts for nuclear spiral arms, ansae at the bar's ends, and spiral arms with the outer two Gaussians, respectively.

\begin{sidewaysfigure}
\includegraphics[clip=true,trim= 11mm 1mm 3mm 5mm,width=0.249\textwidth]{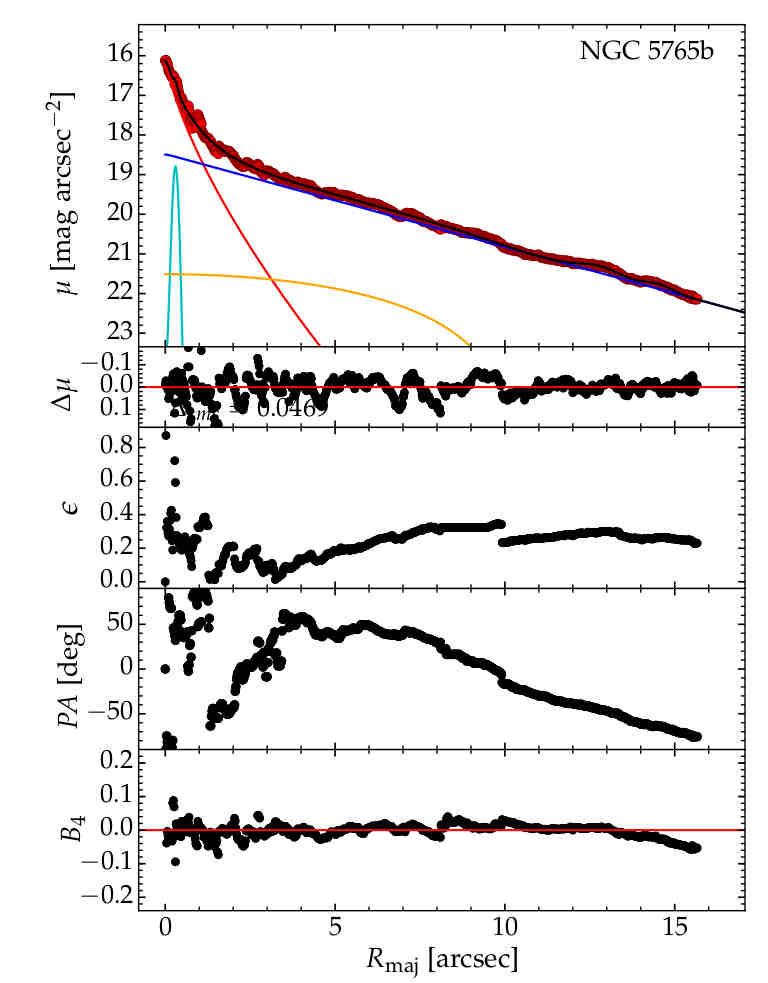}
\includegraphics[clip=true,trim= 11mm 1mm 3mm 5mm,width=0.249\textwidth]{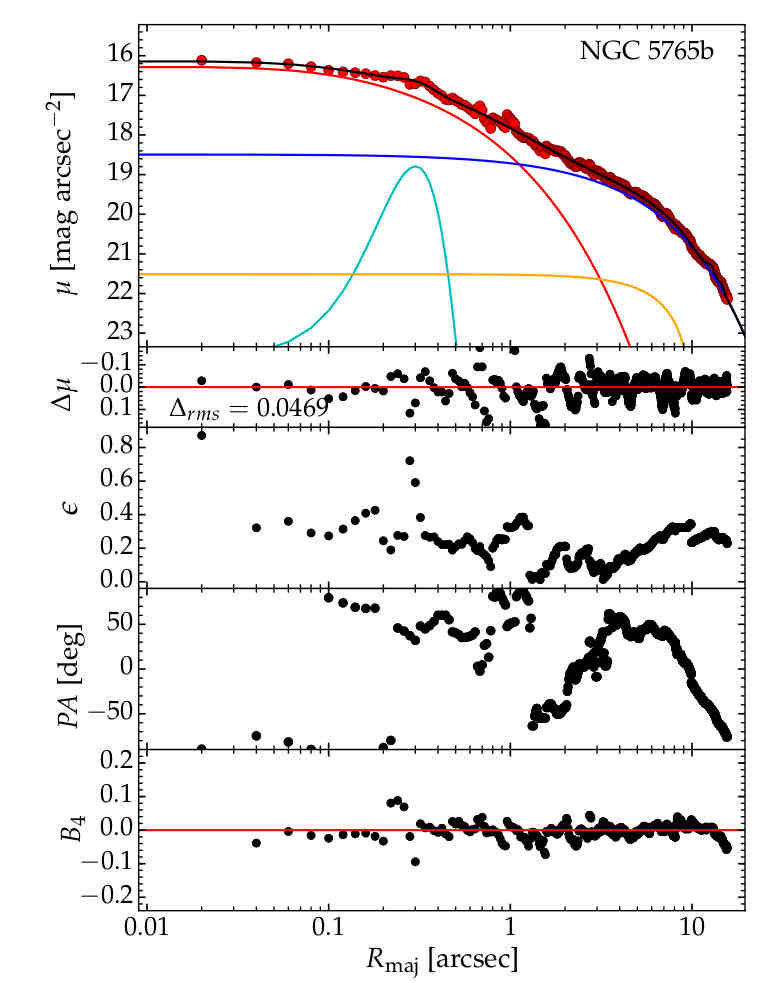}
\includegraphics[clip=true,trim= 11mm 1mm 3mm 5mm,width=0.249\textwidth]{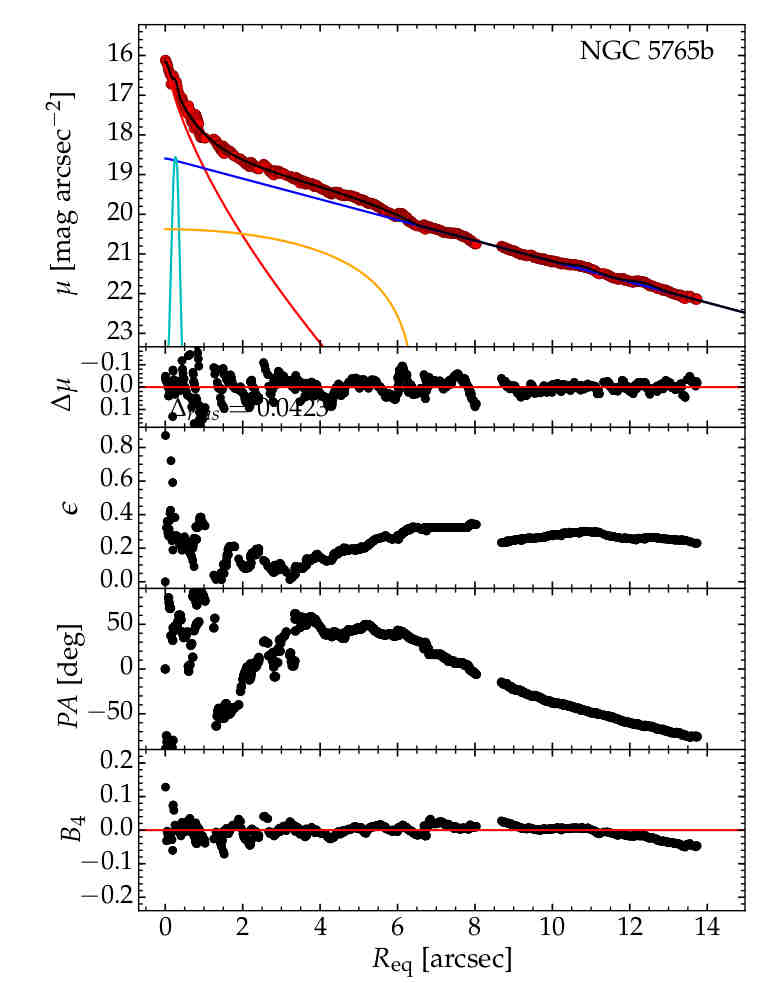}
\includegraphics[clip=true,trim= 11mm 1mm 3mm 5mm,width=0.249\textwidth]{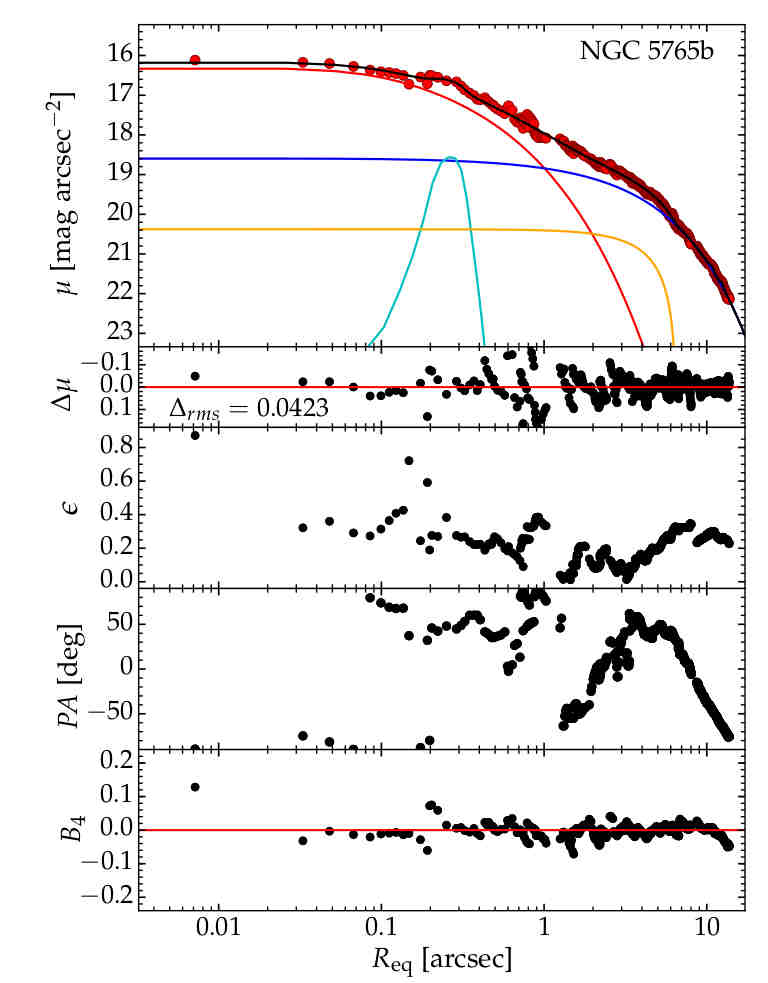}
\caption{\textit{HST} WFC3 UVIS2 F814W surface brightness profile for NGC~5765b, with a physical scale of 0.6491$\,\text{kpc}\,\text{arcsec}^{-1}$. \textbf{Left two panels}---The model represents $0\arcsec \leq R_{\rm maj} \leq 15\farcs66$ with $\Delta_{\rm rms}=0.0469\,\text{mag\,arcsec}^{-2}$. \underline{S{\'e}rsic Profile Parameters:} \textcolor{red}{$R_e=1\farcs11\pm0\farcs05$, $\mu_e=18.72\pm0.06\,\text{mag\,arcsec}^{-2}$, and $n=1.46\pm0.04$.} \underline{Ferrers Profile Parameters:} \textcolor{Orange}{$\mu_0 = 21.51\pm0.20\,\text{mag\,arcsec}^{-2}$, $R_{\rm end} = 10\farcs58\pm0\farcs32$, and $\alpha = 3.28\pm0.56$.} \underline{Exponential Profile Parameters:} \textcolor{blue}{$\mu_0 = 18.48\pm0.04\,\text{mag\,arcsec}^{-2}$ and $h = 4\farcs63\pm0\farcs07$.} \underline{Additional Parameters:} three Gaussian components added at: \textcolor{cyan}{$R_{\rm r}=0\farcs30$, $12\farcs74\pm0\farcs03$, \& $14\farcs47\pm0\farcs04$; with $\mu_0 = 18.41\pm0.31$, $23.36\pm0.07$, \& $24.14\pm0.11\,\text{mag\,arcsec}^{-2}$; and FWHM = $0\farcs11\pm0\farcs04$, $1\farcs31\pm0\farcs10$, \& $0\farcs94\pm0\farcs13$, respectively.} The outer two Gaussians are below the visible portion of the plots. \textbf{Right two panels}---The model represents $0\arcsec \leq R_{\rm maj} \leq 13\farcs74$ with $\Delta_{\rm rms}=0.0423\,\text{mag\,arcsec}^{-2}$. \underline{S{\'e}rsic Profile Parameters:} \textcolor{red}{$R_e=1\farcs00\pm0\farcs05$, $\mu_e=18.83\pm0.07\,\text{mag\,arcsec}^{-2}$, and $n=1.51\pm0.05$.} \underline{Ferrers Profile Parameters:} \textcolor{Orange}{$\mu_0 = 20.38\pm0.07\,\text{mag\,arcsec}^{-2}$, $R_{\rm end} = 6\farcs54\pm0\farcs08$, and $\alpha = 2.74\pm0.25$.} \underline{Exponential Profile Parameters:} \textcolor{blue}{$\mu_0 = 18.58\pm0.01\,\text{mag\,arcsec}^{-2}$ and $h = 4\farcs17\pm0\farcs02$.} \underline{Additional Parameters:} three Gaussian components added at: \textcolor{cyan}{$R_{\rm r}=0\farcs27\pm173\farcs97$, $10\farcs78\pm0\farcs05$, \& $12\farcs36\pm0\farcs05$; with $\mu_0 = 14.86\pm99.99$, $24.26\pm0.16$, \& $24.65\pm0.17\,\text{mag\,arcsec}^{-2}$; and FWHM = $0\farcs01\pm829\farcs82$, $0\farcs69\pm0\farcs12$, \& $0\farcs66\pm0\farcs13$, respectively.} The outer two Gaussians are below the visible portion of the plots.}
\label{NGC5765b_plot}
\end{sidewaysfigure}

\subsection{NGC~5765b (Figure~\ref{NGC5765b_plot})}

The nuclei of the galaxy pair NGC5765a and NGC5765b, are separated by $22\farcs83$. We have added a Gaussian component at $R_{\rm maj} =R_{\rm eq}\approx 0\farcs3$ to account for the presence of near-center star clusters and two Gaussian components at $R_{\rm maj} \approx 13\arcsec$ ($R_{\rm eq}\approx 11\arcsec$) and $R_{\rm maj} \approx 14\arcsec$ ($R_{\rm eq}\approx 12\arcsec$) to account for crossings of the $m=2$ spiral arms.

\begin{sidewaysfigure}
\includegraphics[clip=true,trim= 11mm 1mm 3mm 5mm,width=0.249\textwidth]{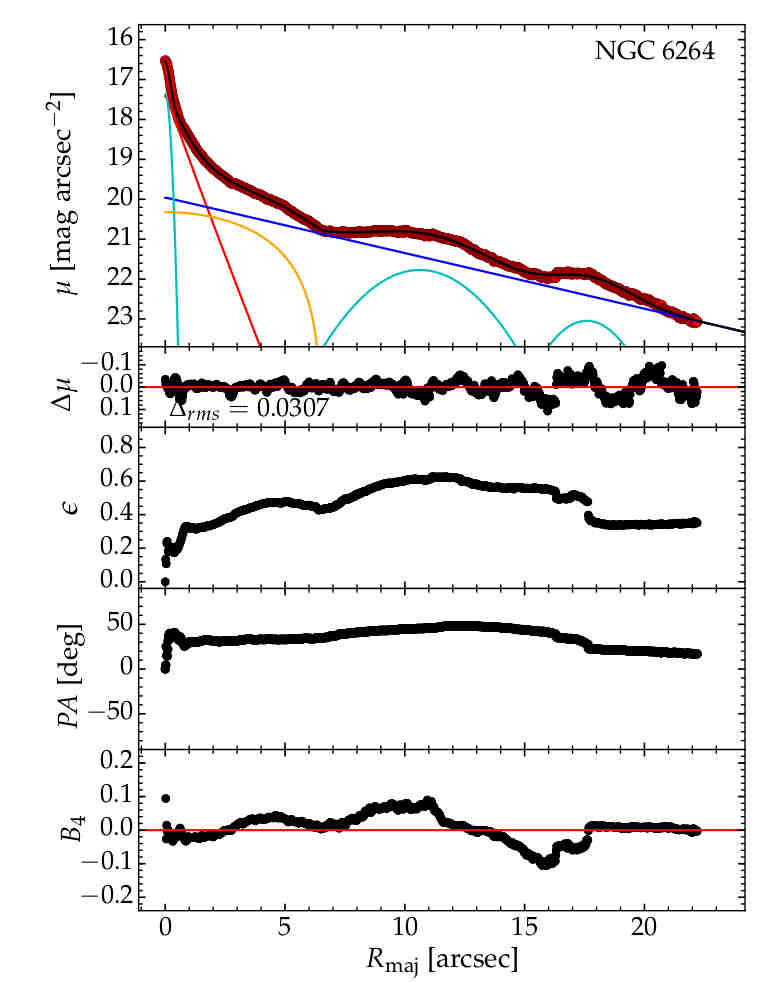}
\includegraphics[clip=true,trim= 11mm 1mm 3mm 5mm,width=0.249\textwidth]{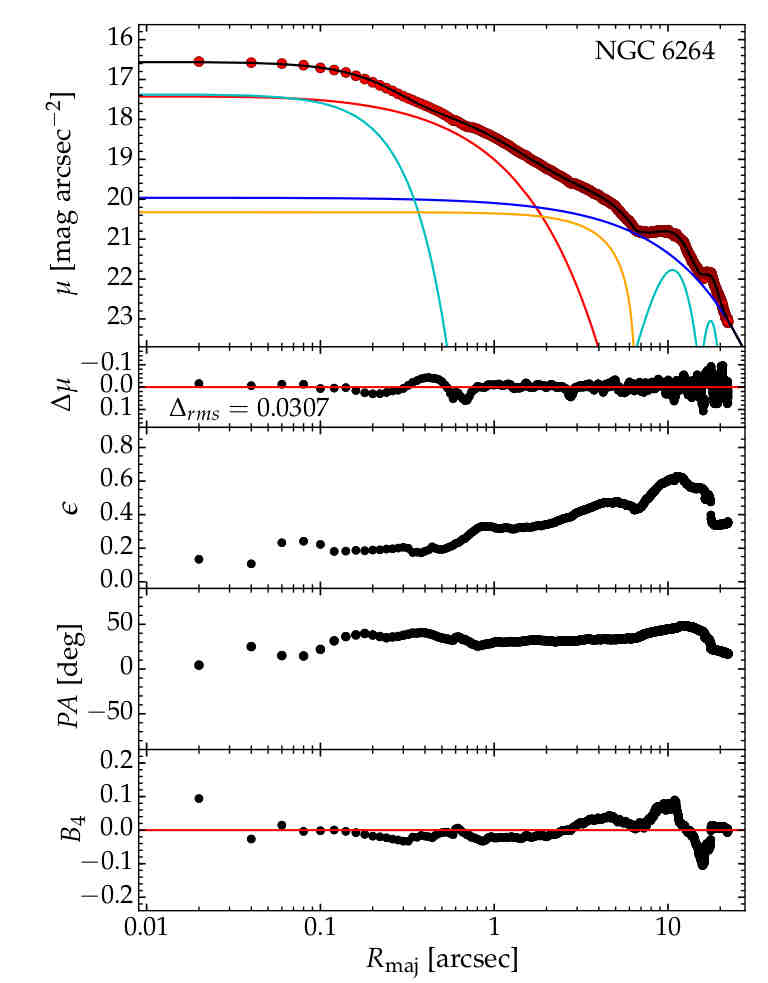}
\includegraphics[clip=true,trim= 11mm 1mm 3mm 5mm,width=0.249\textwidth]{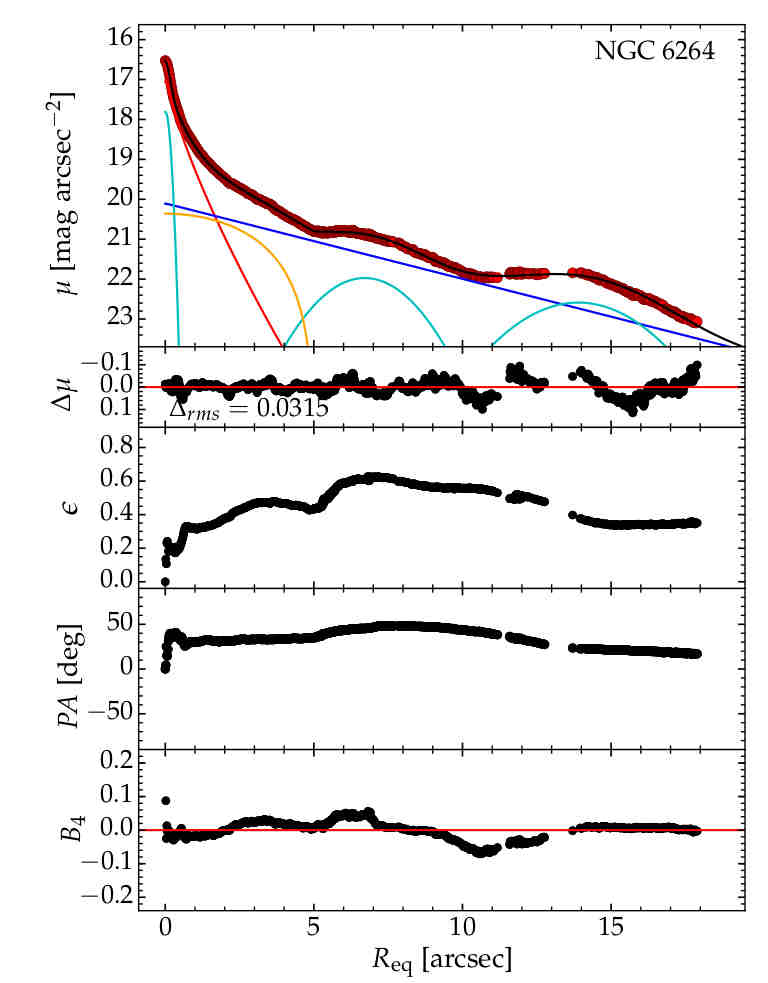}
\includegraphics[clip=true,trim= 11mm 1mm 3mm 5mm,width=0.249\textwidth]{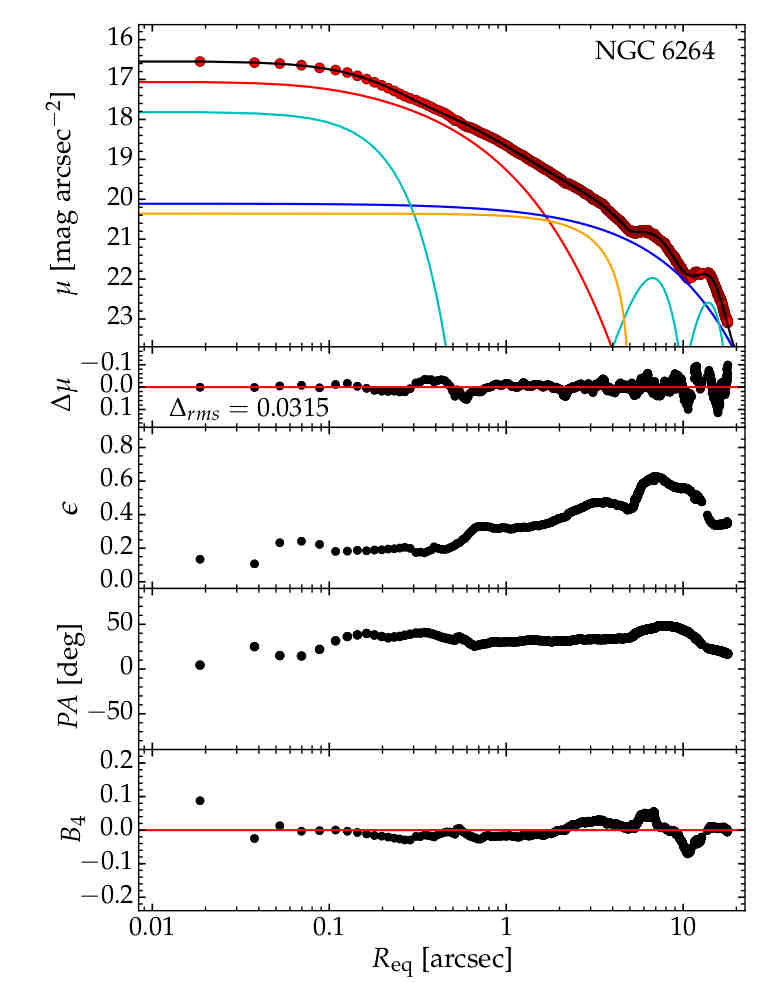}
\caption{\textit{HST} WFC3 UVIS2 F814W surface brightness profile for NGC~6264, with a physical scale of 0.7463$\,\text{kpc}\,\text{arcsec}^{-1}$. \textbf{Left two panels}---The model represents $0\arcsec \leq R_{\rm maj} \leq 22\farcs2$ with $\Delta_{\rm rms}=0.0307\,\text{mag\,arcsec}^{-2}$. \underline{S{\'e}rsic Profile Parameters:} \textcolor{red}{$R_e=1\farcs13\pm0\farcs03$, $\mu_e=19.23\pm0.04\,\text{mag\,arcsec}^{-2}$, and $n=1.04\pm0.05$.} \underline{Ferrers Profile Parameters:} \textcolor{Orange}{$\mu_0 = 20.34\pm0.04\,\text{mag\,arcsec}^{-2}$, $R_{\rm end} = 6\farcs58\pm0\farcs05$, and $\alpha = 2.98\pm0.14$.} \underline{Exponential Profile Parameters:} \textcolor{blue}{$\mu_0 = 19.97\pm0.02\,\text{mag\,arcsec}^{-2}$ and $h = 7\farcs79\pm0\farcs05$.} \underline{Additional Parameters:} three Gaussian components added at: \textcolor{cyan}{$R_{\rm r}=0\arcsec$, $10\farcs62\pm0\farcs03$, \& $17\farcs59\pm0\farcs02$; with $\mu_0 = 17.28\pm0.05$, $21.78\pm0.01$, \& $23.06\pm0.01\,\text{mag\,arcsec}^{-2}$; and FWHM = $0\farcs35\pm0\farcs01$, $5\farcs03\pm0\farcs07$, \& $3\farcs46\pm0\farcs04$, respectively.} \textbf{Right two panels}---The model represents $0\arcsec \leq R_{\rm maj} \leq 17\farcs9$ with $\Delta_{\rm rms}=0.0307\,\text{mag\,arcsec}^{-2}$. \underline{S{\'e}rsic Profile Parameters:} \textcolor{red}{$R_e=1\farcs05\pm0\farcs06$, $\mu_e=19.37\pm0.08\,\text{mag\,arcsec}^{-2}$, and $n=1.35\pm0.09$.} \underline{Ferrers Profile Parameters:} \textcolor{Orange}{$\mu_0 = 20.40\pm0.09\,\text{mag\,arcsec}^{-2}$, $R_{\rm end} = 5\farcs06\pm0\farcs06$, and $\alpha = 3.28\pm0.23$.} \underline{Exponential Profile Parameters:} \textcolor{blue}{$\mu_0 = 20.10\pm0.02\,\text{mag\,arcsec}^{-2}$ and $h = 5\farcs74\pm0\farcs04$.} \underline{Additional Parameters:} three Gaussian components added at: \textcolor{cyan}{$R_{\rm r}=0\arcsec$, $6\farcs73\pm0\farcs03$, \& $13\farcs84\pm0\farcs01$; with $\mu_0 = 17.73\pm0.15$, $22.00\pm0.01$, \& $22.60\pm0.00\,\text{mag\,arcsec}^{-2}$; and FWHM = $0\farcs31\pm0\farcs01$, $3\farcs48\pm0\farcs05$, \& $4\farcs83\pm0\farcs02$, respectively.}}
\label{NGC6264_plot}
\end{sidewaysfigure}

\subsection{NGC~6264 (Figure~\ref{NGC6264_plot})}

The galaxy NGC~6264 possesses a Type 2 Seyfert AGN \citep{Veron-Cetty:2006}, for which we have added a central Gaussian component. Two additional Gaussians have been added to account for spiral arms at $R_{\rm maj} \approx 11\arcsec$ ($R_{\rm eq}\approx 7\arcsec$) and $R_{\rm maj} \approx 18\arcsec$ ($R_{\rm eq}\approx 14\arcsec$).

\begin{sidewaysfigure}
\includegraphics[clip=true,trim= 11mm 1mm 3mm 5mm,width=0.249\textwidth]{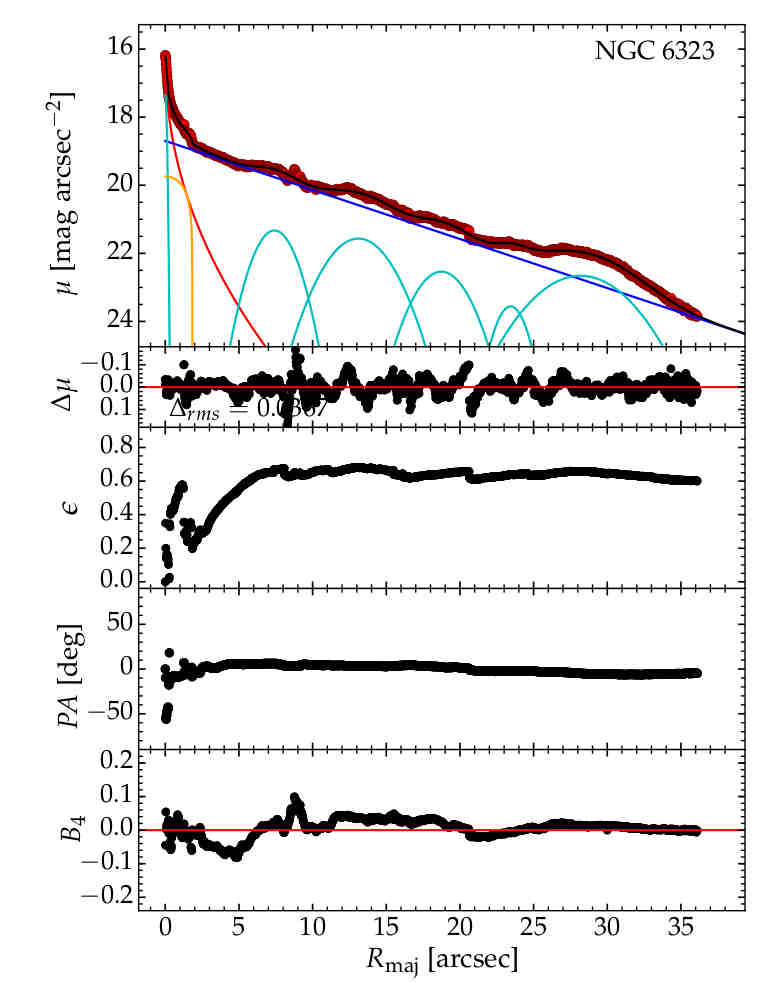}
\includegraphics[clip=true,trim= 11mm 1mm 3mm 5mm,width=0.249\textwidth]{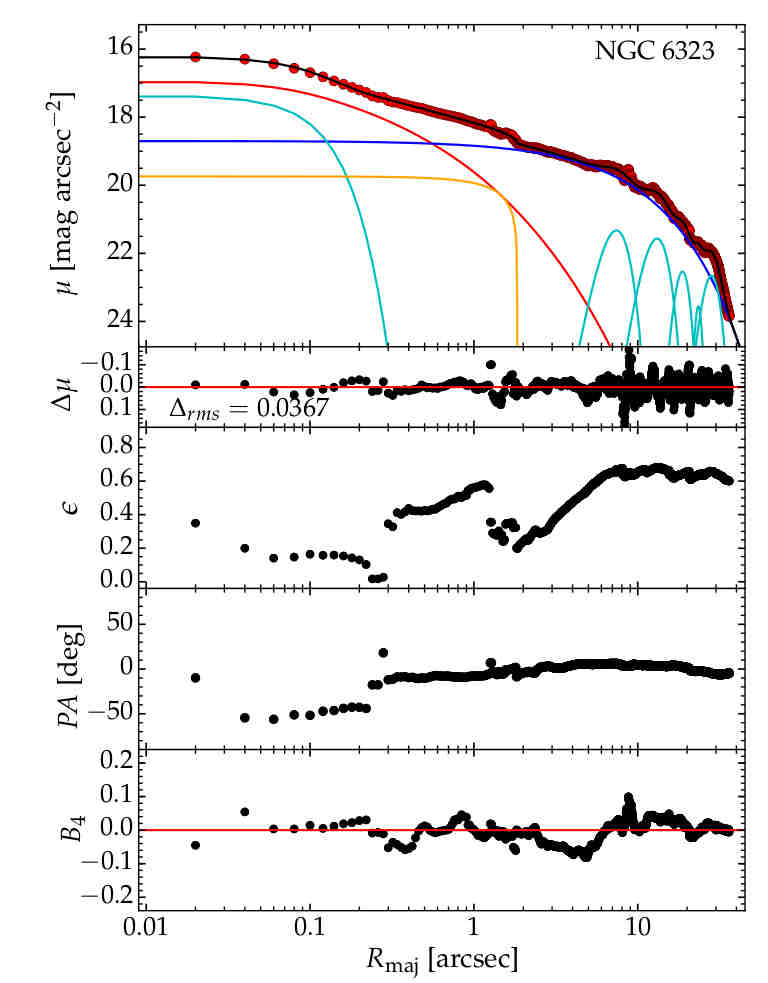}
\includegraphics[clip=true,trim= 11mm 1mm 3mm 5mm,width=0.249\textwidth]{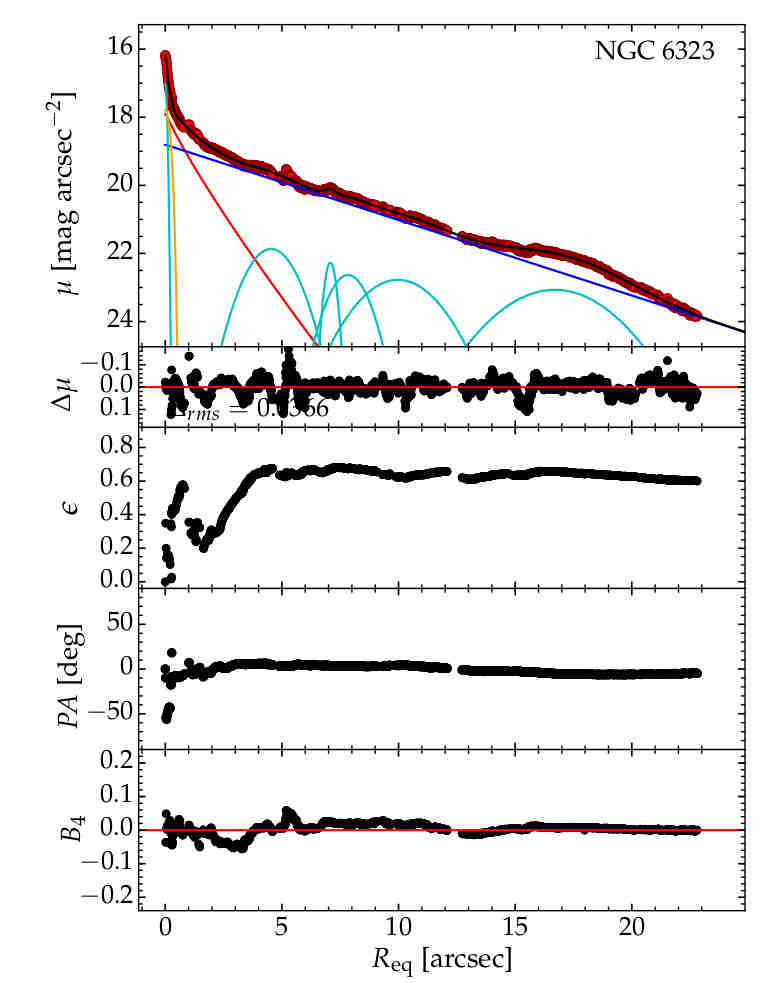}
\includegraphics[clip=true,trim= 11mm 1mm 3mm 5mm,width=0.249\textwidth]{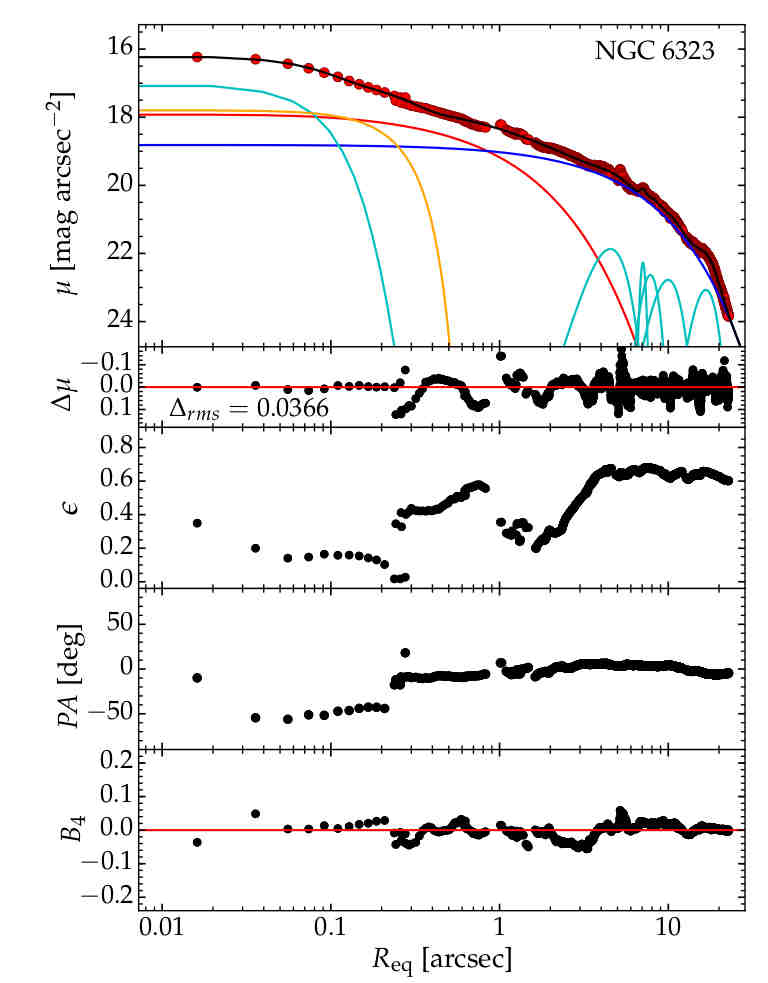}
\caption{\textit{HST} WFC3 UVIS2 F814W surface brightness profile for NGC~6323, with a physical scale of 0.5384$\,\text{kpc}\,\text{arcsec}^{-1}$. \textbf{Left two panels}---The model represents $0\arcsec \leq R_{\rm maj} \leq 36\farcs1$ with $\Delta_{\rm rms}=0.0367\,\text{mag\,arcsec}^{-2}$. \underline{S{\'e}rsic Profile Parameters:} \textcolor{red}{$R_e=1\farcs53\pm0\farcs10$, $\mu_e=20.39\pm0.17\,\text{mag\,arcsec}^{-2}$, and $n=2.09\pm0.20$.} \underline{Ferrers Profile Parameters:} \textcolor{Orange}{$\mu_0 = 19.74\pm0.10\,\text{mag\,arcsec}^{-2}$, $R_{\rm end} = 1\farcs84\pm0\farcs00$, and $\alpha = 1.27\pm0.15$.} \underline{Exponential Profile Parameters:} \textcolor{blue}{$\mu_0 = 18.70\pm0.00\,\text{mag\,arcsec}^{-2}$ and $h = 7\farcs53\pm0\farcs01$.} \underline{Additional Parameters:} six Gaussian components added at: \textcolor{cyan}{$R_{\rm r}=0\arcsec$, $7\farcs41\pm0\farcs03$, $13\farcs11\pm0\farcs03$, $18\farcs74\pm0\farcs04$, $23\farcs43\pm0\farcs04$, \& $28\farcs18\pm0\farcs03$; with $\mu_0 = 17.02\pm0.16$, $21.33\pm0.02$, $21.57\pm0.01$, $22.54\pm0.02$, $23.56\pm0.03$, \& $22.66\pm0.00\,\text{mag\,arcsec}^{-2}$; and FWHM = $0\farcs16\pm0\farcs01$, $2\farcs84\pm0\farcs08$, $4\farcs44\pm0\farcs09$, $3\farcs75\pm0\farcs10$, $2\farcs28\pm0\farcs09$, \& $6\farcs75\pm0\farcs04$, respectively.} \textbf{Right two panels}---The model represents $0\arcsec \leq R_{\rm maj} \leq 22\farcs8$ with $\Delta_{\rm rms}=0.0366\,\text{mag\,arcsec}^{-2}$. \underline{S{\'e}rsic Profile Parameters:} \textcolor{red}{$R_e=1\farcs71\pm0\farcs24$, $\mu_e=19.98\pm0.12\,\text{mag\,arcsec}^{-2}$, and $n=1.15\pm0.13$.} \underline{Ferrers Profile Parameters:} \textcolor{Orange}{$\mu_0 = 17.74\pm0.22\,\text{mag\,arcsec}^{-2}$, $R_{\rm end} = 0\farcs52\pm0\farcs20$, and $\alpha = 10.00\pm26.38$.} \underline{Exponential Profile Parameters:} \textcolor{blue}{$\mu_0 = 18.80\pm0.08\,\text{mag\,arcsec}^{-2}$ and $h = 4\farcs90\pm0\farcs08$.} \underline{Additional Parameters:} six Gaussian components added at: \textcolor{cyan}{$R_{\rm r}=0\arcsec$, $4\farcs54\pm0\farcs07$, $7\farcs08\pm0\farcs02$, $7\farcs83\pm0\farcs16$, $9\farcs98\pm0\farcs22$, \& $16\farcs70\pm0\farcs09$; with $\mu_0 = 16.37\pm0.11$, $21.87\pm0.23$, $22.27\pm0.11$, $22.63\pm0.19$, $22.77\pm0.23$, \& $23.07\pm0.05\,\text{mag\,arcsec}^{-2}$; and FWHM = $0\farcs10\pm0\farcs01$, $2\farcs19\pm0\farcs32$, $0\farcs53\pm0\farcs06$, $1\farcs80\pm0\farcs27$, $3\farcs64\pm0\farcs47$, \& $5\farcs11\pm0\farcs14$, respectively.}}
\label{NGC6323_plot}
\end{sidewaysfigure}

\subsection{NGC~6323 (Figure~\ref{NGC6323_plot})}

A central Gaussian plus five additional off-center Gaussian components have been added to account for spiral arms at $R_{\rm maj} \approx 7\arcsec$ ($R_{\rm eq}\approx 5\arcsec$), $R_{\rm maj} \approx 13\arcsec$ ($R_{\rm eq}\approx 7\arcsec$), $R_{\rm maj} \approx 19\arcsec$ ($R_{\rm eq}\approx 8\arcsec$), $R_{\rm maj} \approx 23\arcsec$ ($R_{\rm eq}\approx 10\arcsec$), and $R_{\rm maj} \approx 28\arcsec$ ($R_{\rm eq}\approx 17\arcsec$).

\begin{sidewaysfigure}
\includegraphics[clip=true,trim= 11mm 1mm 3mm 5mm,width=0.249\textwidth]{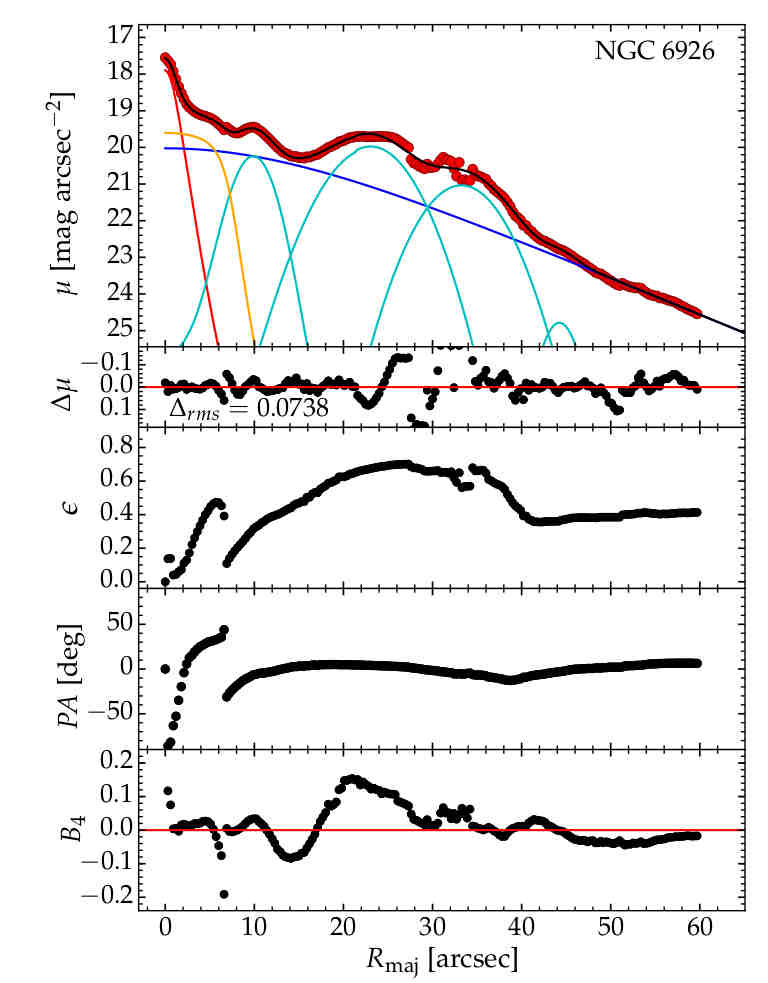}
\includegraphics[clip=true,trim= 11mm 1mm 3mm 5mm,width=0.249\textwidth]{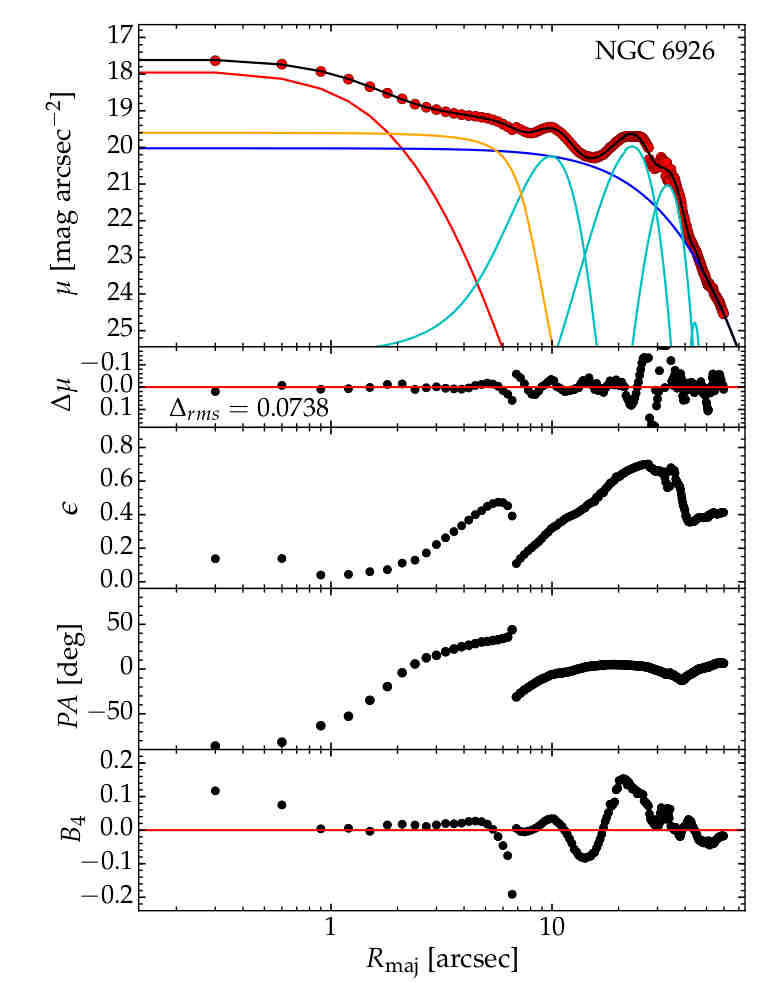}
\includegraphics[clip=true,trim= 11mm 1mm 3mm 5mm,width=0.249\textwidth]{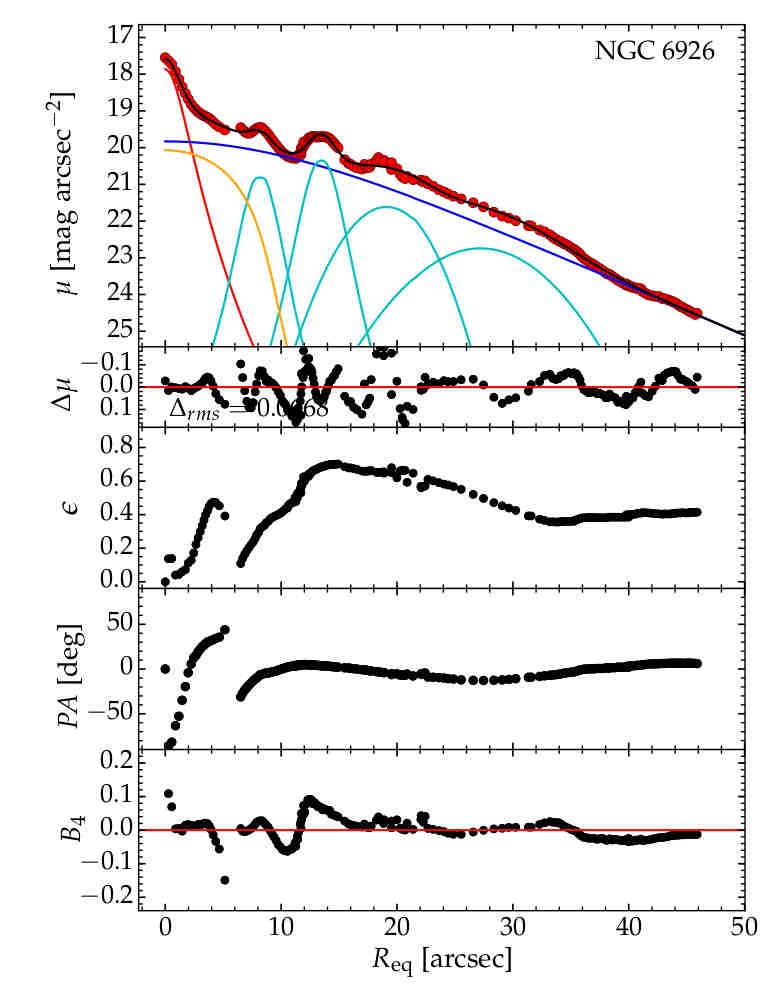}
\includegraphics[clip=true,trim= 11mm 1mm 3mm 5mm,width=0.249\textwidth]{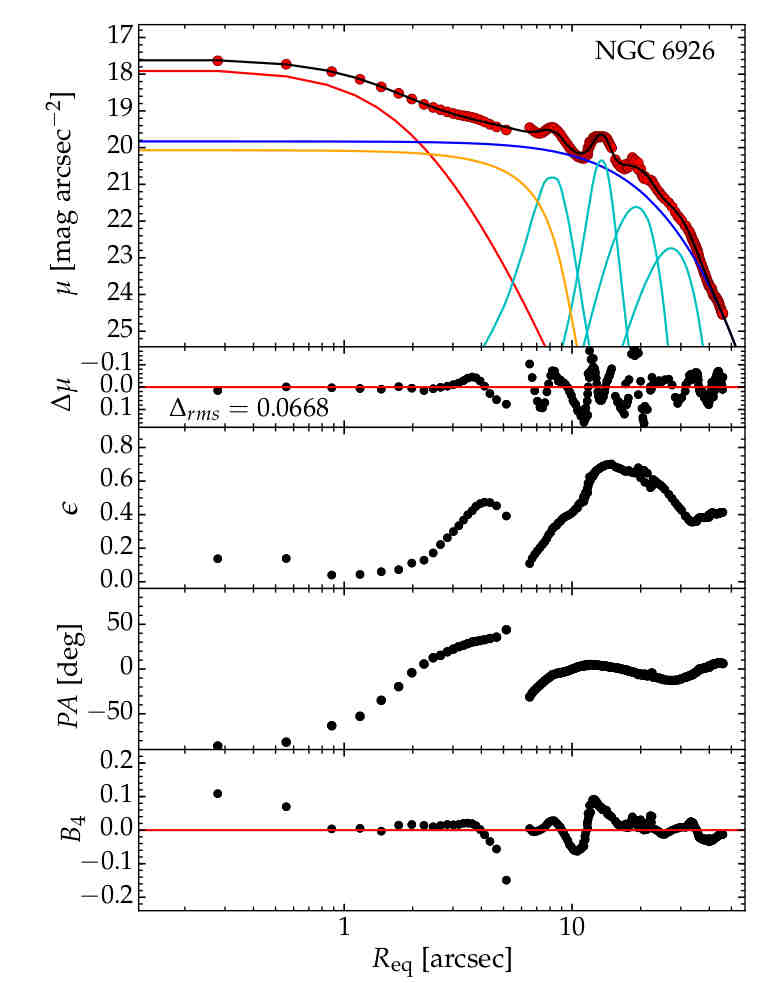}
\caption{\textit{Spitzer} $3.6\,\micron$ surface brightness profile for NGC~6926, with a physical scale of 0.4248$\,\text{kpc}\,\text{arcsec}^{-1}$. \textbf{Left two panels}---The model represents $0\arcsec \leq R_{\rm maj} \leq 60\arcsec$ with $\Delta_{\rm rms}=0.0738\,\text{mag\,arcsec}^{-2}$. \underline{S{\'e}rsic Profile Parameters:} \textcolor{red}{$R_e=0\farcs57\pm0\farcs17$, $\mu_e=17.52\pm1.17\,\text{mag\,arcsec}^{-2}$, and $n=1.60\pm1.01$.} \underline{Ferrers Profile Parameters:} \textcolor{Orange}{$\mu_0 = 19.56\pm0.45\,\text{mag\,arcsec}^{-2}$, $R_{\rm end} = 6\farcs97\pm1\farcs12$, and $\alpha = 0.73\pm1.64$.} \underline{Edge-on Disk Model Parameters:} \textcolor{blue}{$\mu_0 = 20.01\pm0.10\,\text{mag\,arcsec}^{-2}$ and $h_z = 21\farcs55\pm0\farcs40$.} \underline{Additional Parameters:} four Gaussian components added at: \textcolor{cyan}{$R_{\rm r}=9\farcs94\pm0\farcs18$, $23\farcs04\pm0\farcs14$, $33\farcs24\pm0\farcs29$, \& $44\farcs24\pm0\farcs47$; with $\mu_0 = 19.99\pm0.09$, $19.97\pm0.03$, $21.04\pm0.04$, \& $24.78\pm0.22\,\text{mag\,arcsec}^{-2}$; and FWHM = $3\farcs50\pm0\farcs60$, $8\farcs55\pm0\farcs46$, $8\farcs42\pm0\farcs39$, \& $3\farcs97\pm1\farcs01$, respectively.} \textbf{Right two panels}---The model represents $0\arcsec \leq R_{\rm eq} \leq 46\arcsec$ with $\Delta_{\rm rms}=0.0668\,\text{mag\,arcsec}^{-2}$. \underline{S{\'e}rsic Profile Parameters:} \textcolor{red}{$R_e=0\farcs86\pm0\farcs24$, $\mu_e=18.69\pm0.78\,\text{mag\,arcsec}^{-2}$, and $n=2.33\pm0.53$.} \underline{Ferrers Profile Parameters:} \textcolor{Orange}{$\mu_0 = 20.02\pm0.58\,\text{mag\,arcsec}^{-2}$, $R_{\rm end} = 8\farcs97\pm2\farcs06$, and $\alpha = 3.52\pm0.00$.} \underline{Edge-on Disk Model Parameters:} \textcolor{blue}{$\mu_0 = 19.82\pm0.02\,\text{mag\,arcsec}^{-2}$ and $h_z = 15\farcs98\pm0\farcs07$.} \underline{Additional Parameters:} four Gaussian components added at: \textcolor{cyan}{$R_{\rm r}=8\farcs20\pm4\farcs46$, $13\farcs47\pm0\farcs05$, $19\farcs10\pm0\farcs26$, \& $27\farcs20\pm0\farcs63$; with $\mu_0 = 18.07\pm64.80$, $19.09\pm0.46$, $21.52\pm0.13$, \& $22.74\pm0.07\,\text{mag\,arcsec}^{-2}$; and FWHM = $0\farcs23\pm8\farcs67$, $0\farcs83\pm0\farcs40$, $6\farcs38\pm0\farcs61$, \& $10\farcs92\pm0\farcs67$, respectively.}}
\label{NGC6926_plot}
\end{sidewaysfigure}

\subsection{NGC~6926 (Figure~\ref{NGC6926_plot})}\label{NGC6926}

The galaxy NGC~6926 is inclined by $58\fdg0\pm7\fdg8$ with respect to the plane of the sky. Nonetheless, we found that it was preferable to use the edge-on disk model to fit the disk. Additionally, it possesses a Type 2 Seyfert AGN \citep{Veron-Cetty:2006}. We have added a Gaussian for this, plus a Gaussian component at $R_{\rm maj} \approx 10\arcsec$ ($R_{\rm eq}\approx 8\arcsec$), to account for ansae at the ends of the bar and three Gaussian components to account for spiral arms at $R_{\rm maj} \approx 23\arcsec$ ($R_{\rm eq}\approx 13\arcsec$), $R_{\rm maj} \approx 33\arcsec$ ($R_{\rm eq}\approx 19\arcsec$), and $R_{\rm maj} \approx 44\arcsec$ ($R_{\rm eq}\approx 27\arcsec$). The inner S{\'e}rsic model expresses a likely nuclear (disk) component, not a spheroid.

\begin{sidewaysfigure}
\includegraphics[clip=true,trim= 11mm 1mm 3mm 5mm,width=0.249\textwidth]{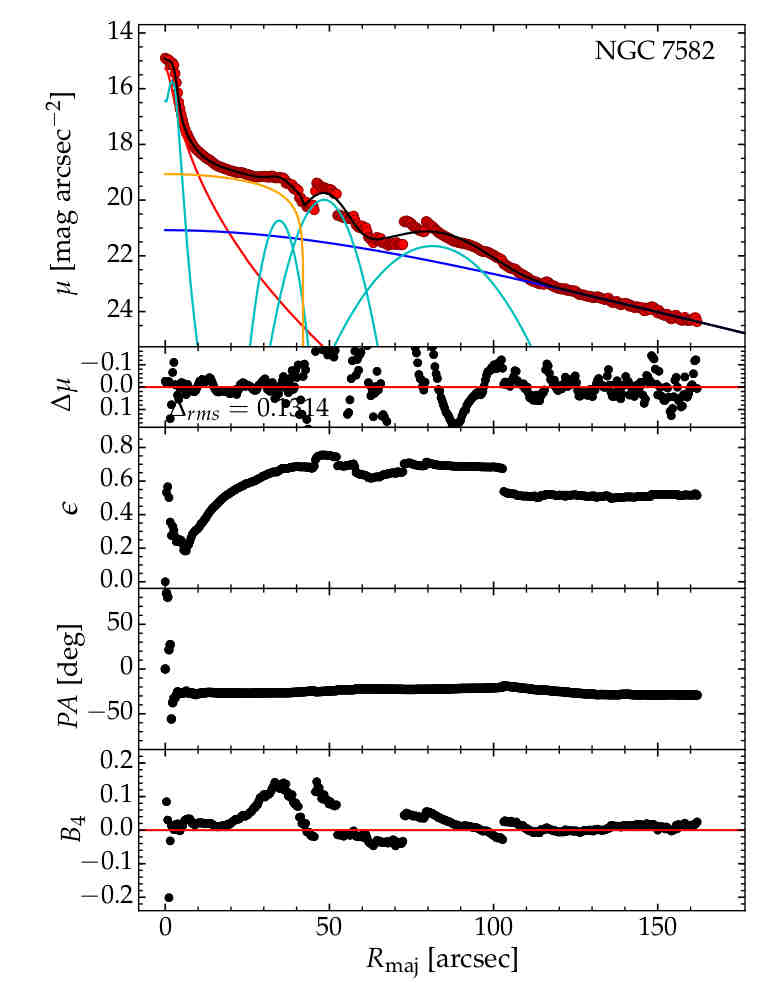}
\includegraphics[clip=true,trim= 11mm 1mm 3mm 5mm,width=0.249\textwidth]{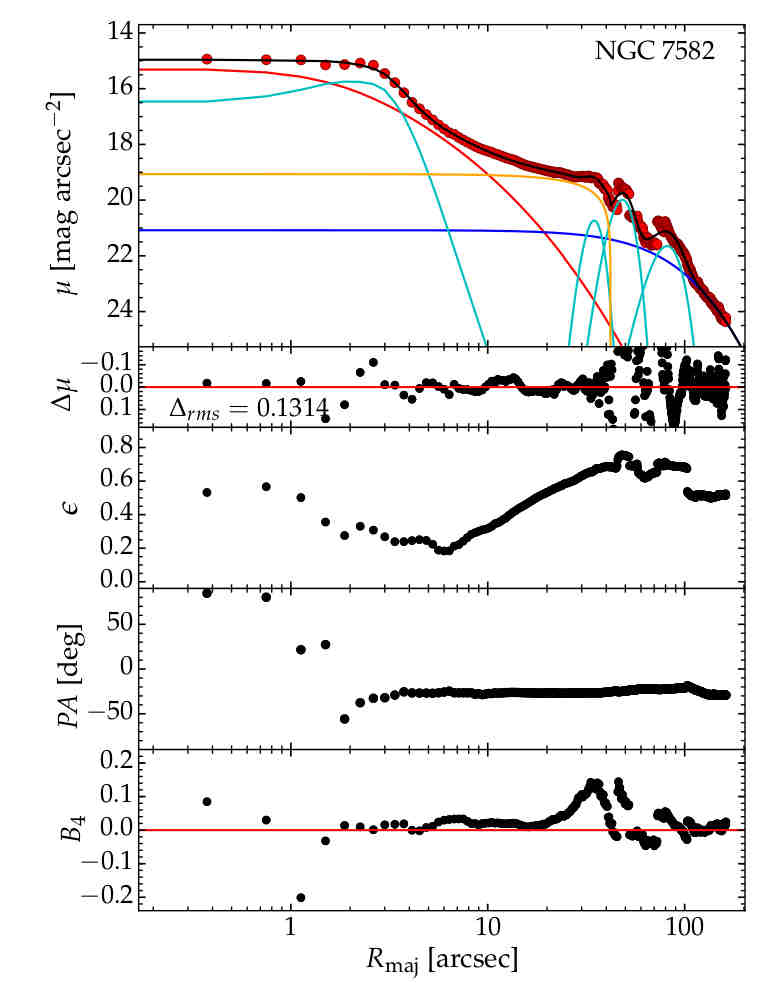}
\includegraphics[clip=true,trim= 11mm 1mm 3mm 5mm,width=0.249\textwidth]{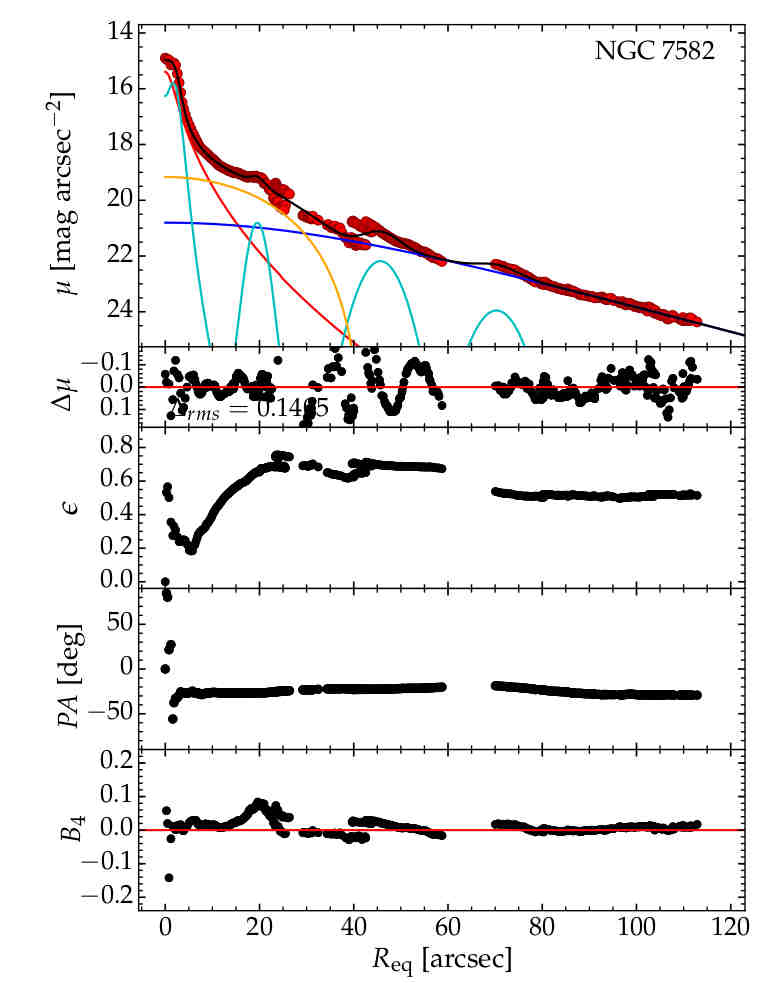}
\includegraphics[clip=true,trim= 11mm 1mm 3mm 5mm,width=0.249\textwidth]{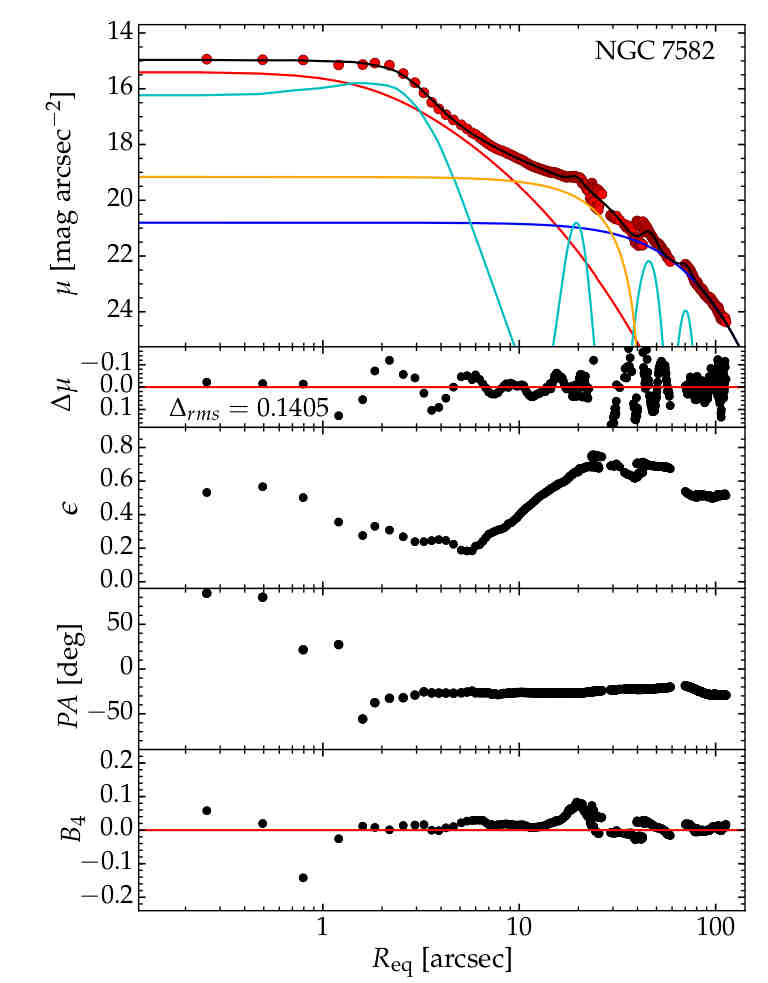}
\caption{\textit{Spitzer} $3.6\,\micron$ surface brightness profile for NGC~7582, with a physical scale of 0.0963$\,\text{kpc}\,\text{arcsec}^{-1}$. \textbf{Left two panels}---The model represents $0\arcsec \leq R_{\rm maj} \leq 162\arcsec$ with $\Delta_{\rm rms}=0.1314\,\text{mag\,arcsec}^{-2}$. \underline{S{\'e}rsic Profile Parameters:} \textcolor{red}{$R_e=5\farcs33\pm1\farcs26$, $\mu_e=17.04\pm0.52\,\text{mag\,arcsec}^{-2}$, and $n=2.20\pm0.54$.} \underline{Ferrers Profile Parameters:} \textcolor{Orange}{$\mu_0 = 19.07\pm0.11\,\text{mag\,arcsec}^{-2}$, $R_{\rm end} = 42\farcs07\pm0\farcs18$, and $\alpha = 1.40\pm0.35$.} \underline{Edge-on Disk Model Parameters:} \textcolor{blue}{$\mu_0 = 21.08\pm0.11\,\text{mag\,arcsec}^{-2}$ and $h_z = 73\farcs88\pm2\farcs11$.} \underline{Additional Parameters:} four Gaussian components added at: \textcolor{cyan}{$R_{\rm r}=2\farcs37\pm21\farcs23$, $34\farcs71\pm0\farcs77$, $48\farcs32\pm0\farcs26$, \& $81\farcs44\pm0\farcs50$; with $\mu_0 = 13.16\pm363.61$, $20.74\pm0.30$, $19.99\pm0.04$, \& $21.65\pm0.04\,\text{mag\,arcsec}^{-2}$; and FWHM = $0\farcs29\pm46\farcs78$, $7\farcs27\pm2\farcs42$, $12\farcs40\pm0\farcs52$, \& $27\farcs42\pm1\farcs14$, respectively.} \textbf{Right two panels}---The model represents $0\arcsec \leq R_{\rm eq} \leq 113\arcsec$ with $\Delta_{\rm rms}=0.1405\,\text{mag\,arcsec}^{-2}$. \underline{S{\'e}rsic Profile Parameters:} \textcolor{red}{$R_e=4\farcs55\pm1\farcs26$, $\mu_e=17.66\pm0.54\,\text{mag\,arcsec}^{-2}$, and $n=2.21\pm0.56$.} \underline{Ferrers Profile Parameters:} \textcolor{Orange}{$\mu_0 = 19.16\pm0.24\,\text{mag\,arcsec}^{-2}$, $R_{\rm end} = 42\farcs60\pm3\farcs12$, and $\alpha = 7.13\pm1.92$.} \underline{Edge-on Disk Model Parameters:} \textcolor{blue}{$\mu_0 = 20.81\pm0.05\,\text{mag\,arcsec}^{-2}$ and $h_z = 48\farcs19\pm0\farcs54$.} \underline{Additional Parameters:} four Gaussian components added at: \textcolor{cyan}{$R_{\rm r}=1\farcs94\pm999\arcsec$, $19\farcs54\pm0\farcs24$, $45\farcs66\pm0\farcs20$, \& $70\farcs29\pm0\farcs38$; with $\mu_0 = 12.69\pm999$, $20.36\pm0.31$, $22.19\pm0.07$, \& $23.96\pm0.10\,\text{mag\,arcsec}^{-2}$; and FWHM = $0\farcs09\pm999\arcsec$, $2\farcs32\pm0\farcs90$, $8\farcs18\pm0\farcs54$, \& $9\farcs14\pm0\farcs96$, respectively.}}
\label{NGC7582_plot}
\end{sidewaysfigure}

\subsection{NGC~7582 (Figure~\ref{NGC7582_plot})}

The galaxy NGC~7582 is inclined by $64\fdg3\pm5\fdg2$ with respect to the plane of the sky. Because of this, we use the edge-on disk model to represent it. We have added four Gaussian components at $R_{\rm maj} = R_{\rm eq} \approx 2\arcsec$, $R_{\rm maj} \approx 35\arcsec$ ($R_{\rm eq}\approx20\arcsec$), $R_{\rm maj} \approx 48\arcsec$ ($R_{\rm eq}\approx46\arcsec$), and $R_{\rm maj} \approx 81\arcsec$ ($R_{\rm eq}\approx70\arcsec$) to account for the inner ring and three spiral-arm crossings, respectively.

\subsection{UGC~3789 (Figure~\ref{UGC3789_plot} and \ref{fig:annotate})}

We have added two Gaussian components at $R_{\rm maj} \approx 18\arcsec$ ($R_{\rm eq}\approx 12\arcsec$) and $R_{\rm maj} \approx 33\arcsec$ ($R_{\rm eq}\approx 35\arcsec$) to account for a ring and spiral arms, respectively.

\begin{sidewaysfigure}
\includegraphics[clip=true,trim= 11mm 1mm 3mm 5mm,width=0.249\textwidth]{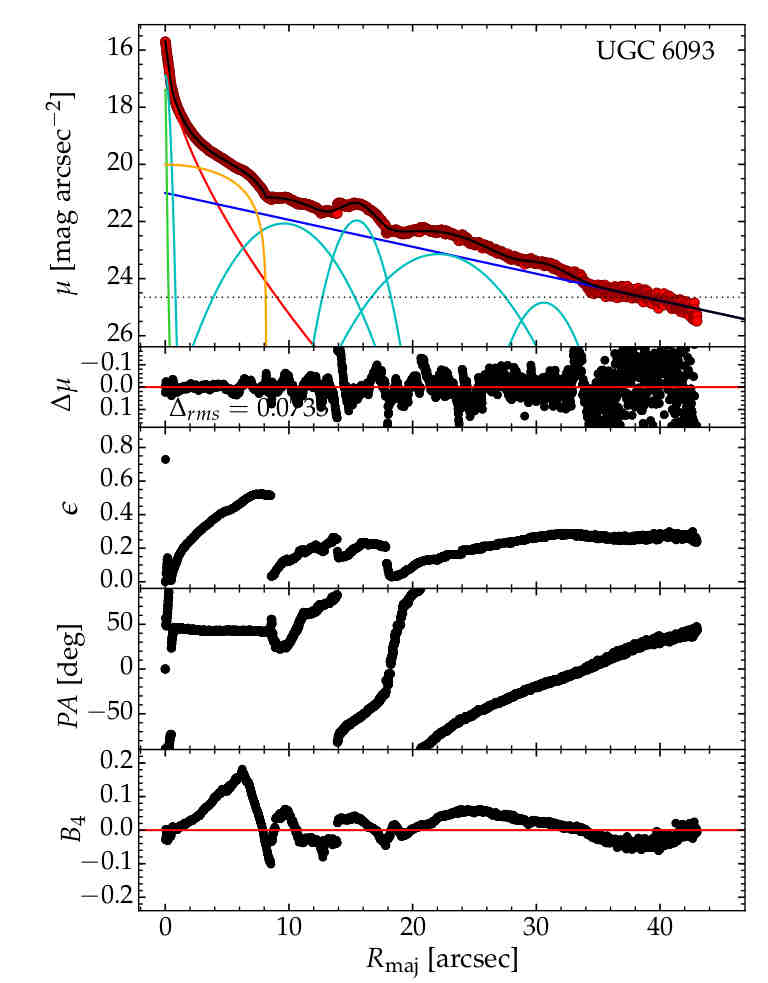}
\includegraphics[clip=true,trim= 11mm 1mm 3mm 5mm,width=0.249\textwidth]{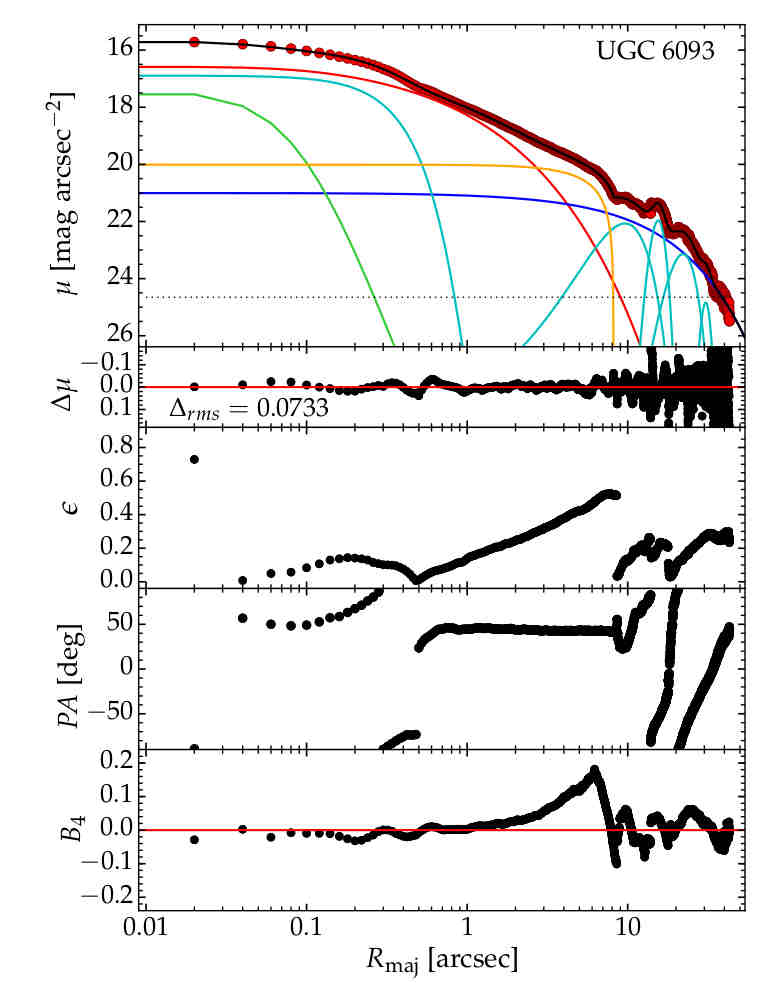}
\includegraphics[clip=true,trim= 11mm 1mm 3mm 5mm,width=0.249\textwidth]{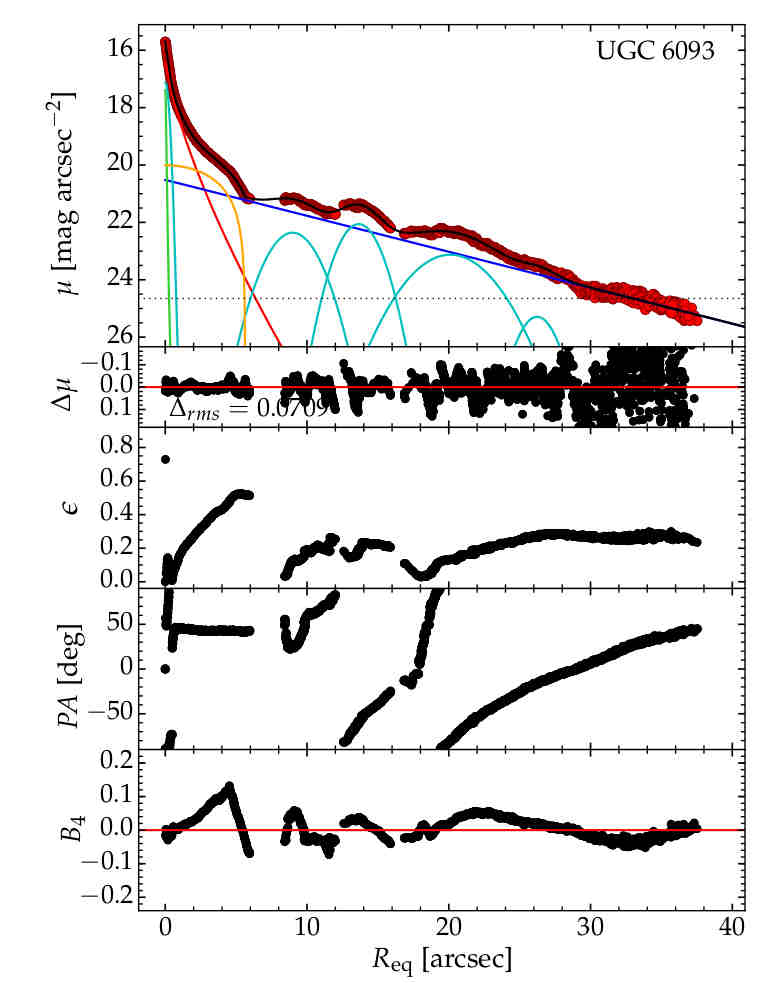}
\includegraphics[clip=true,trim= 11mm 1mm 3mm 5mm,width=0.249\textwidth]{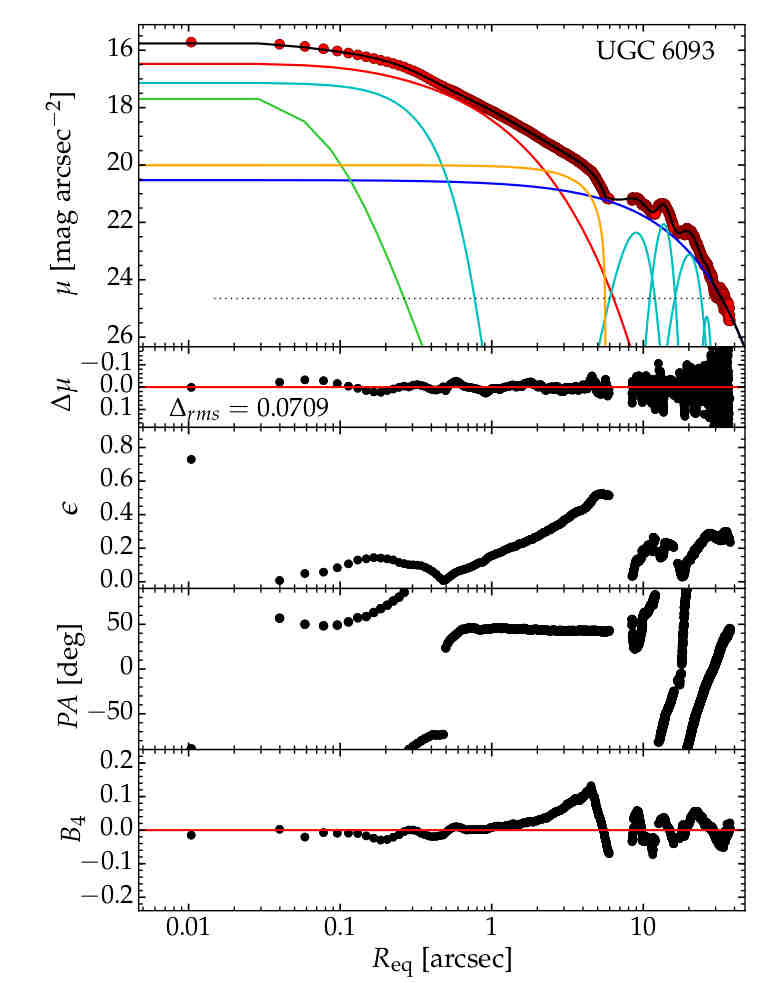}
\caption{\textit{HST} WFC3 UVIS2 F814W surface brightness profile for UGC~6093, with a physical scale of 0.7409$\,\text{kpc}\,\text{arcsec}^{-1}$. \textbf{Left two panels}---The model represents $0\arcsec \leq R_{\rm maj} \leq 43\farcs02$ with $\Delta_{\rm rms}=0.0733\,\text{mag\,arcsec}^{-2}$. \underline{Point Source:} \textcolor{LimeGreen}{$\mu_0 = 17.41\pm0.35\,\text{mag\,arcsec}^{-2}$.} \underline{S{\'e}rsic Profile Parameters:} \textcolor{red}{$R_e=1\farcs84\pm0\farcs17$, $\mu_e=19.27\pm0.14\,\text{mag\,arcsec}^{-2}$, and $n=1.55\pm0.20$.} \underline{Ferrers Profile Parameters:} \textcolor{Orange}{$\mu_0 = 20.02\pm0.09\,\text{mag\,arcsec}^{-2}$, $R_{\rm end} = 8\farcs16\pm0\farcs04$, and $\alpha = 2.26\pm0.15$.} \underline{Exponential Profile Parameters:} \textcolor{blue}{$\mu_0 = 21.01\pm0.06\,\text{mag\,arcsec}^{-2}$ and $h = 11\farcs50\pm0\farcs20$.} \underline{Additional Parameters:} five Gaussian components added at: \textcolor{cyan}{$R_{\rm r}=0\arcsec$, $9\farcs60\pm0\farcs17$, $15\farcs47\pm0\farcs03$, $22\farcs08\pm0\farcs07$, \& $30\farcs60\pm0\farcs07$; with $\mu_0 = 16.81\pm0.16$, $22.08\pm0.07$, $21.97\pm0.02$, $23.15\pm0.03$, \& $24.85\pm0.03\,\text{mag\,arcsec}^{-2}$; and FWHM = $0\farcs49\pm0\farcs03$, $6\farcs17\pm0\farcs39$, $2\farcs94\pm0\farcs06$, $7\farcs65\pm0\farcs18$, \& $3\farcs95\pm0\farcs14$, respectively.} \textbf{Right two panels}---The model represents $0\arcsec \leq R_{\rm maj} \leq 37\farcs53$ with $\Delta_{\rm rms}=0.0709\,\text{mag\,arcsec}^{-2}$. \underline{Point Source:} \textcolor{LimeGreen}{$\mu_0 = 17.40\pm0.38\,\text{mag\,arcsec}^{-2}$.} \underline{S{\'e}rsic Profile Parameters:} \textcolor{red}{$R_e=1\farcs27\pm0\farcs09$, $\mu_e=18.87\pm0.10\,\text{mag\,arcsec}^{-2}$, and $n=1.41\pm0.16$.} \underline{Ferrers Profile Parameters:} \textcolor{Orange}{$\mu_0 = 20.02\pm0.09\,\text{mag\,arcsec}^{-2}$, $R_{\rm end} = 5\farcs63\pm0\farcs04$, and $\alpha = 2.50\pm0.18$.} \underline{Exponential Profile Parameters:} \textcolor{blue}{$\mu_0 = 20.53\pm0.05\,\text{mag\,arcsec}^{-2}$ and $h = 8\farcs66\pm0\farcs10$.} \underline{Additional Parameters:} five Gaussian components added at: \textcolor{cyan}{$R_{\rm r}=0\arcsec$, $8\farcs98\pm0\farcs05$, $13\farcs65\pm0\farcs02$, $20\farcs11\pm0\farcs05$, \& $26\farcs23\pm0\farcs09$; with $\mu_0 = 17.05\pm0.23$, $22.37\pm0.06$, $22.07\pm0.03$, $23.14\pm0.03$, \& $25.30\pm0.06\,\text{mag\,arcsec}^{-2}$; and FWHM = $0\farcs46\pm0\farcs04$, $3\farcs40\pm0\farcs17$, $2\farcs82\pm0\farcs06$, $5\farcs59\pm0\farcs13$, \& $2\farcs71\pm0\farcs18$, respectively.}}
\label{UGC6093_plot}
\end{sidewaysfigure}

\subsection{UGC~6093 (Figures~\ref{UGC6093_plot})}

We have added a PSF and a central Gaussian to model excess nuclear light. Four additional Gaussians have been added to account for spiral arms at $R_{\rm maj} \approx 10\arcsec$ ($R_{\rm eq}\approx 9\arcsec$), $R_{\rm maj} \approx 15\arcsec$ ($R_{\rm eq}\approx 14\arcsec$), $R_{\rm maj} \approx 22\arcsec$ ($R_{\rm eq}\approx 20\arcsec$), and $R_{\rm maj} \approx 31\arcsec$ ($R_{\rm eq}\approx 26\arcsec$).

\begin{deluxetable}{lcccccccc}
\tabletypesize{\footnotesize}
\tablecolumns{9}
\tablecaption{Major-axis Galaxy Components\label{table:comps}}
\tablehead{
\colhead{Galaxy Name} & \colhead{S{\'e}rsic} & \colhead{Core-S{\'e}rsic} & \colhead{Ferrers} & \colhead{Exponential} & \colhead{Broken Exponential\tablenotemark{a}} & \colhead{Edge-on Disk Model} & \colhead{Gaussian} & \colhead{Point Source}
}
\startdata
\object{Circinus} & 1 & \nodata & 1 & 1 & \nodata & \nodata & 4 & \nodata \\
\object{Cygnus~A} & 1 & \nodata & \nodata & 1 & \nodata & \nodata & 1 & \nodata \\
\object{ESO~558-G009} & 1 & \nodata & \nodata & \nodata & \nodata & 1 & \nodata & 3 \\
\object{IC~2560} & 1 & \nodata & 1 & 1 & \nodata & \nodata & 2 & \nodata \\
\object[SDSS J043703.67+245606.8]{J0437+2456}\tablenotemark{b} & 1 & \nodata & 2 & 1 & \nodata & \nodata & 4 & 1 \\
\object{Mrk~1029} & 1 & \nodata & \nodata & 2 & \nodata & \nodata & 1 & \nodata \\
\object{NGC~0224}\tablenotemark{c} & 1 & \nodata & 1 & 1 & \nodata & \nodata & \nodata & \nodata \\
\object{NGC~0253} & 1 & \nodata & 1 & \nodata & \nodata & 1 & 4 & \nodata \\
\object{NGC~1068} & 1 & \nodata & 1 & 1 & \nodata & \nodata & 3 & \nodata \\
\object{NGC~1097} & 1 & \nodata & 1 & 1 & \nodata & \nodata & 5 & 1 \\
\object{NGC~1300} & 1 & \nodata & 1 & \nodata & 1 & \nodata & 3 & \nodata \\
\object{NGC~1320} & 1 & \nodata & \nodata & 1 & \nodata & \nodata & \nodata & \nodata \\
\object{NGC~1398} & 1 & \nodata & 1 & 1 & \nodata & \nodata & 4 & \nodata \\
\object{NGC~2273} & 1 & \nodata & 1 & 1 & \nodata & \nodata & 6 & \nodata \\
\object{NGC~2748} & 1 & \nodata & \nodata & \nodata & \nodata & 1 & 2 & \nodata \\
\object{NGC~2960} & 1 & \nodata & \nodata & 1 & \nodata & \nodata & 1 & \nodata \\
\object{NGC~2974} & 1 & \nodata & 1 & 1 & \nodata & \nodata & 1 & \nodata \\
\object{NGC~3031} & \nodata & 1 & \nodata & 1 & \nodata & \nodata & 4 & \nodata \\
\object{NGC~3079} & 1 & \nodata & 1 & \nodata & \nodata & 1 & 3 & \nodata \\
\object{NGC~3227} & 1 & \nodata & 1 & 1 & \nodata & \nodata & 1 & \nodata \\
\object{NGC~3368} & 1 & \nodata & 2 & 1 & \nodata & \nodata & 3 & 1 \\
\object{NGC~3393} & 1 & \nodata & 1 & \nodata & 1 & \nodata & 2 & \nodata \\
\object{NGC~3627} & 1 & \nodata & 1 & 1 & \nodata & \nodata & \nodata & \nodata \\
\object{NGC~4151} & 1 & \nodata & 1 & 1 & \nodata & \nodata & 2 & \nodata \\
\object{NGC~4258} & 1 & \nodata & 1 & 1 & \nodata & \nodata & \nodata & 1 \\
\object{NGC~4303} & 1 & \nodata & 1 & \nodata & 1 & \nodata & 4 & \nodata \\
\object{NGC~4388} & 1 & \nodata & 1 & \nodata & \nodata & 1 & 2 & \nodata \\
\object{NGC~4395} & \nodata & \nodata & 2 & 1 & \nodata & \nodata & \nodata & 1 \\
\object{NGC~4501} & 1 & \nodata & \nodata & \nodata & 1 & \nodata & 5 & 1 \\
\object{NGC~4594} & \nodata & 1 & \nodata & 1 & \nodata & \nodata & \nodata & \nodata \\
\object{NGC~4699} & \nodata & 1 & 1 & 1 & \nodata & \nodata & 2 & \nodata \\
\object{NGC~4736} & 1 & \nodata & \nodata & 1 & \nodata & \nodata & 4 & \nodata \\
\object{NGC~4826} & 1 & \nodata & \nodata & 1 & \nodata & \nodata & 5 & \nodata \\
\object{NGC~4945} & 1 & \nodata & \nodata & \nodata & 1 & \nodata & \nodata & \nodata \\
\object{NGC~5055} & 1 & \nodata & \nodata & 1 & \nodata & \nodata & 1 & \nodata \\
\object{NGC~5495} & 1 & \nodata & 1 & 1 & \nodata & \nodata & 4 & 1 \\
\object{NGC~5765b} & 1 & \nodata & 1 & 1 & \nodata & \nodata & 3 & \nodata \\
\object{NGC~6264} & 1 & \nodata & 1 & 1 & \nodata & \nodata & 3 & \nodata \\
\object{NGC~6323} & 1 & \nodata & 1 & 1 & \nodata & \nodata & 6 & \nodata \\
\object{NGC~6926}  & 1 & \nodata & 1 & \nodata & \nodata & 1 & 4 & \nodata \\
\object{NGC~7582} & 1 & \nodata & 1 & \nodata & \nodata & 1 & 4 & \nodata \\
\object{UGC~3789} & 1 & \nodata & 1 & 1 & \nodata & \nodata & 2 & \nodata \\
\object{UGC~6093} & 1 & \nodata & 1 & 1 & \nodata & \nodata & 5 & 1 \\
\enddata
\tablecomments{}
\tablenotetext{a}{All of our broken exponential models are ``truncated,'' where the scale length decreases (the light falloff becomes steeper), i.e., $h_1>h_2$.}
\tablenotetext{b}{SDSS J043703.67+245606.8}
\tablenotetext{c}{From \citet{Savorgnan:2016}.}
\end{deluxetable}

\section{Surface Brightness Profiles}\label{App3}

\end{document}